*To be submitted*

# Photoinduced charging and spatially selective photochemistry on ferroelectric surfaces

Sergei V. Kalinin

Department of Materials Science and Engineering, University of Tennessee, Knoxville, Tennessee 37920, United States

E-mail: sergei2@utk.edu

**Abstract**

Ferroelectric surfaces exhibit polarization-dependent photoreactivity and provide a route to program chemical patterns through local electrical control of polarization. Observations accumulated over several decades show reaction products in domain interiors, at domain boundaries, and in other selected regions, indicating that the mechanisms extend beyond a single polarization-induced band-bending picture. Here we analyze light-induced charging and interfacial reactions on uniformly polarized ferroelectrics and domain patterns, for dry surfaces and surfaces in contact with an electrochemically active solution. A common semiconductor transport and charge-conservation framework distinguishes five cases: an isolated dry ferroelectric, a dry ferroelectric connected to a carrier reservoir, a blocking ferroelectric–electrolyte junction, a reactive interface with an external carrier return, and two opposite domains coupled through the solid and electrolyte. Analytical solutions and numerical evaluations connect stationary potentials versus illumination to differential optical spectra and finite light-step responses. We further examine the doping effects on static potentials in all five cases and dynamic photovoltage in the isolated film and a fixed-polarization, bare-surface contacted-film variant. For reactive interfaces, the same framework determines reduction and oxidation rates, their spatial distributions, and accumulated reaction amount. The analysis covers surface charging and photochemical conversion beyond deposition, while retaining the electrical and chemical boundaries that distinguish finite charging from sustained turnover, and explains the observed range of photodeposition morphologies. Finally, this framework builds the theoretical foundation for the modulated photovoltage imaging methods that can readily be realized in SPM under illumination.

## 1. Introduction

Ferroelectric polarization can change the chemical and photochemical properties of a surface through the coupling to electronic charge, adsorbates, and the surrounding medium. A biased scanning probe can write polarization locally on the nanometer scale, while piezoresponse force microscopy maps the resulting domain structure. This combination makes it possible to design a spatial pattern electrically and then use it to direct surface reactions. Early demonstrations of polarization-dependent local reactivity and ferroelectric templates for metal photodeposition established this connection between domain writing and chemical patterning. Studies of adsorption and water-mediated screening further showed that the surface environment participates in the state being patterned. [1, 2, 3, 4]

Classical semiconductor photoelectrochemistry provides the basic framework for understanding light-induced surface charging and chemical reactions. Depletion-layer and diffusion theories relate absorbed light to carrier collection, while semiconductor–liquid transfer theory connects surface carrier populations to chemical activities and reaction driving forces. Quantitative open-circuit studies further established the importance of bulk transport in determining the attainable junction photovoltage. [5, 6, 7, 8] Surface states contribute by storing charge, mediating interfacial transfer, or promoting recombination. Their occupation can depart from equilibrium with the semiconductor when exchange with the liquid competes with carrier capture and emission. [9, 10] Intensity-modulated photocurrent and impedance theories connect these processes to stationary and dynamic measurements, including conditions under which charging contributes substantially to the terminal current. [11, 12, 13] A quantitative description therefore requires the electron and hole continuity equations, electrostatics, and interfacial kinetic laws to be solved with consistent electrical and carrier-exchange boundary conditions. Band bending describes the electrostatic contribution, but carrier transport, recombination, and exchange determine how that potential translates into photovoltage and operating current. [14, 15]

Ferroelectric polarization adds an electrically switchable contribution to the internal field and interfacial charge, together with polarization-dependent surface compensation and adsorption. Its spatial patterning provides a means to redistribute carrier collection and surface reactions, allowing an electrically written domain structure to control local chemistry. These effects extend the semiconductor framework through the coupling of polarization to carrier transport and interfacial kinetics. [1, 3, 4, 16, 17, 18] Under illumination, the resulting behavior includes domain-selective reduction and oxidation, changes in surface-potential contrast, photocurrent near domain walls, and polarization-dependent catalytic conversion. As in conventional semiconductors, sustained chemical conversion requires charge balance: electron consumption by reduction must be accompanied by hole consumption or replenishment of the electronic reservoir. The complementary process can occur elsewhere on the same object, through an external circuit and counterreaction, or through oxidation of a chemical donor. Without such a path, net transfer remains transient and is limited by the accompanying change in stored charge and interfacial potential. Polarization can therefore control where reactions occur, while the available charge-return paths determine whether they can be sustained. [19, 20, 21, 22, 23, 24, 25]

Here we examine surface charging, reduction and oxidation, and spatially selective photochemical conversion, including but not restricted to the formation of deposited products. We compare the five cases in Figure 1, progressing from charge redistribution to reservoir exchange, ionic screening, sustained conversion, and lateral coupling. Each case is examined through its stationary potential as a function of illumination, its response to a small periodic change in light about an illuminated state, and its evolution after increasing and decreasing light steps. Polarization reversal and a nonpolar reference accompany this comparison. We also examine how donor and acceptor populations change screening and stationary potentials in all five cases, and the dynamic photovoltage in the isolated and bare contacted films. In the reactive cases, the same carrier populations determine partial reaction rates and accumulated reaction amount. Analytical reductions identify the mechanism and its limiting behavior; numerical solutions evaluate the retained nonlinear or coupled equations and test their consistency. The full derivations, parameter sets, depth profiles, transient families, and extended literature analysis are provided in the Supplementary Material.

## 2. Prior Work

Ferroelectric photoassisted chemistry has a substantial history preceding nanoscale deposition patterns. Nasby and Quinn examined $BaTiO_3$ photoelectrodes, while Inoue and co-workers studied water decomposition on poled PZT and on semiconductor–metal–ferroelectric structures prepared with opposite polar faces. These experiments connected the electrical state of the material to a chemical output and established the relevance of the electrode or composite architecture. [26, 27, 28] Later domain-resolved experiments made the spatial relationship explicit. Complementary markers on $BaTiO_3$ studied by Giocondi and Rohrer and visible-light-induced reduction on $BiFeO_3$ establish domain-dependent chemical activity. Other experiments find reduction on both polarization orientations or changes in the preferred region after thermal treatment of periodically poled $LiNbO_3$. Notably, domain polarity alone does not specify the sign or location of every reaction. [22, 29, 30, 31] Experiments on ultrathin PZT films demonstrate that the bottom-electrode choice changes the preferred polarization and photochemical response. Titania coatings on $BaTiO_3$ can retain a reaction pattern inherited from the ferroelectric substrate, with the effect depending on coating thickness and donor doping. [32, 33] The domain wall can be a region of distinct electrostatics or transport, but a wall-adjacent product does not by itself determine which contribution is responsible.

The range of these polarization coupled chemical effects extends beyond deposition. Work by Kakekhani and Ismail-Beigi establishes the role of polarization in adsorption and the catalytic energy landscape, complementing its role in electronic separation. Polarization-dependent mediator adsorption has been used to control competing reactions in overall water splitting. [16, 34] Single-domain and spatially modified ferroelectric photocatalysts have been examined for hydrogen evolution, water oxidation, and paired gas production. Domain-engineered $BiFeO_3$ photoanodes and poled $BaTiO_3$ nanocrystals demonstrate the relevance of polarization to contacted photoelectrochemistry as well as particulate photocatalysis. [35, 36, 37, 38, 39, 40, 41] In $PbTiO_3$, selective modification of polarized facets can suppress electron-trapping defects and improve charge utilization. First-principles calculations further show that oxygen vacancies can change the water-oxidation pathway, supplying a chemical input that must accompany a continuum carrier model. [42, 43] Carbon conversion, oxygen activation, and selective organic oxidation provide further examples, with their own requirements for reactant, product, and complementary-carrier balances. [44, 45, 46]

Light also changes the surface without producing a deposited material. Ultraviolet-generated charge and water adsorption contribute to $BaTiO_3$ domain screening; illumination and interfacial adsorbates can modify switchable polarization, and above-bandgap light can stabilize written domains. [47, 48, 49] Focused illumination of lithium niobate has been used to inhibit chemical etching during exposure or to modify the response to subsequent etching. These experiments provide direct examples of surface chemistry controlled by a light beam on a selected polar face. [50, 51] Note that photovoltage, wettability, particle motion in a photoinduced field, and dye decolorization probe different observables and cannot all be read as measures of the same physics. The additional literature synthesis is given in the Supplementary Material. [23, 24, 52, 53, 54, 55]

Quantitative ferroelectric models already address important parts of this problem. Glickstein and co-workers coupled electrostatics, optical generation, carrier drift and diffusion, and recombination to calculate reaction currents in coated ferroelectrics, and subsequently extended the calculation to lateral collection in multidomain structures. These models use specified solid and solution boundaries, including a rear carrier contact. [56, 57] Two-dimensional semiconductor-particle calculations likewise relate conversion to bulk recombination and the distribution of surface catalyst regions. [58] Explicit semiconductor–electrolyte calculations by He and co-workers and Harmon and co-workers resolve the liquid and reactive interface more directly. Equilibrium ionic-compensation and ferroionic theories supply a complementary description of chemical screening and its feedback on polarization. The opportunity is to connect these established descriptions through a deliberate comparison of physical boundaries and measured quantities. [17, 18, 59, 60]

Local studies have clearly elucidated the role of non-local current transport necessitated by charge balance. Scanning-probe studies on silver ion conductors reveal that a local excitation can be coupled to transport and reaction outside the immediate tip region, with the environment affecting that coupling. Analyses of biased-tip measurements on ferroelectric films further identify ionic surface charging and electrochemical changes that can accompany preparation of the local state. [61, 62, 63] The same architectural issue arises under local illumination. A dry in-plane conducting path to an equilibrating electrode can provide the carrier-reservoir function of a rear contact. This reduction requires a sufficiently small collection impedance over the illuminated region and the frequencies examined. Redistribution into a neighboring, uncontacted part of the sample instead preserves the total charge of the isolated body. Recent photo-scanning electrochemical microscopy maps electron transfer and hydrogen-evolution activity with nanoscale resolution; combined activity and photovoltage measurements on $MoS_2$ also show that an insulating support can participate in lateral charge separation. Such observations reinforce the need to state the complete illuminated object and its exchange boundaries. [64, 65]

## 3. Common Model and Boundary Conditions

The surface normal points from the ferroelectric into the exterior, and positive polarization points toward the exposed face. In a one-dimensional film, $x=0$ denotes the rear and $x=L$ the front. We use $q>0$, $V_T=k_B T/q$, electron and hole number densities $n$ and $p$, and potential $\phi$. The fields and thermalized carrier fluxes satisfy

$$D=\epsilon_b E+P, \tag{1}$$

$$E=-\nabla\phi, \tag{2}$$

$$\nabla\cdot D=q\left(p-n+N_D^+-N_A^-\right), \tag{3}$$

$$\Gamma_n=-D_n\nabla n+\mu_n n\nabla\phi, \tag{4}$$

$$\Gamma_p=-D_p\nabla p-\mu_p p\nabla\phi, \tag{5}$$

$$\partial_t n=-\nabla\cdot\Gamma_n+G-R, \tag{6}$$

$$\partial_t p=-\nabla\cdot\Gamma_p+G-R. \tag{7}$$

The Einstein relation is $D_i=\mu_i V_T$ in the nondegenerate approximation. The evaluated models use pair generation followed by carrier drift, diffusion, and interfacial exchange. A uniform source permits the analytical isolated and lateral reductions; a depth-dependent source $G(y)=G_0\exp(-\alpha_{\text{abs}}y)$ is retained in the contacted calculations, where $y=L-x$ is inward depth. For monochromatic light, $G_0=\eta_{\text{pair}}(1-R_{\text{opt}})\alpha_{\text{abs}}I/(h\nu)$. Absorption, reflectance, and mobile-pair yield are needed to convert a normalized source derivative into a derivative with respect to incident optical intensity. An intrinsic bulk photovoltaic current is a separate constitutive contribution and is not generated automatically by writing a nonzero pair source in the continuity equations. Shift-current and bulk-photovoltaic theories describe that additional contribution. [66, 67, 68] Switchable diode behavior and above-bandgap photovoltages in $BiFeO_3$ motivate distinguishing bulk optical-current generation from domain-wall potential and conductivity effects. [69, 70, 71]

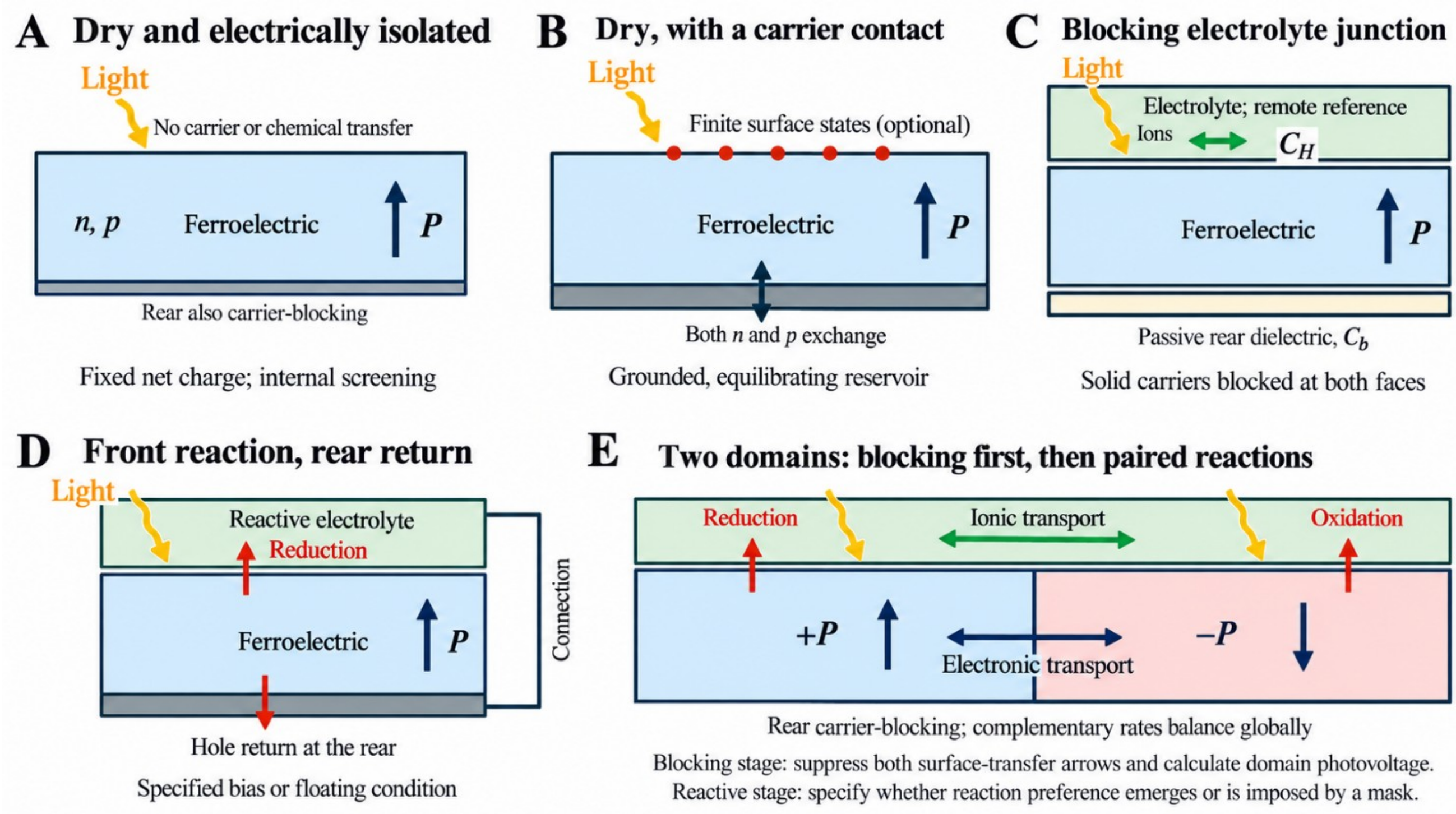

***Figure 1.*** *The five physical cases and the paths available to charge. A: an isolated dry slab, with both electronic faces blocking. B: a dry surface with a rear contact that exchanges electrons and holes; finite electronic surface states are included as a declared variant. C: a blocking solid–electrolyte junction with a passive rear dielectric capacitance; neither rear nor front exchanges solid carriers. D: front reduction with a hole-conducting rear and an external complementary reaction; exchanging the carrier and reaction labels gives the oxidation counterpart. E: opposite domains with a blocking rear, considered first without reactions and then with complementary surface transfer and lateral solid/liquid transport. A remote lateral contact can serve the reservoir role shown at the rear in B or D when its collection impedance is sufficiently small. Arrows show permitted paths, rather than equal instantaneous fluxes.*

The polarization approximation is stated for each case. In the finite-depth contacted models, $\Gamma_P \partial_t P = -aP - bP^3 - cP^5 + g_P \nabla^2 P + E_z$, with natural boundaries $\partial_n P = 0$. The background permittivity in (1) excludes the susceptibility already represented by this polarization variable. The analytical fixed-domain models instead prescribe the polarization source and use a constant differential permittivity once. A prescribed residual source can incorporate immobile compensation present before illumination. Its paired compensating sheets are removed with the source in the corresponding nonpolar control.

This distinction also determines what is held fixed in a semiconductor reference. For A, C, and the analytical domain calculation, the nonpolar reference retains the fixed dielectric and transport coefficients while setting the residual polarization source to zero. For the contacted B and D calculations, the reference retains the same background permittivity and carrier, surface, and contact laws, but removes the entire ferroelectric polarization variable. It therefore also removes its differential susceptibility and relaxation. It is a semiconductor control with the same background permittivity, rather than a control with the same total frequency-dependent dielectric response. Within an intrinsic electron–hole-symmetric blocking geometry, that control has zero photovoltage and a nonzero photocarrier population. Contact asymmetry, unequal transport, doping, or surface chemistry can produce a semiconductor photovoltage without ferroelectricity.

The continuity equations impose a constraint shared by all five cases. With $J_n = -q\Gamma_n$ and $J_p = q\Gamma_p$, equal generation and recombination terms cancel from the free-charge balance. Thus $\partial_t \rho_f + \nabla\cdot(J_n + J_p) = 0$, and Gauss's law gives $\nabla\cdot(J_n + J_p + \partial_t D) = 0$. The displacement term includes changing polarization. In one dimension this total current is independent of depth even during charging. A potential reference fixes a gauge, whereas a contact law specifies exchange of carriers; neither condition can replace the other.

Dynamic surface storage denotes a finite population of explicitly counted charge-bearing states. For separate electron and hole sites, $\sigma_s = q\left(N_s^+ \theta_+ - N_s^- \theta_-\right)$, with occupations determined by capture and emission balances. The same capture rate appears as a carrier flux leaving the solid, a change in the corresponding site population, and a contribution to the displacement jump through $\sigma_s$. These three appearances enforce charge conservation. A capacitance specifies a relation between charge and voltage between named nodes; a finite site population additionally specifies how that charge can be acquired and released. Surface-state and chemical-compensation theories provide established forms of these balances. [9, 10, 17, 18, 72]

At a reacting front, let $r_r$ and $r_o$ be signed rates of reduction and oxidation in electron equivalents per unit area and time. Without separate site capture, $\Gamma_n \cdot \hat{n} = r_r$ and $\Gamma_p \cdot \hat{n} = r_o$. The conventional outward Faradaic current is $j_F = q(r_o - r_r)$. Integrating over the solid and any modeled electronic surface inventory gives

$$\frac{dQ}{dt} = I_{\mathrm{rear,in}} + q\int_{A_f} \left(r_r - r_o\right)\, dA. \tag{8}$$

Here $I_{\mathrm{rear,in}}$ is conventional current entering through the rear. With a blocking rear, stationary operation requires equal integrated reduction and oxidation in electron equivalents. Their local rates need not coincide, and their integrated rates need not be equal while the object charges. A held-potential rear can sustain a front half-reaction only within a complete device that includes the complementary external process. A floating metal sets zero external total current but can still exchange carriers with the adjoining semiconductor as its potential changes. It is therefore distinct from a boundary that blocks both species.

Reaction kinetics must also retain the chemical reference. For reversible carrier transfer, define the local solid potential relative to the reaction plane as $U$. Reference densities can be written $n_r(U) = n_r^0 \exp(U/V_T)$ and $p_o(U) = p_o^0 \exp(-U/V_T)$. A useful generic law is $r_r = S_n(U)\left[n_s - n_r(U)\right]$ and $r_o = S_p(U)\left[p_s - p_o(U)\right]$, with positive transfer velocities and the specified forward/reverse reference. The affinities are $A_r = k_B T \ln\left[n_s / n_r(U)\right]$ and $A_o = k_B T \ln\left[p_s / p_o(U)\right]$. Their sum is the electron–hole quasi-Fermi-level separation minus the chemical free-energy demand per electron equivalent. These relations distinguish electrostatic band bending from chemical driving force. Equal partial rates for a common dissolved redox pair recycle its species; distinct reduction and oxidation couples can convert separate reservoirs at the same balanced electron-equivalent rate. Quantitative multistep chemistry requires its own species, activities, and intermediates. [6, 7, 73]

Spectroscopic rate-law studies of hematite water oxidation connect the current to accumulated surface holes, and combined spectroscopy and atomistic calculations have been used to propose multihole pathways. [74, 75] Continuum analyses additionally show that carrier accumulation can change the Helmholtz-layer voltage and thereby the interfacial transfer rate, contributing to an apparent higher kinetic order. [76, 77] Operando experiments have resolved a crossover from population-controlled to potential-driven water-oxidation kinetics as the surface charge and compensation change. [78] These results motivate retaining both carrier populations and electrochemical potential in a rate law, with any proposed intermediate population counted explicitly when its dynamics are resolved.

The transport in electrolyte is analyzed where its ionic dynamics are needed. Its fluxes are $\Gamma_i^l = -D_i^l [\nabla c_i + (z_i c_i / V_T) \nabla \phi_l]$, accompanied by species continuity and $-\nabla \cdot (\epsilon_l \nabla \phi_l) = q \sum_i z_i c_i$. The compact-layer voltage and the diffuse liquid potential are separate quantities. Case C retains the time-dependent linearized charge behavior of a symmetric electrolyte. The analytical lateral model uses the limit of an electroneutral conducting liquid with rapidly equilibrating diffuse charge and retains a modal spreading resistance. Its temporal range must satisfy the ionic relaxation requirement, assessed below. The nonlinear thick-crystal benchmark fixes the liquid reaction-plane potential. These liquid models are different reductions of the same transport problem and are identified in each figure. [59, 60, 79, 80, 81]

The charge carrier populations set the timescales. Bimolecular loss, $R = k_b (np - n_i^2)$, permits the analytical A/C reduction. The contacted reactive model uses the effective law $R = (np - n_i^2) / [\tau_0 (n + p + 2 n_i)]$. This has the stationary form of intrinsic Shockley–Read–Hall recombination; using it instantaneously defines a memoryless pair-loss approximation. Physical elimination of a trap occupation additionally requires $\omega \tau_{\text{occ}} \ll 1$ and negligible omitted trapped-charge feedback. The underlying occupation kinetics and this condition are given in the Supplementary Material. The lifetime symbols of different recombination laws are consequently not interchangeable. [82]

All optical spectra use one measurement definition. We define $\theta = t / t_{\text{ref}}$ and $\Omega = \omega t_{\text{ref}}$, with the time scale defined by the rate law of each case. For $g(\theta) = g_0 + \text{Re}[\hat{g} \exp(i\Omega\theta)]$, we define $H_V = \widehat{(V / V_T)} / \hat{g}$ about the stationary state at $g_0$, with the electrical and chemical controls held fixed. The main figures show this absolute derivative and phase relative to a positive incident-light modulation. In particular, $H_V(0; g_0) = d(V_{\text{ss}} / V_T) / d g_0$ on the same stable branch. A negative static slope has a phase of $180^\circ$ in this convention; a zero response has no defined phase. A physical sinusoid requires a positive mean greater than its small modulation amplitude. The dark derivative is a one-sided static limit or a formal limiting susceptibility. Intensity-dependent amplitudes should not be confused with the response to fractional source modulation, which is $g_0 H_V$. [11, 12, 13, 77]

For a modal or discretized state satisfying $M \dot{u} = f(u, g)$, the evaluated small-signal solution obeys $(i\Omega M - J) \hat{u} = f_g \hat{g}$. Finite light steps retain the nonlinear change of the mean carrier population and any evolving polarization, sites, or chemical states. Their endpoints are compared with stationary solutions under identical boundaries. The examples below use dimensionless, explicitly specified parameters to isolate mechanisms. They are not a fit to one named oxide across all five geometries; material-specific conversion requires consistent optical, defect, transport, dielectric, and interfacial parameters. The intrinsic concentration is retained as a thermodynamic quantity distinct from the ionized donor/acceptor density.

## 4. Photocharging without Chemical Turnover

The first three cases describe charging without sustained chemical conversion. We compare an electrically isolated dry film, a dry film connected to a carrier reservoir, and a blocking junction with an electrolyte. Their stationary and dynamic potentials show how electronic and ionic compensation depend on the available charge paths.

### 4.1 Case A - Isolated Dry Ferroelectric

Case A establishes the response that requires neither a carrier reservoir nor a chemical transfer channel. Both faces of the dry slab block electrons and holes, the net mobile charge is fixed at its neutral initial value, and the exterior normal displacement vanishes. Photogeneration changes the electron–hole population and its spatial separation while preserving this charge constraint. The observable is the front–rear dipole voltage. Setting the midpoint potential to zero fixes a gauge and does not ground the specimen.

An exact nonlinear stationary solution and its depth profiles are given in Figure S1. For an analytical treatment of optical dynamics, we take a weak-potential limit with fixed residual polarization, equal electron and hole diffusivities, uniform generation, and bimolecular recombination. Length is normalized by $L_0$, time by $\tau_R = (k_b n_i)^{-1}$, potential by $V_T$, and permittivity by $\epsilon_{\text{ref}}$. Charge per area is divided by $\epsilon_{\text{ref}} V_T / L_0$; thus polarization and displacement share that scale, and an areal capacitance is scaled by $\epsilon_{\text{ref}}/L_0$. The volume-charge and current-density scales are $\epsilon_{\text{ref}} V_T / L_0^2$ and $\epsilon_{\text{ref}} V_T / (L_0 \tau_R)$, respectively. In the following analytical expressions, $\phi$, $P$, and the transport coefficients denote these normalized quantities. The carrier population is $m = N/n_i$, the optical source is $g = G/(k_b n_i^2)$, and $a_s^0 = 2 q n_i L_0^2 / (\epsilon_{\text{ref}} V_T)$.

To zeroth order in the residual source, the neutral population obeys $\dot{m} = g - (m^2 - 1)$, with $\overline{m} = \sqrt{1 + g_0}$ and $\hat{m}/\hat{g} = (s + 2\overline{m})^{-1}$ for $s = i\Omega$. The first-order charge current is $J_s = -D_s(\partial_x \rho + a_s^0 m \partial_x \phi)$, with $\partial_t \rho = -\partial_x J_s$ and $\partial_x(P - \epsilon_s \partial_x \phi) = \rho$. The blocking boundaries impose zero charge current at both faces, while their zero displacement ensures $\int_0^L \rho \, dx = 0$. Defining $\lambda_s(m) = \sqrt{\epsilon_s / (a_s^0 m)}$, the stationary solution is

$$\phi(x) = \frac{P \lambda_s}{\epsilon_s} \frac{\sinh[(x - L/2)/\lambda_s]}{\cosh[L/(2\lambda_s)]}, \tag{9}$$

$$\Delta\phi = PF(\overline{m}), \tag{10}$$

$$F(m) = \frac{2\lambda_s}{\epsilon_s} \tanh\left(\frac{L}{2\lambda_s}\right). \tag{11}$$

Positive polarization produces a positive potential at the exposed face, with electrons displaced toward that face and holes toward the rear. Increasing light raises the pair population, reduces the screening length, and lowers the voltage magnitude. The integrated charges in the two halves remain equal and opposite. The derivative is $d\Delta\phi / d g_0 = P F_m / (2\overline{m})$, where $F_m = -\lambda_s [\tanh a - a \operatorname{sech}^2 a] / (\epsilon_s m)$ and $a = L/(2\lambda_s)$. Its sign is opposite to $P$. In a film much thinner than its screening length the voltage approaches $PL/\epsilon_s$, with weak optical sensitivity. For well-separated screening layers it approaches $2P\lambda_s/\epsilon_s$ and decreases as $(1 + g_0)^{-1/4}$ under the specified bimolecular loss law.

Figure 2 compares the two polarization orientations and the nonpolar semiconductor. The illustrative parameters are $L = \epsilon_s = 1$, $a_s^0 = 4$, $|P| = 0.08$, and $D_s = 0.25$. The largest midpoint-referenced potential is only $0.0305 V_T$, keeping this dynamic example in the controlled weak-potential regime. The voltage decreases from $0.0609275 V_T$ in darkness to $0.0251465 V_T$ at $g = 100$. The nonpolar control retains the changing carrier population but has zero voltage in this symmetric model. The finite-potential stationary family in Figure S1 uses its own stated parameters; it is not presented as the same parameter family continued through these weak-potential dynamic curves.

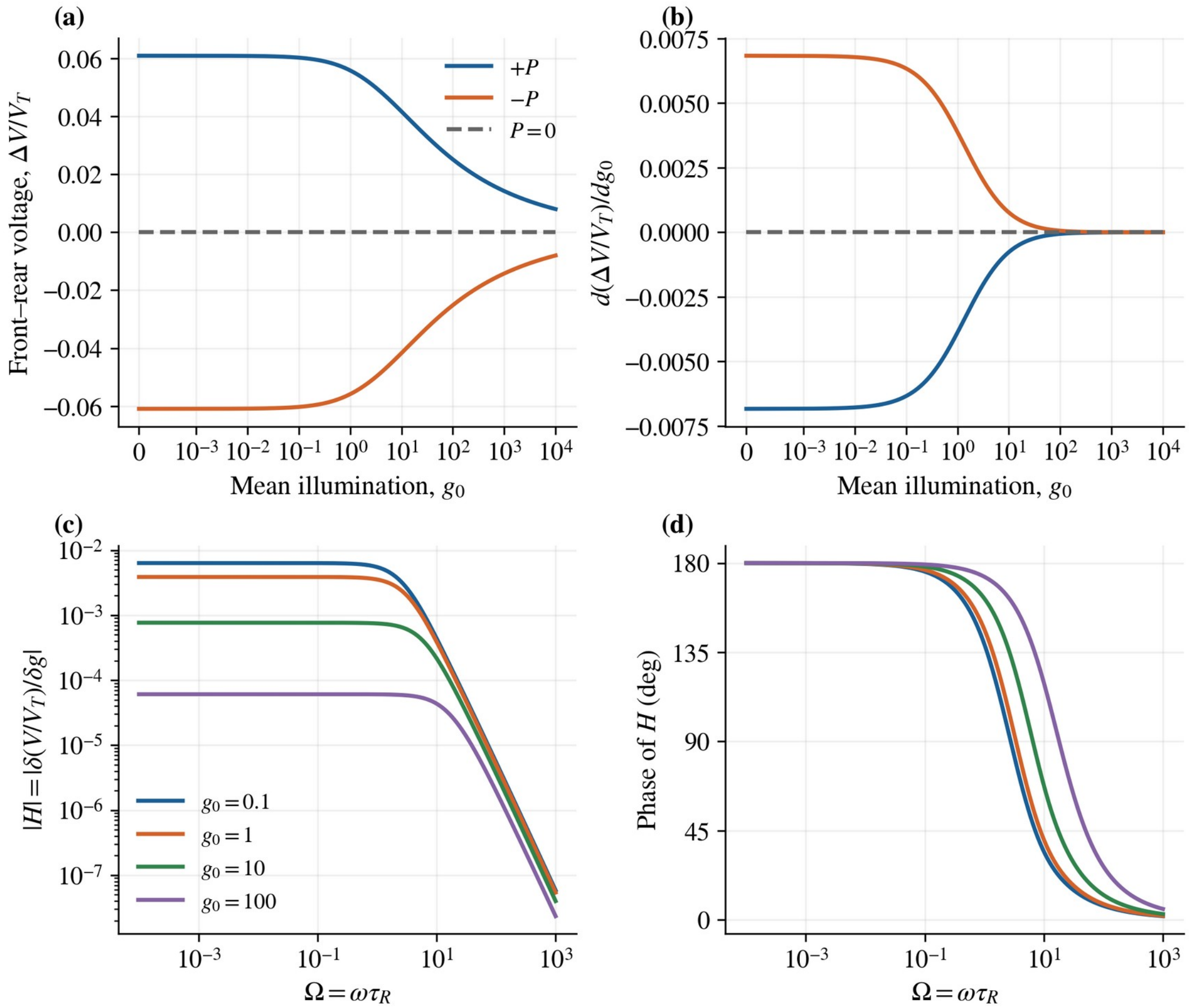

***Figure 2.*** *Isolated dry-film charging (A). (a,b) Stationary front–rear potential and its exact optical derivative for opposite residual polarizations and the symmetric nonpolar control. (c,d) Absolute optical amplitude and phase for* $+P$ *at four mean light levels. The phase is referred to positive incident-light modulation;* $-P$ *has the same amplitude and a* $180^\circ$ *phase shift. The control has zero voltage response and undefined phase. Parameters are* $L=\epsilon_s=1$, $a_s^0=4$, $|P|=0.08$, *and* $D_s=0.25$. *The analytical modal spectrum resolves both pair-population and screening relaxation.*

The donor/acceptor dependence of this isolated-film response, including its dynamic photovoltage, is examined in Section 7.

The optical response contains a population relaxation and a set of spatial screening relaxations. Defining $k_j=(2j+1)\pi/L$, with $j=0,1,2,\ldots$, $\kappa_0^2=a_s^0/\epsilon_s$, and $\bar{\kappa}^2=\kappa_0^2\bar{m}$, projection onto the odd charge modes yields the optical voltage response given by Equation (12).

$$H_A(s;g_0)=\frac{-8PD_s\kappa_0^2}{\epsilon_s L(s+2\bar{m})}\sum_{j=0}^{\infty}\frac{1}{(k_j^2+\bar{\kappa}^2)\left[s+D_s(k_j^2+\bar{\kappa}^2)\right]}. \quad (12)$$

The static limit of this convergent series is the derivative of (10). Each term has the same static sign and combines the population time with a diffusive screening time. Consequently $|H_A(i\Omega)|\le|H_A(0)|$: the fixed-source symmetric isolated model cannot exhibit a finite-frequency voltage amplification above its own static sensitivity. Increasing the mean light level reduces that sensitivity and moves the population and screening relaxation rates upward. Reversing polarization reverses the complex response without changing its magnitude. These are properties of the declared reduction, rather than general restrictions on every isolated semiconductor or every ferroelectric photovoltage.

Finite light steps can also be treated analytically. Following a change to $g_f$, the neutral population is $m(t)=m_f\left[m_i+m_f\tanh(m_f t)\right]/\left[m_f+m_i\tanh(m_f t)\right]$, with $m_i=\sqrt{1+g_i}$ and $m_f=\sqrt{1+g_f}$. Each charge mode satisfies a first-order equation with the time-dependent screening coefficient and has an integrating-factor solution requiring only a smooth one-dimensional quadrature. Its instantaneous equilibrium moves monotonically after an increasing or decreasing light step; starting from the preceding stationary state, every mode follows in the same direction. The voltage therefore approaches its new value monotonically, with a lag behind instantaneous screening equilibrium. The light-on, light-off, and nonzero-mean step families and the corresponding population and charge balances are retained in Figure S2.

### 4.2 Case B - Dry Ferroelectric with a Carrier Reservoir

Allowing a dry ferroelectric to exchange carriers with an electrical reservoir changes the charging problem even when the exposed surface supports no chemical conversion. In case B, both electrons and holes equilibrate at the rear contact, whose potential is fixed. The exposed potential remains an output of the electrostatic problem. The contact can supply or remove compensating charge, whereas the total charge of the isolated specimen in case A is constrained. A lateral conducting connection to an electrode provides the same reservoir limit when its equilibration time and voltage loss are negligible on the measurement scales; a finite lateral resistance would require a spatially resolved contact model. This distinction is relevant to surface-potential measurements of illuminated ferroelectrics, in which the observed charge relaxation depends on the electrical connection as well as on the prepared polarization [23, 24, 83].

For the contacted calculations, let $y$ denote inward depth divided by the film thickness, with the exposed face at $y=0$ and the rear at $y=1$. We write $v=\phi/V_T$, divide the carrier densities by the intrinsic concentration $n_i$, and use $g_0=G_0\tau_R/n_i$ for the surface generation. The absorption profile is $g_0\exp(-\beta y)$, with $\beta=2$ in the examples. Bimolecular net recombination fixes the time scale $\tau_R=(k_b n_i)^{-1}$. The rear conditions are $v(1)=0$ and $n(1)=p(1)=1$, together with natural polarization conditions. The Landau equations for polarization, both carrier populations, and electrostatic potential are solved together. The dimensionless material parameters and the complete equations are given in the supplementary material; these are illustrative parameter sets rather than a calibration of a particular oxide.

Two exposed-surface limits are compared. Without electronic sites, both carrier fluxes vanish at the exposed face and the exterior normal displacement is neglected. With electronic sites, fixed areal populations $N_+$ and $N_-$ can store positive and negative charge. Their charged fractions $\theta_+$ and $\theta_-$ evolve by reversible capture and emission: holes charge the positive family and electrons charge the negative family. The stored charge is $\sigma_s=q(N_+\theta_+-N_-\theta_-)$. Each capture term appears in the corresponding carrier boundary flux, so charge added to a site is removed from the carrier population. The occupancies remain between zero and one, and the signed charge transferred into each family is bounded by its capacity. At stationary occupancy the two selected front currents vanish individually. Thus these sites describe finite electronic charge storage; they do not constitute a sustained source of reaction product.

Writing $K_\pm=k_{\mathrm{cap},\pm}n_i\tau_\pm$ and $T_\pm=\tau_\pm/\tau_R$, the occupation equations are

$$\frac{d\theta_+}{d(t/\tau_R)}=\frac{K_+p_s(1-\theta_+)-\theta_+}{T_+}, \tag{13}$$

$$\frac{d\theta_-}{d(t/\tau_R)} = \frac{K_- n_s(1-\theta_-)-\theta_-}{T_-}. \tag{14}$$

The stationary charged fractions are therefore $K_+ p_s/(1+K_+ p_s)$ and $K_- n_s/(1+K_- n_s)$. These expressions specify a finite population with saturating occupation. They do not add a freely adjustable surface capacitance to an otherwise unchanged carrier boundary: the same capture and release rates must determine both the stored charge and the carrier flux.

An integration of Gauss's law separates the contributions to the measured voltage. Define $\pi(y)=P_\perp(y)/P_*$, where positive polarization points out of the exposed surface, $\epsilon^*=\epsilon_b V_T/(P_* L)$, and $\delta=q n_i L/P_*$. Then

$$v_s = \frac{\int_0^1 \pi(y)\, dy + \sigma_s/P_* + \delta \int_0^1 (1-y)[p(y)-n(y)]\, dy}{\epsilon^*}. \tag{15}$$

The three terms are the polarization contribution, the electronic site charge, and the first spatial moment of mobile charge. The moment weights charge according to its distance from the grounded rear. Consequently, neither the total number of photocarriers nor the mean polarization alone determines the surface voltage. The identity remains valid during the nonlinear transient and provides the interpretation of the stationary and dynamic responses below.

Figure 3(a,b) compares the stationary exposed potentials with and without sites. For outward polarization, the finite-site voltage decreases from approximately $1.859\,V_T$ in darkness to $0.170\,V_T$ at $g_0=10^6$. Without sites, the voltage first increases from $1.208\,V_T$, reaches approximately $1.800\,V_T$ near $g_0=108$, and then decreases to $0.173\,V_T$ at $10^6$. The opposite prepared polarization gives the reflected voltage curve. The high-generation voltages are similar, although the dark states and intermediate intensity dependence differ appreciably. Illumination therefore need not produce a monotonic voltage change simply because it increases the supply of screening carriers.

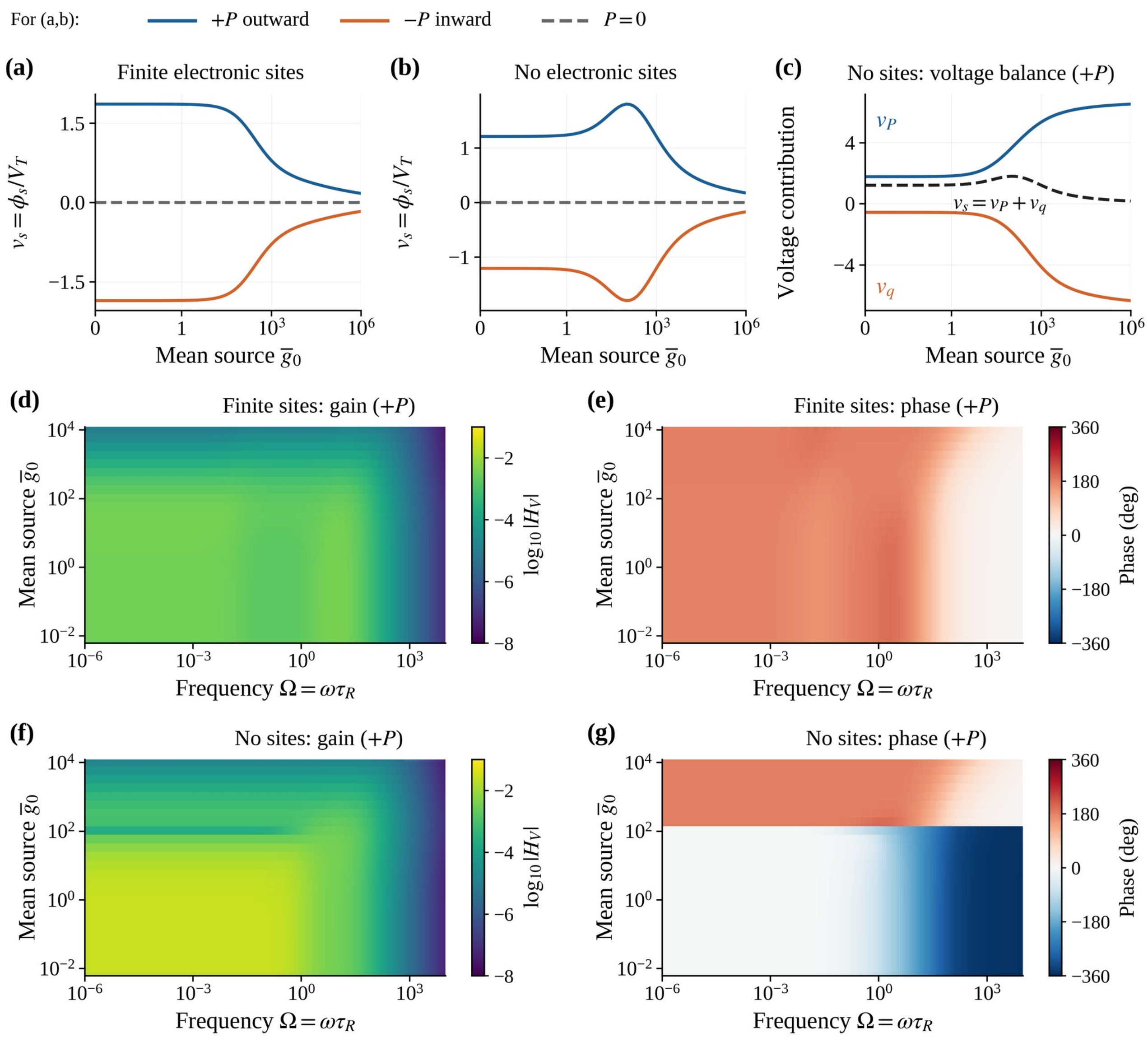

***Figure 3.*** *Grounded dry-film charging (case B). (a,b) Stationary surface potential with finite reversible electronic sites and without sites, respectively, for the two prepared polarities and the nonpolar reference. The polarity key above the figure applies to (a,b). (c) Polarization contribution $v_P$, mobile-charge contribution $v_q$, and their sum $v_s = v_P + v_q$ for the no-site, outward-polarized film. The dashed black curve in (c) is the total voltage. (d,e) Optical gain and phase with finite sites; (f,g) gain and phase without sites. The maps show the outward preparation and the absolute derivative $H_V = \delta(\phi_s / V_T)/\delta g_0$, without a mean-generation multiplier. Phase is referenced to positive incident-light modulation and unwrapped along frequency at each mean. The low-frequency branch approaches $0^\circ$ or $180^\circ$ according to the sign of the stationary derivative. Reversing polarization reverses the complex response. The symmetric $P = 0$ voltage response vanishes and has no defined phase. Both rear carrier populations are equilibrated, and $\Omega = \omega\tau_R$. The $P = 0$ control removes the polar mode while retaining the background permittivity. The stationary source axes are linear below $10^{-2}$ and logarithmic above it; the stationary sweep extends to $10^6$, while the optical maps extend to $10^4$. Parameters are specified in the Supplementary Material.*

The origin of the voltage maximum is resolved in Figure 3(c). In the no-site film, photocarriers screen the depolarizing field and permit the polarization magnitude to increase. This raises the polarization contribution in equation (15). Simultaneously, the mobile-charge moment increasingly compensates that contribution. The measured voltage increases while polarization recovery dominates and decreases when the additional screening dominates. The maximum is a cancellation in a particular observable, rather than the disappearance of photocarrier generation or a loss of the polarized branch. With sites, the additional compensating charge changes this balance and removes the stationary maximum over the illustrated range.

The semiconductor reference is defined by setting the explicit polarization field to zero and removing its dynamical equation, while retaining the background permittivity, optical generation, carrier transport, recombination, and contact and site laws. It does not retain the differential susceptibility of the removed polar mode. It therefore measures the combined effect of spontaneous polarization and its relaxation within the chosen ferroelectric model, rather than isolating spontaneous polarization at an exactly matched dielectric response. For equal electron and hole mobilities, identical site families, and an intrinsic unbiased rear, the reference preserves $n=p$ and equal occupancies under arbitrary illumination. Its potential is identically zero even though its carrier and site populations change. This symmetry result is specific to these constitutive laws and boundaries; the reactive semiconductor reference considered below has a nonzero photovoltage.

A controlled analytical limit makes that distinction explicit. For uniform weak generation in the intrinsic no-site reference, the equal carrier excess $c$ obeys $\dot{c}=Mc''-2c+g$, with $c'(0)=0$ and $c(1)=0$. The exact surface carrier response is

$$\frac{\delta c_s}{\delta g}=\frac{1-\operatorname{sech}\sqrt{(2+i\Omega)/M}}{2+i\Omega}. \tag{16}$$

Thus carrier transport and recombination give a nonzero optical population response while symmetry keeps the voltage zero. The finite-depth ferroelectric calculation additionally requires the field and polarization response; it cannot be replaced by this scalar limit. The analytical expression and its independent discretized evaluation are compared in Section S7.

The accumulation/depletion asymmetry and dynamic response of a doped, fixed-polarization bare-film variant are examined in Section 7.

The periodic response in Figure 3(d–g) is evaluated about the stationary state at each mean illumination. We use $H_V=\delta v_s/\delta g_0$ and phase relative to a positive incident-light modulation with time dependence $\exp(i\Omega t/\tau_R)$, where $\Omega=\omega\tau_R$. The maps show the absolute derivative per unit generation, without multiplication by the mean illumination. Reversing the polarization reverses the complex response and adds $180^\circ$ to its phase, while preserving its magnitude. The semiconductor voltage response is zero and its phase is undefined. Both the gain and the characteristic frequencies change with the operating intensity because the carrier distributions, recombination derivatives, and polarization stiffness change with the stationary state.

A useful consequence of the separate voltage contributions is illustrated in supplementary Figure S4. At the no-site stationary maximum, the zero-frequency derivative vanishes, yet the finite-frequency response remains nonzero. The sampled gain reaches approximately 0.00262 per unit generation near $\Omega=7.50$. Polarization and mobile screening cancel in the static derivative but acquire different amplitudes and phases when illumination is modulated. The voltage maximum therefore does not represent optical insensitivity at all frequencies. This interpretation follows from the coupled linearized equations and is consistent with the more general need to distinguish observable relaxation from an assumed microscopic rate in optical modulation measurements [11, 13].

Figure 4 examines the charge and transport behind these voltage curves. Both constitutive examples permit an increase of mean polarization under illumination, but the exposed polarization need not track the mean monotonically. The net site charge is also nonmonotonic. This behavior is compatible with fixed site capacities because the two independent families can both become substantially occupied; their difference determines the net charge. The full depth profiles in Figure S3 show why a single surface or bulk carrier density is insufficient. The rear continues to impose its equilibrium carrier values, while the illuminated interior and the exposed boundary develop different populations and screening fields.

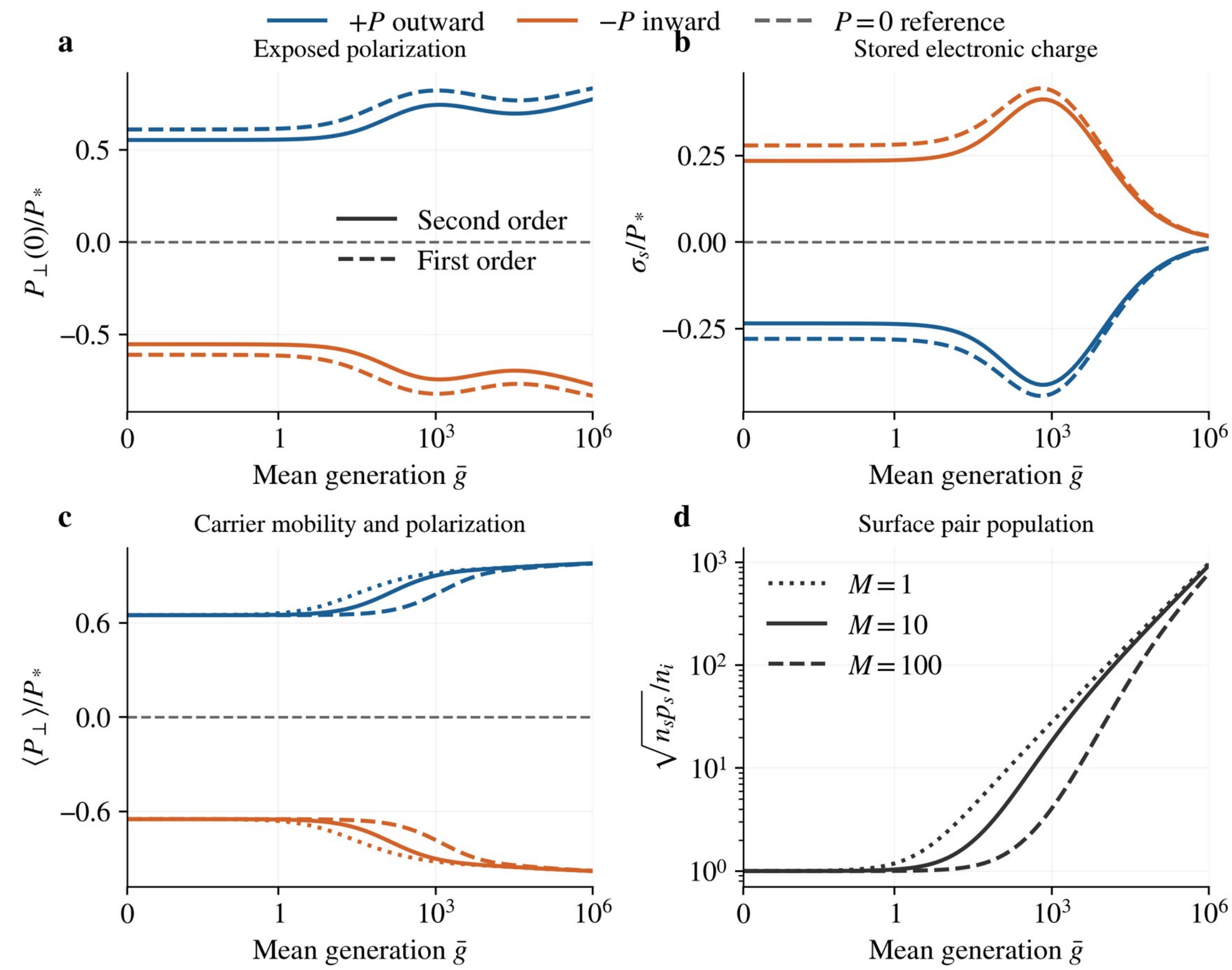

***Figure 4.*** *Surface charge and carrier supply in the grounded dry film. (a,b) Exposed polarization and net electronic site charge for second-order (solid) and first-order (dashed) Landau parameter sets. (c,d) Mean polarization and geometric mean of the surface carrier densities for equal transport parameters* $M_n = M_p = M = 1,10,100$, *with the second-order constitutive law. The line styles in (c) match the mobility legend in (d). The surface pair population is identical for the two polarities by electron–hole symmetry. The nonpolar reference has zero polarization and, for the symmetric site families, zero net site charge; it is not an additional mobility curve in (d). Both carrier species equilibrate at the rear. Faster extraction can reduce the steady photocarrier population and the polarization recovery at fixed generation. The nonmonotonic site charge is the difference of two bounded occupancies. All polarization components use the outward convention.*

The mobility dependence in Figure 4(c,d) further identifies the role of that reservoir. Raising both mobilities increases the rate at which generated carriers can reach the rear contact. At a fixed optical source, their stationary excess can therefore decrease, which weakens the illumination-induced recovery of polarization. The geometric mean $\sqrt{n_s p_s}$ measures the surface pair population and is unchanged by exchanging the two polarization orientations in this symmetric example. Mobility affects both collection and recombination through the full spatial solution; assigning a larger steady screening charge solely because transport is faster would miss the extraction boundary. The first- and second-order constitutive comparisons and the additional asymmetric cases are retained in Figure 4(a,b) and Figure S6.

The full light-on and light-off trajectories in Figure 5(a,b) provide an especially direct demonstration. In the no-site film, switching the source from zero to $10^3$ changes the outward stationary voltage only from approximately $1.208\,V_T$ to $1.215\,V_T$. During the transition, however, the voltage falls to approximately $0.526\,V_T$ before recovering. On switch-off, it rises to approximately $2.456\,V_T$ before returning to the dark state. The close stationary endpoints conceal a large evolving redistribution of polarization and mobile charge. Since these trajectories contain no electronic site variables, the overshoot cannot be attributed uniquely to trapping. Its sign is reversed for the opposite preparation, and the symmetric semiconductor remains at zero voltage.

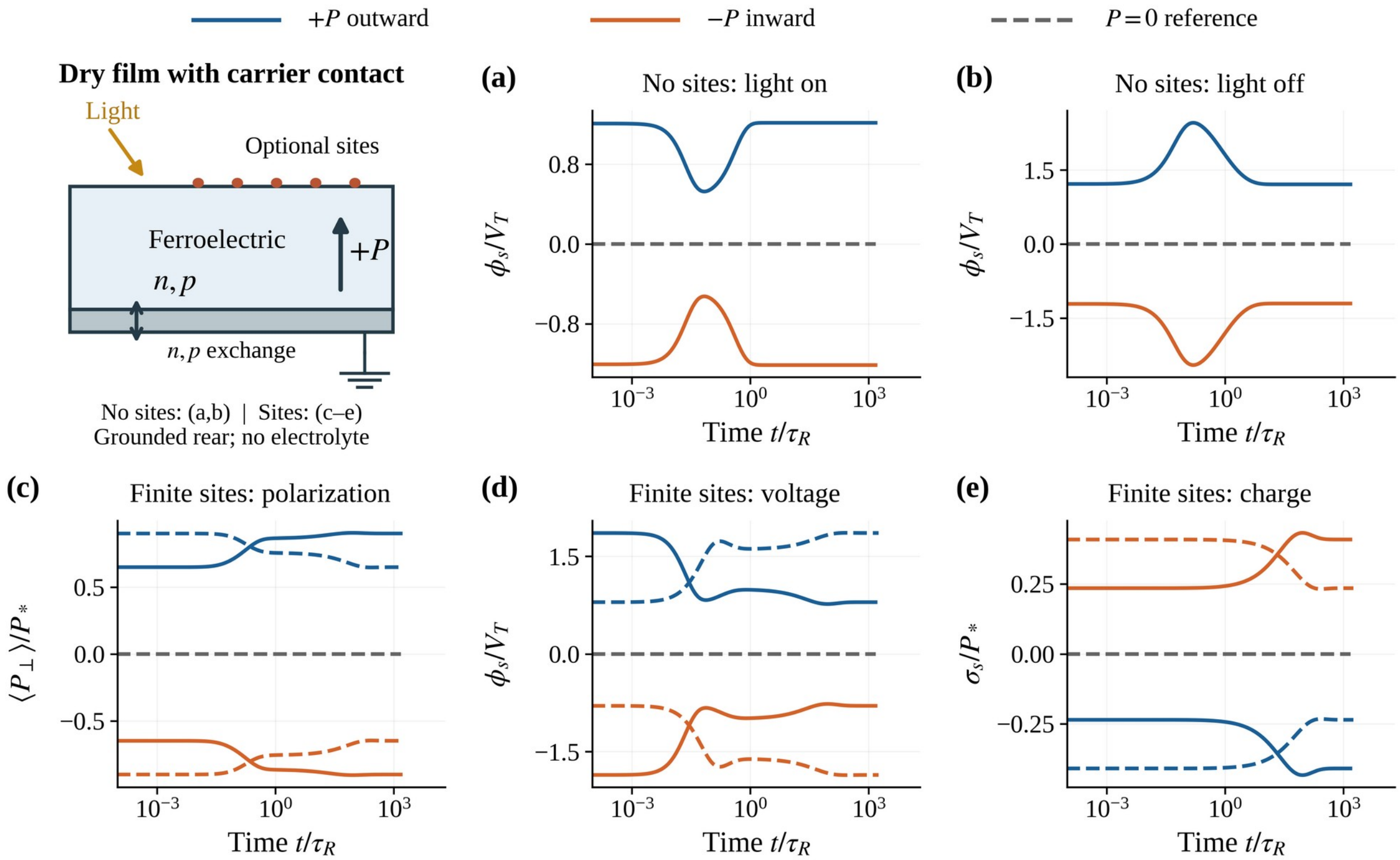

***Figure 5.*** *Physical origin of grounded-film transients. The unlettered upper-left schematic shows the dry ferroelectric and its grounded rear contact, which exchanges both electrons and holes. Surface sites are absent in (a,b) and included in (c–e); the arrow illustrates the outward-polarized preparation. (a,b) Full nonlinear light-on and light-off voltage responses without electronic sites for source changes between* $g_0 = 0$ *and* $10^3$*. Almost equal stationary endpoint voltages coexist with a large intervening excursion. (c–e) Mean polarization, exposed potential, and net electronic site charge for the finite-site model. For the polar curves in (c–e), solid and dashed lines denote light-on and light-off, respectively, each with its own time origin. Color identifies polarization throughout: blue is outward* $+P$*, orange is inward* $-P$*, and gray is the nonpolar reference. Both polarities recover their initial dark branch, and the prepared polarization does not reverse. The* $P = 0$ *reference remains at zero polarization, voltage, and net site charge despite changing carrier and individual site populations.*

The finite-site trajectories in Figure 5(c–e) connect the voltage to the evolving polarization and stored charge. Light-induced carrier redistribution, polarization recovery, and site occupation proceed on coupled time scales, producing shoulders and a modest voltage overshoot. After the source is removed, all three observables recover their dark values for this symmetric parameter set. The polarization retains its sign throughout. Thus the transient is a charging response, and its shape alone does not demonstrate a polarization reversal. The unequal-mobility example that does switch polarization is treated separately in Figure S5; its branch fold and finite-duration pulse depend on the specified transport asymmetry and preparation.

These time scales cannot generally be assigned one-to-one to isolated capture or recombination constants. A local site perturbation has a denominator $i\Omega + \left(1 + K_{\pm}\bar{c}_s\right)/T_{\pm}$, but it is driven by a carrier population that itself responds to field, transport, and polarization. The measured voltage combines all of those state changes through equation (15). The complete linearized operator, nonlinear steps between nearby illuminated states, and independently integrated sinusoidal perturbations are compared in the supplementary material. The results establish a practical distinction: a large transient surface-potential excursion can accompany little stationary voltage change, and it can occur with either a finite site inventory or solely with mobile charge and polarization.

### 4.3 Case C - Blocking Ferroelectric-Electrolyte Junction

Case C introduces the electrolyte while retaining carrier isolation of the solid. Neither electrons nor holes cross the front interface or the rear boundary. There is no Faradaic reaction and no electronic surface-site population in this case. The solid space charge responds to illumination, the electrolyte redistributes ionic charge, and the measured potential depends on how these changes divide among the solid, compact layer, diffuse layer, and rear environment.

A passive rear dielectric of areal capacitance $C_b$ couples the specimen to a reference plane held at the remote liquid potential. It permits displacement charging of that plane without electron or hole exchange through the rear. The front compact layer has capacitance $C_H$. The symmetric electrolyte has thickness $H$, fixed concentrations and potential at its remote boundary, and zero ionic flux at the compact layer. This remote reservoir can supply ionic current and displacement during redistribution. It is not assigned a second, incompatible zero-current condition.

Using the normalization of A and $u$ as the distance from the compact layer into the liquid, the electrical boundaries are $D_s(0)=-C_b\phi_s(0)$, $D_l(0)=D_s(L)$, and $\phi_s(L)-\phi_l(0)=D_s(L)/C_H$. Both solid faces block charge current, the liquid front blocks ionic current, and $\phi_l(H)=\rho_l(H)=0$. The neutral initial solid charge remains fixed, so the displacement has the same value at its two faces at every time. The liquid charge sector obeys $J_l=-D_l^{\text{ion}}(\partial_u\rho_l+a_l\partial_u\phi_l)$, $\partial_t\rho_l=-\partial_u J_l$, and $-\epsilon_l\partial_u^2\phi_l=\rho_l$, with $\lambda_l=\sqrt{\epsilon_l/a_l}$. These equations retain diffuse-charge dynamics explicitly. [59, 60, 80]

The stationary solution follows from the same solid function $F$ as A. If $d$ is the common front and rear solid displacement, the solid voltage is $(P-d)F$. The liquid has differential capacitance $C_D=(\epsilon_l/\lambda_l)\coth(H/\lambda_l)$. Defining $S=C_b^{-1}+C_H^{-1}+C_D^{-1}$ gives

$$d=\frac{PF}{F+S}, \tag{17}$$

$$\Delta\phi_s=dS=(P-d)F, \tag{18}$$

$$\phi_s(L)=d\left(C_H^{-1}+C_D^{-1}\right), \tag{19}$$

$$\phi_s(0)=-d/C_b. \tag{20}$$

The compact and diffuse drops are $d/C_H$ and $d/C_D$, respectively. The entire voltage partition therefore follows from a single charge balance and electrostatic loop. In the illustrative calculation, $C_b=2$, $C_H=5$, $H=1$, $\epsilon_l=0.5$, and $a_l=50$, so the liquid is ten screening lengths thick and $C_D\simeq5$. The remaining solid parameters are those of A. In darkness the front potential is $0.0147\,V_T$ relative to the liquid, while the voltage across the solid is $0.033\,V_T$. At $g=100$ they decrease to $0.0083\,V_T$ and $0.0186\,V_T$, respectively. The solid potential relative to the remote liquid is thus different from the internal dipole voltage and from the compact-layer voltage.

Figure 6 shows this stationary division and the importance of the passive rear boundary. As $C_b\to0$, the neutral planar solid has zero displacement at both faces. The front and liquid potentials in this chosen reference remain zero, while the internal solid photovoltage remains finite and equals the isolated result. This limiting statement concerns the laterally uniform, neutral planar problem with the specified compact law; it is not a general statement that an isolated particle in liquid cannot change its exterior potential. Finite $C_b$ produces a measurable front-to-liquid optical signal without converting the rear into a carrier contact. Reversing polarization reverses all first-order potentials, and the symmetric nonpolar control has zero charge-sector signal.

The donor/acceptor dependence of the stationary voltage partition is examined in Section 7.

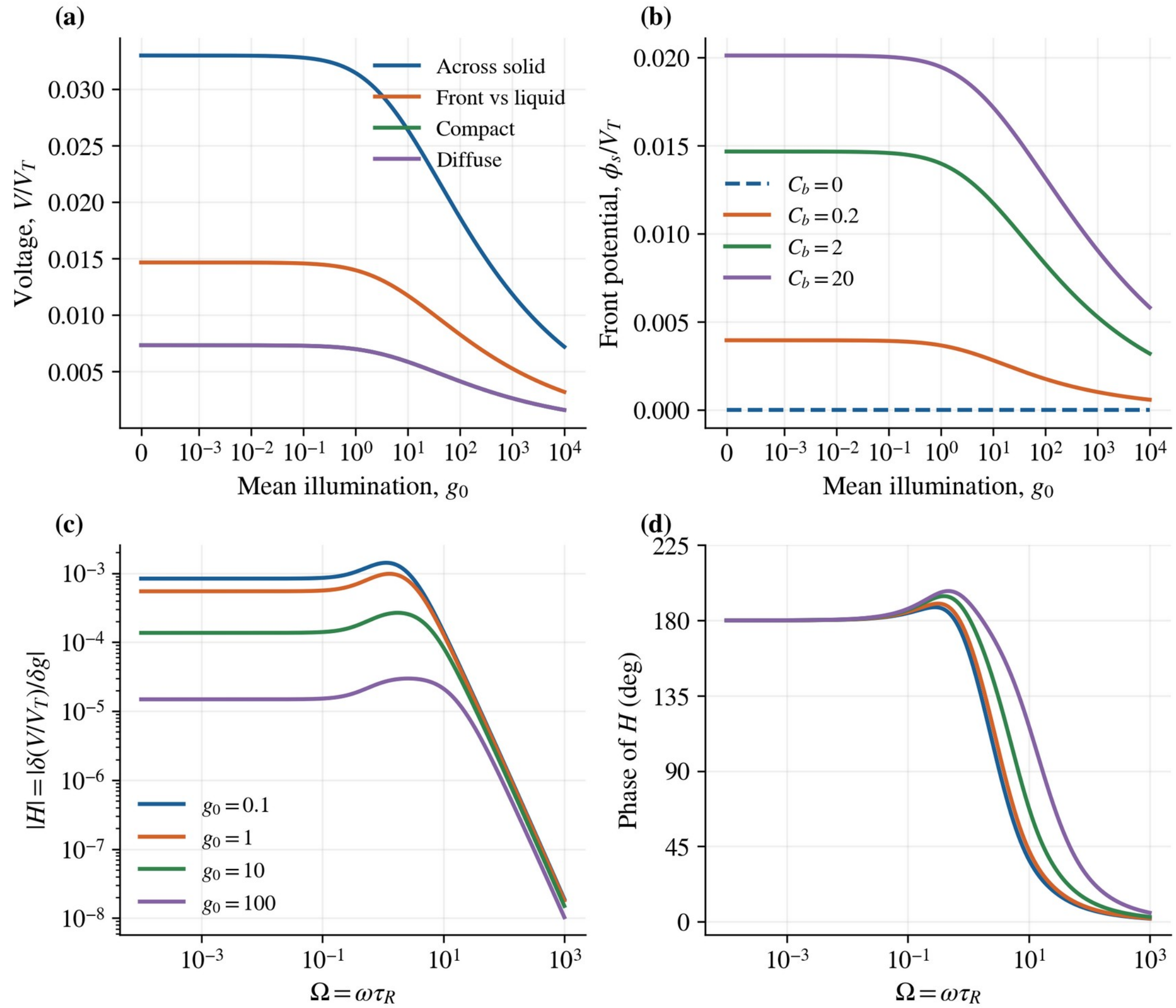

***Figure 6.*** *Blocking ferroelectric–electrolyte junction (C). (a) Stationary voltage partition for* $+P$*; the compact and diffuse drops overlap because their capacitances are equal. (b) Front potential relative to remote liquid as the passive rear capacitance varies. In the strict* $C_b = 0$ *planar limit the front signal vanishes while the internal dipole voltage remains. (c,d) Absolute optical response at* $C_b = 2$ *and* $D_l^{\text{ion}} = 0.02$*. The other liquid parameters are* $H = 1$*,* $\epsilon_l = 0.5$*,* $a_l = 50$*, and* $C_H = 5$*, with the solid parameters of Figure 2. Reversing* $P$ *reverses the voltage and complex response. All solid carrier boundaries remain blocking.*

The frequency response can be obtained without assigning an empirical ionic time constant. In each homogeneous layer, the potential basis is $\hat{\phi}_h = c_0 + c_1 x + A\exp(-\kappa_s x) + B\exp[-\kappa_s(1-x)]$, where $\kappa_s^2 = a/\epsilon + s/D_q$ and the square root has nonnegative real part. The constant and linear terms retain the potential reference and current-carrying solution, while the bounded exponentials represent screening modes. An illumination-driven particular solution is added in the solid through $\hat{m}/\hat{g} = (s + 2\overline{m})^{-1}$. The electrical, current, and charge constraints then determine eight coefficients in an algebraic system. Its zero-frequency limit is the derivative of (19); the full construction and its nonsingular evaluation near zero frequency are given in the Supplementary Material.

Unlike case A, this blocking junction can have a finite-frequency response larger than its own static slope. At $g_0 = 1$, the baseline ionic diffusivity $D_l^{\text{ion}} = 0.02$ gives a peak front-potential amplitude approximately 1.79 times the static value. Decreasing the ionic diffusivity by a factor of 100 increases this ratio to approximately $2.57$, whereas increasing it by 100 removes the enhancement over the evaluated range. The stationary solution is unchanged in this variation because the screening length and capacitances are fixed. Figure 7 places these spectra beside the light-on and light-off trajectories and the charge balance.

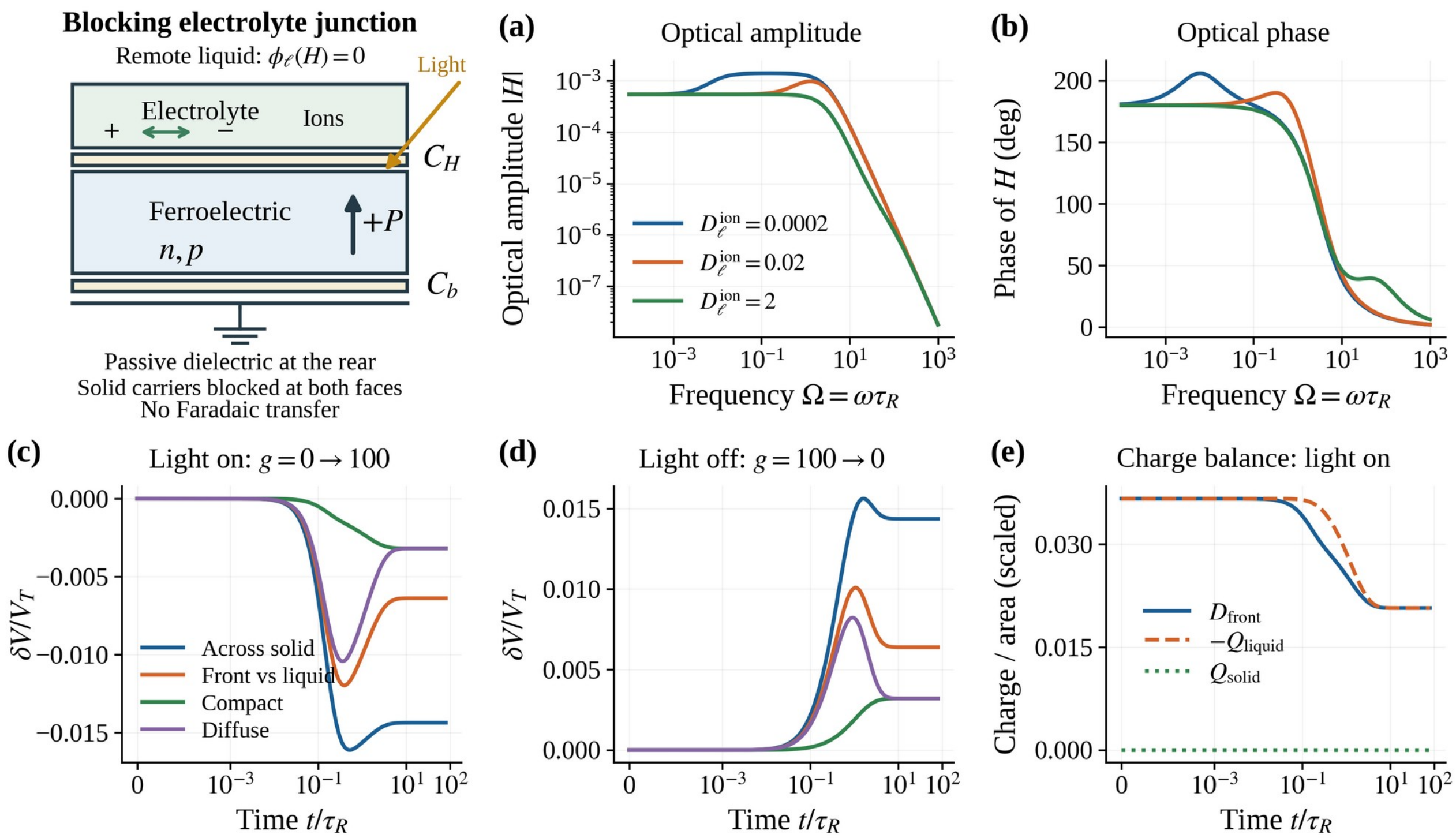

***Figure 7.*** *Ionic origin of the blocking-junction dynamics. The unlettered upper-left schematic shows the blocking ferroelectric–electrolyte junction, compact-layer capacitance $C_H$, and passive rear dielectric capacitance $C_b$. The rear reference plane is separated from the solid by the dielectric; neither solid face exchanges carriers. The remote liquid fixes the potential reference, and the interface has no Faradaic transfer or electronic surface sites. (a,b) Absolute front-potential optical response at mean illumination $g_0=1$ for three ionic diffusivities. The stationary state is identical in all three calculations. (c,d) Voltage changes from the initial state following light-on, $g:0\rightarrow100$, and light-off, $g:100\rightarrow0$, for $D_l^{ion}=0.02$. The colors in (c) also identify the voltage contributions in (d). (e) Charge balance during light-on: interfacial displacement $D_{front}$, minus the integrated liquid charge $-Q_{liquid}$, and total solid charge $Q_{solid}$. Their relation is $D_{remote}=D_{front}+Q_{liquid}$, while the solid remains neutral. Electronic redistribution precedes ionic equilibration and produces an overshoot. The illustrated preparation is outward polarized. The full nonzero-mean steps and depth-resolved responses are in the Supplementary Material.*

The enhancement has a direct physical origin. Immediately after increased illumination, the solid carrier population and screening profile can change before the ionic distribution has relaxed. The front potential initially reflects this unequal redistribution between the two media. Subsequent ionic motion shifts the compact and diffuse contributions and partly reverses the initial voltage excursion. The corresponding phase advance relative to the negative static derivative and the step overshoot are two measurements of the same coupled dynamics. No chemical transfer, electronic surface trap, or reversal of the prescribed polarization is required.

For the computed step from $g=0$ to $100$, the front potential moves from approximately $0.015\,V_T$ to $0.0030\,V_T$ near $t=0.40\,\tau_R$, then recovers to approximately $0.0083\,V_T$. The maximum excursion is about 1.87 times the final change. After light-off it rises temporarily to approximately $0.01837\,V_T$ near $1.09\,\tau_R$ before returning to darkness. Steps between two nonzero light levels also overshoot (Figure S8). The small difference between the finite-volume endpoints quoted here and the continuum stationary values is separately quantified by spatial refinement.

The last panel of Figure 7 identifies the relevant charge path. During the transient, the interfacial displacement need not equal minus the total diffuse liquid charge. Gauss's law gives $D_l(H)=D_s(L)+\int_0^H \rho_l \, du$; their difference is the displacement at the remote liquid boundary. The solid's total mobile charge nevertheless remains zero. Simultaneously setting this remote displacement to zero and fixing both the reservoir potential and an independently evolving liquid charge would overconstrain the intended experiment.

The analytical stationary and harmonic solutions satisfy the stated boundary equations to numerical precision. Refining a conservative discretization from 48 solid/80 liquid cells to 96/160 cells reduces the largest selected static front-voltage error to $0.0285\,\%$ and the complex optical-response error to $0.158\,\%$. The retained modes are stable at the examined operating points. A further comparison that restores nonlinear electron, hole, and ionic species equations on the same electrostatic map changes the selected stationary values by at most $0.00045\,\%$ and the step response by less than $0.0036\,\%$ of the full change. The overshoot therefore persists beyond the first-order charge reduction at these weak potentials. Its interpretation remains ionic voltage redistribution under a blocking boundary, distinct from the finite electronic-state mechanism in B.

## 5. Reactive Photoelectrochemistry

We next allow carriers to cross the ferroelectric–solution interface through electrochemical reactions. The analysis first treats a single surface reaction sustained by a rear carrier return, and then paired reactions with a floating contact. These comparisons connect surface potential, partial reaction rates, and carrier transport under fixed and floating electrical conditions.

### 5.1 Case D - Reactive Interface with a Rear Carrier Return

Opening an electrochemical channel changes the purpose of the transport path. In cases A–C, illumination establishes and perturbs a distribution of compensating charge. In case D, carriers can also be transferred to maintained chemical species, and a stationary reaction requires their complementary return. The primary example has one reduction channel at the exposed ferroelectric–liquid interface and a rear contact that equilibrates holes while blocking electrons. The rear potential is fixed relative to the liquid. Zero imposed bias is an electrical operating condition; it does not make the specimen isolated. The external connection and its complementary counterreaction complete the stationary circuit. The corresponding oxidation-only problem follows by exchanging electrons and holes, reversing polarization and bias, and exchanging the contact and chemical parameters when the material statistics have the required symmetry.

The liquid in this one-dimensional reactive limit has fixed reaction-plane activities and negligible resolved transport loss. Its electrostatic contribution is represented by a compact capacitance $C_H$ to the maintained liquid potential. Finite ionic transport is treated explicitly in the blocking junction and later in the domain calculations. The reactive carrier populations use the time scale $\tau_0$ and the instantaneous pair-loss law $R=(np-1)/(n+p+2)$. This has the stationary form of a symmetric Shockley–Read recombination law [82]. Its use in dynamics assumes sufficiently rapid eliminated trap occupation and negligible omitted trap-charge feedback. The frequency results below are solutions of that memoryless constitutive reduction; a physical interpretation as trap-mediated recombination additionally requires $\omega\tau_{\mathrm{occ}} \ll 1$. The trap-occupation equation and the associated validity conditions are given in the supplementary material.

The interfacial rates retain forward and reverse transfer. Using the normalization of the contacted-film equations, the two electron-equivalent rates are

$$r_r = s_{n0}\, e^{-\alpha_n v_s}\left(n_s - e^{v_s+\rho_r}\right), \tag{21}$$

$$r_o = s_{p0}\, e^{\alpha_p v_s}\left(p_s - e^{-v_s-\rho_o}\right). \tag{22}$$

Here $s_{i0}=S_{i0}\tau_0/L$ are transfer prefactors, $\alpha_i$ are compact-layer activation fractions, and $\rho_i$ specify the maintained redox electron electrochemical potentials relative to the intrinsic solid reference, in units of $k_BT$. The examples use $\alpha_n=\alpha_p=1/2$. For the primary single-channel calculation, $s_{p0}=0$. In the inward coordinate, the front carrier conditions are $j_n(0)=r_r$ and $j_p(0)=0$, and the rear conditions are $j_n(1)=0$, $p(1)=1$, and $v(1)=0$. The exposed displacement obeys $d(0)=-c_H v_s$, where $c_H=C_H V_T/P_*$. These conditions retain finite-depth carrier supply and determine the exposed potential through the instantaneous electrostatic state.

Figure 8 separates four observables that are frequently combined in qualitative band diagrams: surface potential, reaction current, reaction affinity, and dissipative transport loss. Their distinction follows directly from the constitutive laws. The electron and hole quasi-Fermi energies are $f_n=\ln n-v$ and $f_p=-\ln p-v$. Reduction has the dimensionless affinity $\eta_r=\ln n_s-v_s-\rho_r$, and oxidation has $\eta_o=\ln p_s+v_s+\rho_o$. Equations (21)–(22) give each signed rate the sign of its corresponding affinity. The electrostatic potential enters that affinity together with the local carrier population and the chemical reservoir. Consequently, the optical voltage alone does not specify either the reaction driving force or the rate. This separation is also central to continuum descriptions of semiconductor–electrolyte transfer [14, 15, 59, 60, 77, 84, 85].

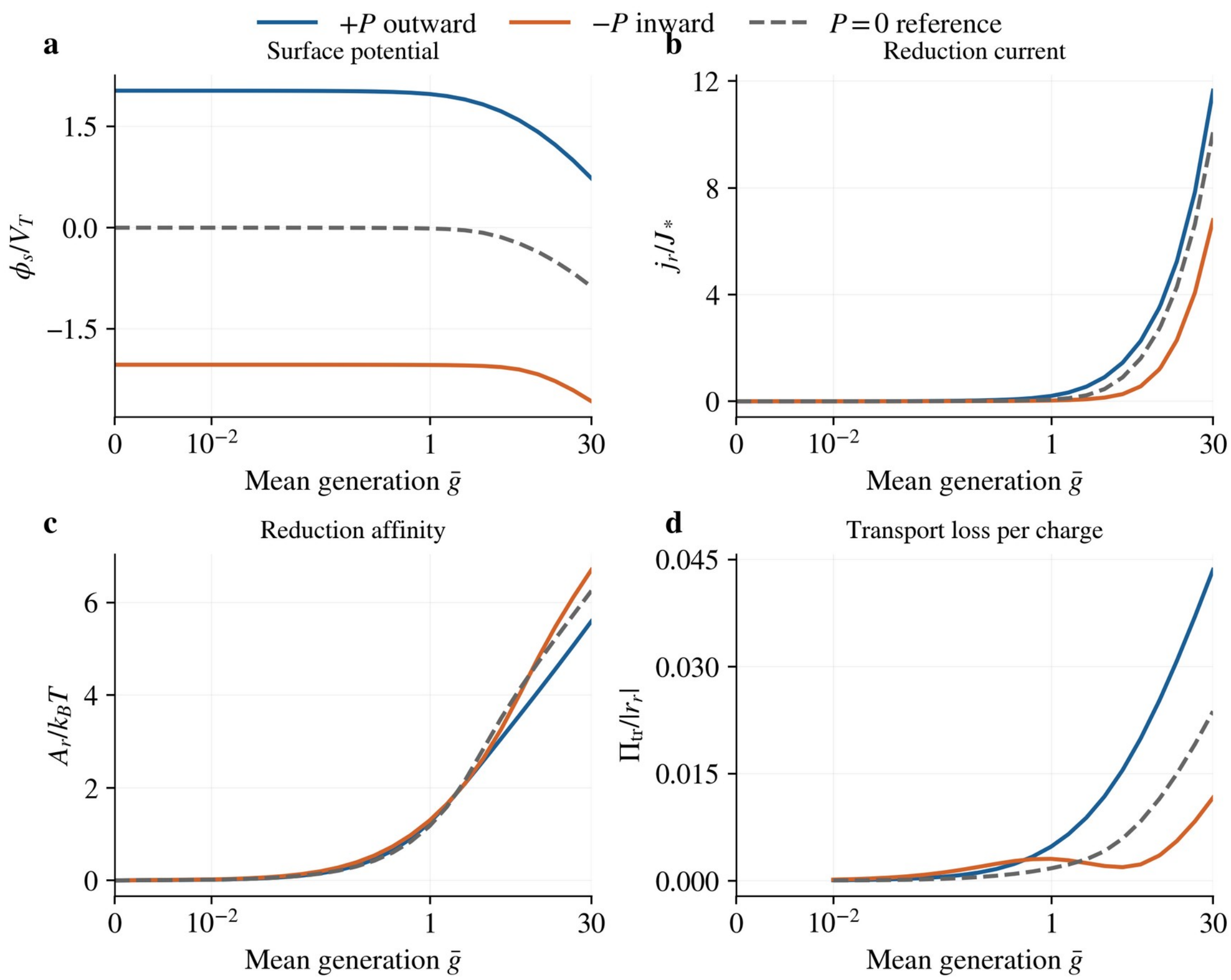

***Figure 8.*** *Sustained front reduction with selective rear return (case D). (a–d) Surface potential, front reduction current, local reaction affinity, and nonnegative transport dissipation divided by the nonzero terminal current. The rear potential is fixed at zero relative to the maintained liquid, the rear equilibrates holes and blocks electrons, and the front admits one reversible reduction channel. The electrostatic potential, reaction affinity, and transport loss are distinct quantities. The ratio in (d), in thermal-voltage units, is undefined at zero current and is omitted there. The nonpolar reference retains the background permittivity and the same selective contact and chemistry, so it has a nonzero photovoltage.* $J_*=q n_i L/\tau_0$ *and* $r_r=j_r/J_*$*; the reactive parameter set is specified separately from case B.*

For the illustrated source $g_0=10$, the outward-polarized film carries a reduction current of approximately $3.646\,J_*$, the semiconductor reference $2.845\,J_*$, and the inward-polarized film $1.281\,J_*$, with $J_*=q\,n_i\,L/\tau_0$. Their exposed potentials are approximately $1.405\,V_T$, $-0.369\,V_T$, and $-2.180\,V_T$, respectively. The absorbed pair-generation integral is $4.323\,J_*$ in electron-equivalent current units. The corresponding collected fractions are therefore approximately 84.3%, 65.8%, and 29.6% for this parameter set. The semiconductor now develops a finite optical potential because front reduction and rear hole exchange break the electron–hole symmetry of the dry reference. Reversing polarization while retaining the same chemical channel and contact does not reverse the entire solution.

The doping dependence of the reduced stationary interfacial potential is examined in Section 7, with the contact and solution references stated explicitly.

The depth-resolved fields and quasi-Fermi profiles are shown in Figure S9. Electron and hole currents vary separately as carriers are generated and recombine, but their sum is spatially constant at stationarity. Integrating the electron balance gives $r_r=\int_0^1 (g-R)\,dy$ because the rear blocks electrons. The equal rear hole current supplies the complementary return. This is stronger information than the sign of the near-surface field: it identifies how the full specimen supports the reaction. A rear contact that equilibrates both carrier species gives a different solution. In the checked zero-bias example, most pairs then leave through the rear and the front reduction is substantially smaller. The complete carrier ledgers and the contact comparison are provided in the supplement.

The transport diagnostic in Figure 8(d) follows from the quasi-Fermi gradients. Since $j_n=M_n n f_n'$ and $j_p=M_p p f_p'$, the dimensionless dissipation is

$$\Pi_{\mathrm{tr}}=\int_0^1\left[\frac{j_n^2}{M_n n}+\frac{j_p^2}{M_p p}\right]dy\geq 0\,. \tag{23}$$

Dividing by the nonzero single-channel current gives a transport loss per charge in units of $V_T$. At $g_0=10$, this quantity is approximately 0.0257 for outward polarization, 0.0118 for the semiconductor, and 0.00369 for inward polarization. These values differ from the corresponding electrostatic potentials. Equilibrium band bending is therefore not an Ohmic voltage loss. Generation, recombination, reaction affinity, and reservoir work complete the carrier free-energy balance; the supplementary calculation verifies that identity separately from the charge balance.

Figure 9 shows the differential optical voltage about each stationary illuminated state. All three preparations have their own amplitude and phase surfaces because their carrier populations, field distributions, and transfer derivatives differ. The displayed gain is $|H_V|=|\delta v_s/\delta g_0|$, and phase is referenced to positive light modulation. A negative stationary slope contributes a $180^\circ$ low-frequency sign change; the plotted phase should not be read as a delay without accounting for that sign. The frequencies are normalized by $\tau_0$, so they cannot be assigned the dry-film time scale or a physical frequency without a consistent material parameter set. The highest displayed frequencies indicate the behavior of the stated reduced equations, subject to the trap-elimination condition given above.

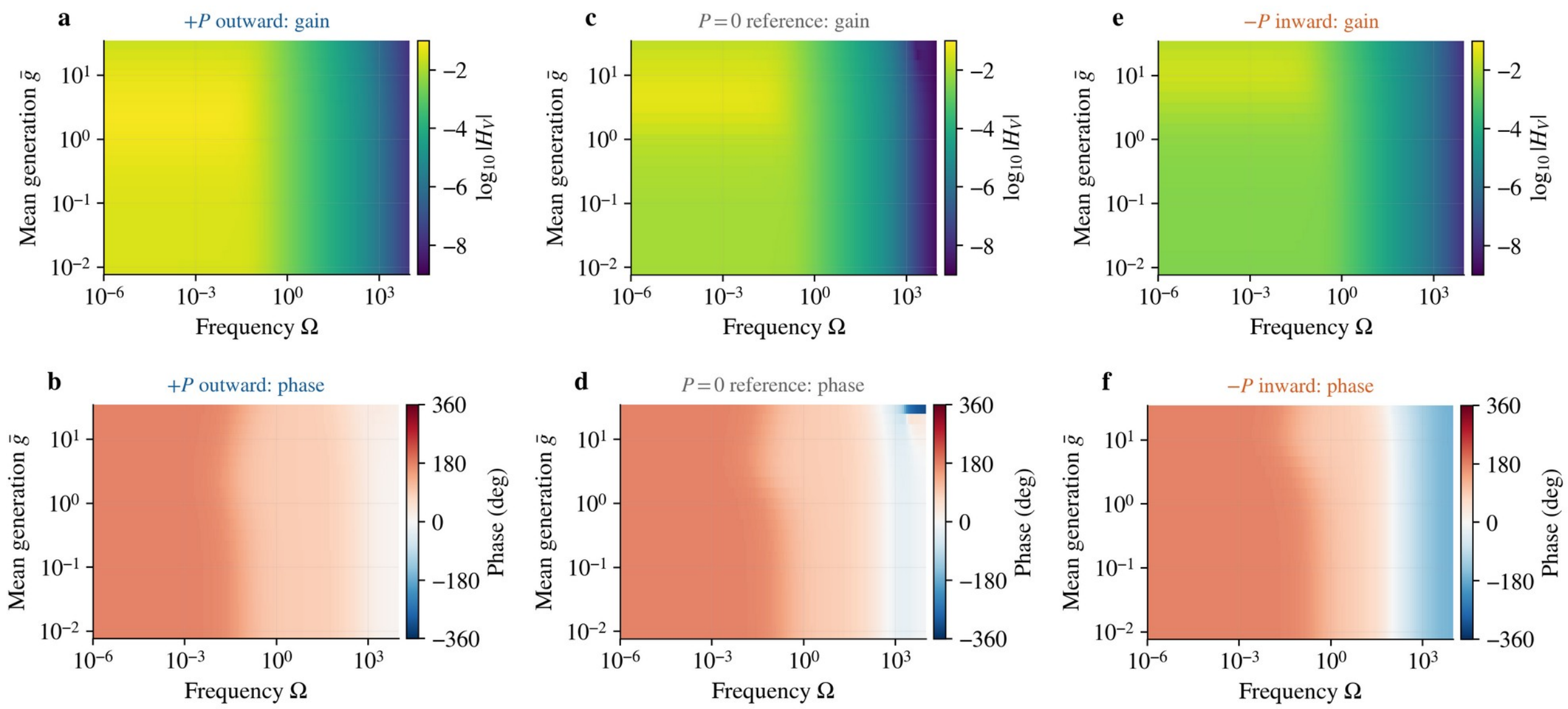

***Figure 9.*** *Optical voltage response of the reactive film at fixed rear bias. The three columns show outward polarization, the nonpolar reference, and inward polarization. Upper panels give the absolute generation derivative* $|H_V| = |\delta(\phi_s/V_T)/\delta g_0|$*; lower panels give incident-light-referenced phase, unwrapped along frequency separately at each mean intensity. The low-frequency phase branch is near* $0^\circ$ *or* $180^\circ$ *according to the sign of the stationary derivative. Branch winding is retained rather than folded into* $[-180^\circ, 180^\circ]$*. Every spectrum is evaluated about its own stationary state with one front reduction channel, a hole-equilibrating/electron-blocking rear, and fixed rear potential.* $\Omega = \omega\tau_0$*. The full displayed band is the response of the stated instantaneous pair-loss model; identifying that law with eliminated trap-mediated recombination additionally requires* $\omega\tau_{\mathrm{occ}} \ll 1$ *and negligible omitted trap-charge feedback. No dimensional frequency calibration is assigned.*

Potential, affinity, and current have different optical transfer functions. In supplementary Figure S10, all three are evaluated from one linearized state and compared with directly integrated nonlinear sinusoidal perturbations. The change of reduction affinity is $\delta\eta_r = \delta n_s/\bar{n}_s - \delta v_s$, which makes the additional carrier contribution explicit. The light-on/off current and potential trajectories, together with steps between nearby illuminated states, are retained in Figure S11; the applied-bias and chemical-threshold comparisons are in Figure S12. The step results bridge the finite-amplitude experiment to the differential response, while keeping the active front channel and rear return unchanged. The chemically transferred amount follows the time integral of the appropriate partial reaction rate with its electron stoichiometry; displacement and dry-site capture currents do not contribute to that amount.

The separate roles of bulk supply and surface transfer are also apparent in a weak-field minority-carrier limit. With a maintained majority population, a reflecting minority-carrier rear, linear recombination coefficient $k$, and finite front transfer $S$, uniform generation gives

$$\frac{\delta j_s}{\delta g} = \frac{SM\,q_\Omega \tanh q_\Omega}{(k+i\Omega)\left[S + M\,q_\Omega \tanh q_\Omega\right]}, \tag{24}$$

$$q_\Omega = \sqrt{\frac{k+i\Omega}{M}}. \tag{25}$$

The root has positive real part. The denominator separates the finite interfacial transfer from the diffusion–recombination supply. In the stationary fast-transfer limit, the collection factor becomes $l\tanh(1/l)$, where $l = \sqrt{M/k}$. Maintaining the majority carriers requires the declared complementary return. This analytical limit predicts the collection current under its specified assumptions; the potential and affinity still require the coupled electrostatics and both carrier populations. Its derivation and numerical verification are retained in Sections S5.3 and S7.

### 5.2 Floating Potential and Paired Reactions

A useful closure comparison retains the contacted geometry but opens both front channels and permits the rear metal to float. The rear remains hole-permeable and electron-blocking, while its potential is no longer prescribed. Zero external total current includes displacement and gives

$$\frac{c_H}{\delta}\frac{d v_s}{d\left(t/\tau_0\right)}=r_r-r_o. \tag{26}$$

The electrostatic solution then determines the rear voltage. At stationarity, reduction and oxidation balance and the rear hole conduction current vanishes. During charging, the two partial rates need not be equal, and transient hole exchange at the floating metal is compatible with zero external total current because its displacement changes. This remains a carrier-permeable boundary, distinct from the inactive rear of the blocking liquid and isolated domain cases. It provides a direct comparison of externally maintained and floating operation without introducing a sixth primary physical case.

Figure 10 uses a common maintained redox pair, with $\rho_r=\rho_o=0$. Equal positive stationary reduction and oxidation rates recycle its two chemical forms and remove photocarriers; they do not imply net accumulation of one reduced product. After light-on, the partial-rate imbalance changes the interfacial charge until their balance is recovered. The outward-polarized floating voltage at $g_0=10$ is approximately $2.709\,V_T$, with each partial rate near $0.6205\,J_*$. On switch-off, residual carriers initially maintain an imbalance that raises the voltage to approximately $2.749\,V_T$ before it returns to its dark value. The rate traces in Figure 10(c) and the voltage traces in (a,b) describe the same charge inventory through equation (26). No finite electronic site population is present in this example.

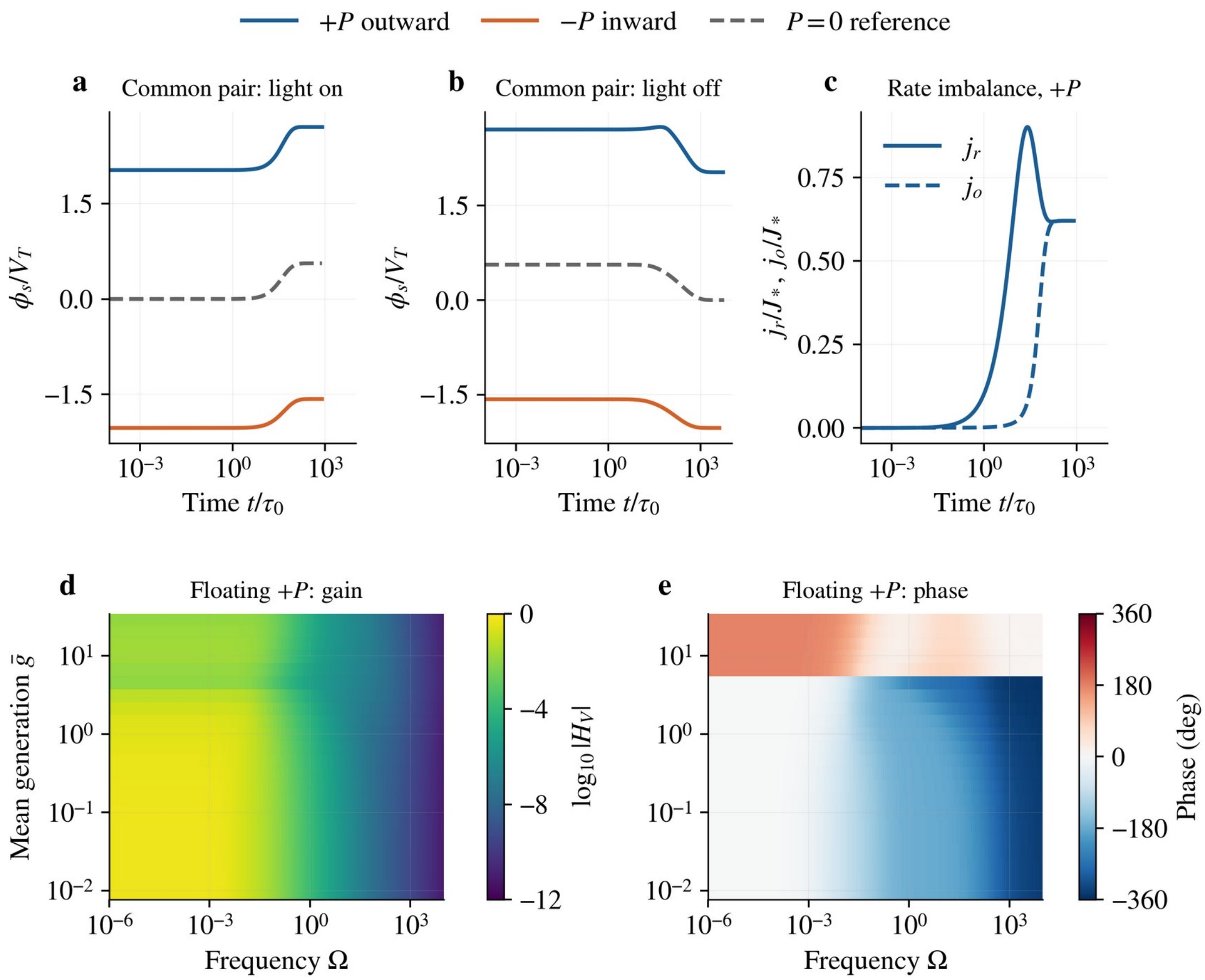

*__Figure 10.__ Floating-metal response with a common maintained redox pair. (a,b) Surface voltage after steps from $g_0 = 0$ to 10 and back, for both polarities and the nonpolar reference. Each step uses its own time origin. (c) Reduction and oxidation partial rates during the outward-polarized light-on step. Their difference changes compact interfacial charge according to $\left(c_H/\delta\right) d v_s/d\left(t/\tau_0\right) = r_r - r_o$. (d,e) Absolute optical gain and incident-reference phase for outward polarization over mean generation and frequency. Phase is unwrapped along frequency; the complete polarity/reference spectra are in the supplement. The rear metal remains hole-permeable and electron-blocking, while its potential adjusts to zero external total current. It is not an inactive rear. Equal stationary partial rates recycle the same redox species and do not establish net product accumulation. The same memoryless recombination and fast-liquid restrictions as in the reactive fixed-bias model apply.*

The periodic common-pair response in Figure 10(d,e) is linearized about its own floating stationary states. Its mean-intensity dependence differs from the fixed-bias single-reaction response because the rear voltage and both interfacial rates now participate in the relaxation. The common-pair frequency response and its controls are specified in Section S5.4. When the two front channels instead connect distinct maintained redox couples, equal electron-equivalent rates can represent net chemical conversion. The chemical gap then enters through $\eta_r + \eta_o = \ln\left(n_s p_s\right) - \left(\rho_r - \rho_o\right)$. The retained distinct-couple calculations show both reverse dark conversion and illumination-driven reversal, with polarity-dependent crossing intensities; their stationary, transient, and periodic results are given in Figure S13, Figure S14, and Figure S15, with the chemical crossings in Figure S12.

Taken together, the contacted results identify the electrical and chemical conditions behind a measured optical potential. A fixed-bias reactive film can sustain a single front channel because the rear supports a complementary current. A floating film with two front channels instead relaxes toward a balance of its partial rates. Both can exhibit nonmonotonic voltage transients, but their relevant inventories and return conditions differ from those of a dry film with finite electronic sites.

These distinctions are necessary for using surface-potential dynamics to infer reaction behavior and for relating a predicted current to accumulated chemical amount.

## 6. Spatially Selective Chemistry in Domain Patterns

Oppositely polarized domains allow complementary reactions to communicate through lateral transport in the solid and electrolyte. We first examine the domain potentials and reaction rates, then use the nonlinear finite-depth benchmark to resolve the current paths. The final part relates lateral transport to reaction extent and accumulated product.

### 6.1 Case E - Two Opposite Domains

The two-domain geometry introduces lateral redistribution of photogenerated carriers and a lateral ionic return through the solution. These processes determine whether a polarization pattern produces a detectable chemical pattern, and whether spatial selectivity is accompanied by a change in total conversion. The distinction is relevant to domain-resolved photochemical markers, which establish preferential reaction at selected surface locations without directly measuring the complete operating current path [22, 86, 87, 88]. Earlier multidomain calculations likewise demonstrate the importance of lateral collection and domain width [57]. Semiconductor-particle models also identify the distribution of catalyst regions and bulk recombination as controls on collection [58]. Here, the rear face remains impermeable to electrons and holes. Its fixed compensation and passive dielectric response are retained. The blocking and reactive calculations use the same domain geometry and differ in the front transfer coefficient.

We first consider a frozen, smooth stripe pattern of period $2w$. The outward-polarized domain occupies $0<x<w$, and the inward-polarized domain occupies $w<x<2w$. The assigned residual surface charge is positive over the outward domain. This assignment includes the specified fixed compensation; polarization orientation alone does not determine the sign of a partially compensated surface charge. A vertical 180° wall is electrically neutral in the bulk because the varying polarization component is perpendicular to the direction of variation. No additional wall conductance, wall trap population, or wall-specific reaction channel is introduced. Both redox couples are available at every surface position in the analytical reactive model. They are distinct buffered couples with equal thermoneutral reference carrier densities, so balanced reduction and oxidation can convert separate chemical reservoirs without a rear-electrode current.

In the thin-film limit, the carrier populations and electrostatic potential are represented by sheet variables. The compact layer stores charge, and the outer liquid conducts laterally. For a nonzero lateral wave number $k$, a deep Ohmic liquid has area-specific resistance $R_1(k)=1/(\kappa_1|k|)$. The same conductivity determines the resistance of every spatial harmonic. The passive backing contributes $\epsilon_{\text{back}}|k|$, while the lateral film electrostatics contributes $\epsilon_x L k^2$. These terms are included in $C_g(k)=C_0+\epsilon_{\text{back}}|k|+\epsilon_x L k^2$, where $C_0$ is the compact-dominated interfacial capacitance. A rear boundary that blocks particles therefore need not have zero displacement field for each lateral mode.

The analytical reference uses equal electron and hole diffusivities, equal transfer velocities $S$, and bimolecular pair loss. We define $t_0=(k_b n_i)^{-1}$, $\theta=t/t_0$, $g=G/(k_b n_i^2)$, $m=N/n_i$, and $s=S t_0/L$. The leading spatially uniform population obeys

$$\frac{dm}{d\theta}=g-(m^2-1)-s(m-1), \tag{27}$$

$$\bar{m}=1+\frac{2\bar{g}}{2+s+\sqrt{(2+s)^2+4\bar{g}}}. \tag{28}$$

Blocking corresponds to $s=0$; the paired-reaction example uses $s=0.2$. These quantities have a definite conversion to incident intensity through the absorption, generation yield, photon energy, and film thickness. The present comparisons use normalized parameters, rather than assigning the same dimensional material to all five boundary cases. The complete parameter sets and conversions are given in the Supplementary Material.

For a weak surface-charge pattern, write $n/n_i=m+b(x)$, $p/n_i=m-b(x)$, $u=\phi_s/V_T$, $v=\phi_l/V_T$, and $\zeta=u-v$. The dimensionless Gauss relation for each mode is $b_k=\delta_k-c_g u_k+c_0 v_k$, with $C_*=2qLn_i/V_T$, $c_0=C_0/C_*$, $c_g=C_g/C_*$, and $\delta_k=\sigma_k/(C_* V_T)$. The interfacial rates follow the separate carrier populations and the solid–liquid drop: $r_r=S\left[n_s-n_i e^{\zeta}\right]$ and $r_o=S\left[p_s-n_i e^{-\zeta}\right]$. Their first-order difference is controlled by $c_k=b_k-\zeta_k$. Thus the electrostatic potential at the solid surface and the driving force for the reaction remain distinct observables.

The stationary solution makes the two lateral transport controls explicit. Let $d_k=D_x k^2 t_0$, $f_k=d_k/(d_k+s)$, $h_k=t_0/\left[R_1(k)C_*\right]$, $\chi_k=s f_k/h_k$, and $\Theta_k=\chi_k(\bar{m}-1)/(1+\chi_k)$. Then

$$\bar{u}_k=\frac{\delta_k}{c_g+f_k\bar{m}+1-f_k+(c_0+1-f_k)\Theta_k}, \tag{29}$$

$$\bar{v}_k=-\Theta_k\bar{u}_k, \tag{30}$$

$$\bar{c}_k=\frac{f_k(\bar{m}-1)\bar{u}_k}{1+\chi_k}. \tag{31}$$

The bare transport crossover length is $l_{tr}=(\frac{D_x L}{S})^{\frac{1}{2}}$. Thus the transport coordinate is $\frac{D_x L k^2}{S}=(k\, l_{tr})^2$, and $f_k=\frac{1}{2}$ when $k\, l_{tr}=1$. The reaction-decay length $l_c$ also includes the electrostatic screening factor given in Equation (S331).

For a blocking surface, the stationary liquid modes vanish and $\bar{u}_k=\delta_k/(c_g+\bar{m})$. Illumination therefore reduces the magnitude of the domain potential by increasing carrier screening. The full stationary intensity dependences are shown in Figure S18; the spatial comparison in Figure 11a shows the dark, illuminated blocking, and illuminated reactive states. At $g=3$, the positive-domain average changes from $0.07848\,V_T$ in darkness to $0.05891\,V_T$ under blocking conditions. The opposing domain has the negative of this potential, so the domain contrast is twice the positive-domain average. The spatial mean remains zero by symmetry. The corresponding unpatterned reference also has zero potential under these symmetric constitutive and boundary assumptions, although it develops a nonzero uniform reaction rate when the two distinct chemical channels are open.

Opening the reaction channels changes the stationary carrier inventory as well as the lateral current. At $g=3$, the mean population becomes $\bar{m}=1.95183$. For an ideally conducting liquid, the positive-domain potential is $0.07498\,V_T$, larger than in the illuminated blocking state because interfacial consumption reduces screening. Both partial reactions remain active over both domains, with complementary local enhancements (Figure 11b). At first order, 50.56% of the total reduction occurs on the positive domain. This is a weak preference, despite the recognizable rate pattern. Increasing the fundamental liquid resistance parameter $\rho_1=R_1(k_1)C_*/t_0$ from zero to 500 reduces the solid potential to $0.04262\,V_T$, while the liquid potential becomes $-0.03575\,V_T$. The compact drop remains approximately $0.07836\,V_T$. Consequently, a reduced solid-potential contrast does not by itself establish a proportional reduction in the local electrochemical drop.

The doping-dependent shift of the two domain potentials relative to the solution is examined in Section 7.

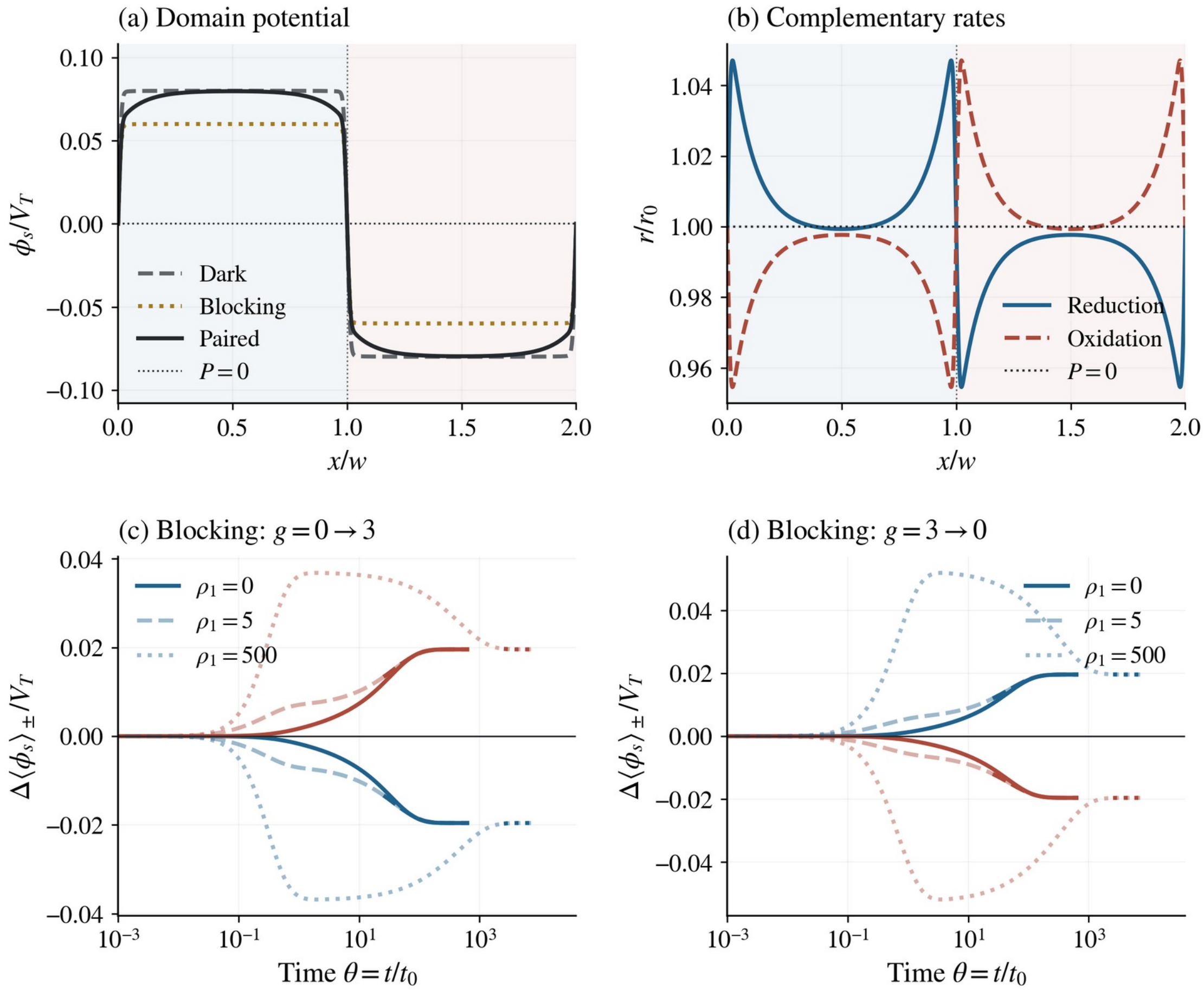

***Figure 11.*** *Domain potentials, complementary reactions, and blocking illumination steps in Case E. (a) Dark, illuminated blocking, and illuminated paired-reaction potential profiles; the unpatterned reference has zero potential in this symmetric model. (b) Both partial rates on both domains at g=3 for an ideally conducting liquid, including the complete local quadratic neutral-density correction;* $r_0=S(N-n_i)$*. Blue shading denotes the outward domain,* $0<x/w<1$*, and red shading the inward domain,* $1<x/w<2$*. (c,d) Changes of the domain-averaged potential following* $g{:}0\to3$ *and* $3\to0$*, respectively, for* $\rho_1=0$*, 5, and 500. Blue/red curves represent the outward/inward domain; line style identifies liquid resistance. These are solutions of the fast-diffuse, Ohmic-liquid reduction. For the explicitly illustrative ratio* $\varepsilon\ell/\varepsilon_x=0.8$*, saturated curve portions satisfy the observation-time diagnostic* $t\geq10\tau M$*; pale portions retain the formal limiting solution at earlier times. This diagnostic does not establish a full ionic error bound. The strong early overshoot at* $\rho_1=500$ *is consequently not presented as a resolved electrolyte transient. The static panels are long-time solutions and do not use this temporal cutoff.*

Uniform optical modulation perturbs the mean carrier population rather than the fixed surface-charge source. Its susceptibility is $H_m=(2\overline{m}+s+i\Omega)^{-1}$, where $\Omega=\omega t_0$. The finite-liquid charge problem has the form $M_k\dot{x}_k=B_k-K_k(m)x_k$, with $x_k=(u_k,v_k)^T$. The matrices follow from the carrier inventory, Gauss law, and the interfacial current condition, and are given in the Supplementary Material. Differentiating these same equations gives

$$H_{x,k}=\frac{-d_k\overline{u}_k}{2\overline{m}+s+i\Omega}\left[K_k+i\Omega M_k\right]^{-1}\begin{pmatrix}1\\0\end{pmatrix}, \tag{32}$$

$$H_{u,k}^{(R_1=0)}=\frac{-(d_k/c_g)\overline{u}_k}{(2\overline{m}+s+i\Omega)(\omega_{q,k}+i\Omega)}, \tag{33}$$

where $\omega_{q,k}=d_k\left(1+\overline{m}/c_g\right)+s\left(1+1/c_g\right)$. The ideal-liquid expression separates the neutral carrier relaxation from charge redistribution. For finite liquid resistance, the solid and liquid charge states introduce two coupled electrostatic relaxation rates. Their real parts are positive for the specified positive transport and capacitance coefficients. Stable relaxation nevertheless permits an overshoot in a measured voltage because that voltage combines charge contributions with different temporal weights.

The gain and phase in Figure 12 are referenced to an absolute modulation $\delta g$, with phase measured relative to positive illumination. The positive-domain stationary derivative is negative, so its phase approaches 180° as frequency approaches zero. Reversing the domain reverses the potential response and preserves its magnitude. The global mean potential does not detect this antisymmetric response. Increasing mean illumination changes both the carrier susceptibility and the electrostatic relaxation rates, which is why a single dark-state relaxation time cannot describe the full intensity–frequency maps. The response of the compact drop can also be smaller than the responses of the solid and liquid potentials because both potentials can move together. The component-resolved calculations are retained in Figure S19 and Figure S20.

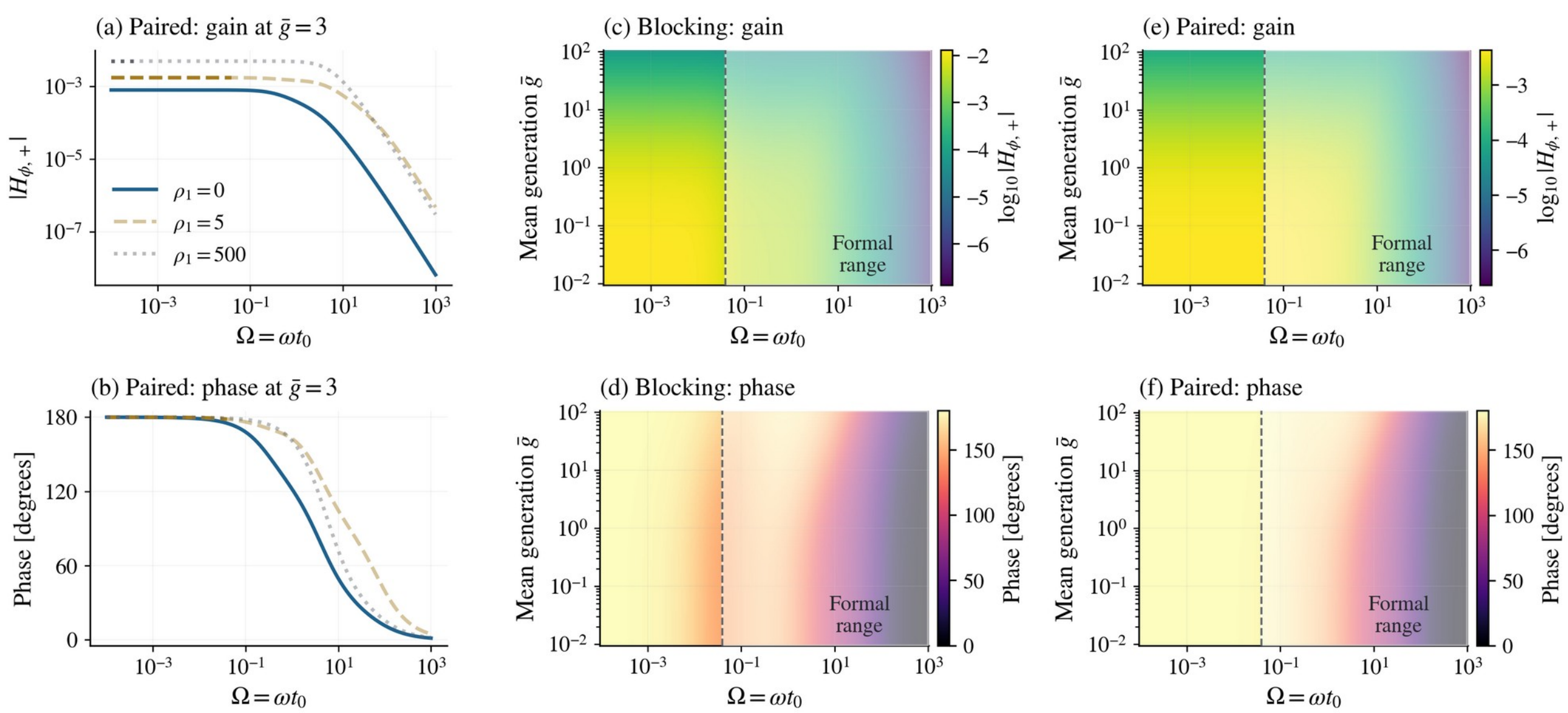

*Figure 12. Periodic optical response of the two-domain thin film. Each column pairs gain (top) with the corresponding phase (bottom). (a,b) Gain and phase of Hφ,+=δ(⟨φs⟩+/VT)/δg at mean generation $\bar{g}$=3, comparing $\rho_1$=0, 5, and 500 with paired reactions. (c,d) Blocking-surface gain and phase over mean generation and temporal frequency for $\rho_1$=5. (e,f) The corresponding paired-reaction maps at the same liquid load and geometry. Phase is measured relative to positive incident-light modulation; the negative stationary potential slope gives the branch approaching 180° at low frequency. The opposite domain has the same gain and the sign-reversed complex response. Pale curves and pale map regions mark ωτM>0.1 for the explicitly illustrative dielectric ratio εℓ/εx=0.8, using τM/$t_0$=0.632136$\rho_1$εℓ/εx. The dashed map boundary is Ω=0.03955. The spectra remain the mathematical fast-diffuse-limit solutions throughout; their formal continuation is displayed separately from the portion passing this necessary time-separation diagnostic. The example is not a material calibration.*

The temporal interpretation requires the liquid approximation to be stated quantitatively. The present modal equations take the limit of an equilibrated diffuse layer and an electroneutral Ohmic outer liquid, while retaining compact charging. For a symmetric dilute electrolyte, the omitted bulk charge-relaxation time is $\tau_M=\epsilon_1/\kappa_1=\lambda_D^2/D_i$. Diffuse charging and ionic transport introduce additional temporal structure when this elimination is not justified [80]. For the geometry used here, $\tau_M/t_0=0.632136\,\rho_1\left(\epsilon_1/\epsilon_x\right)$. The resistance and the omitted ionic time therefore cannot be assigned independently. As an explicit diagnostic, Figure 11 and Figure 12 use $\epsilon_1/\epsilon_x=0.8$ to distinguish the portions satisfying $\omega \tau_M \leq 0.1$, or $t \geq 10\tau_M$, from the formal continuation of the limiting equations. This ratio is illustrative and is not a material calibration. Passing the displayed Maxwell-time check is necessary but not sufficient to validate the complete liquid reduction. A late observation time also does not remove the unresolved contribution of an earlier ionic transient to accumulated amount. The necessary spatial and capacitance limits, including $k\lambda_D \ll 1$ and $C_D \gg C_0$, are also specified in the Supplementary Material.

For this diagnostic, the frequency cutoffs are $\Omega = 0.03955$ at $\rho_1 = 5$ and $3.955 \times 10^{-4}$ at $\rho_1 = 500$. The pronounced early overshoot of the most resistive blocking example occurs before the omitted liquid charge has been shown to equilibrate. It remains a well-defined solution of the stated limiting model; it is not a resolved prediction of ionic dynamics for that dielectric ratio. Its origin within the model is informative. The fixed residual charge is partitioned among mobile semiconductor charge, compact charge, and the film/backing displacement. After a light step, the solid and liquid potentials initially move together while the compact drop changes more slowly. Subsequent redistribution recovers the stationary blocking endpoint, which is independent of liquid conductivity. The complete trajectories are retained, with the regime distinction visible, rather than using their early extrema to infer a material relaxation time.

### 6.2 Nonlinear Finite-Depth Benchmark

The finite-depth calculation in Figure 13 provides a complementary nonlinear spatial benchmark. It uses a 500 nm domain width, exponential absorption, separate electron and hole transport, and the high-injection pair-loss law $R = 2np/[\tau(n+p)]$. Reduction-active and oxidation-active regions are prescribed on complementary domains, with transfer velocities of 0.1 and 10 $\mathrm{m\,s^{-1}}$, respectively. The liquid is ideally conducting and the deep boundary is carrier-blocking. These chemical masks are retained in the semiconductor control. Thus the calculation tests carrier supply and electrostatic redistribution within an assigned chemical pattern; it does not infer exclusive chemical accessibility from polarization. The unmasked weak-pattern calculation and this masked nonlinear benchmark have different kinetic and parameter assumptions.

The depth maps show how those assumptions produce unequal electron and hole populations, a surface potential varying between approximately $-19.4$ and 19.2 mV, and a conventional-current path through the solid between the two reaction regions. The streamlines show the solid portion of the path; its liquid continuation is represented by the ideal-conduction boundary. At the reference lifetime of 1 µs, the collected-pair fraction is 24.80%, compared with 22.00% for the unpolarized control with the same chemical masks. The mean reduction current is 1.134 $\mathrm{A\,m^{-2}}$ over its active domain, or 0.5668 $\mathrm{A\,m^{-2}}$ averaged over the entire period. Integrated reduction and oxidation agree. Their sum must not be counted as twice the converted pair flux.

Although the reduction rate is enhanced near the domain boundaries, the reaction remains broad. The nearest-wall distance containing 90% of the reduction is 217.8 nm, compared with 225 nm for a uniform profile over the same domain (Figure 13d). Increasing the lifetime from 10 ns to 10 µs raises collection from 2.71% to 53.99%, while this distance changes only from 214.6 to 220.8 nm (Figure 13e,f). Carrier survival therefore strongly affects amount without producing an equivalent change in integrated spatial extent in this example. Grid refinement changes collection by 0.17% and the extent by 0.0034%. These checks support the reported surface observables; the high-injection approximation is not a calibration of equilibrium carrier and defect populations deep in a named oxide.

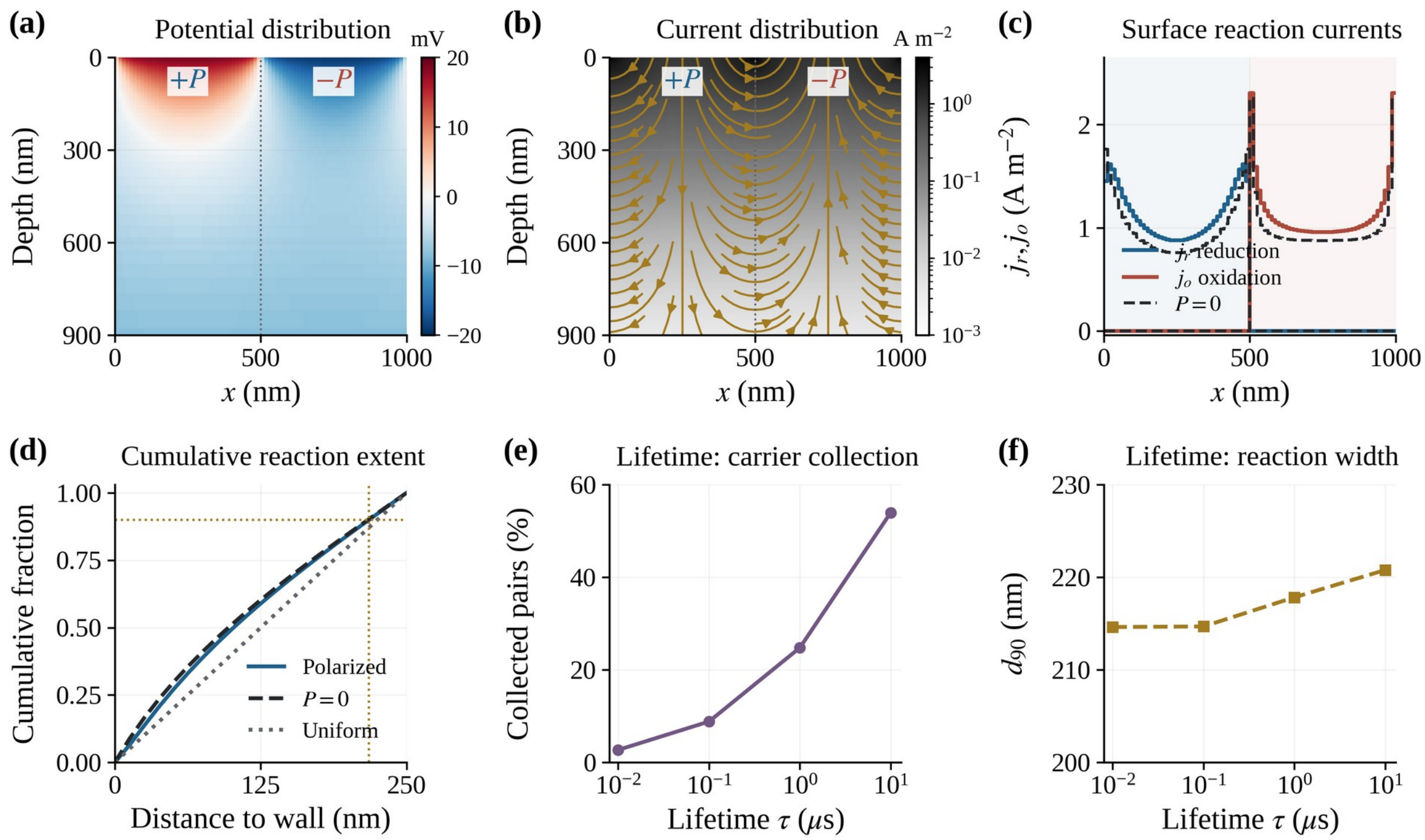

***Figure 13.*** *Finite-depth nonlinear benchmark with complementary chemical masks. (a) Electrostatic potential within the upper 900 nm of the 1.5 μm calculation depth. (b) Conventional-current magnitude and streamlines within the solid. The liquid continuation is represented by the ideal-conduction boundary and is not an explicitly calculated liquid current field. (c) Surface reduction and oxidation currents; dashed curves show the nonpolar semiconductor with the same prescribed reaction masks. (d) Cumulative reduction versus distance from the nearest domain wall, including the uniform-profile comparison. The dotted guides identify the distance* $d_{90}$ *containing 90% of reduction. (e,f) Collected-pair fraction and* $d_{90}$*, respectively, versus the carrier lifetime. These two plots separate the amount collected from the spatial extent of the reaction. Electron and hole density maps remain in Supplementary Figure S22(b,c). The period is 1 μm, temperature 300 K, mobilities* $\mu_n = \mu_p = 3 \times 10^{-7}$ *m² V⁻¹ s⁻¹, residual surface-charge amplitude 0.002 C m⁻², and interfacial capacitance 0.1 F m⁻². Illumination is 10 W m⁻² at 3.5 eV, with absorption coefficient* $10^7$ *m⁻¹ and mobile-pair yield 0.8. Reduction and oxidation transfer velocities on their assigned domains are 0.1 and 10 m s⁻¹. The high-injection recombination law is* $R = 2np / [\tau(n+p)]$*. This stationary benchmark tests transport within prescribed chemical masks; it does not infer exclusive chemical accessibility from polarization or validate the temporal thin-film reduction.*

### 6.3 Lateral Transport, Reaction Extent, and Accumulated Product

The transport maps in Figure 14a–c return to the unmasked analytical model and separate local contrast, domain allocation, and total conversion. Weak solid lateral diffusion suppresses redistribution between domains. Large liquid resistance suppresses the complementary current even when the solid readily redistributes carriers. Uniform local paired turnover can still occur because both couples are available everywhere. For a zero-mean weak pattern, the first-order rate perturbations average to zero. The first change in mean yield is quadratic. With $b$ and $\zeta$ denoting the first-order fields,

$$\delta \bar{R} = \frac{s}{2\bar{m} + s}\left[\langle b^2 \rangle - \bar{m}\langle \zeta^2 \rangle\right], \tag{34}$$

$$\bar{R} = \frac{\bar{r}\, t_0}{L n_i}. \tag{35}$$

The two terms represent the change in pair recombination associated with carrier separation and the change in reversible interfacial transfer associated with the compact drop. They compete. For one spatial harmonic, the sign changes when the effective transport factor $f_{\text{eff}} = f/\left(1+\chi\,\overline{m}\right)$ crosses $1/\left(1+\sqrt{\overline{m}}\right)$. Each harmonic of a stripe has its own factor. Figure 14c therefore uses a single harmonic to display the sign boundary explicitly, while the stripe results use the complete mode sum. A stronger spatial pattern can accompany either increased or decreased total conversion under the stated kinetics.

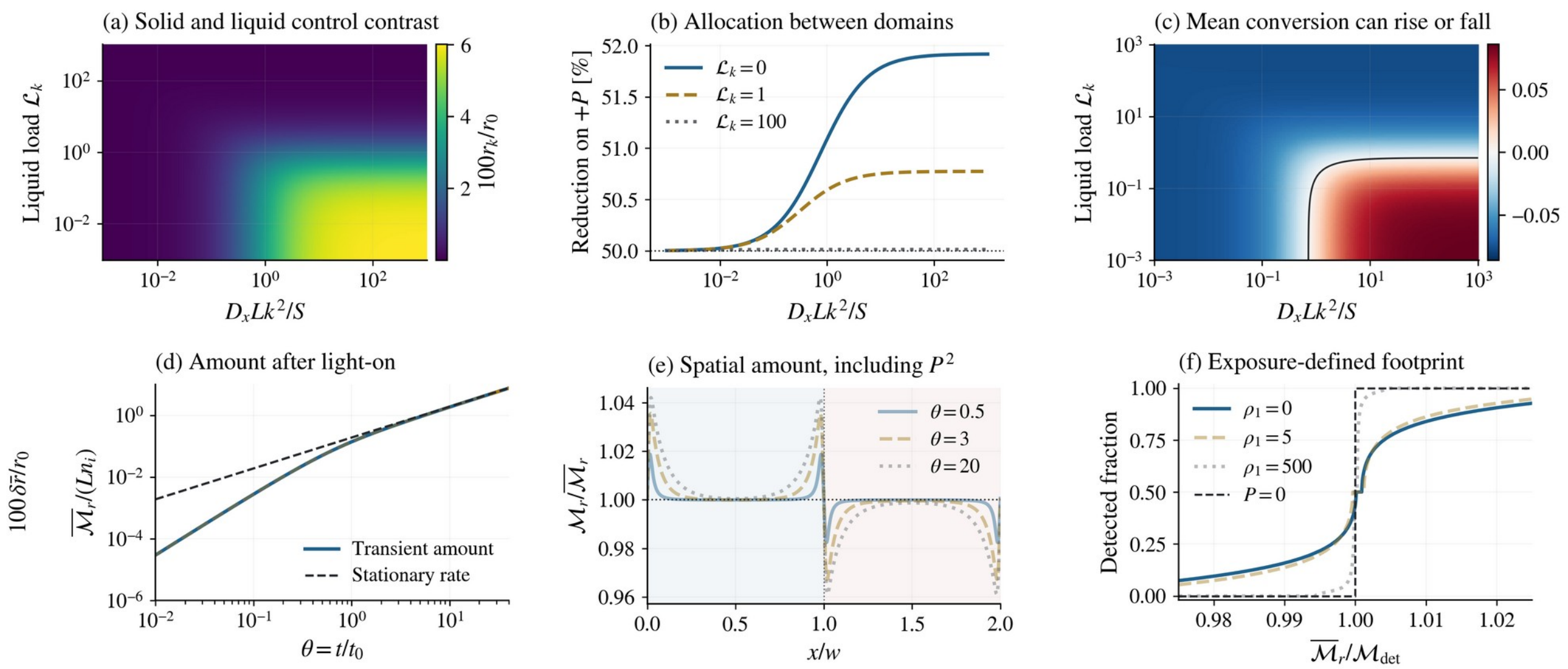

***Figure 14.*** *Lateral transport, chemical allocation, and accumulated product. (a) Relative partial-rate contrast versus solid transport $D_xLk^2/S$ and liquid load $\mathcal{L}k=2qSn_iR\ell(k)/VT$ for one spatial harmonic. (b) Fraction of total reduction allocated to the outward half-period. (c) Leading mean-yield correction; the contour marks zero change in total conversion. Panels (a–c) use $k\ell_0=1$, $\bar{g}=3$, and the same normalized electrostatic and transfer parameters as the analytical reference. (d) Mean amount following $g{:}0\to3$ in the stripe model, compared with extrapolating the final stationary rate to zero exposure time. (e) Spatial amount divided by its mean at $\theta=0.5$, 3, and 20 for $\rho_1=5$. (f) Detected fraction at threshold $M_{det}=3Ln_i$ as a function of mean amount; $P=0$ is spatially uniform. Panels (d–f) integrate the actual transient rates, including all local even neutral harmonics through quadratic order. Their fundamental stripe wave number is $k_1\ell_0=0.158034$. Blue/red background shading in (e) denotes outward/inward domains. Pale finite-liquid portions retain the fast-diffuse limiting result where the $\varepsilon\ell/\varepsilon_x=0.8$ observation-time diagnostic is not satisfied. A later observation-time criterion alone does not recover an omitted early ionic contribution to accumulated amount. Threshold positions also retain the stated weak-pattern approximation.*

To determine a local accumulated amount consistently through the same order, we also retain the full quadratic neutral-density field, including its nonzero even harmonics. Its forced diffusion equation and temporal solution are derived in the Supplementary Material. The additional nonzero even harmonics leave the established spatial mean and the integrated allocation over each equal-width domain unchanged through quadratic order, but they modify the local rate and the threshold-defined footprint. They are especially relevant when liquid resistance has made the first-order contrast small. All twelve paired-reaction steps have been evaluated with this correction, covering light-on, light-off, and changes between nonzero mean illuminations for the three liquid loads.

The electron-equivalent amount is obtained by integrating the actual transient rate. For a product requiring $v$ electron equivalents, the molar amount is $M_r/(v N_A)$. An experimental detection threshold defines a separate spatial observable:

$$M_r(x,T)=\int_0^T r_r(x,t)\,dt\,, \tag{36}$$

$$f_{\det}(T)=\frac{1}{2w}\int_0^{2w} 1_{\{M_r(x,T)\geq M_{\det}\}}\,dx\,. \tag{37}$$

Following a step from thermoneutral darkness, the normalized mean amount $\overline{M}_r/(L n_i)$ initially scales as $s g_f \theta^2/2$, because the carrier population and reaction rate must first develop. Extrapolating the final stationary rate back to the instant of illumination instead gives a linear increase and overestimates early conversion (Figure 14d). The spatial amount profile also evolves during the charging interval; it is not obtained by multiplying a final rate map by exposure time. Figure 14e includes the complete quadratic spatial correction. The finite-liquid temporal examples retain the same fast-diffuse interpretation and regime markings as the voltage calculations.

At a specified amount threshold, modest rate contrasts produce a continuous change in detected area as exposure increases (Figure 14f). The uniform semiconductor reference crosses the threshold everywhere simultaneously, giving a step in detected fraction. The displayed patterned curves depend on the threshold, exposure, transport, and perturbation order; they are not a unique intrinsic reaction width. The 90% extent of a domain-specific excess, the 90% extent of the total rate, and the detected footprint answer different experimental questions. Their separation allows surface-potential maps, modulated photovoltage, partial-current measurements, and accumulated chemical patterns to test the same declared electrical and chemical boundaries without identifying any one of these observables with all the others.

## 7. Doping Effects on Static and Dynamic Photovoltage

This section examines how donor and acceptor populations change the five-case comparison. We first consider stationary potentials and then the dynamic photovoltage of the isolated film and the bare contacted film with fixed polarization.

The analytical cases above use intrinsic, electron–hole-symmetric carrier statistics because that choice gives closed-form solutions and makes polarization the only source of asymmetry. Oxide ferroelectrics are instead donor- or acceptor-dominated. $BaTiO_3$ conducts as an *n*-type semiconductor under reducing conditions through oxygen-vacancy-related donors and as a *p*-type semiconductor under oxidizing or acceptor-doped conditions [89, 90, 91]. We therefore separate the intrinsic concentration $n_i$ from a reference density $N_{ref}$ and a signed net dopant density. With $d=(N_D^+-N_A^-)/N_{ref}$ and $r_i=n_i/N_{ref}$, neutrality and mass action fix the dark bulk populations

$$n_0=\frac{1}{2}[d+\sqrt{d^2+4r_i^2}], \tag{38}$$

$$p_0=\frac{r_i^2}{n_0}, \tag{39}$$

where the majority hole density is evaluated first for $d<0$ to avoid cancellation, and the minority electron density follows from the mass-action relation. Uniform generation $g=G/(k_b N_{ref}^2)$ adds equal electron and hole excesses $\Delta$, giving the illuminated quasi-neutral populations

$$\Delta=\frac{1}{2}[-(n_0+p_0)+\sqrt{(n_0+p_0)^2+4g}], \tag{40}$$

$$n_b = n_0 + \Delta, \quad (41)$$

$$p_b = p_0 + \Delta. \quad (42)$$

The examples use $r_i = 10^{-6}$, so that the intrinsic material is nearly carrier-free in darkness and its screening is supplied by photocarriers, and $d = \pm 10$ for the doped states. Lengths are scaled by $L_D = (\epsilon V_T / q N_{ref})^{1/2}$, charge per area by $(\epsilon k_B T N_{ref})^{1/2}$, and time by $(k_b N_{ref})^{-1}$. These are mechanism tests of carrier statistics, not a calibration of a particular oxide.

### 7.1 Static photovoltage

Doping enters the five cases in two distinct ways (Table 1, Figure 15a). In the blocking geometries A and C the solid screens the fixed residual polarization through its total mobile density, so the voltage depends only on $S_b = n_b + p_b$. Equal donor and acceptor densities then give identical magnitudes, and doping reduces the front–rear voltage of A from $0.861 V_T$ to $0.577 V_T$ at $g = 1$; C follows the same reduction after the passive capacitive division. Once a carrier reservoir is present, the screening is nonlinear and depends on which carrier accumulates. For a nondegenerate semiconductor with surface potential $u = \phi_s / V_T$, the space-charge integral and the dry (B) or compact-layer (D) boundary give

$$Q_s(u) = -sgn(u)\{2[n_b(e^u - u - 1) + p_b(e^{-u} + u - 1)]\}^{1/2}, \quad (43)$$

$$\sigma + Q_s(u) - c_H u = 0, \quad (44)$$

with $c_H = 0$ for B. In an n-type film, positive exposed polarization is screened by electron accumulation and negative polarization by electron depletion. An *n*-type film therefore screens $+P$ more effectively than $-P$: the two orientations give $+0.568 V_T$ and $-0.697 V_T$ in B and $+0.450 V_T$ and $-0.506 V_T$ in D, compared with $\pm 1.317 V_T$ and $\pm 0.815 V_T$ in the intrinsic film. A *p*-type film reverses this asymmetry exactly. In the two-domain case E with equal transfer coefficients, paired turnover requires equal surface electron and hole densities, which shifts the common mean potential by

$$m = \frac{1}{2} \ln \frac{p_b}{n_b} \quad (45)$$

while the polarization contrast between the domains is set by interfacial and semiconductor screening. Doping thus translates both domain potentials relative to the solution, by $-2.31 V_T$ for the *n*-type film, while the half-contrast falls only from $0.828 V_T$ to $0.477 V_T$. Above $|d| \approx 1.7$ the carrier-type shift exceeds the polarization contrast, and both domains lie on the same side of the solution reference (Figure 15a). Heavy doping drives the A–D potentials toward zero, whereas the E shift continues to grow logarithmically.

***Table 1.*** *Static surface potentials, in units of $V_T$, at $g = 1$ for $d = -10$, 0 and $+10$ with $r_i = 10^{-6}$. A and C are blocking screening limits, B the dry contacted space-charge limit, D adds compact-layer screening ($c_H = 1$), and E is the wide-domain paired-reaction limit. The residual polarization source is $\sigma = 2$ in B, D and E and $P_{res} L / \epsilon V_T = 1$ in A and C.*

| **Case** | ***p*-type +P** | ***p*-type −P** | **intrinsic +P** | **intrinsic −P** | ***n*-type +P** | ***n*-type −P** |
|---|---|---|---|---|---|---|
| **A** | 0.577 | −0.577 | 0.861 | −0.861 | 0.577 | −0.577 |
| **B** | 0.697 | −0.568 | 1.317 | −1.317 | 0.568 | −0.697 |
| **C** | 0.289 | −0.289 | 0.431 | −0.431 | 0.289 | −0.289 |
| **D** | 0.506 | −0.450 | 0.815 | −0.815 | 0.450 | −0.506 |
| **E** | 2.789 | 1.836 | 0.828 | −0.828 | −1.836 | −2.789 |

**7.2 Intensity dependence and dynamic photovoltage**

The illumination dependence separates the two carrier regimes more sharply than the potential at a single intensity. In the contacted film B, the intrinsic photovoltage changes logarithmically over the whole range, and its static sensitivity scales approximately as $g^{-1}$ (Figure 15b,c). A doped film instead has a low-injection plateau. Its photovoltage is nearly constant, with a sensitivity of $4.2\times10^{-3}$ and $9.6\times10^{-3}$ per unit $g$ for the two orientations at $g=1$, about 30–70 times below the intrinsic value of 0.29. It rejoins the intrinsic curve only when the photogenerated density approaches the majority density, $\Delta\approx g/|d|\sim|d|$, that is, near $g\approx d^2$. The depleted orientation, $-P$ in an $n$-type film, has the larger response because the photovoltage there is carried by minority carriers in the depletion field. The doping level and carrier type therefore appear directly as an injection threshold and as an orientation-dependent sensitivity.

The dynamic response follows from the same populations. Linearizing the pair balance about the illuminated state gives a population relaxation rate $k_b(n_b+p_b)$. In a doped film this is the inverse minority-carrier lifetime, which is controlled by the majority density. For the isolated film A with equal diffusivities, Equation (12) is expressed in the doping units defined above. The screening coefficient is $\bar{\kappa}^2=S_b$ in units of $L_D^{-2}$, and the total-population response is $\dfrac{\delta S_b}{\delta g}=\dfrac{2}{s+S_b}$, with $s=i\Omega$. These definitions replace the screening coefficient and population response in Equation (12), giving the factor 16 in the numerator below; $\epsilon_s$ denotes the solid permittivity in both expressions:

$$H_A(s)=\frac{-16\,P\,D_s}{\epsilon_s L(s+S_b)}\sum_j\frac{1}{(k_j^2+\bar{\kappa}^2)[s+D_s(k_j^2+\bar{\kappa}^2)]}. \tag{46}$$

Doping therefore lowers the static sensitivity and raises both the population and screening relaxation rates. At $g=1$, $|H_A(0)|$ falls from $5.8\times10^{-2}$ to $4.1\times10^{-3}$ and the corner frequency rises from $\Omega\approx1.5$ to 4.3 (Figure 15f). Every term retains the same static sign, so $|H_A(i\Omega)|\le|H_A(0)|$ continues to hold, and donor and acceptor doping remain indistinguishable in this blocking geometry.

The contacted film breaks that symmetry dynamically as well. We solved the nonlinear Poisson–drift–diffusion equations for B (Supplementary Section S3.4, Table S11) with the fixed residual source $\sigma=\pm2$, uniform generation, bimolecular recombination, equal normalized mobilities $M=1$, a blocking dry front, and a rear contact $20\,L_D$ deep that equilibrates both carriers at their dark doped values. The equations were then linearized about each illuminated state. The stationary potentials reproduce the space-charge limit of Equations (43)–(44) to $10^{-4}\,V_T$, and the zero-frequency transfer functions agree with finite-difference slopes. In the intrinsic film the two orientations have the same gain and a corner near $\Omega\approx1.4$. In the $n$-type film the accumulated $+P$ orientation responds with a gain of $4.2\times10^{-3}$ and a corner near $\Omega\approx7.8$, whereas the depleted $-P$ orientation has a 2.3-fold larger gain and a lower corner near $\Omega\approx6.1$ (Figure 15d,e). Oppositely polarized domains of one doped film therefore produce different photovoltage spectra, not only spectra of opposite sign. The domain with the larger and slower response identifies the depleted orientation and hence the carrier type. Combined with the injection threshold, this gives an intensity-modulated surface-photovoltage measure of carrier type and density that does not require an electrical contact to the domain. The full Landau, reactive and lateral dynamics with doping remain to be evaluated; the static results above indicate that doping will shift the corresponding transfer functions through the same majority-carrier screening and lifetime.

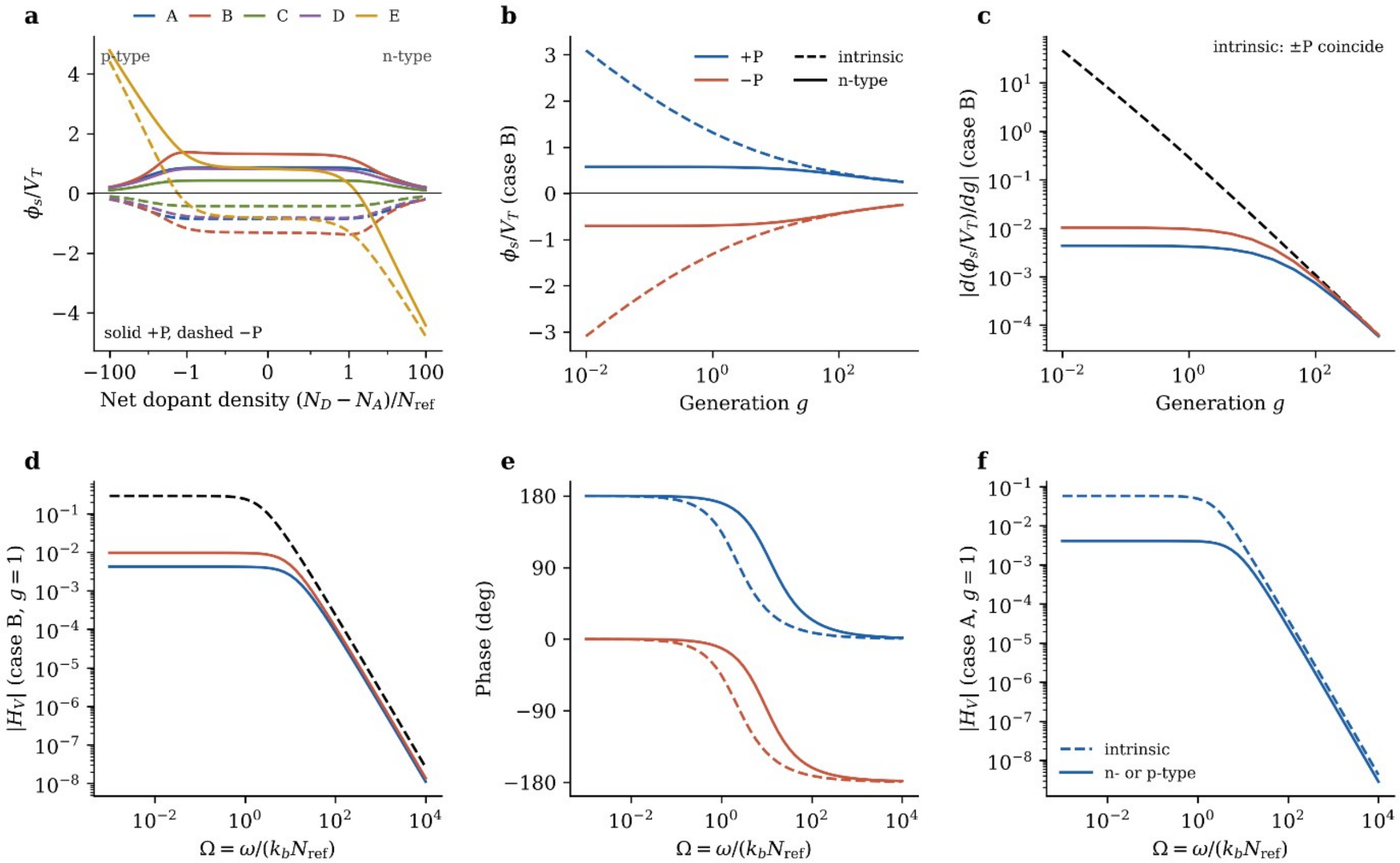

***Figure 15.*** *Doping effects on static and dynamic photovoltage. (a) Static potentials of the five cases versus signed net dopant density at $g=1$; solid curves are $+P$ and dashed curves $-P$. (b) Stationary surface potential of the contacted dry film B versus generation for intrinsic (dashed) and n-type (solid) material; blue denotes $+P$ and red $-P$. (c) Static optical sensitivity $|d(\phi_s/V_T)/dg|$ of B; the intrinsic orientations coincide. (d,e) Gain and phase of the B photovoltage at $g=1$, referenced to positive light modulation, from the linearized drift–diffusion model. (f) Gain of the isolated film A at $g=1$ from Equation (46); donor and acceptor doping give identical curves. $r_i=10^{-6}$, $|d|=10$; $p$-type results follow from the n-type ones by reversing $P$. Frequencies are normalized by $k_b N_{ref}$. Line style identifies polarization in (a), but carrier type in (b).*

## 8. Discussion

The five cases connect a common carrier-generation process to observables with different physical meanings. In A, light changes an internal dipole distribution under a fixed total-charge constraint. B permits exchange with a dry reservoir and adds a separately identifiable contribution when finite surface states participate. C preserves electronic isolation but couples the solid to ionic screening and passive electrostatic boundaries. D sustains a front reaction through a specified carrier return. E replaces that rear exchange by coupled regions that consume complementary carriers and communicate through the solid and liquid. The surface potential, a terminal current, a local partial reaction rate, and a cumulative reaction amount consequently probe different parts of the same charge and species balances. Doping preserves the donor–acceptor symmetry of the isolated screening limit, introduces accumulation/depletion asymmetry in the contacted film, and shifts the common potential of the two domains in the reduced paired-reaction model (Section 7).

This distinction explains why the optical spectrum is useful in addition to the stationary light dependence. A stationary slope identifies the response after all retained populations have equilibrated. A periodic perturbation changes their relative participation according to frequency, while a finite light step reveals how one population evolves before another. The monotonic isolated response, the competing polarization and screening contributions in the dry contacted film, and the ionic overshoot in the blocking junction cannot be assigned one universal relaxation time. A potential maximum or phase reversal alone is likewise insufficient to identify a chemical intermediate. Interpreting the spectrum together with site charge, ionic charge, polarization, and partial currents constrains the mechanism much more directly.

The domain calculations extend this reasoning spatially. The domain-potential contrast is set by electrostatics and charge relaxation, whereas reaction localization also depends on carrier consumption and lateral transport. A transport length inferred from a reaction profile need not equal a screening length or the width of a potential feature. Accumulated conversion integrates a rate over exposure, and an experimentally thresholded image adds a detection criterion. The local second-order neutral population retained here further distinguishes a small correction to the average conversion from a correction that is appreciable relative to a weak spatial contrast. These effects explain why the total rate, preferential allocation, and apparent footprint should be reported separately.

The derivations specify which parts of this comparison are exact within a reduction and which require a restricted physical regime. The A/C optical solutions use weak potentials with a symmetric neutral population; restoring nonlinear species equations at the illustrated parameters gives small corrections. B/D use finite-depth carrier transport and an explicitly chosen polarization and recombination model; their nonpolar reference removes the polarization mode while holding the background permittivity fixed. The E temporal calculations retain the fast-diffuse-layer, conducting-liquid limit and distinguish a necessary time-separation check from full validation. The thick-crystal nonlinear examples resolve transport beneath prescribed reaction regions and provide a complementary benchmark, rather than a matched nonlinear continuation of every weak-pattern calculation. Numerical conservation, refinement, stationary derivatives, and independently evaluated limiting solutions test these specified problems; they do not by themselves calibrate an oxide or establish stability to variables excluded from a reduction.

Application to a particular material requires the same approach at the constitutive level. Poling, composition changes, heating, and electrode preparation can alter defect populations, dielectric susceptibility, adsorption, and transfer laws together with polarization. The carrier density, absorption profile, mobility, loss kinetics, residual compensation, contact selectivity, and reaction activities must therefore be assigned consistently. An additional bulk photovoltaic current, wall-specific transport, or chemically evolving surface can be included through its own constitutive equation and charge balance. The present results establish the consequences of the explicitly retained mechanisms and identify the measurements needed to separate these additional contributions.

## 9. Conclusions

We analyzed light-induced charging and redox chemistry on ferroelectric surfaces within one semiconductor transport and charge-conservation framework, applied to five electrical topologies: an isolated dry film, a dry film connected to a carrier reservoir, with optional finite surface states, a blocking ferroelectric–electrolyte junction, a reactive interface with a rear carrier return, and oppositely polarized domains coupled through the solid and the electrolyte. In each case the same carrier populations determine the stationary photovoltage, its small-signal optical spectrum and its response to finite light steps. In the reactive cases they also determine the partial reduction and oxidation rates and the accumulated product, with polarization reversal and nonpolar and doped references as controls. The comparison shows that the charge-return path matters as much as the polarization itself. An isolated or blocking body can only redistribute a conserved charge, a carrier reservoir or floating metal permits charge exchange, and sustained conversion requires a complementary reaction or an external return. Surface potential, terminal current, partial rate and deposited amount therefore measure different parts of the same charge balance and cannot be substituted for one another.

Within this framework the range of photodeposition morphologies on domain patterns follows without invoking wall-specific conductance or catalytic activity. On a two-domain surface both reactions remain available everywhere, and the polarization-dependent enhancement of reduction on the outward domain is confined to within a lateral transfer length $l_c \propto \left(D_x L/S\right)^{1/2}$ of each wall, set by the competition between lateral carrier redistribution and interfacial transfer. When lateral transport is limited, so that $l_c$ is smaller than the domain width, the reduction rate peaks next to the walls (Figure 11b). We therefore expect the formation of reaction products at the boundaries. Once the accumulated amount crosses a detection threshold, nanoparticles can decorate the domain walls, and deposition extends toward the domain interiors with longer exposure (Figures 13 and 14e,f). Faster lateral transport fills the domain interiors instead, whereas a resistive electrolyte or high injection suppresses the contrast and yields deposition on both polarization orientations. Donor and acceptor populations shift both domain potentials relative to the solution while retaining a polarization-dependent contrast; sufficiently strong doping places both potentials on the same side of the solution reference in the illustrated reduced model (Figure 15a). Interior, wall-decorating and orientation-independent deposition therefore emerge as transport and charge-closure regimes of a single model.

The calculations also predict new classes of photovoltage behavior in the time and frequency domains that a stationary band diagram cannot capture. A contacted film can show a stationary photovoltage maximum at which the static optical derivative vanishes while the modulated response remains finite, because polarization recovery and mobile screening cancel only at zero frequency. Large transient excursions can separate nearly identical stationary endpoints without any trap population. A blocking ferroelectric–electrolyte junction can respond above its static slope at finite frequency and overshoot after a light step through ionic lag alone. Floating paired reactions show transient partial-rate imbalance, and illumination can reverse the direction of conversion between distinct redox couples at polarity-dependent thresholds. Doping adds an injection threshold, a corner frequency controlled by the majority density and, in contacted films, different photovoltage amplitudes and corner frequencies on oppositely polarized domains of the same sample. These signatures make intensity-modulated surface-photovoltage microscopy, photocurrent spectroscopy and light-step measurements quantitative probes. They can separate electronic storage, ionic screening, polarization relaxation and Faradaic turnover, and they can identify the carrier type and the lateral transport length. They also yield design rules for programming light-driven chemistry with written polarization patterns: the domain width relative to $l_c$, the electrolyte conductivity, the doping, and the charge-return architecture. These rules can be applied directly in automated scanning-probe experiments that write, illuminate and read the same surface.

## Author and AI Contributions

Sergei V. Kalinin conceived the study, defined the physical cases and boundary-condition comparisons, directed the analysis, provided scientific interpretation, and guided the manuscript organization and revision. ChatGPT (OpenAI) and Claude (Anthropic) assisted with literature searches, analytical derivations, preparation of numerical-analysis and plotting code, and manuscript drafting and revision under the author's direction. The reported calculations and figures are based on the stated equations and archived numerical data. Responsibility for the scientific content, source attribution, and approval for submission rests with the author.

## Acknowledgments

This work (SVK) was supported by the DOE BES project DE-SC0026253, Deciphering Electrochemical Transformation Pathways on the Nanometer Scale: Advancing Fundamental Discovery for Material Innovation.

## Data and Code Availability

The analytical derivations, parameter values, numerical methods, and extended results are provided in the Supplementary Material. The accompanying analysis and figure archive contains the calculation scripts, saved arrays, figure sources, and bibliographic evidence used in this study, together with instructions for reproducing the reported calculations.

**Supplementary Material**

Photoinduced charging and spatially selective photochemistry on ferroelectric surfaces

**Sergei V. Kalinin***

*Department of Materials Science and Engineering, University of Tennessee, Knoxville, Tennessee 37920, United States*

*Correspondence: sergei2@utk.edu

ORCID: 0000-0001-5354-6152

This supplement develops the five boundary cases used in the manuscript. The formulation, carrier statistics, optical observables, and nonpolar controls are defined once; each case then gives its boundary-specific derivation and supporting results. Numerical methods and verification are collected in S7, and parameter sets and literature context in S8. Figures, equations, tables, and references have independent supplementary numbering.

Contents: S1, common formulation and semiconductor statistics; S2, isolated film; S3, contacted dry film; S4, blocking electrolyte; S5, reactive contacted film; S6, domain patterns; S7, numerical methods and verification; S8, parameters, comparative doping results, and model context.

## S1. Common formulation, carrier statistics, and observables

The five cases differ in the inventories and paths available to charge, not in the existence of photogeneration. A describes a dry isolated film; B a dry film coupled to a carrier reservoir; C an electronically blocking solid–electrolyte junction; D a reactive interface with a specified rear return; and E a fixed pair of opposite domains. The illustrative constitutive sets are complementary controlled examples, not a single calibrated material with only its boundary changed.

### S1.1 Coordinates, electrostatics, and carrier conservation

Positive polarization points toward the exposed surface. The outward coordinate is $-L<z<0$, with the front at $z=0$; the isolated-film coordinate is $x=z+L$. Finite-depth contacted calculations use inward depth $y=-z$ and $P_{¿}=-P_z$. Captions identify retained inward-positive ordinates explicitly; a change of coordinate is not a reversal of the prepared physical state.

The material is described by the electrostatic potential $\phi$, polarization $P$, electron number density $n$, and hole number density $p$. Here $q>0$ is the elementary charge; dopant and ionic concentrations are also number densities unless a molar unit is explicitly stated. In the models that resolve polarization relaxation, the constitutive equations are

$$\Gamma_P \partial_t P = -aP - bP^3 - cP^5 + g_P \nabla^2 P + E_z, \tag{S1}$$

$$D = \epsilon_b E + P\hat{z}, \tag{S2}$$

$$E = -\nabla \phi, \tag{S3}$$

$$\nabla \cdot D = q\left(p - n + N_D^+ - N_A^-\right). \tag{S4}$$

The background permittivity $\epsilon_b$ excludes the polarization susceptibility already represented by the first equation. Natural polarization boundaries, $\partial_n P = 0$, are used in the finite-depth examples. The prepared positive and negative states need not retain identical magnitudes after illumination in an asymmetric device. In the analytical fixed-domain reductions, polarization is prescribed and the dielectric response is represented once by a constant differential permittivity. A fixed residual polarization source can include pre-existing, immobile compensation. Removing this source in a passive reference also removes its paired compensation sheets; otherwise the proposed reference retains a polar surface charge.

For thermalized, nondegenerate carriers, the particle fluxes and continuity equations are

$$\Gamma_n = -D_n \nabla n + \mu_n n \nabla \phi, \tag{S5}$$

$$\Gamma_p = -D_p \nabla p - \mu_p p \nabla \phi, \tag{S6}$$

$$\partial_t n = -\nabla \cdot \Gamma_n + G - R, \tag{S7}$$

$$\partial_t p = -\nabla \cdot \Gamma_p + G - R. \tag{S8}$$

Here $D_i = \mu_i V_T$ and $V_T = k_B T / q$. The conventional electronic currents are $J_n = -q \Gamma_n$ and $J_p = q \Gamma_p$. A uniform pair source is used where it permits an exact reduction. The finite-depth contacted examples instead use $G(y) = G_0 \exp(-\alpha_{\text{abs}} y)$. For monochromatic incident intensity, $G_0 = \eta_{\text{pair}} (1 - R_{\text{opt}}) \alpha_{\text{abs}} I_{\text{inc}} / (h\nu)$, with the mobile-pair yield and optical reflectance stated separately. Thus a calculated derivative with respect to $G_0$ or its dimensionless counterpart can be converted to a derivative with respect to incident intensity once the absorption and pair yield are known.

Generation creates electrons and holes equally. With the ionized bulk dopant charge held fixed, define $\rho_f = q(p - n + N_D^+ - N_A^-)$. Subtracting the carrier balances and using Gauss's law gives

$$\partial_t \rho_f + \nabla \cdot (J_n + J_p) = 0, \tag{S9}$$

$$\nabla \cdot (J_n + J_p + \partial_t D) = 0. \tag{S10}$$

The displacement current includes the time-dependent polarization. It does not represent a second source of electronic carriers. In one dimension the total current is independent of depth, even while the conduction current varies during charging. This identity supplies a direct numerical check and prevents a surface reaction from being sustained by an unaccounted source of charge.

The optical source creates thermalized pairs. An intrinsic bulk photovoltaic current would require a separate constitutive term; it is not generated by setting the polarization to a nonzero value in the pair-source law [S1–S3]. Fixed-source models A, C, and E represent the stable dielectric response once, whereas B and D resolve a Landau polarization mode explicitly.

**S1.2 Semiconductor statistics and the meaning of doping**

The analytical limits of Sections S2–S6 use intrinsic, electron–hole-symmetric carrier statistics because that choice exposes several closed-form solutions. Real oxide ferroelectrics can instead be donor- or acceptor-dominated. In $BaTiO_3$ the carrier type depends on defect chemistry and oxygen chemical potential; oxygen-vacancy-related defects favor n-type conduction under reducing conditions, whereas sufficiently oxidizing or acceptor-doped conditions can produce p-type transport [S4–S6]. The stationary extension below isolates the consequence of this carrier asymmetry without identifying a particular chemical dopant with a single free-carrier density.

Use an independent normalization density $N_{ref}$, signed ionized dopant density $d = (N_D^+ - N_A^-) / N_{ref}$, and true intrinsic-density ratio $r_i = n_i / N_{ref}$. Densities in the following equations are divided by $N_{ref}$. The neutral dark reference and illuminated quasi-neutral populations obey

$$n_0 - p_0 = d, \tag{S11}$$

$$n_0 p_0 = r_i^2, \tag{S12}$$

$$n_b = n_0 + \Delta, \tag{S13}$$

$$p_b = p_0 + \Delta. \tag{S14}$$

$$\Delta = \frac{-(n_0 + p_0) + \sqrt{(n_0 + p_0)^2 + 4g}}{2}, \tag{S15}$$

$$g = \frac{G}{k_b N_{ref}^2}. \tag{S16}$$

These uniform bimolecular pair-balance relations distinguish the true intrinsic concentration from both the chosen unit and the dopant-controlled majority population. They are the reference state for the reduced static comparisons, not a replacement for the spatial carrier equations in a contacted reactive device.

For numerical evaluation, use the cancellation-free dark root: for $d \geq 0$, calculate $n_0 = \left(d + \sqrt{d^2 + 4r_i^2}\right)/2$ and $p_0 = r_i^2/n_0$; for $d < 0$, calculate $p_0 = \left(-d + \sqrt{d^2 + 4r_i^2}\right)/2$ and $n_0 = r_i^2/p_0$. The static comparisons use this stable form. The dynamic doping examples are limited to A and a fixed-polarization, bare-surface variant of B; C–E are extended only in their stated static reductions.

The recombination law is part of each reduction. Bimolecular recombination, $R = k_b\left(np - n_i^2\right)$, permits the exact isolated stationary construction and the weak-polarization dynamical solution. The finite-depth reactive calculation uses $R = \left(np - n_i^2\right)/\left[\tau_0\left(n + p + 2n_i\right)\right]$, an explicitly stated intrinsic Shockley–Read–Hall form. [S7] These laws have different intensity dependences and different definitions of their time parameters. For equal small pair perturbations in the dark, the first has lifetime $\left(2k_b n_i\right)^{-1}$ and the second has lifetime $2\tau_0$. A high-injection lifetime used in a thick-crystal transport estimate is a third quantity. Comparing numerical values of these symbols without their defining rate laws would change the physical problem.

### S1.3 Surface storage, reaction affinity, and charge closure

Dynamic surface storage means an explicitly counted population of charge-bearing states whose occupation changes with time. For electron and hole capture sites, one convenient description is $\sigma_s = q\left(N_s^+\theta_+ - N_s^-\theta_-\right)$. The occupancies obey separate capture/emission balances. An electron captured from the solid reduces its free-electron inventory and makes the corresponding surface inventory more negative; a captured hole has the opposite charge effect. The normal displacement jump contains $\sigma_s$, while the normal carrier flux contains the same capture rate. This makes the electronic and site balances consistent. An unspecified additional surface charge or an independently added capacitance would not supply the missing kinetic law. Chemical adsorption states can also store charge, but require the activities and reactions responsible for changing their occupation. [S8–S11]

At a reactive front interface, let $r_r$ and $r_o$ be signed electron-equivalent rates per unit area, positive for forward reduction and forward oxidation, respectively. Negative values describe their specified reverse reactions. Without capture into separate sites, the outward particle fluxes are $\Gamma_n \cdot \hat{n} = r_r$ and $\Gamma_p \cdot \hat{n} = r_o$. Their conventional outward Faradaic current is $j_F = q\left(r_o - r_r\right)$. If $Q$ includes the solid mobile charge and all explicitly modeled surface-state charge, its integrated balance is

$$\frac{dQ}{dt} = I_{\text{rear,in}} + q\int_{A_f}\left(r_r - r_o\right)\,dA\,. \tag{S17}$$

Here $I_{\text{rear,in}}$ is conventional charge current entering through the rear contact. In a rear-blocking object, unequal total reduction and oxidation charge the object. Equal integrated electron-equivalent rates are consequently necessary for sustained stationary paired conversion, although the two reactions can occur in different places. A single channel can operate transiently using a finite initial charge inventory; that inventory and its potential change must be counted. A grounded contact can support a single front reaction only as part of a circuit with a complementary charge-consuming process. Specifying an electrode potential supplies an electrical boundary for the solid, while the external counterreaction completes the physical device.

The chemical reference also matters. For a reversible reduction couple, the reference carrier density at the reaction plane can be written $n_r(U)=n_r^0\exp(U/V_T)$; for oxidation, $p_o(U)=p_o^0\exp(-U/V_T)$, where $U$ is the solid potential relative to that plane. The products satisfy $n_r p_o=n_i^2\exp(\Delta\mu/k_B T)$ with $\Delta\mu=\mu_r-\mu_o$. The electron-scale quasi-Fermi energies and reaction affinities are

$$E_{Fn}=E_C^0-q\phi+k_B T\ln(n/N_C), \tag{S18}$$

$$E_{Fp}=E_V^0-q\phi-k_B T\ln(p/N_V), \tag{S19}$$

$$A_r=E_{Fn}-\mu_r=k_B T\ln(n/n_r), \tag{S20}$$

$$A_o=\mu_o-E_{Fp}=k_B T\ln(p/p_o), \tag{S21}$$

$$A_r+A_o=E_{Fn}-E_{Fp}-\Delta\mu. \tag{S22}$$

An activated reversible transfer law consistent with these references is

$$r_r=S_{n0}e^{-\alpha_n U/V_T}\left[n_s-n_r^0 e^{U/V_T}\right], \tag{S23}$$

$$r_o=S_{p0}e^{\alpha_p U/V_T}\left[p_s-p_o^0 e^{-U/V_T}\right]. \tag{S24}$$

Here $U=\phi_s-\phi_l^{\mathrm{rp}}$ is the solid potential relative to the liquid reaction plane, $n_s$ and $p_s$ are the solid carrier densities at that boundary, $S_{n0}$ and $S_{p0}$ are positive transfer velocities, and $0\le\alpha_n,\alpha_p\le 1$. With number densities as inputs, $r_r$ and $r_o$ have units of electron equivalents per area per time. The constant-transfer analytical limit sets $\alpha_n=\alpha_p=0$; the normalized reactive-film examples retain their stated activation factors.

Rate laws proportional to $n-n_r$ and $p-p_o$, multiplied by positive transfer coefficients, have signs consistent with these affinities. A forward and reverse branch must share the same reference if detailed balance is to be retained. For a common dissolved redox pair, equal positive reduction and oxidation rates recycle the same species and produce no net amount of its reduced form. Distinct reduction and oxidation couples can instead undergo chemical conversion at equal electron-equivalent rates. Both situations close the electrical balance, but their product balances differ. Generic carrier-transfer laws establish this distinction; they do not by themselves determine all intermediates or product selectivities of water oxidation, carbon conversion, or another multistep reaction. [S12–S14]

The liquid is described, where it is resolved dynamically, by ionic concentrations $c_i$, charge numbers $z_i$, and potential $\phi_l$:

$$\Gamma_i^l=-D_i^l\left(\nabla c_i+\frac{z_i c_i}{V_T}\nabla\phi_l\right), \tag{S25}$$

$$\partial_t c_i=-\nabla\cdot\Gamma_i^l, \tag{S26}$$

$$-\nabla\cdot(\epsilon_l\nabla\phi_l)=q\sum_i z_i c_i. \tag{S27}$$

Case C resolves the symmetric electrolyte charge dynamics. The analytical model in E instead eliminates diffuse-layer relaxation and retains the Ohmic outer-liquid resistance; its timescale restriction is stated before the temporal results in S6. D and the nonlinear thick-crystal benchmark use maintained, ideally conducting liquid references. Conductivity, reactant delivery, and a compact voltage drop are distinct inputs [S15–S19].

The carrier-exchange boundaries, charge-return paths, and measured voltage for each case are summarized in Table S1.

*Table S1. Cases, exchange conditions, and voltage references.*

| Case | Exposed boundary | Rear and charge closure | Primary voltage |
|---|---|---|---|
| A | Dry; electrons and holes blocked | Both faces carrier-blocking; fixed net solid charge; no exterior normal field | Front–rear dipole voltage; midpoint potential is a gauge |
| B | Dry; no chemical transfer; finite sites optional | Grounded rear exchanges both electrons and holes | Front potential relative to rear metal |
| C | Solid–electrolyte transfer blocked; compact and diffuse charge resolved | Rear carrier-blocking; passive dielectric capacitance to a reference plane | Solid-front potential relative to remote liquid, and separate solid/compact/diffuse drops |
| D | Declared front reduction or oxidation kinetics | Selective rear carrier return and a complete complementary circuit; floating comparisons explicitly identified | Front/reaction-plane and front/rear potentials at held bias or floating terminal condition |
| E | Two domains: blocking first, paired reactions second | Rear carrier-blocking; lateral solid and liquid transport; global paired balance | Domain contrast and spatial solid/liquid potentials relative to the specified mean reference |

### S1.4 Optical measurements, controls, and constitutive scope

Dry experiments need not place the carrier reservoir directly behind the illuminated region. An in-plane conducting path to a remote equilibrating contact can play the same reservoir role. Its reduction to an ideal rear contact requires that the lateral path maintain approximately the same electrochemical boundary over the region and frequencies examined. For a strip of length l, sheet conductance $\Sigma$, and areal differential charge storage $C_a$, the linear contact-driven relaxation obeys $C_a \partial_t v = \Sigma \partial_x^2 v$ away from sources. A reservoir at one end and zero lateral current at the other give the longest charging time $\tau_{\parallel} = 4 C_a l^2 / (\pi^2 \Sigma)$. More generally, the sheet conductance and storage are replaced by the relevant coupled carrier and site response. The ideal-contact limit requires both a small voltage loss at the operating current and $\omega \tau_{\parallel} \lll 1$ for the measured modulation. An uncontacted in-plane path redistributes charge inside the specimen and retains the isolated charge inventory of A. This distinction is consistent with the role of spreading and environmental transport in local electrochemical measurements. [S20, S21]

Every optical response is evaluated around a specified mean illumination $g_0$. We write $g(t) = g_0 + \mathrm{Re}[\hat{g} \exp(i\Omega t)]$ in normalized time and define $H_V(\Omega; g_0) = \hat{V}/\hat{g}$. A physical sinusoidal intensity modulation is taken with $g_0 > |\hat{g}|$ and a sufficiently small amplitude for linear response. A derivative at $g_0 = 0$ is the formal limiting susceptibility or the one-sided static derivative from nonnegative illumination; a finite sinusoid centered at zero would not be a physical nonnegative light source. The zero-frequency limit of $H_V$ equals $dV_{ss}/dg_0$ on the same stable branch with the same held electrical and chemical conditions. The reported phase is either the signed phase of $H_V$ or, where stated, the phase of $H_V/H_V(0)$. The latter removes the static sign reversal for ease of comparison; it must not be used when the static slope vanishes. A zero response has no defined phase. Spatial Fourier wave number and temporal frequency remain independent variables in E.

For a discretized or modal state $u$ with mass matrix $M$ and evolution $M\dot{u}=f(u,g)$, linearization at a stationary state gives $(i\Omega M-J)\hat{u}=f_g\hat{g}$. We use this relation to evaluate complete amplitude/phase spectra and illumination–frequency maps, and check selected responses against nonlinear time integration. It is a calculation method, rather than a substitute for the evaluated spectra. Large light steps retain the change of mean carrier density and any nonlinear surface or polarization kinetics. Their endpoints, charge balance, and the populations responsible for each transient feature are examined together. Analytical intensity-modulation theory provides an established precedent for this organization. [S22–S24]

For A, C, and the fixed-domain reduction E, removing the residual source also removes its fixed paired compensation and leaves the assigned dielectric response unchanged. The B/D nonpolar reference instead removes the explicit Landau field and its dynamics while retaining the background permittivity and the stated optical, carrier, and interface laws. It therefore does not match the full differential dielectric response of the polar branch.

For a locally uniform stationary polarization $\overline{P}$, linearization of $\Gamma_P\partial_t P=-aP-bP^3-cP^5+g_P\nabla^2 P+E$ gives

$$\chi_P(k,\omega)=\frac{\delta P}{\delta E}=\frac{1}{a+3b\overline{P}^2+5c\overline{P}^4+g_P k^2+i\omega\Gamma_P}. \tag{S28}$$

The corresponding differential permittivity is $\epsilon_b+\chi_P$ in this local uniform limit. Both terms have dimensions of absolute permittivity. In a nonuniform film, the response is an operator with the actual polarization boundary conditions, rather than a single scalar. Removing $P$ leaves $\epsilon_b$ alone. A comparison intended to isolate spontaneous polarization from polar softness would instead have to specify and retain a matched differential response at the operating state, wave vector, and frequency of interest. Such a comparison is not identical to the nonpolar reference displayed here. This limitation does not alter the exact zero-voltage symmetry of case B, but it constrains a quantitative attribution of differences in case D solely to the static spontaneous field.

The nonlinear stationary family and weak-potential temporal family of A use different declared source amplitudes. Their results are not one strong-field dynamic continuation. Likewise, the contact, thin-film, and thick-crystal examples retain their own normalization and constitutive laws. Comparisons between cases concern the stated physical mechanisms unless a common parameter set is explicitly identified.

## S2. Case A: dry, electrically isolated film

### S2.1 Stationary closure and nonlinear screening

The dry isolated film establishes what illumination can change before either a contact or a surface reaction is introduced. Both faces of the intrinsic film, $-L<z<0$, block electrons and holes. The surrounding dry dielectric carries zero normal field far from the planar film. Its total mobile charge is zero, and the fixed residual source $P_{\mathrm{res}}$ has the sign of the prepared polarization. Fixed equal and opposite compensation sheets, if present, are included in this residual source. The electrostatic boundary is consequently $\epsilon\,\partial_z\phi=P_{\mathrm{res}}$ at both faces. Here $\epsilon$ is a constant differential permittivity on the chosen stable polarization branch; the residual source is held fixed. These assumptions isolate carrier screening from changes in the polarization itself. Setting $\phi(-L/2)=0$ fixes the potential reference without introducing an electrode.

For uniform absorbed generation and bimolecular recombination, the stationary carrier problem has an exact reduction. Write $G-R=-k_b\left(np-N_G^2\right)$, where

$$N_G=\sqrt{n_i^2+G/k_b}, \tag{S29}$$

$$g = \frac{G}{k_b n_i^2}, \tag{S30}$$

$$\lambda_G = \sqrt{\frac{\epsilon V_T}{2 q N_G}}, \tag{S31}$$

$$a_G = \frac{L}{2\lambda_G} = a_0 (1+g)^{1/4}. \tag{S32}$$

The dielectric approximation must be specified before its boundary conditions are imposed. Let $P_0$ be a constant polarization source and write $D_z = P_0 - \epsilon\phi'$. With a strictly frozen polarization, $\epsilon$ is the background permittivity. Alternatively, $\epsilon$ can include the constant differential response of a stable branch, with $P_0$ the corresponding intercept. In the latter reduction the same differential susceptibility is not represented a second time by a separate polarization equation. Illumination does not change the prescribed residual source in this case.

If fixed compensation sheets $\sigma_f$ and $\sigma_b = -\sigma_f$ are included at the front and rear, Gauss's interfacial jump condition gives $D_z(0) = -\sigma_f$ and $D_z(-L) = \sigma_b = -\sigma_f$ because both exterior fields vanish. Consequently,

$$\epsilon\phi'(0) = \epsilon\phi'(-L) = P_0 + \sigma_f \equiv P_{\text{res}}, \tag{S33}$$

$$\Gamma_n(-L) = \Gamma_n(0) = \Gamma_p(-L) = \Gamma_p(0) = 0, \tag{S34}$$

$$\int_{-L}^{0} q(p-n)\, dz = D_z(0) - D_z(-L) = 0. \tag{S35}$$

Without separate sheets, $P_{\text{res}} = P_0$ and the internal displacement vanishes at both boundaries. A potential reference, $\phi(-L/2) = 0$, completes the description without imposing a conducting contact. Fixed total charge refers to the difference between electron and hole inventories; the sum can change because illumination creates pairs. For the comparison of the two polarities, the residual sources have equal magnitudes and opposite signs. The matched passive reference removes that residual source. Retaining a compensation charge left by one polar state would define a different passive comparison.

The particle fluxes and stationary continuity equations are

$$\Gamma_n = -D_n (n' - n\phi'/V_T), \tag{S36}$$

$$\Gamma_p = -D_p (p' + p\phi'/V_T), \tag{S37}$$

$$\Gamma_n' = \Gamma_p' = G - k_b (np - n_i^2) = -k_b (np - N_G^2), \tag{S38}$$

$$N_G^2 = n_i^2 + G/k_b. \tag{S39}$$

These signs follow the outward $+z$ coordinate. Conventional currents are $J_n = -q\Gamma_n$ and $J_p = q\Gamma_p$. Define the dimensionless particle electrochemical potentials

$$u_n = \ln(n/N_G) - \phi/V_T, \tag{S40}$$

$$u_p = \ln(p/N_G) + \phi/V_T. \tag{S41}$$

The hole particle potential $u_p$ carries the opposite sign to the hole quasi-Fermi energy expressed on the electron-energy scale. Since $\Gamma_n = -D_n n u_n'$ and $\Gamma_p = -D_p p u_p'$, multiplying each stationary balance by its corresponding $u_i$ and integrating by parts gives

$$\int_{-L}^{0} \left[ D_n n (u_n')^2 + D_p p (u_p')^2 + k_b (np - N_G^2) \ln\left(\frac{np}{N_G^2}\right) \right] dz = 0. \tag{S42}$$

For positive densities, every term is nonnegative: $(x-1)\ln x \geq 0$. Thus $u_n' = u_p' = 0$, $np = N_G^2$, and both particle fluxes vanish throughout the film. This is an exact stationary identity for the combined generation–recombination operator. It is not a statement that the illuminated material is in thermal equilibrium or that its physical entropy production vanishes. Optical absorption and recombination continue at equal nonzero rates, $R = G$. The proof also explains why unequal positive mobilities do not change this stationary solution. Nonuniform generation, a different recombination law, carrier-permeable faces, or additional capture channels generally removes this particular closure.

The constant electrochemical potentials give $n = A e^{\phi/V_T}$ and $p = B e^{-\phi/V_T}$, with $AB = N_G^2$. A potential shift absorbs the ratio $A/B$. The neutral, reflection-symmetric solution then has $A = B = N_G$ in the midplane gauge. Using the intrinsic band reference $E_i^0$,

$$E_{Fn} = E_i^0 - q\phi + k_B T \ln(n/n_i) = E_i^0 + k_B T \ln(N_G/n_i), \tag{S43}$$

$$E_{Fp} = E_i^0 - q\phi - k_B T \ln(p/n_i) = E_i^0 - k_B T \ln(N_G/n_i), \tag{S44}$$

$$E_{Fn} - E_{Fp} = k_B T \ln(1+g), \tag{S45}$$

$$g = G/(k_b n_i^2). \tag{S46}$$

Both quasi-Fermi energies are spatially uniform, while the band edges and carrier densities remain nonuniform.

Set $v = \phi/V_T$, $\zeta = 2(z + L/2)/L$, $\lambda_G^2 = \epsilon V_T/(2 q N_G)$, $a_G = L/(2\lambda_G)$, and $s_P = P_{\text{res}} L/(2\epsilon V_T)$. The full-slab Poisson equation is

$$v_{\zeta\zeta} = a_G^2 \sinh v, \tag{S47}$$

$$v_\zeta(-1) = v_\zeta(1) = s_P. \tag{S48}$$

The reduced Poisson–Boltzmann functional is strictly convex, which gives a unique solution for the specified derivative boundaries. Reflection symmetry makes the potential odd about the midplane. For positive residual source define $v_s = v(1)$ and $m_G = v_\zeta(0)$. Multiplication by the potential derivative and integration gives the first integral and the half-slab quadrature:

$$v_\zeta^2 = m_G^2 + 4 a_G^2 \sinh^2(v/2), \tag{S49}$$

$$s_P^2 = m_G^2 + 4 a_G^2 \sinh^2(v_s/2), \tag{S50}$$

$$1 = \int_0^{v_s} \frac{du}{\sqrt{m_G^2 + 4 a_G^2 \sinh^2(u/2)}}. \tag{S51}$$

The two scalar relations determine the midplane slope and surface voltage. The profile follows from the same quadrature with its upper limit replaced by the local potential. The Gauss inventory gives the front-half charge directly:

$$Q_{\text{front}} = \frac{2\epsilon V_T}{L} m_G - P_{\text{res}}, \tag{S52}$$

$$Q_{\text{rear}} = -Q_{\text{front}}, \tag{S53}$$

$$\frac{|Q_{\text{front}}|}{|P_{\text{res}}|} = 1 - \frac{|m_G|}{|s_P|} < 1 \ (P_{\text{res}} \neq 0). \tag{S54}$$

Its limiting magnitude is the prescribed residual source, even though the number of electron–hole pairs need not be bounded as the idealized generation is increased. The passive limit has $Q_{\text{front}} = Q_{\text{rear}} = 0$, $n = p = N_G$, and $v = 0$. Increasing $a_G$ decreases the positive solution: differentiating its half-slab equation with respect to $a_G$ and applying the maximum principle gives $\partial v / \partial a_G < 0$. Thus illumination reduces the signed electrostatic voltage magnitude smoothly; this model has no abrupt illumination threshold.

**S2.2 Controlled weak-potential limit and its error**

For $|v| \ll 1$, linearizing $\sinh v$ gives

$$v^{\text{lin}}(\zeta) = \frac{s_P}{a_G} \frac{\sinh(a_G \zeta)}{\cosh a_G}, \tag{S55}$$

$$v_s^{\text{lin}} = \frac{s_P}{a_G} \tanh a_G, \tag{S56}$$

$$\Delta\phi_{\text{iso}}^{\text{lin}} = 2 V_T v_s^{\text{lin}} = \frac{2 P_{\text{res}} \lambda_G}{\epsilon} \tanh\left(\frac{L}{2\lambda_G}\right). \tag{S57}$$

The sign and magnitude of the approximation error can be bounded without fitting. On the positive half-slab, $1 \le \sinh v / v \le K_s \equiv \sinh(v_s)/v_s$. Comparison with the two linear problems of inverse screening lengths $a_G$ and $a_G \sqrt{K_s}$ gives

$$\frac{s_P}{a_G \sqrt{K_s}} \tanh\left(a_G \sqrt{K_s}\right) \le v_s \le \frac{s_P}{a_G} \tanh a_G, \tag{S58}$$

$$0 \le \frac{v_s^{\text{lin}} - v_s}{v_s} \le \sqrt{K_s} - 1. \tag{S59}$$

The same relative bound holds for the negative polarity. The directly evaluable sufficient condition $|v_s^{\text{lin}}| \le 0.2$ implies $|v_s| \le 0.2$. Then the relative voltage error is at most $0.00333445$ $(0.333445\,\%)$, while the local source linearization obeys $\sinh(0.2)/0.2 - 1 = 0.00668001$. This is a controlled small-potential condition; agreement at larger potentials can also arise from weak screening over the thickness and does not establish the linearization generally.

For well-separated screening layers, the first integral retains the nonlinear surface-voltage correction:

$$v_s \simeq 2 \ \text{arsinh}\left(\frac{s_P}{2 a_G}\right), \tag{S60}$$

$$\Delta\phi_{\text{iso}} \simeq 4 V_T \ \text{arsinh}\left(\frac{P_{\text{res}} \lambda_G}{2 \epsilon V_T}\right). \tag{S61}$$

For $|s_P| / a_G \ll 1$, this reduces to $\Delta\phi_{\text{iso}} \simeq 2 P_{\text{res}} \lambda_G / \epsilon$. Since $\lambda_G = \lambda_0 (1+g)^{-1/4}$, the high-injection decay is proportional to $G^{-1/4}$. For strong surface potential, the inverse-hyperbolic-sine form retains the nonlinear screening that the linear expression omits.

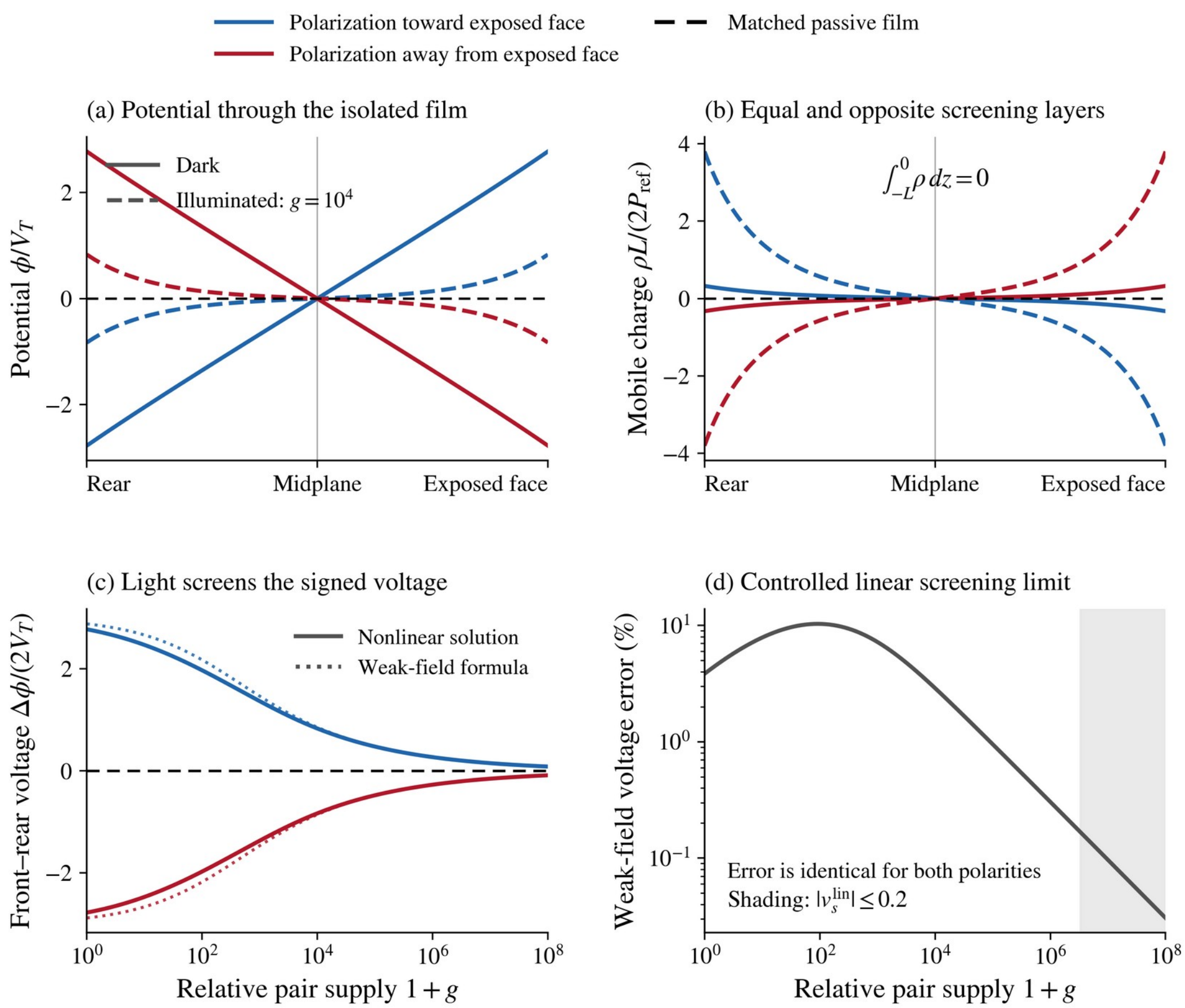

***Figure S1.*** *Stationary screening in a dry isolated film with both faces blocking electrons and holes. Blue denotes polarization toward the exposed face, red denotes polarization away from it, and the black dashed line is the nonpolar semiconductor control. The exact nonlinear calculation uses $a_0 = 0.35$ and $|s_P| = 3$. (a,b) Electrostatic potential and mobile-charge profiles in darkness (solid) and at $g = 10^4$ (dashed). The potential reference is the midplane; the integrated mobile charge is zero. The charge normalization uses the common nonzero residual magnitude $P_{\mathrm{ref}}$ of the two polar films. (c) Signed front–rear electrostatic voltage versus relative pair supply $1+g$, with the exact quadrature result shown solid and the small-potential formula dotted. The photoinduced voltage is the difference from the dark ordinate. (d) Relative error of the linear voltage magnitude, identical for the two polarities. Shading identifies the sufficient controlled regime $|v_s^{\mathrm{lin}}| \le 0.2$. The passive film has zero potential and mobile-charge density but a light-dependent carrier population.*

The nonlinear benchmark uses $a_0 = 0.35$ and $|s_P| = 3$. Its dimensional mapping is joint: intrinsic density, residual charge, film thickness, dielectric response, and generation must satisfy the definitions of the dimensionless groups. The temporal examples below use a much smaller residual source. Parameter values are collected in S8; verification of the quadrature and its independent transport solution is in S7.

### S2.3 Optical modulation and finite light steps

The nonlinear stationary construction establishes what the isolated film can do without a carrier contact: the total electronic charge remains fixed while the electron–hole inventory and its spatial separation change. To calculate the temporal response analytically, we now take a controlled weak-potential limit of the same geometry. The polarization source is fixed, photogeneration is uniform, electron and hole diffusivities are equal, and recombination is bimolecular. The sum of the carrier densities is spatially uniform to zeroth order in the polarization source. The charge density, potential, and carrier difference enter at first order. Terms coupling charge to a nonuniform neutral population appear at higher order and are not retained in this reduction.

For these calculations, lengths are divided by the physical solid thickness $L_0$, time by $\tau_R = (k_b n_i)^{-1}$, potential by $V_T$, permittivity by a reference $\epsilon_{\mathrm{ref}}$, areal charge by $\epsilon_{\mathrm{ref}} V_T / L_0$, and volume charge by $\epsilon_{\mathrm{ref}} V_T / L_0^2$. Polarization and electrical displacement use the areal-charge scale; current density uses $\epsilon_{\mathrm{ref}} V_T / (L_0 \tau_R)$. Thus $D_s$ below denotes $D_s^{\mathrm{phys}} \tau_R / L_0^2$, and an areal capacitance denotes $C^{\mathrm{phys}} L_0 / \epsilon_{\mathrm{ref}}$. The dimensionless intrinsic screening coefficient is $a_s^0 = 2 q n_i L_0^2 / (\epsilon_{\mathrm{ref}} V_T)$. This notation keeps the potential normalization identical in A and C while allowing a different liquid permittivity and diffusivity.

Let $m = N / n_i$, $g = G / (k_b n_i^2)$, and $\overline{m} = \sqrt{1 + g_0}$. The neutral pair population satisfies

$$\dot{m} = g - (m^2 - 1), \tag{S62}$$

$$\overline{m} = \sqrt{1 + g_0}, \tag{S63}$$

$$\frac{\hat{m}}{\hat{g}} = \frac{1}{s + 2\overline{m}}, \tag{S64}$$

$$s = i\Omega. \tag{S65}$$

The stationary population changes as the square root of the total dark-plus-optical supply. Its differential relaxation becomes faster as illumination increases. This population time is not yet the surface-voltage time, because the carriers must also redistribute in space. For a finite change from $g_i$ to $g_f$, with the initial population $m_i = \sqrt{1 + g_i}$ and $m_f = \sqrt{1 + g_f}$, the exact solution is

$$m(t) = m_f \frac{m_i + m_f \tanh(m_f t)}{m_f + m_i \tanh(m_f t)}, \tag{S66}$$

$$\int_0^t m(u)\, du = \ln\left[\cosh(m_f t) + \frac{m_i}{m_f} \sinh(m_f t)\right]. \tag{S67}$$

The charge sector follows directly by subtracting the linearized electron and hole fluxes. The physical mobile-charge density is $q(p - n)$; its dimensionless value below is $\rho = q(p - n) L_0^2 / (\epsilon_{\mathrm{ref}} V_T)$, with $n$ and $p$ here denoting physical number densities:

$$J_s = -D_s \left(\partial_x \rho + a_s^0 m\, \partial_x \phi\right), \tag{S68}$$

$$\partial_t \rho = -\partial_x J_s, \tag{S69}$$

$$D_e = P - \epsilon_s \partial_x \phi, \tag{S70}$$

$$\partial_x D_e = \rho. \tag{S71}$$

Both carrier-blocking boundaries give $J_s(0) = J_s(L) = 0$ in this charge sector; the neutral-population flux also vanishes. The zero exterior field gives $D_e(0) = D_e(L) = 0$. Consequently $\int_0^L \rho\, dx = 0$ at every time. Setting the midpoint potential to zero determines the gauge, while the measurable internal response is $\Delta\phi = \phi(L) - \phi(0)$.

Writing $\lambda_s(m)=\sqrt{\epsilon_s/(a_s^0 m)}$ and evaluating it at $m=\overline{m}$ for a stationary state, the weak-potential profile and voltage are

$$\phi(x)=\frac{P\lambda_s}{\epsilon_s}\frac{\sinh[(x-L/2)/\lambda_s]}{\cosh[L/(2\lambda_s)]}, \tag{S72}$$

$$\Delta\phi=PF(\overline{m}), \tag{S73}$$

$$F(m)=\frac{2\lambda_s}{\epsilon_s}\tanh\left(\frac{L}{2\lambda_s}\right). \tag{S74}$$

The exact derivative of this weak-potential stationary voltage is obtained without numerical differentiation. Set $a=L/(2\lambda_s)$; then

$$F_m=\frac{-\lambda_s}{\epsilon_s m}\left[\tanh a-a\,\mathrm{sech}^2 a\right], \tag{S75}$$

$$\frac{d\Delta\phi}{d g_0}=\frac{P F_m}{2\overline{m}}. \tag{S76}$$

The voltage decreases in magnitude as the pair population increases. The nonpolar symmetric reference has a changing carrier population but zero potential. To retain the spatial screening times rather than assigning an instantaneous screening length, expand the odd charge modes as follows.

The isolated geometry permits a stronger analytical dynamical result than a single lumped screening time. Let $k_j=(2j+1)\pi/L$ and write the antisymmetric charge distribution as $\rho(x,t)=\sum_{j\geq 0}\rho_j(t)\sin[k_j(x-L/2)]$. The constant normal-field boundary imposes $\rho'(0)=\rho'(L)=-\kappa^2(t)P$, where $\kappa^2(t)=a_s^0 m(t)/\epsilon_s$. Projecting the diffusion equation onto the odd sine modes, including this boundary term, gives

$$\dot{\rho}_j=-D_s\left[k_j^2+\kappa^2(t)\right]\rho_j-\frac{4D_sP}{L}(-1)^j\kappa^2(t), \tag{S77}$$

$$b_j=\frac{-L(-1)^j\rho_j}{4P}, \tag{S78}$$

$$\dot{b}_j=-D_s\left[k_j^2+\kappa^2(t)\right]b_j+D_s\kappa^2(t), \tag{S79}$$

$$\Delta\phi(t)=\frac{PL}{\epsilon_s}-\frac{8P}{\epsilon_s L}\sum_{j=0}^{\infty}\frac{b_j(t)}{k_j^2}. \tag{S80}$$

The sine expansion is a weak expansion of the driven Neumann problem; the nonzero boundary derivative is represented by the projected boundary forcing. Omitting that term would incorrectly remove the optical screening response. At fixed illumination, $\overline{b}_j=\overline{\kappa}^2/(k_j^2+\overline{\kappa}^2)$. Substitution into the voltage sum reproduces the hyperbolic-tangent solution.

Linearizing each scalar mode and inserting the population transfer function yields an explicit analytical optical spectrum:

$$H_A(s;g_0)=\frac{-8PD_s\kappa_0^2}{\epsilon_s L(s+2\overline{m})}\sum_{j=0}^{\infty}\frac{1}{(k_j^2+\overline{\kappa}^2)\left[s+D_s(k_j^2+\overline{\kappa}^2)\right]}, \tag{S81}$$

$$\kappa_0^2 = a_s^0/\epsilon_s, \tag{S82}$$

$$\bar{\kappa}^2 = \kappa_0^2 \bar{m}. \tag{S83}$$

Each term combines the pair-population relaxation with a diffusive screening mode. All static weights have the same sign. Therefore the magnitude cannot exceed its zero-frequency value in this reduction: $|H_A(i\Omega)| \le |H_A(0)|$. The zero-frequency sum equals the derivative obtained from the stationary hyperbolic-tangent solution. The spectrum has a low-frequency plateau, an illumination-dependent roll-off, and a phase lag from the combined carrier-population and charge redistribution dynamics. Polarization reversal changes the sign of the transfer function while preserving its magnitude. This result is specific to the declared symmetric, fixed-source isolated film; it is not a general restriction on contacted or ionically coupled photovoltages.

A finite illumination step also has an analytical integrating-factor solution for each mode. Define $I(t) = \int_0^t m(u)\, du$ from the exact population expression and $A_j(t) = D_s k_j^2 t + D_s \kappa_0^2 I(t)$. Then

$$b_j(t) = e^{-A_j(t)} b_j(0) + D_s \kappa_0^2 \int_0^t m(u) e^{-[A_j(t) - A_j(u)]}\, du. \tag{S84}$$

Only a smooth one-dimensional quadrature remains. For an increasing light step, the instantaneous stationary value $\kappa^2(t)/\left(k_j^2 + \kappa^2(t)\right)$ increases monotonically. Starting from the preceding stationary state, every $b_j$ follows it from below. The positive-$P$ voltage therefore decreases monotonically. The decreasing-step argument reverses these signs. Thus the absence of overshoot in Figure S2 follows from the mode equations and is not merely an observation from one numerical trace.

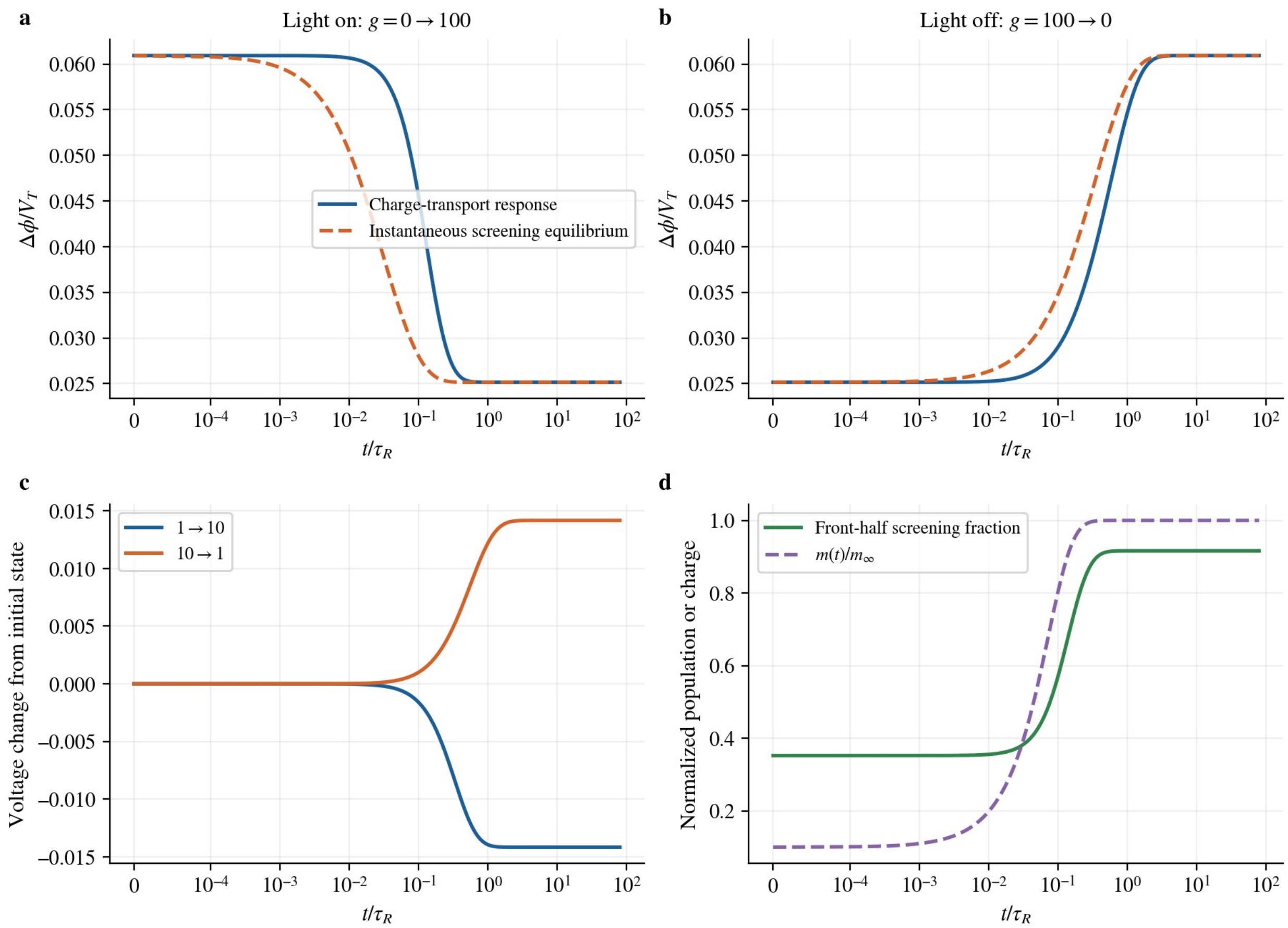

***Figure S2.*** *Large and nonzero-mean illumination steps in A. (a,b) Light on and off between* $g=0$ *and* $100$. *The dashed curve assumes instantaneous charge equilibration at the exact time-dependent pair population; the solid curve resolves charge transport. (c) Steps between* $g=1$ *and* $10$ *in both directions. (d) Pair population and the fraction of the residual source compensated in the front half of the slab during light-on. The charge response lags the population while the total slab charge remains zero. The independently evaluated analytical mode quadratures reproduce these conservative transport traces within the stated spatial-discretization accuracy.*

### S2.4 Donor and acceptor extension

For the blocking linearized reductions A and C, the differential screening length is $\lambda_s=\left[\epsilon k_B T/\left(q^2(n_b+p_b)\right)\right]^{1/2}$, with dimensional carrier populations in this expression. The isolated front–rear voltage is $\Delta\phi_A=2P_{res}\lambda_s\tanh\left[L/(2\lambda_s)\right]/\epsilon$. In C this solid dipole enters the passive capacitance division specified in S4.1. These relations depend on the total population and are invariant under $d\rightarrow -d$ at equal donor and acceptor magnitudes.

This donor–acceptor symmetry belongs to the stated blocking reduction and is not imposed on contacted or chemically asymmetric cases. The common static comparison is given in S8. For the dynamic normalization below, the screening-length unit corresponding to S1.2 is $L_D=\left[\epsilon k_B T/\left(q^2 N_{ref}\right)\right]^{1/2}$.

For the isolated film with equal diffusivities, the linearized pair balance about the illuminated quasi-neutral state gives a population response with rate $S_b=n_b+p_b$, in units of $k_b N_{ref}$, and the screening coefficient $\bar{\kappa}^2=S_b$ in units of $L_D^{-2}$:

$$\frac{\hat{S}}{\hat{g}}=\frac{2}{s+S_b}, \tag{S85}$$

$$H_A(s)=\frac{-16\,P\,D_s}{\epsilon L(s+S_b)}\sum_j \frac{1}{(k_j^2+S_b)[s+D_s(k_j^2+S_b)]}. \tag{S86}$$

The evaluation uses $L=\epsilon=P=1$ in the units of Table S10, $D_s=0.25$, and 4000 odd modes. Its zero-frequency limit equals the derivative of $\Delta\phi=2P\lambda\tanh(L/2\lambda)/\epsilon$ with $\lambda=S_b^{-1/2}$ to $10^{-8}$ relative. In this blocking geometry the transfer function depends on doping only through $S_b$, so donor and acceptor densities of equal magnitude give identical spectra.

## S3. Case B: dry film with a carrier reservoir

### S3.1 Finite-depth problem and optional electronic sites

Case B connects the dry ferroelectric to a carrier reservoir. The exposed surface remains chemically blocking, while the rear contact fixes the electron and hole electrochemical potentials. We first specify the continuum model and its optical linearization, then compare the same contact with and without a finite population of electronic surface sites. The detailed dry-film examples include illumination, mobility, surface occupation, depth-profile, and polarization-stability comparisons. Their electrical rear boundary remains a carrier reservoir throughout.

The exposed face is at inward depth $x=0$ and the rear at $x=1$. In physical coordinates, $x=y/L=-z/L$, with the paper's outward coordinate satisfying $-L<z<0$. Define $u=P_{\text{in}}/P_*=-P_z/P_*$, $v=\phi/V_T$, $d=D_{\text{in}}/P_*$, and $\hat{t}=t/\tau_*$. Electron and hole densities are divided by $n_i$. Conventional currents point in the inward direction and are divided by $J_*=q\,n_i\,L/\tau_*$. Thus blue, outward polarization corresponds to $u<0$, and red, inward polarization to $u>0$.

The dimensionless parameters are

$$\epsilon^*=\frac{\epsilon_b V_T}{P_* L}, \tag{S87}$$

$$\delta=\frac{q\,n_i\,L}{P_*}, \tag{S88}$$

$$M_i=\frac{D_i\tau_*}{L^2}, \tag{S89}$$

$$g_0=\frac{G_0\tau_*}{n_i}, \tag{S90}$$

$$\beta=\alpha_{\text{abs}}L. \tag{S91}$$

Here $\epsilon_b$ is the absolute background permittivity. If the dimensional polarization equation is $\Gamma\,\partial_t P=-a_L P-b_L P^3-c_L P^5+g_P\partial_y^2 P+E_y$, then $A=a_L P_* L/V_T$, $B=b_L P_*^3 L/V_T$, $C=c_L P_*^5 L/V_T$, $\kappa=g_P P_*/(L V_T)$, and $\gamma=\Gamma P_* L/(V_T\tau_*)$. These relations identify the parameters required to attach physical lengths, intensities, and times to an illustrative dimensionless set.

Dots and primes denote $\hat{t}$ and $x$ derivatives. The full equations are

$$\gamma\dot{u}=-Au-Bu^3-Cu^5+\kappa u''-v', \tag{S92}$$

$$d=u-\epsilon^* v', \tag{S93}$$

$$d'=\delta(p-n), \tag{S94}$$

$$j_n=M_n(n'-nv'), \tag{S95}$$

$$j_p=-M_p(p'+pv'), \tag{S96}$$

$$\dot{n}=j_n'+g_0(\hat{t})e^{-\beta x}-R(n,p), \tag{S97}$$

$$\dot{p} = -j_p' + g_0(\hat{t})e^{-\beta x} - R(n, p). \tag{S98}$$

The background permittivity excludes the soft-mode polarization represented by $u$. Natural polarization conditions are $u'(0) = u'(1) = 0$. The nonpolar semiconductor control sets $u = 0$ and removes its dynamic equation, retaining all other parameters of the corresponding case. It does not retain an unstable Landau extremum as a putative semiconductor state.

The charging set uses $\tau_* = (k_b n_i)^{-1}$ and $R = np - 1$. The reactive set uses $\tau_* = \tau_0$ and $R = (np - 1)/(n + p + 2)$. These two illustrative material parameter sets have different recombination laws and time scales. Numerical values of $g_0$ or angular frequency cannot be identified across them with the same physical irradiance or seconds without a material calibration.

The rear conditions are $v(1) = 0$ and $n(1) = p(1) = 1$. The exterior normal displacement above the dry front is neglected. A front site charge $\sigma_s$ therefore gives $d(0) = \sigma_s / P_*$. This boundary specifies displacement, not the exposed potential. There is no electrolyte, resolved compact-layer voltage, or continuously replenished chemical reactant at the front.

With no sites the complete front conditions are simply

$$d(0) = 0, \tag{S99}$$

$$j_n(0) = j_p(0) = 0. \tag{S100}$$

In the finite-site version, fixed areal populations $N_+$ and $N_-$ have charged fractions $\theta_+$ and $\theta_-$. A positive site captures a hole and a negative site captures an electron. Emission returns that same carrier to the solid. There is no cross-capture recombination channel or movement of the sites. The emission times are $\tau_\pm$, and $K_\pm = k_{\mathrm{cap},\pm} n_i \tau_\pm$ are dimensionless capture-to-emission ratios. With $s_\pm = q N_\pm / P_*$ and $T_\pm = \tau_\pm / \tau_*$,

$$\dot{\theta}_+ = \frac{K_+ p_s (1 - \theta_+) - \theta_+}{T_+}, \tag{S101}$$

$$\dot{\theta}_- = \frac{K_- n_s (1 - \theta_-) - \theta_-}{T_-}, \tag{S102}$$

$$d(0) = s_+ \theta_+ - s_- \theta_-, \tag{S103}$$

$$j_n(0) = \frac{s_-}{\delta} \dot{\theta}_-, \tag{S104}$$

$$j_p(0) = \frac{-s_+}{\delta} \dot{\theta}_+. \tag{S105}$$

The front exchanges carriers with this finite inventory during charging and release. It is carrier-blocking only at stationary occupancy in these two selected channels. No-site calculations remove the site variables and apply zero front flux directly; they do not set a denominator or a capacity to zero in the site evolution equations.

For nonnegative carrier densities and positive kinetic coefficients, the occupancy interval $0 \le \theta_\pm \le 1$ is invariant. Therefore $-q N_- \le \sigma_s \le q N_+$, and the net number of carriers stored in either family between any two times is bounded by its areal capacity $N_\pm$. Repeated capture and emission can exchange carriers many times, but the signed time integral of each selected front current equals the finite change of its occupancy.

### S3.2 Voltage decomposition and finite charge inventory

An exact integration of Gauss's law and the grounded rear condition gives an instructive voltage decomposition, valid for the full nonlinear state:

$$v_s = \frac{-\int_0^1 u\, dx + d(0) + \delta \int_0^1 (1-x)(p-n)\, dx}{\epsilon^*}. \tag{S106}$$

Polarization, site charge, and the first spatial moment of mobile charge enter separately. This identity explains why the sign of a differential optical voltage need not be inferred from the sign of the change in polarization or the total number of photocarriers. A grounded film can change its total mobile-plus-site charge through the rear. Its exact balance, in units of $q n_i L$, is

$$\frac{d}{d\hat{t}}\left[\int_0^1 (p-n)\, dx + \frac{s_+\theta_+ - s_-\theta_-}{\delta}\right] = -\left[j_n(1) + j_p(1)\right]. \tag{S107}$$

For the no-site case the site term is absent. At every depth,

$$j_n + j_p + \delta^{-1}\dot{d} = 0. \tag{S108}$$

This zero *total* current does not set the two conduction currents to zero during a transient. At stationary illumination the rear can remove generated pairs through opposite electron and hole currents. Integrating the front capture current measures changing occupancy; it does not give an indefinitely accumulated chemical product.

For equal mobilities and identical site families, simultaneous electron–hole exchange and $(u, v, d) \mapsto (-u, -v, -d)$ relates the two polar preparations. In the passive control, a symmetric state obeys $n = p$, $\theta_+ = \theta_-$, and $v = 0$ at every time for arbitrary prescribed illumination. The carrier and occupancy dynamics persist even though the voltage signal is identically zero. This is an exact symmetry of these boundaries and constitutive choices, not a general statement that a nonferroelectric semiconductor cannot produce photovoltage.

The two surface families store charge independently; their occupations can both rise while their difference falls. They are not a single trap that alternately captures electrons and holes. A stationary illuminated rear contact can remove pairs through equal and opposite conventional carrier currents even when the terminal total current is zero.

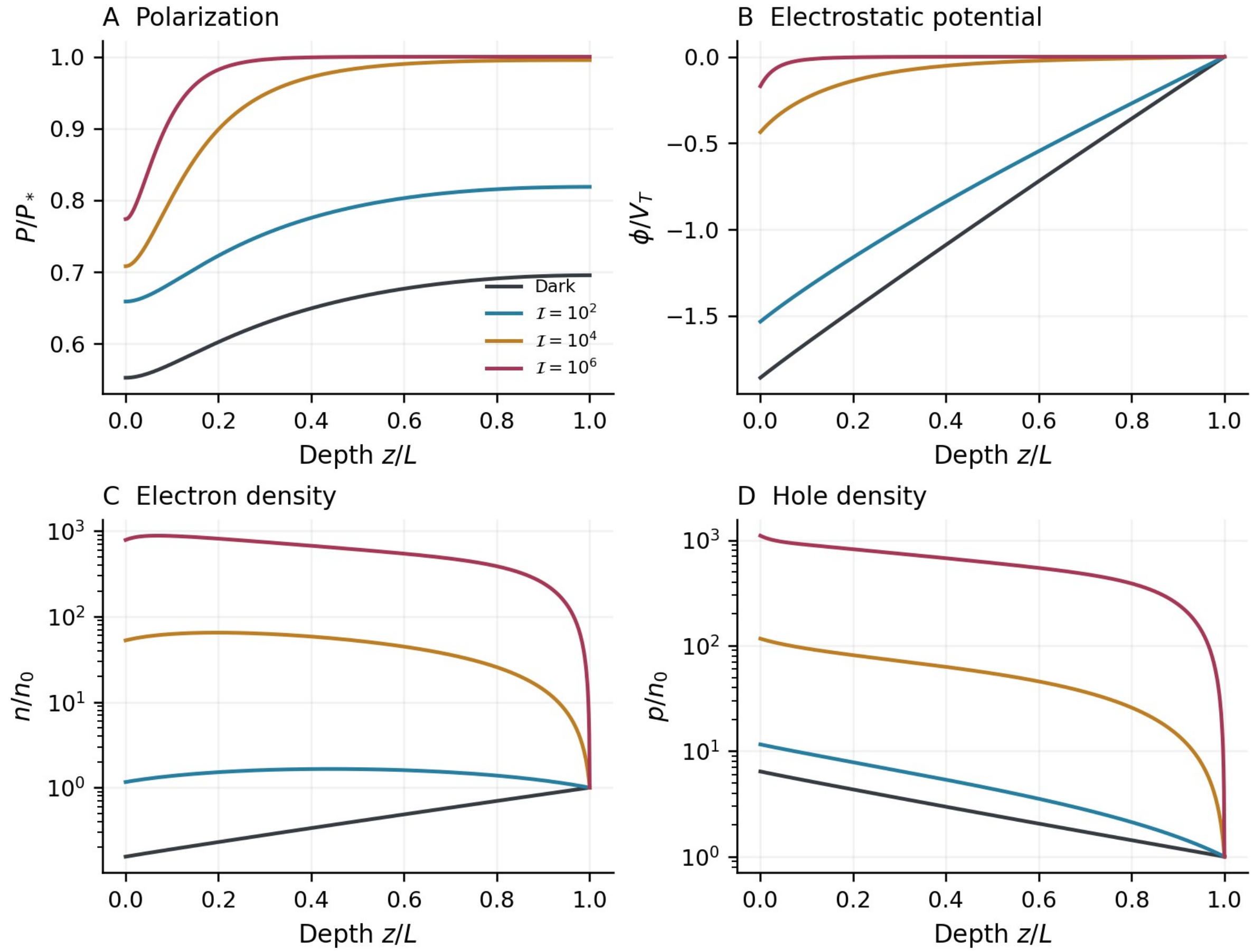

***Figure S3.*** *Depth profiles of the stationary charge-site states. The panels show inward polarization, electrostatic potential, electron density, and hole density for source intensities* $I=0,100,10^4,10^6$ *on the positive inward second-order branch. The exposed face is at* $y/L=0$ *and the rear electrode at* $y/L=1$*. Carrier-density axes are logarithmic. The rear conditions remain* $n=p=n_i$ *and* $\phi=0$ *at every intensity, so appreciable depth gradients remain even when carrier transport is relatively rapid. The positive inward dark state has negative exposed bound charge, negative surface potential, electron depletion, and hole accumulation. Illumination changes both carrier populations and their space-charge distribution. The local band-edge shift has the opposite sign to the electrostatic potential.*

## S3.3 Dynamic response and mechanism-resolving controls

Let a barred state be a stationary solution at a specified mean $\bar{g}$. Perturb the incident source by $\mathrm{R}\left[\tilde{g}\, e^{i\Omega \hat{t}}\right]$, with $|\tilde{g}| \ll \bar{g}$ when a nonzero illuminated state is studied. Here $\Omega=\omega_{\mathrm{phys}}\tau_*$ is dimensionless angular frequency. Tildes denote first-order complex amplitudes. Linearizing the bulk equations gives

$$i\Omega\gamma\,\tilde{u}=-\left(A+3B\bar{u}^2+5C\bar{u}^4\right)\tilde{u}+\kappa\,\tilde{u}''-\tilde{v}', \tag{S109}$$

$$\tilde{d}=\tilde{u}-\epsilon^*\tilde{v}', \tag{S110}$$

$$\tilde{d}'=\delta\left(\tilde{p}-\tilde{n}\right), \tag{S111}$$

$$\tilde{j}_n=M_n\left(\tilde{n}'-\bar{v}'\tilde{n}-\bar{n}\tilde{v}'\right), \tag{S112}$$

$$\tilde{j}_p=-M_p\left(\tilde{p}'+\bar{v}'\tilde{p}+\bar{p}\tilde{v}'\right), \tag{S113}$$

$$i\Omega\,\tilde{n}=\tilde{j}_n'+e^{-\beta x}\tilde{g}-R_n\tilde{n}-R_p\tilde{p}, \tag{S114}$$

$$i\Omega\tilde{p}=-\tilde{j}_p'+e^{-\beta x}\tilde{g}-R_n\tilde{n}-R_p\tilde{p}. \tag{S115}$$

For B, $(R_n,R_p)=(\bar{p},\bar{n})$. For D,

$$R_n=\frac{(\bar{p}+1)^2}{(\bar{n}+\bar{p}+2)^2}, \tag{S116}$$

$$R_p=\frac{(\bar{n}+1)^2}{(\bar{n}+\bar{p}+2)^2}. \tag{S117}$$

At B's rear, $\tilde{v}(1)=\tilde{n}(1)=\tilde{p}(1)=0$. Without sites, $\tilde{d}(0)=\tilde{j}_n(0)=\tilde{j}_p(0)=0$. With sites,

$$(i\Omega+b_+)\tilde{\theta}_+=a_+\tilde{p}_s, \tag{S118}$$

$$(i\Omega+b_-)\tilde{\theta}_-=a_-\tilde{n}_s, \tag{S119}$$

$$a_+=\frac{K_+(1-\bar{\theta}_+)}{T_+}, \tag{S120}$$

$$b_+=\frac{1+K_+\bar{p}_s}{T_+}, \tag{S121}$$

$$a_-=\frac{K_-(1-\bar{\theta}_-)}{T_-}, \tag{S122}$$

$$b_-=\frac{1+K_-\bar{n}_s}{T_-}, \tag{S123}$$

$$\tilde{d}(0)=s_+\tilde{\theta}_+-s_-\tilde{\theta}_-, \tag{S124}$$

$$\tilde{j}_n(0)=\frac{s_-}{\delta}i\Omega\tilde{\theta}_-, \tag{S125}$$

$$\tilde{j}_p(0)=\frac{-s_+}{\delta}i\Omega\tilde{\theta}_+. \tag{S126}$$

These boundary admittances contain the local capture/emission times. The full optical response additionally contains carrier redistribution and polarization feedback; $b_\pm^{-1}$ cannot be assigned to a measured voltage mode without evaluating its coupling and observable weight.

The discretized observable response and the checks relating it to stationary derivatives, small steps, and direct sinusoidal integration are collected in S7. The surface-state capture time is a coefficient in the coupled system; the measured voltage mode also contains carrier redistribution and polarization feedback.

A particularly useful prediction occurs at the no-site stationary voltage maximum (Figure S4). The 81-cell response gives $\bar{g}=108.04745$, $v_s=1.80025765$, and a static optical slope smaller than $4\times10^{-13}$ in magnitude. Nevertheless, the evaluated finite-frequency gain reaches $|H_v|=0.00261539$ near $\Omega=7.50$. The operating state is stable. The derivatives of the polarization and mobile-charge terms cancel at zero frequency, but their different dynamic phases prevent cancellation at finite frequency. Consequently, a nearly flat potential-versus-intensity curve does not imply that the same state is insensitive to periodically modulated illumination. This maximum is a property of the coupled observable, not evidence for an additional unstable state or an independently fitted electrical circuit.

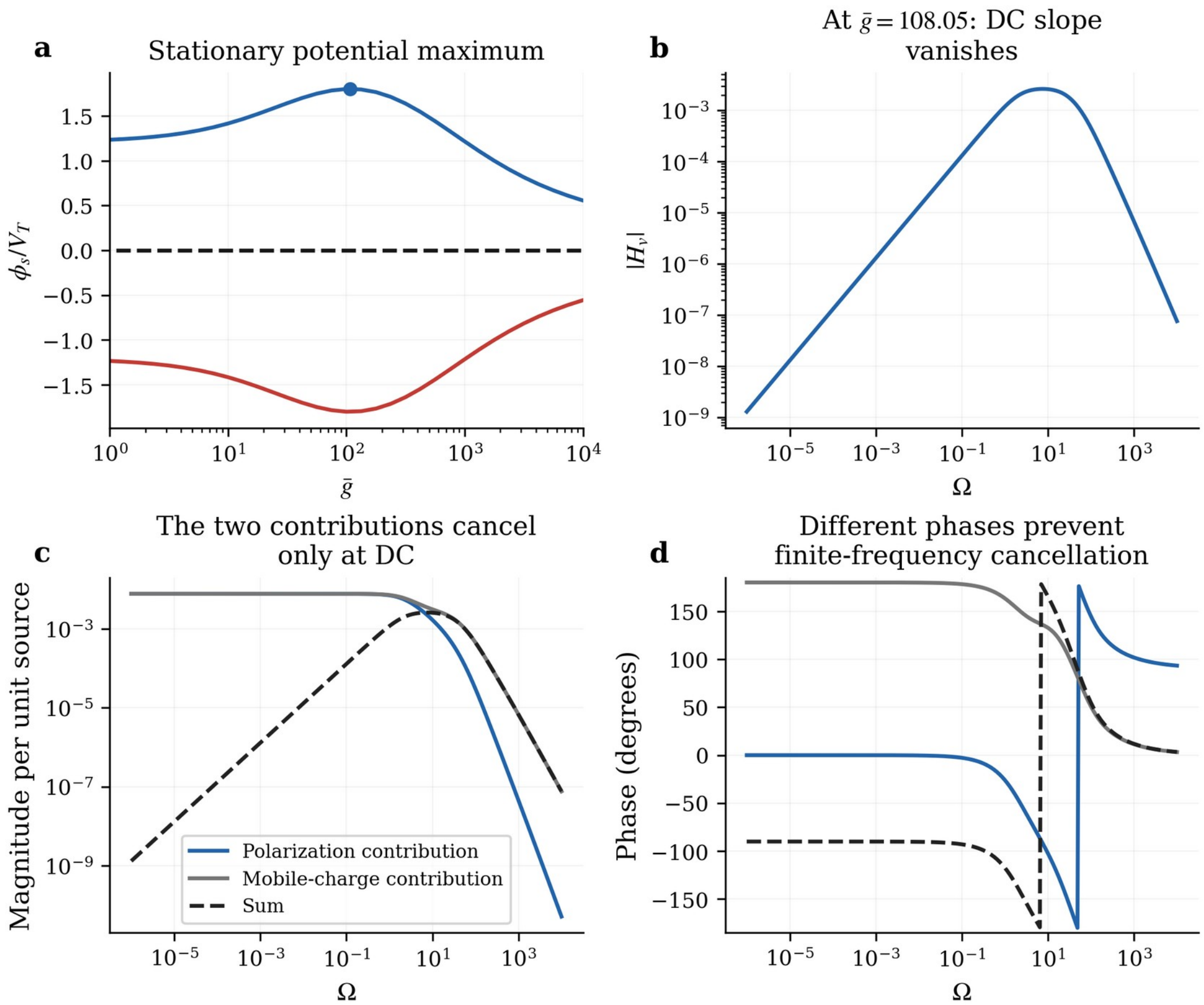

***Figure S4.*** *Finite optical response at a stationary voltage maximum in the grounded film without traps. At the 81-cell stationary extremum* $\bar{g} = 108.047445$*, the DC surface-voltage derivative has magnitude below* $4 \times 10^{-13}$ *while the stable dynamic response is nonzero. The sampled maximum* $|H_v|$ *is* $0.00261539$ *at* $\Omega = 7.49894$*. The polarization and mobile-charge moments cancel in the static derivative but acquire different phases at finite frequency. This is an evaluated response of the coupled model, not a new instability or a fitted relaxation circuit. The opposite polar preparation has the negative voltage response; the passive control remains zero.*

Removing the sites still gives a large nonmonotonic transient. In the no-site outward preparation, the same light step starts at $v_s = 1.20756$, reaches approximately $0.52578$ near $0.070\,\tau_R$, and ends at $1.21492$. Its stationary start and finish are close, while its intervening voltage change is substantial. Switch-off raises the potential to approximately 2.45551 before it returns to the original branch. Polarization evolves from approximately $0.2658\,P_*$ to $0.8007\,P_*$ during illumination, and the charge-moment identity assigns the evolving voltage to polarization and mobile carriers without introducing any hidden site population. The equal-mobility passive semiconductor remains at zero potential throughout both pulses.

Two simple limits help interpret, but do not replace, the coupled voltage solution. In an intrinsic passive no-site B film with equal mobilities, the carrier pair perturbation around the uniform dark equilibrium satisfies $\dot{c} = Mc'' - 2c + g$ for uniform generation, with $c'(0) = 0$ and $c(1) = 0$. Its exact surface transfer function is

$$\frac{\tilde{c}_s}{\tilde{g}} = \frac{1 - \operatorname{sech}\sqrt{(2 + i\Omega)/M}}{2 + i\Omega}. \tag{S127}$$

The surface voltage is nevertheless zero by symmetry. This explicitly separates a nonzero carrier response from a voltage observable. The result assumes weak injection about $n = p = 1$, uniform generation, no sites, and the two-carrier rear contact; it is not the finite-field ferroelectric spectrum.

A conditional switching example is shown in Figure S5. Changing the mobility pair from $(10,10)$ to $(30,3)$ while keeping the symmetric site parameters breaks illuminated electron–hole symmetry. A mean-constrained continuation locates the branch fold at $g_0=194.07491687$, and a source-250 pulse moves the initial branch to the opposite polarization, which persists after the light is removed (Figure S5). Selected finite-volume stability eigenvalues change sign across the fold. This result concerns the specified unequal-transport preparation and should not be used to explain the symmetric pulse, whose polarization retains its sign. Both belong to the same dry grounded geometry; their different behavior follows from explicitly changed transport parameters.

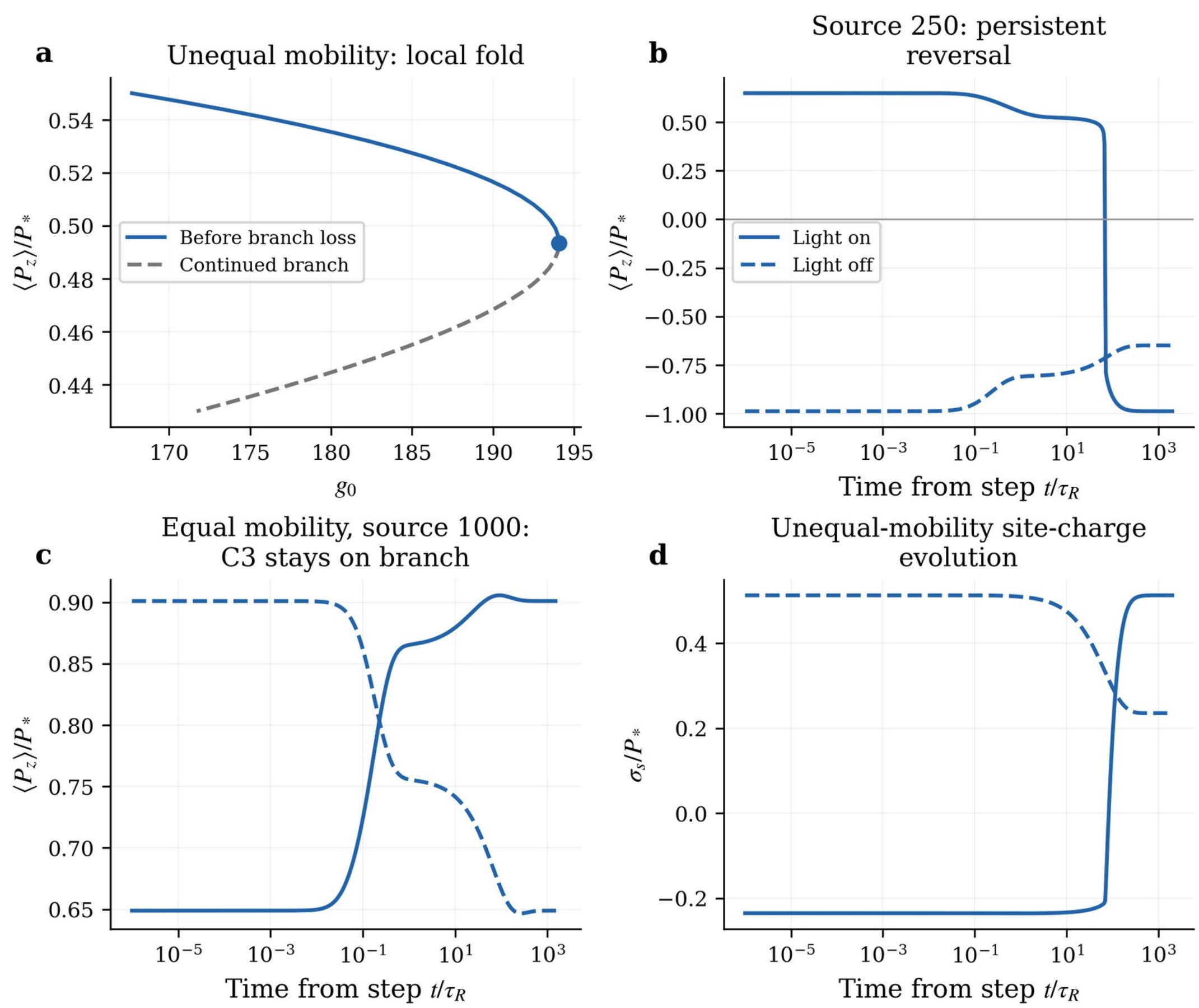

***Figure S5.*** *Distinct finite-site mechanisms under the same grounded geometry. The equal-mobility pulse gives coupled shoulders and relaxation without reversing the prepared polarization. Changing the dimensionless electron and hole mobilities from* $M_n=M_p=10$ *to* $M_n=30$ *and* $M_p=3$ *gives the fold at* $g_0=194.07491687$. *A source-250 pulse lasting* $2000\,\tau_R$ *moves the initially outward-polarized branch to the opposite branch, which persists for the subsequent* $2000\,\tau_R$ *dark interval. Solid and dashed traces distinguish on/off segments, each with its own time origin; their blue color labels the initial preparation. The finite-volume stability samples straddle the local fold. No general optical-switching threshold is inferred for other contacts or kinetics.*

The mobility and surface-kinetic comparisons hold the rear reservoir fixed. Increasing both mobilities can drain the photoexcited population more efficiently and weaken illumination-induced screening. Unequal mobilities or unequal site activities break the paired-carrier symmetry. The switching example is a conditional branch-loss calculation, not a general prediction for dry ferroelectrics.

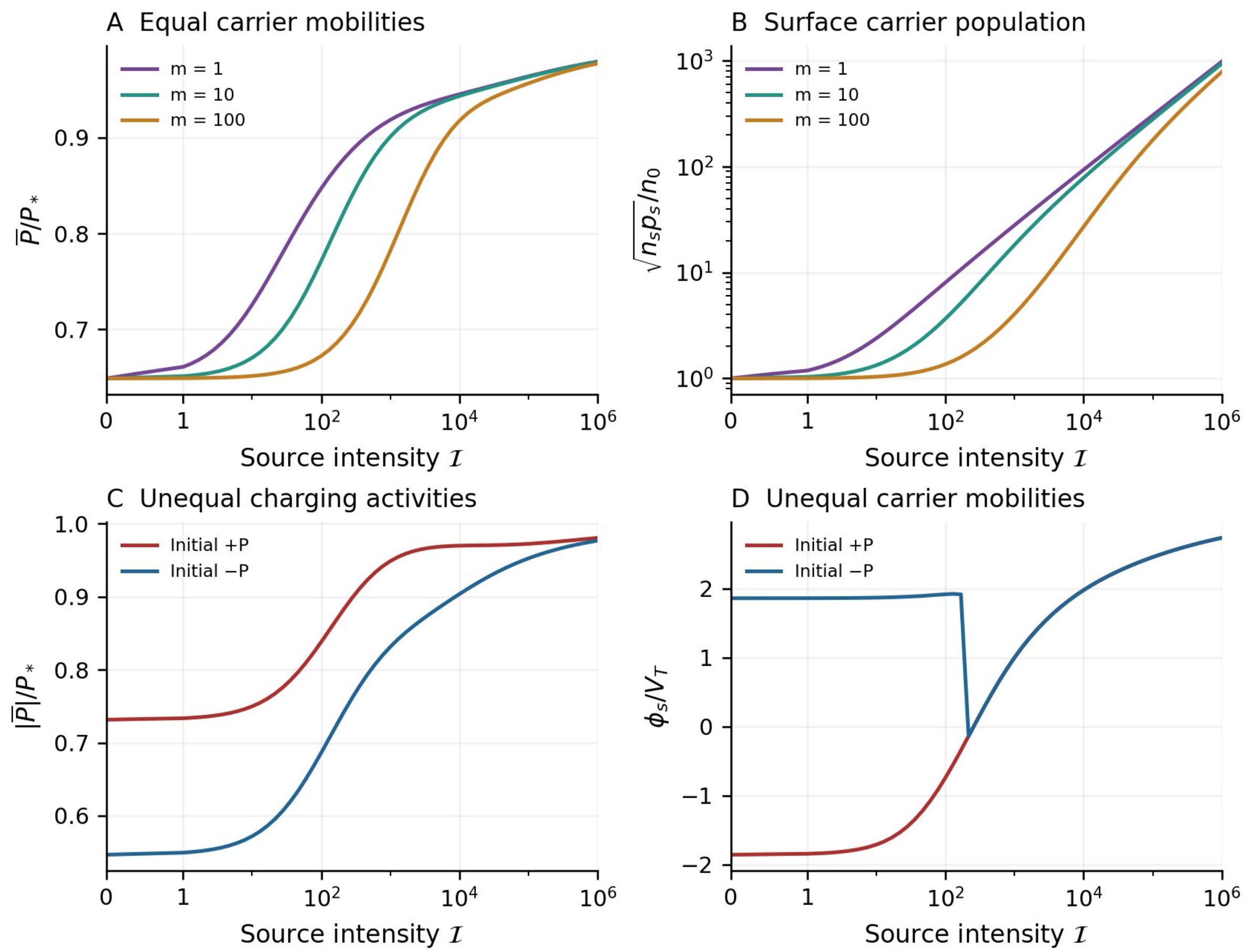

***Figure S6.*** *Carrier transport and surface-kinetic asymmetry in the charge-site model. Panels (A,B) show mean inward polarization and the geometric mean of the two surface carrier densities for equal transport parameters* $M_n = M_p = 1,10,100$. *The ideal rear electrode equilibrates both carrier species, so faster transport can drain photocarriers and weaken the illumination-induced screening at fixed generation. Panel (C) shows the magnitude of mean polarization when the charging activities are* $K_+ = 0.06$ *and* $K_- = 0.015$, *with* $M_n = M_p = 10$. *Panel (D) shows surface potential for* $M_n = 30$, $M_p = 3$ *and symmetric charging activities. The initially negative inward branch terminates and joins the positive branch after its fold. This conditional switching is resolved by the continuation and time evolution in Figure S5. All curves retain the same grounded, carrier-equilibrating rear boundary.*

### S3.4 Doping-dependent static potential and fixed-polarization dynamics

The reduced contacted-space-charge comparison uses the one-dimensional charge integral, with the bulk populations from S1.2. In dimensional form, for $u = q\psi_s / k_B T$,

$$Q_s(u) = -sgn(u)\left\{2\epsilon k_B T\left[n_b\left(e^{u} - u - 1\right) + p_b\left(e^{-u} + u - 1\right)\right]\right\}^{1/2}, \tag{S128}$$

$$P_{res} + Q_s = 0. \tag{S129}$$

Here the populations in the charge integral are dimensional; multiplication by $N_{ref}$ restores them from the normalized doping calculation. Electron accumulation screens one polarity differently from depletion/inversion, and exchanging donor for acceptor reverses that asymmetry. This electrostatic limit is separate from the full relaxing-polarization, finite-site model above.

For the contacted dry film B we solve the nonlinear Poisson–drift–diffusion problem without a Landau relaxation variable, so that the doping dependence of the space-charge response is isolated. With depth $x$ measured inward from the exposed face in units of $L_D$:

$$-v''=p-n+d, \tag{S130}$$

$$\Gamma_n=-M(n'-nv'), \tag{S131}$$

$$\Gamma_p=-M(p'+pv'), \tag{S132}$$

$$\partial_t n=-\Gamma_n'+g-(np-r_i^2), \tag{S133}$$

$$\partial_t p=-\Gamma_p'+g-(np-r_i^2), \tag{S134}$$

with $v'(0)=-\sigma$ and $\Gamma_n(0)=\Gamma_p(0)=0$ at the dry front, and $v=0$, $n=n_0$, $p=p_0$ at a rear contact placed $20\,L_D$ deep. The parameters are $\sigma=\pm 2$, $M=1$ and uniform $g$. The equations are discretized with Scharfetter–Gummel fluxes on a 261-node geometric grid with a first cell of $0.005\,L_D$, using logarithmic carrier densities, and solved by Newton iteration. Continuation proceeds first in illumination at $\sigma=0$ and then in $\sigma$. Small-signal responses follow from $(i\Omega M-J)\hat{y}=f_g\hat{g}$, where $M$ contains the carrier storage terms and the Poisson rows are algebraic. The stationary surface potentials reproduce the reduced space-charge balance of Table S10 to within $1.3\times 10^{-4}V_T$ for all six combinations of $d\in\{-10,0,10\}$ and $\sigma=\pm 2$. The zero-frequency transfer functions agree with central-difference slopes to $10^{-4}$ relative. Table S11 summarizes the gains and corner frequencies, defined as the $-3$ dB points of $|H_V|$.

The dynamic doping calculation fixes polarization and omits finite surface sites. It therefore isolates the space-charge effect of doping and does not extend the site-driven or Landau-switching spectra. The static gains and corner frequencies reported for this variant are retained in the comparative doping table in S8.

## S4. Case C: blocking ferroelectric–electrolyte junction

### S4.1 Electrical ensemble and stationary voltage partition

Case C retains the carrier isolation of the solid but introduces a polarizable liquid. The distinction from B is electrical: the rear does not exchange electrons or holes. The distinction from D is chemical: neither carrier transfers across the exposed interface. Charge can nevertheless redistribute inside the solid and inside the electrolyte. To calculate the measured potential relative to the liquid, the electrostatic environment on both sides of the solid must be specified.

We place a dielectric spacer of areal capacitance $C_b$ between the rear face and a reference plane. This is passive capacitive coupling, with no carrier exchange through the spacer. The reference plane is at zero potential relative to the remote electrolyte reservoir. A compact layer of capacitance $C_H$ separates the solid front from a symmetric electrolyte of thickness $H$. The electrolyte has equal positive and negative ionic diffusivities, a fixed bulk concentration at its remote boundary, and no Faradaic flux at the compact layer. At first order in the residual polarization source, its salt concentration is unchanged while its charge density responds. Both solid carrier fluxes vanish at both solid faces; the solid's total mobile charge remains its neutral initial value.

Using $u$ as the distance into the liquid from the compact layer and $D$ for electrical displacement, the electrical and charge boundaries are

$$D_s(0)=-C_b\phi_s(0), \tag{S135}$$

$$J_s(0)=J_s(L)=0, \tag{S136}$$

$$D_l(0)=D_s(L), \tag{S137}$$

$$\phi_s(L)-\phi_l(0)=D_s(L)/C_H, \tag{S138}$$

$$J_l(0)=0, \tag{S139}$$

$$\phi_l(H)=0, \tag{S140}$$

$$\rho_l(H)=0, \tag{S141}$$

$$\int_0^L \rho_s\, dx=0. \tag{S142}$$

Here $D_s=P-\epsilon_s\phi_s'$ in the solid and $D_l=-\epsilon_l\phi_l'$ in the liquid. No independently occupied charged surface sheet is present. If such a sheet is introduced, its displacement jump and kinetic balance must be added together. The last condition makes the front and rear solid displacements equal, although either can vary with time through the passive external capacitance. The limit $C_b=0$ describes a strictly isolated rear with zero rear displacement, rather than an Ohmic rear of high electrical resistance.

The weak-potential solid charge equation is the same as in A. The liquid obeys its symmetric linearized Poisson–Nernst–Planck charge equation,

$$J_l=-D_l^{\text{ion}}\left(\partial_u\rho_l+a_l\partial_u\phi_l\right), \tag{S143}$$

$$\partial_t\rho_l=-\partial_u J_l, \tag{S144}$$

$$-\epsilon_l\partial_u^2\phi_l=\rho_l, \tag{S145}$$

$$\lambda_l=\sqrt{\epsilon_l/a_l}. \tag{S146}$$

The superscript on $D_l^{\text{ion}}$ distinguishes ionic diffusivity from electrical displacement. A thermalized symmetric electrolyte with bulk concentration $c_0$ has $a_l=2qc_0L_0^2/\left(\epsilon_{\text{ref}}V_T\right)$ in the normalization already defined. The remote concentration and potential specify a large reservoir; the normal ionic current there is determined by the solution and is not artificially set to zero. This explicit liquid response retains diffuse-charge dynamics that are removed by an equilibrium capacitance approximation. [S15, S16, S18]

At stationarity the solid displacement at its two faces has the common value $d$, and its internal voltage is $(P-d)F$, where $F$ is the isolated-film function. The liquid solution gives $\phi_l(0)=d/C_D$, with

$$C_D=\frac{\epsilon_l}{\lambda_l}\coth\left(H/\lambda_l\right), \tag{S147}$$

$$S=\frac{1}{C_b}+\frac{1}{C_H}+\frac{1}{C_D}, \tag{S148}$$

$$d=\frac{PF}{F+S}, \tag{S149}$$

$$\Delta\phi_s=dS=(P-d)F, \tag{S150}$$

$$\phi_s(0)=-d/C_b, \tag{S151}$$

$$\phi_s(L)=d\left(1/C_H+1/C_D\right), \tag{S152}$$

$$\Delta\phi_H=d/C_H, \tag{S153}$$

$$\phi_l(0)=d/C_D. \tag{S154}$$

These expressions are a closed analytical solution for the stationary voltage partition. The rear dielectric drop, solid screening response, compact drop, and diffuse drop all enter the same electrostatic loop. The quantity $1/F$ acts as the differential capacitance of the screened solid for this fixed-source, neutral-charge problem; it is not a second independent surface-state capacitance.

Increasing illumination reduces $F$ and therefore reduces $d$ for a positive source. The positive solid-front, compact, diffuse, and solid-internal voltages decrease, while the negative rear potential approaches zero. The derivative is

$$\frac{dd}{d\,g_0}=\frac{P\,S\,F_m}{2\,\overline{m}\left(F+S\right)^2}, \tag{S155}$$

$$\frac{d\,\phi_s\left(L\right)}{d\,g_0}=\left(\frac{1}{C_H}+\frac{1}{C_D}\right)\frac{dd}{d\,g_0}. \tag{S156}$$

The polarity-reversed result changes all these signs. The nonpolar reference retains the pair-population response but has no first-order charge or voltage response in this symmetric blocking geometry.

The rear-isolation limit is especially informative. As $C_b \to 0$, $d \to 0$, the compact and diffuse drops vanish, and the front potential relative to the prescribed bulk liquid remains zero. However, the internal solid voltage remains $PF$ and the rear potential becomes $-PF$. Thus a changing internal photovoltage does not necessarily create a changing mean interfacial liquid voltage. This result follows from the initially neutral solid, fixed total charge, and the specified planar zero-field rear. A fixed nonzero total charge would instead give a nonzero constant front potential set by the interfacial charge; illumination alone would still not change that potential in the same blocking topology. More generally, with zero rear displacement, Gauss's law fixes the liquid displacement by the combined charge of the solid and its interfacial sheets. Internal capture and release conserve this combined inventory and therefore do not change the mean liquid displacement. A changing interfacial dipole could alter the measured solid-front potential only through its explicitly modeled additional voltage drop. The finite-$C_b$ calculation changes the electrical coupling and allows both solid and liquid charge to redistribute.

Figure S7 compares the stationary voltage partition with the corresponding potential and ionic-charge profiles across the solid and liquid.

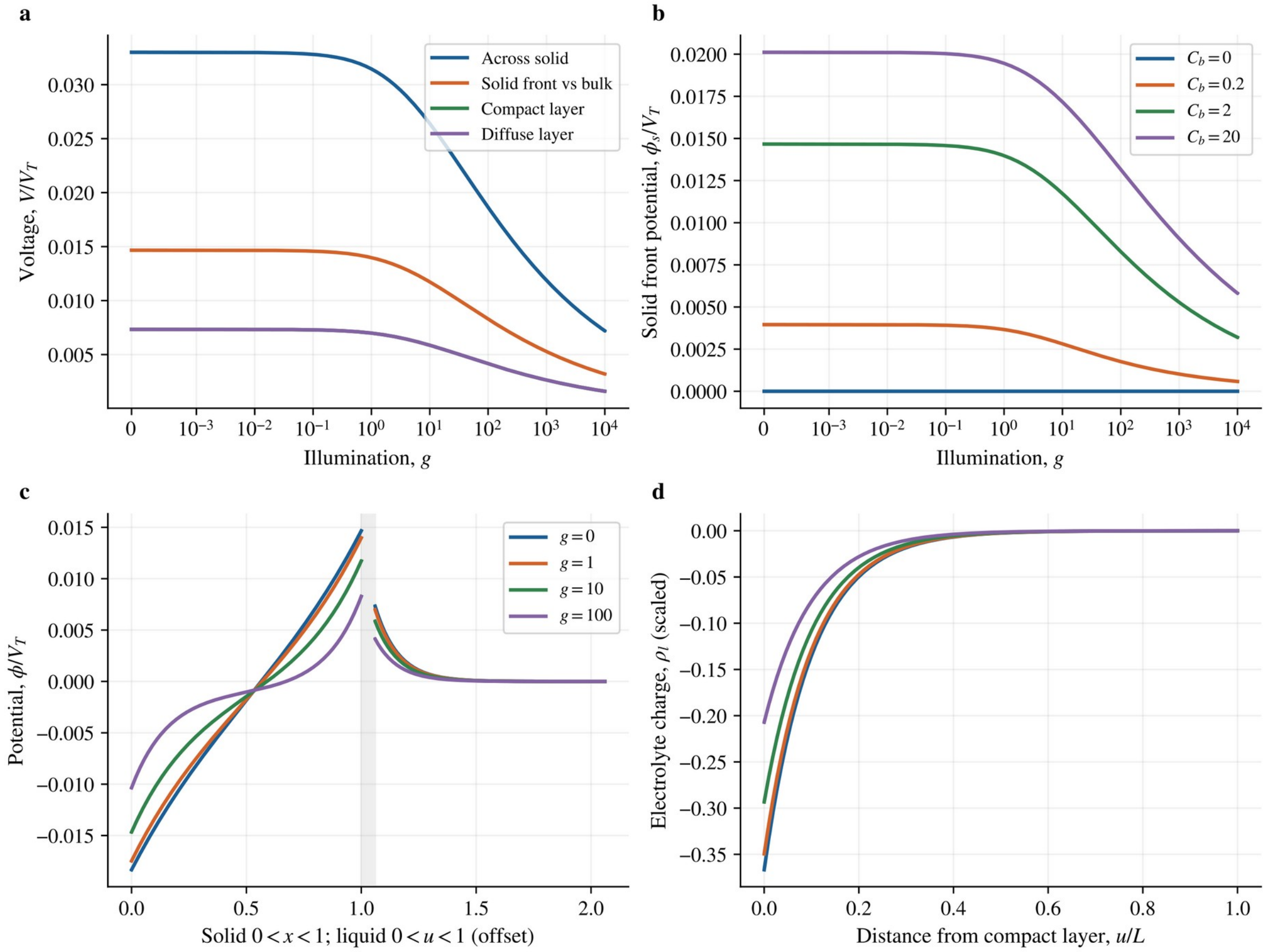

***Figure S7.*** *Stationary blocking-junction solution in C. (a) Illumination dependence of the solid-internal, solid-front, compact-layer, and diffuse-layer voltages. (b) Dependence of the measured front potential on passive rear capacitance, including strict rear isolation. (c) Solid and liquid depth profiles; the narrow shaded interval indicates the compact voltage jump, not a spatially resolved layer. (d) Equilibrium liquid charge profiles. Parameters are those of A together with* $H=1$, $\epsilon_l=0.5$, $\lambda_l=0.1$, $C_H=5$, *and baseline* $C_b=2$. *The negative source gives the sign-reversed profiles; the matched nonpolar control gives zero potential and charge.*

## S4.2 Coupled solid and ionic response

The temporal problem can also be reduced analytically without replacing the liquid by a fitted relaxation time. For a homogeneous charge-transport layer with permittivity $\epsilon$, diffusivity $D_q$, and charge coefficient $a$, let $\kappa_s^2=a/\epsilon+s/D_q$. A bounded basis for its homogeneous potential response is

$$\hat{\phi}_h=c_0+c_1 x+A e^{-\kappa_s x}+B e^{-\kappa_s(1-x)}. \tag{S157}$$

Its accompanying displacement, charge, and current are obtained from $\hat{D}=-\epsilon\hat{\phi}_h{}'$, $\hat{\rho}=-\epsilon\hat{\phi}_h{}''$, and $\hat{J}=-D_q\left(\hat{\rho}\,'+a\hat{\phi}_h{}'\right)$. The constant and linear terms must be retained: the linear term carries an Ohmic charge current and the exponentials carry the screening modes. Writing only a decaying diffuse-layer potential would remove an allowed part of the dynamic solution. The square root is chosen with nonnegative real part, so neither basis exponential grows through the layer.

In the solid, illumination additionally changes $a = a_s^0 m$. Let $\psi_0$ denote the decaying part of the stationary solid potential, whose inverse length is $\bar{\kappa} = \sqrt{a_s^0 \bar{m} / \epsilon_s}$. Since the stationary charge current vanishes, no stationary linear-potential current term is present. A particular optical solution for unit $\hat{g}$ is

$$\hat{\phi}_p = \frac{-D_s a_s^0}{\epsilon_s s(s+2\bar{m})} \psi_0, \tag{S158}$$

$$\hat{D}_p = -\epsilon_s \hat{\phi}_p', \tag{S159}$$

$$\hat{\rho}_p = -\epsilon_s \hat{\phi}_p'', \tag{S160}$$

$$\hat{J}_p = -s \hat{D}_p. \tag{S161}$$

The liquid has no direct optical source in this model. Four coefficients in the solid and four in the liquid are fixed by the linearized electrical boundaries, zero rear solid current, zero front liquid current, and solid charge neutrality. Neutrality replaces one redundant solid blocking-current equation because $\hat{J}_s + s\hat{D}_s$ is depth independent. The resulting eight-by-eight linear system gives all requested complex voltages and charge responses. Although the chosen particular solution contains $1/s$, the complete solution has a finite zero-frequency limit. At zero frequency we use the exact stationary derivative directly; the evaluated finite-frequency solution approaches it continuously.

The gain enhancement in C differs from the monotonic modal response in A. During a rapid increase in carrier density, the solid can change its screening profile while the liquid charge is still close to its preceding distribution. The measured front voltage initially reflects that delayed ionic compensation. As the ions redistribute, part of the excursion recovers toward the smaller stationary change. The phase advance relative to the signed static derivative and the light-step overshoot are two representations of this same voltage-partition dynamics. They do not require a change of polarization sign, a chemical reaction, or electronic surface traps.

For $g_0 = 1$, the front-potential modulation gain peaks at approximately 1.79 times its static value for the baseline ionic diffusivity. Slowing ionic motion by a factor of 100 increases this factor to approximately 2.57; making it 100 times faster removes the enhancement over the evaluated range. The diffuse-layer potential is more sensitive to this separation of timescales, with corresponding peak factors of approximately 3.17 and 4.68 for the baseline and slow liquids. The stationary curves are unchanged because ionic diffusivity affects the relaxation, rather than the equilibrium screening length, in this parameter variation.

For finite steps, the neutral population retains the exact solution used in A. The remaining charge equations form a linear system with time-dependent coefficients because $m(t)$ changes the solid screening coefficient. The full evolution may be written with its linear propagator; finite steps do not generally reduce to a single exponential because the solid and liquid operators are coupled. We evaluate this charge evolution with a conservative finite-volume method and an implicit time integrator, using the analytical population at every time. Electrostatics is solved as a linear charge-to-potential map, and one dependent solid charge cell is eliminated so that the prescribed net charge is maintained exactly. These calculations resolve the noncommuting solid/ionic relaxation rather than assuming instantaneous capacitances for all times.

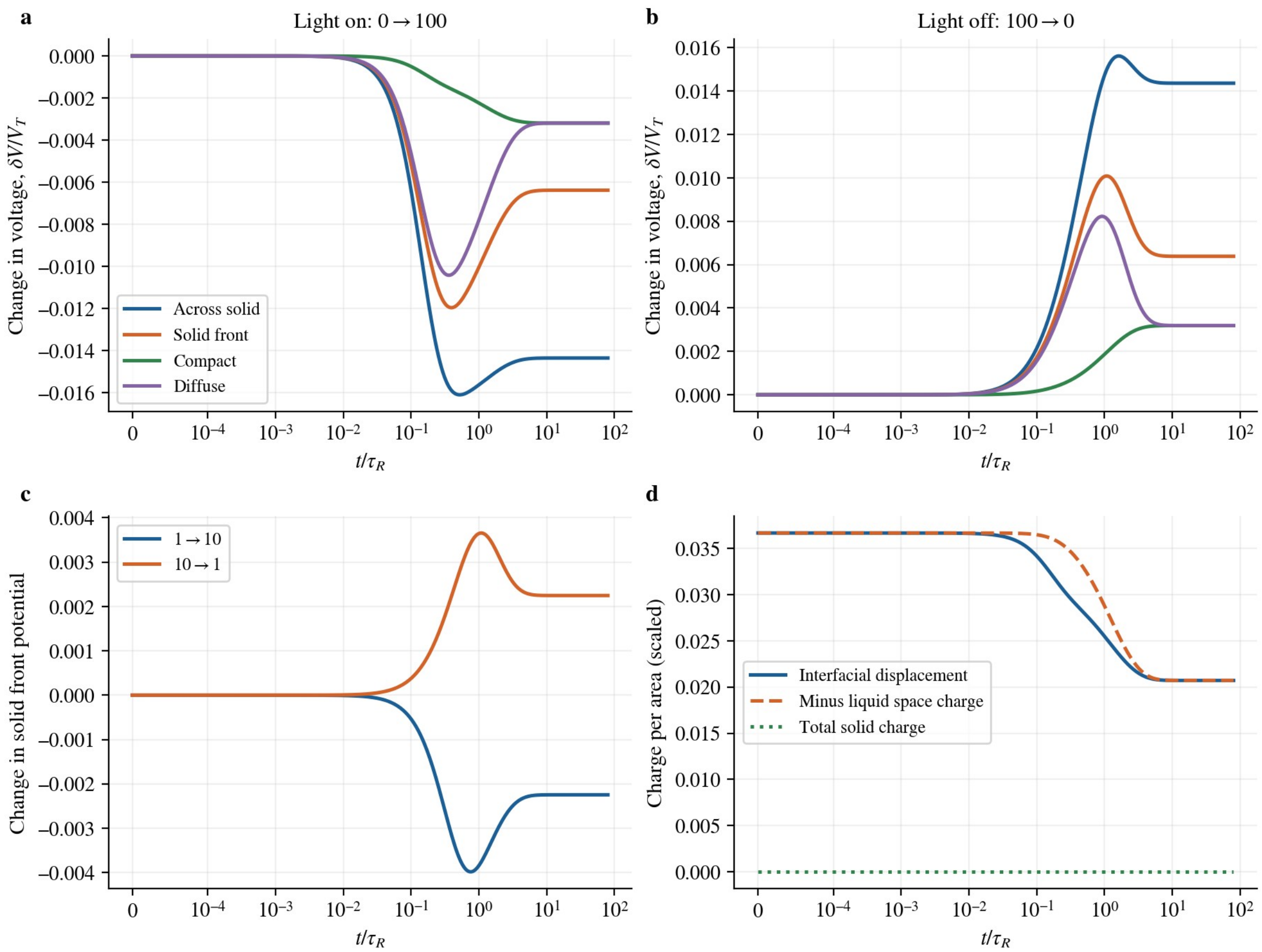

***Figure S8.*** *Finite illumination steps in the blocking solid–liquid junction. (a,b) Changes in each voltage after light-on and light-off between* $g=0$ *and* $100$. *(c) Front-potential changes for steps between two nonzero mean intensities. (d) Interfacial displacement, the negative of the integrated liquid space charge, and the conserved total solid charge. During the transient the first two need not coincide: their difference is the displacement at the remote liquid boundary. The compact voltage changes more slowly than the initial diffuse and solid-front excursion in this example. All traces approach the analytically determined stationary endpoints.*

The charge ledger explains why an integrated diffuse-layer charge cannot always be identified with the front displacement during a transient. Gauss's law gives $D_l(H)=D_s(L)+\int_0^H \rho_l\,du$. The remote displacement is approximately zero at stationarity for the chosen ten-screening-length liquid. During redistribution it can be nonzero and accompanies the current supplied by the remote ionic reservoir. A model that simultaneously sets that displacement to zero, fixes the reservoir potential, and retains an independently evolving diffuse charge would impose an extra constraint and change the response.

The nonmonotonic front voltage is produced by delayed ionic compensation and changing voltage partition, with frozen polarization and no Faradaic channel. Its amplitude and timing belong to the specified liquid height, diffusivity, and passive rear capacitance. The strict rear-isolation limit fixes the mean liquid-facing displacement; it can retain an internal solid photovoltage without the finite-capacitance front-to-liquid optical signal.

### S4.3 Static doping extension

In the passive planar doping reduction, the screened solid dipole is multiplied by the stated passive capacitance division. The solid screening length is set by $n_b+p_b$, so equal-magnitude donor and acceptor densities give equal voltage magnitudes under the same capacitances. This is the static extension of the blocking topology. The doping extension in C is restricted to stationary electrostatics. The nonlinear electrolyte and finite liquid-height problem is not replaced by this static summary.

## S5. Case D: reactive interface with a rear carrier return

### S5.1 Surface kinetics, electrical control, and contact selectivity

Case D opens a front-surface reaction and specifies the electronic return that sustains it. The primary example pairs front reduction with a hole-equilibrating, electron-blocking rear contact at fixed bias. Its depth profiles, reaction affinities, and optical response follow from the same carrier populations. A floating rear metal and two available front reactions are subsequently examined as explicit operating variants of the contacted geometry.

Use the inward coordinate, normalization, common polarization/carrier equations, and stationary-state linearization defined for Case B above. The reactive set has SRH recombination, time scale $\tau_0$, and the different parameter values specified in the contact parameter table. Its actual front and rear conditions are given here.

The conducting liquid is represented by fixed reaction-plane activities and a bulk potential chosen as zero. A compact capacitance $c_H = C_H V_T / P_*$ supplies the interfacial electrostatic condition. This reduced model assumes negligible resolved liquid voltage loss and no independently evolving diffuse-layer or reagent-transport field; it must not be presented as a solution of finite-conductivity electrolyte transport.

The dimensionless transfer prefactors are $s_{n0} = S_{n0} \tau_0 / L$ and $s_{p0} = S_{p0} \tau_0 / L$. The maintained couples have electron electrochemical potentials $\mu_r$ and $\mu_o$, with $\rho_r = (\mu_r - E_i^0)/(k_B T)$ and $\rho_o = (\mu_o - E_i^0)/(k_B T)$ relative to the solid's intrinsic energy reference $E_i^0$. Their reaction-plane state and the compact activation fractions are fixed constitutive inputs. The normalized reduction and oxidation rates are

$$r_r = s_{n0} e^{-\alpha_n v_s} \left( n_s - e^{v_s + \rho_r} \right), \tag{S162}$$

$$r_o = s_{p0} e^{\alpha_p v_s} \left( p_s - e^{-v_s - \rho_o} \right). \tag{S163}$$

All executed reactive calculations use $\alpha_n = \alpha_p = 1/2$. The front conditions are

$$d(0) = -c_H v_s, \tag{S164}$$

$$j_n(0) = r_r, \tag{S165}$$

$$j_p(0) = -r_o. \tag{S166}$$

The primary D calculation sets $s_{p0} = 0$ and imposes

$$v(1) = 0, \tag{S167}$$

$$j_n(1) = 0, \tag{S168}$$

$$p(1) = 1. \tag{S169}$$

The rear is a hole reservoir and blocks electrons. The fixed rear potential relative to the maintained liquid is an operating electrical constraint, with its external connection and complementary counterreaction understood. Zero applied bias does not mean isolation. The retained imposed-bias maps use the cathodic convention $v(1) = -V/V_T$, where $V$ is the dimensional applied bias. A contact exchanging both species would be a different boundary law; it is not substituted here because the metal is described as conducting.

For a sole front oxidation channel, the corresponding selective return instead equilibrates electrons and blocks holes: $n(1) = 1$, $j_p(1) = 0$, and the specified rear potential is retained. Under intrinsic electron–hole-symmetric statistics, exchange $n \leftrightarrow p$, $M_n \leftrightarrow M_p$, $(u, v, d) \mapsto (-u, -v, -d)$, and $(j_n, j_p) \mapsto (-j_p, -j_n)$. Exchange the transfer prefactors and activation fractions and set $(\rho_r, \rho_o) \mapsto (-\rho_o, -\rho_r)$; a nonzero applied bias also changes sign. This maps the front-reduction solution exactly to the front-oxidation counterpart, including its optical amplitude and phase. The zero-bias calculation therefore supplies the conjugate oxidation result without a redundant numerical run when those symmetry assumptions hold. Unequal material statistics, changed chemistry, or a nonselective contact require their actual equations instead.

The exact grounded-voltage reduction is

$$(\epsilon^* + c_H) v_s = -\int_0^1 u \, dx + \delta \int_0^1 (1-x)(p-n) \, dx. \tag{S170}$$

The compact charge is consequently coupled to the finite-depth carrier and polarization state, not prescribed as an extra independent voltage relaxation.

The reduced liquid is a maintained reaction-plane reservoir: it supplies activities and an electrical reference but does not resolve a finite liquid voltage loss or evolving diffuse layer. The primary rear is electron-blocking and hole-equilibrating. Replacing it with a contact that exchanges both carriers changes the operating problem.

Table S2 compares the stationary reduction and rear carrier currents for selective and nonselective rear contacts.

*Table S2. Effect of rear carrier selectivity on stationary reduction and return currents.*

| **Prepared state** | **Reduction with selective rear** | **Reduction with both-carrier rear** | **Rear electron current, both-carrier case** | **Rear hole current, both-carrier case** |
|---|---:|---:|---:|---:|
| Outward polarization | 3.64557502 | 0.00122472 | −4.31483225 | 4.31605697 |
| Nonpolar reference | 2.84524601 | 0.00084956 | −4.31370539 | 4.31455495 |
| Inward polarization | 1.28126977 | 0.00075416 | −4.31530348 | 4.31605764 |

When both carriers equilibrate at the rear, the large opposite rear currents remove most generated pairs, leaving much less front transfer in this example. The integrated electron balance is $j_n(0) - j_n(1) = \int_0^1 (g-R) \, dx$, while the hole balance is $j_p(1) - j_p(0) = \int_0^1 (g-R) \, dx$. Both hold for each contact. The current difference between contacts is therefore a consequence of the specified transport and exchange laws rather than a missing source or uncompensated charge.

## S5.2 Reaction affinity, carrier return, and transport loss

Dimensionless carrier quasi-Fermi energies relative to the intrinsic dark reference are

$$f_n = \ln n - v, \tag{S171}$$

$$f_p = -\ln p - v. \tag{S172}$$

The local forward affinities in units of $k_B T$ are

$$\eta_r = \ln n_s - v_s - \rho_r, \tag{S173}$$

$$\eta_o = \ln p_s + v_s + \rho_o, \tag{S174}$$

$$\eta_r + \eta_o = \ln(n_s p_s) - (\rho_r - \rho_o). \tag{S175}$$

The specified positive prefactors make each rate have the sign of its corresponding affinity. Electrostatic band bending lowers electron energies by $q\phi$; it is not itself a reaction affinity. The linearized optical affinity response is $\tilde{\eta}_r = \tilde{n}_s / \bar{n}_s - \tilde{v}_s$ and $\tilde{\eta}_o = \tilde{p}_s / \bar{p}_s + \tilde{v}_s$. Both are evaluated from the same state resolvent as the voltage.

The exact transport laws can also be written $j_n = M_n n f_n'$ and $j_p = M_p p f_p'$. The nonnegative stationary transport dissipation, normalized by $J_* V_T$, is therefore

$$\Pi_{\mathrm{tr}} = \int_0^1 \left[ \frac{j_n^2}{M_n n} + \frac{j_p^2}{M_p p} \right] dx . \tag{S176}$$

For the single-channel steady D calculation, $j_n + j_p = r_r$ throughout the film. The reported ratio $\Pi_{\mathrm{tr}} / |r_r|$ is a transport-loss-per-terminal-current diagnostic in units of $V_T$, where the current is nonzero. It records irreversible transport rather than assigning the complete electrostatic front–rear variation to an Ohmic loss. Generation, recombination, chemical transfer, and any imposed bias require their own terms for a complete energy budget. The ratio is not used for common-pair floating operation, whose terminal current is zero while internal dissipation can remain finite.

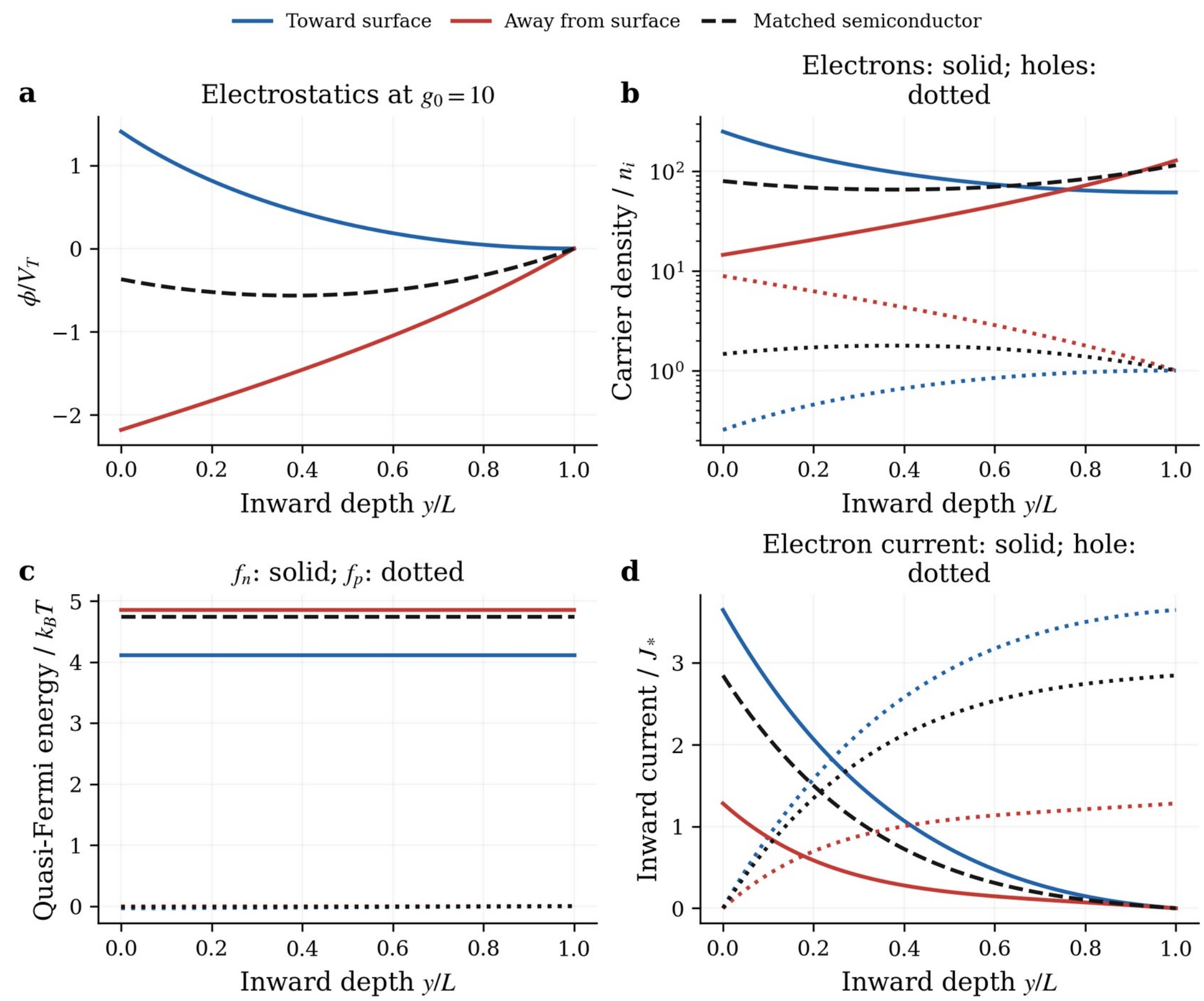

***Figure S9.*** *Finite-depth reaction and return in case D at generation 10. The same polarization and passive controls are shown in electrostatic potential, electron and hole densities, quasi-Fermi energies, and conventional current. The electron current is zero at the rear, while the hole density is fixed there. Electron and hole currents can vary separately through generation/recombination, but their stationary sum is constant. Solid and dotted styles distinguish carrier species; the passive electron trace remains dashed. The profiles are independent collocation solutions, not a uniform-carrier approximation.*

A separate carrier free-energy identity checks the interpretation of the transport loss. Define $f_n = \ln n - v$, $f_p = -\ln p - v$, and $l = \ln(np) = f_n - f_p$ in the inward normalized coordinate. The stationary equations imply

$$\Pi_{\mathrm{tr}} = \left[ j_n f_n + j_p f_p \right]_0^1 + \int_0^1 (g - R) l \, dx . \tag{S177}$$

Writing the rear boundary contribution as $W_b = \left[ j_n f_n + j_p f_p \right]_{x=1}$ and using the two reversible front rates yields

$$\int_0^1 g \ln(np) \; dx + W_b = \Pi_{\mathrm{tr}} + \int_0^1 R \ln(np) \; dx + r_r \eta_r + r_o \eta_o + r_r \rho_r - r_o \rho_o . \tag{S178}$$

The transport term, recombination term, and each interfacial dissipation term are nonnegative for the specified constitutive laws. The final terms describe signed chemical work. For equal stationary partial rates between distinct couples they reduce to $r_r(\rho_r - \rho_o)$, which is negative for reverse conversion driven by a maintained chemical gap. A common-pair floating stationary state can have zero terminal current and positive internal dissipation; dividing by the zero terminal current is not meaningful there. The gain term $\int g \ln(np) \; dx$ is the carrier free-energy input in this model. It is not a complete incident-photon energy balance or a solar-to-chemical efficiency.

**S5.3 Optical response and finite carrier collection**

For D define $a_r = s_{n0} e^{-\alpha_n \bar{v}_s}$, $a_o = s_{p0} e^{\alpha_p \bar{v}_s}$, $\bar{n}_r = e^{\bar{v}_s + \rho_r}$, and $\bar{p}_o = e^{-\bar{v}_s - \rho_o}$. Then

$$\tilde{r}_r = a_r \tilde{n}_s - \left[ \alpha_n \bar{r}_r + a_r \bar{n}_r \right] \tilde{v}_s , \tag{S179}$$

$$\tilde{r}_o = a_o \tilde{p}_s + \left[ \alpha_p \bar{r}_o + a_o \bar{p}_o \right] \tilde{v}_s . \tag{S180}$$

The front perturbations obey $\tilde{d}(0) = -c_H \tilde{v}_s$, $\tilde{j}_n(0) = \tilde{r}_r$, and $\tilde{j}_p(0) = -\tilde{r}_o$. The primary rear conditions are $\tilde{v}(1) = 0$, $\tilde{j}_n(1) = 0$, and $\tilde{p}(1) = 0$. Floating operation instead determines the rear voltage and imposes

$$i\Omega \frac{c_H}{\delta} \tilde{v}_s = \tilde{r}_r - \tilde{r}_o . \tag{S181}$$

Natural polarization perturbations apply at both faces. All mean-dependent coefficients are evaluated at the same stationary illuminated state as the frequency response.

A minority-carrier limit for D instead has a reflecting rear, finite front transfer, negligible drift at the retained order, a clamped majority background, and constant linear recombination coefficient $k$. Writing $q_\Omega = \sqrt{(k + i\Omega)/M}$ with positive real part, the front-current response to a depth-dependent source is

$$\tilde{j}_s = \int_0^1 \frac{S \cosh\left[ q_\Omega (1 - x) \right]}{M q_\Omega \sinh q_\Omega + S \cosh q_\Omega} \tilde{G}(x) \; dx . \tag{S182}$$

For uniform source amplitude,

$$\frac{\tilde{j}_s}{\tilde{g}} = \frac{S M q_\Omega \tanh q_\Omega}{(k + i\Omega)\left[ S + M q_\Omega \tanh q_\Omega \right]} . \tag{S183}$$

The steady fast-transfer limit gives $l \tanh(1/l)$ with $l = \sqrt{M/k}$, and finite $S$ adds the interface resistance. The carrier-blocking rear in this limit matches the electron boundary of the primary D problem. Maintaining the majority population requires the declared hole return. The kernel predicts current collection under its assumptions; it does not predict the local potential without electrostatics and the complementary carrier response. It is therefore presented beside, rather than substituted for, the full numerical D voltage and affinity spectra.

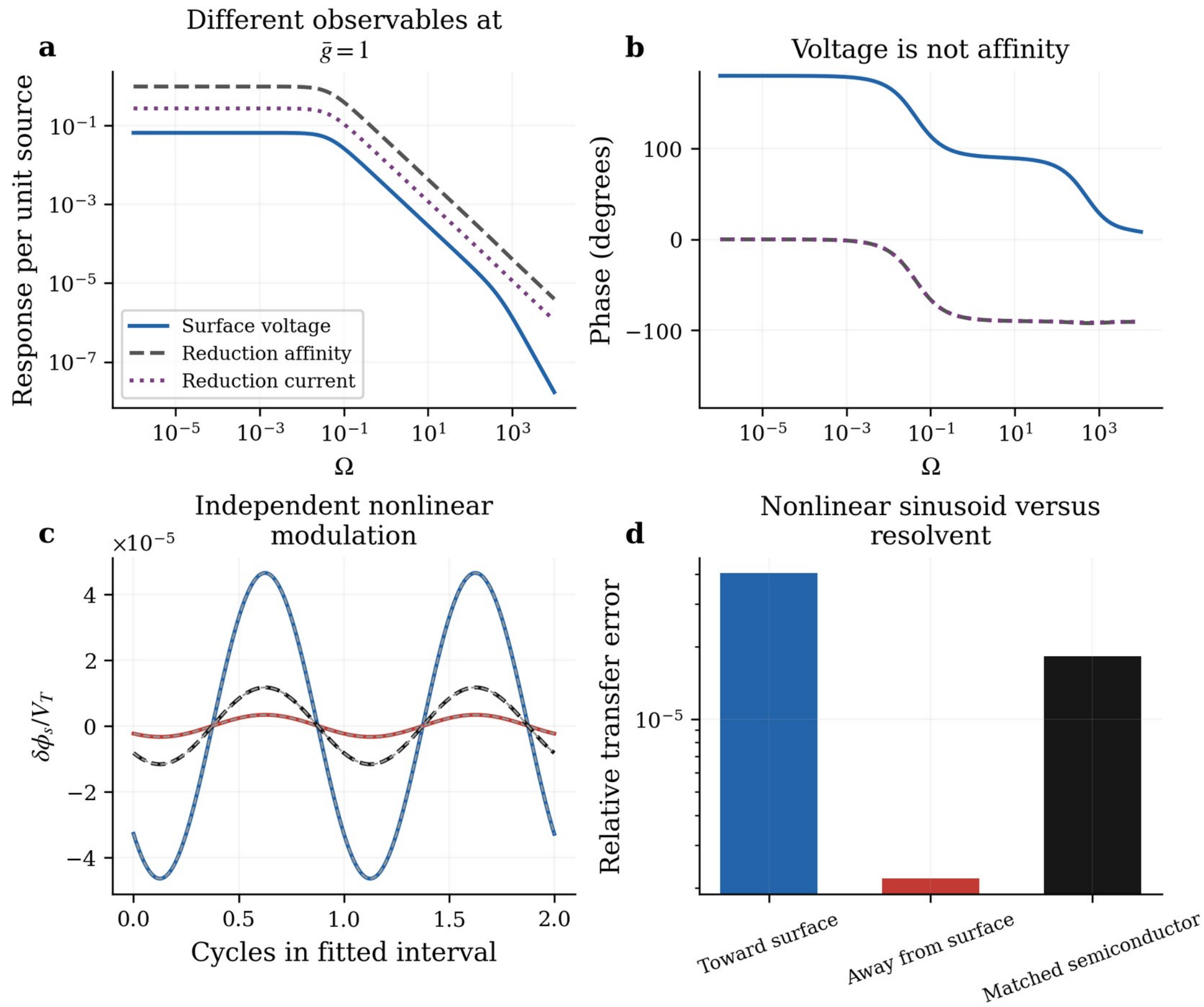

***Figure S10.*** *Potential, reduction affinity, and reduction-current transfer functions for the outward-polarized fixed-bias reactive film at mean generation* $\overline{g}=1$. *Each is differentiated from the same carrier/electrostatic state, and their phases differ. The lower panels compare selected independently integrated nonlinear sinusoidal responses with their linear predictions for both polarities and the passive control. The source modulation is* $0.1\%$ *of the mean; each polarity uses the frequency recorded in the checks. Gray dashed traces show the linear prediction. The largest fitted relative complex voltage-transfer discrepancy is approximately* $4.1\times10^{-5}$.

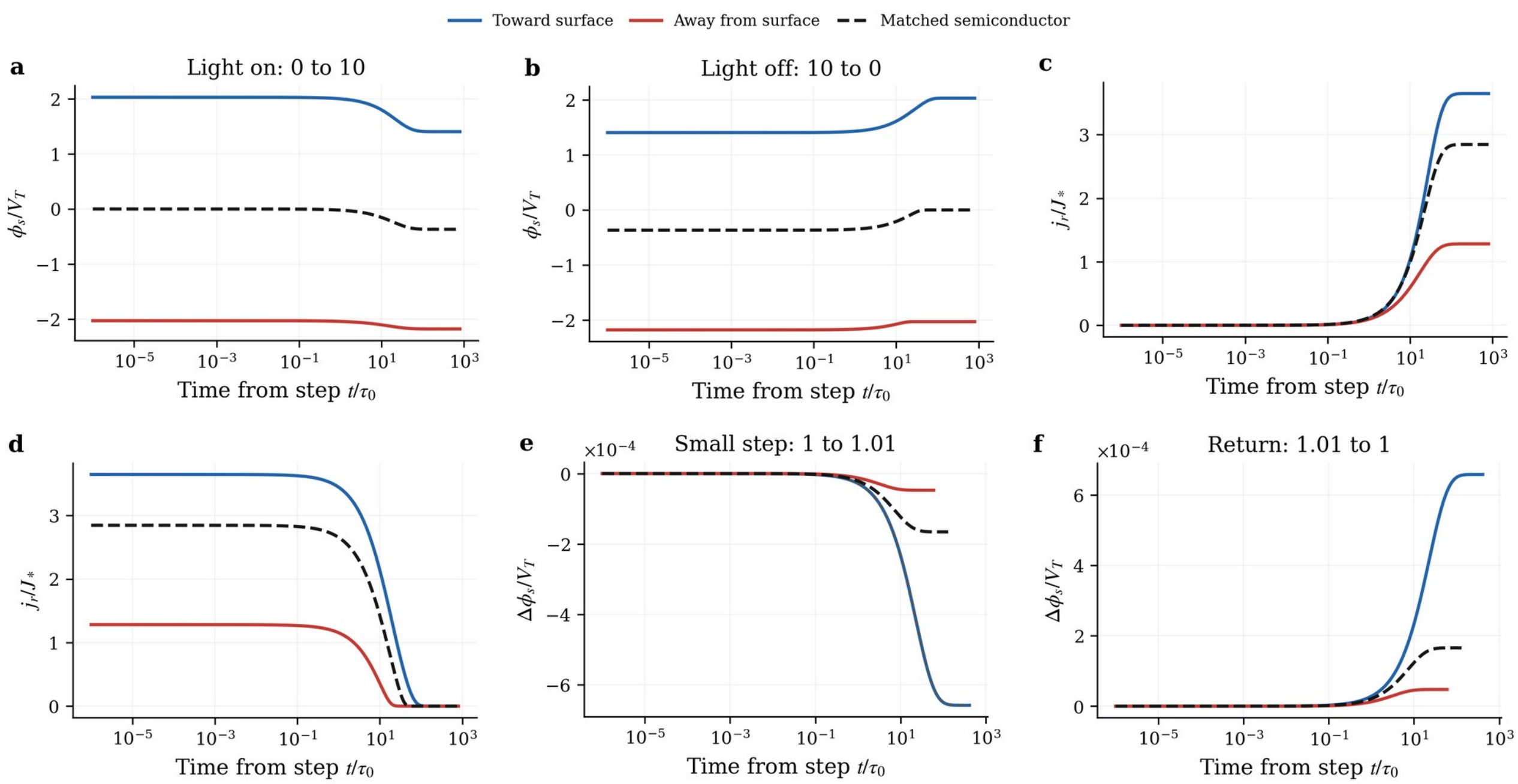

*__Figure S11.__ Full optical step comparison for the primary fixed-bias reactive interface. Panels a,b show the potential during light-on and light-off; c,d show the corresponding reduction current, whose rear return is carried by holes. Each large step has its own time origin and is evolved through the stationary endpoint. Panels e,f show a 1% increase about a nonzero mean and its return; dashed gray in e is the linear-response prediction for outward polarization. The passive control remains reactive and has a nonzero optical potential because the carrier-selective boundary is retained.*

Potential, affinity, and current are differentiated from the same operating state, but have different transfer functions. Finite light-on/off and nonzero-mean steps retain the active reaction and rear return. The product amount is obtained from the appropriate partial Faradaic rate and stoichiometry, not by integrating displacement current or the dry-site capture current.

## S5.4 Floating metal, paired reactions, and chemical conversion

In the floating-metal comparison, $v(1)$ becomes an unknown while $j_n(1)=0$ and $p(1)=1$ remain. Zero external total current gives the additional dynamic closure

$$\frac{c_H}{\delta}\dot{v}_s=r_r-r_o, \tag{S184}$$

and the rear voltage follows from

$$v(1)=v_s+\frac{\int_0^1 u\;dx+c_H v_s-\delta\int_0^1(1-x)(p-n)\;dx}{\epsilon^*}. \tag{S185}$$

The compact charge changes when the partial rates are unequal. At stationarity $r_r=r_o$ and the rear hole flux vanishes. During evolution the rear metal still exchanges holes and its displacement changes; it is not the inactive rear used in cases C or E.

For the common pair, $\rho_r=\rho_o=0$, equal positive rates cycle the same dissolved species and remove photocarriers without net accumulation of that reduced species. For distinct couples, the executed gap is $\rho_r-\rho_o=4$, with separately maintained inventories. Equal partial rates then represent conversion between those reservoirs. The dark state can carry reverse chemical conversion and is not automatically equilibrium. Signed rates are reported rather than treating every nonzero rate as an optical collection current.

The same redox pair and two distinct couples have different chemical meanings even when their electron-equivalent rates balance. For the common pair, equal positive reduction and oxidation rates recycle its reduced and oxidized species and remove photocarriers. For distinct maintained couples with $\rho_r-\rho_o=4$, equal rates instead represent conversion between the specified reservoirs. Their dark state supports reverse conversion. The refined illumination crossings are $g_0=2.1580615$ for polarization toward the surface, 5.3540787 for polarization away, and 4.5950721 for the passive control. At each zero-rate crossing the surface quasi-Fermi splitting equals $4k_BT$. The finite-depth carrier populations and SRH kinetics determine those different intensities; a uniform-carrier threshold from another limit is not substituted for them.

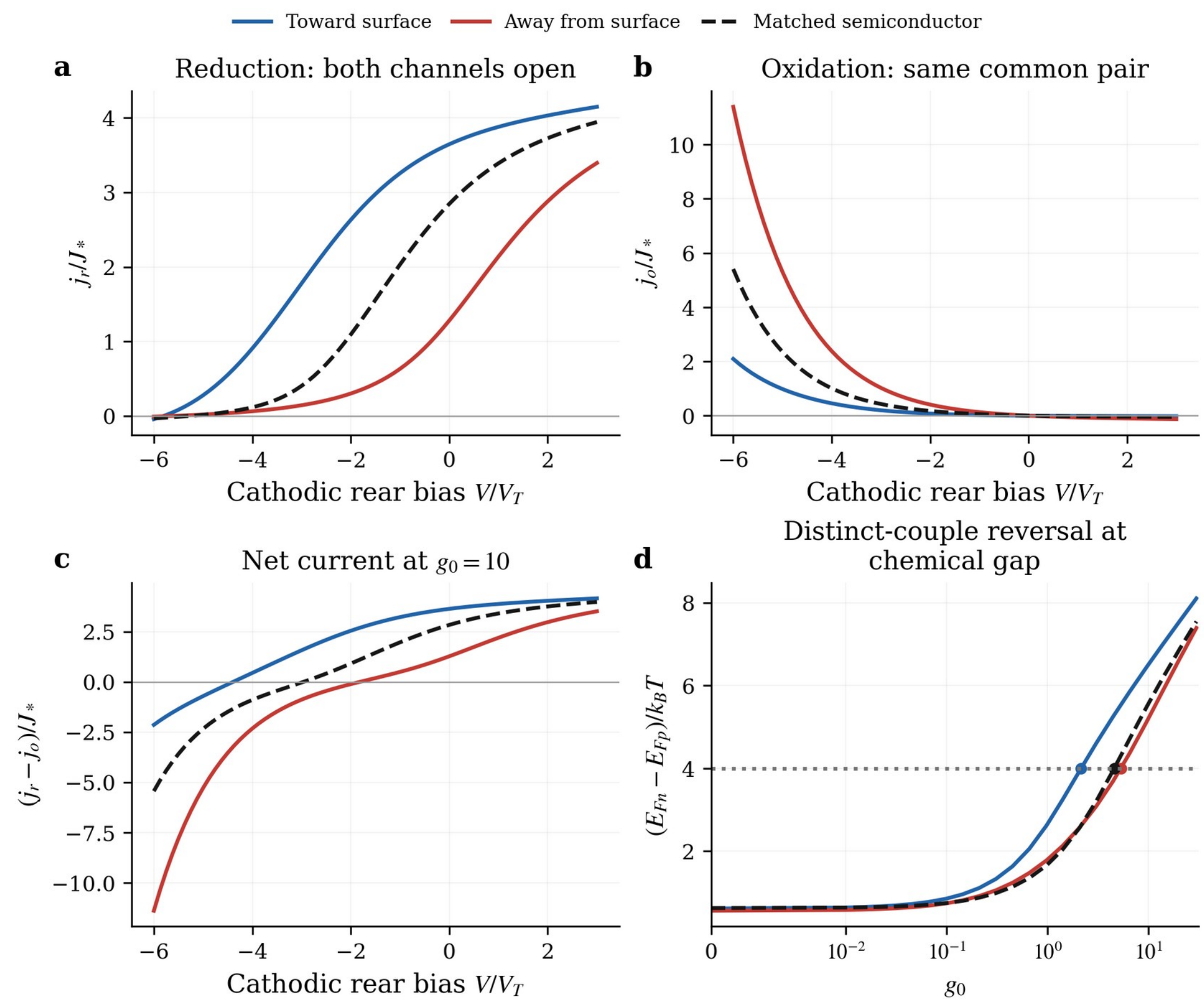

***Figure S12.*** *Extended contacted-interface analyses. Panels a–c show both common-pair partial currents and their net current versus imposed cathodic rear bias at $g_0=10$, with the actual electron-blocking and hole-equilibrating rear. The passive control has the same selective boundary. Panel d uses floating operation and distinct couples with chemical gap $4k_BT$; the independently refined zero-rate intensities occur when the surface quasi-Fermi splitting reaches that gap. Applied bias, floating closure, partial current, and chemical conversion remain separately identified.*

The floating stationary voltage need not follow the fixed-bias voltage (Figure S13). For example, the outward-polarized common-pair surface at source 10 has $v_s\simeq 2.70867$ and equal partial rates near $0.62049$, while the fixed-bias single-channel surface has $v_s\simeq 1.40519$ and a reduction rate near $3.64560$. The carrier boundary, electrostatic closure, and number of active front channels all differ. Calling both conditions open circuit or simply describing both rear objects as metal would conceal that distinction.

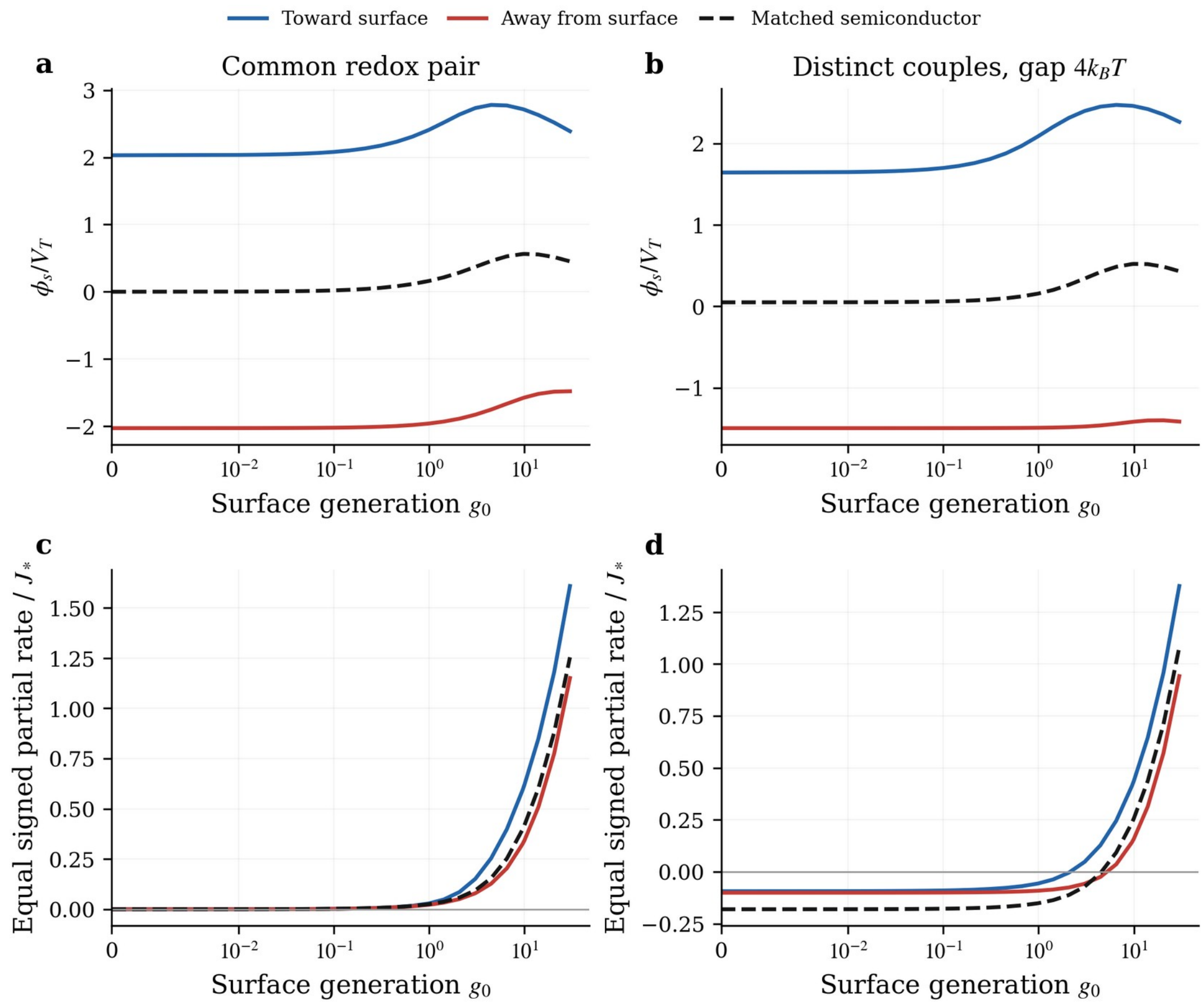

***Figure S13.*** *Uniform floating-metal closure comparisons within the contacted reactive geometry. Both front channels are open and the hole-equilibrating rear remains carrier-permeable; its total external current is zero. The left column uses the same redox pair, so equal positive partial rates recycle its species. The right column uses distinct maintained couples with a* $4\,k_B T$ *chemical gap; the dark reverse conversion changes sign under sufficient illumination. The floating voltage follows both the charge balance and finite-depth electrostatics. The passive control retains the same selective rear and chemistry.*

The floating metal remains a hole reservoir; zero external total current determines its evolving potential and does not turn it into the inactive rear of C or E. Unequal transient partial rates charge the interface. Equal stationary rates recycle a common mediator, whereas two distinct maintained couples can support net chemical conversion.

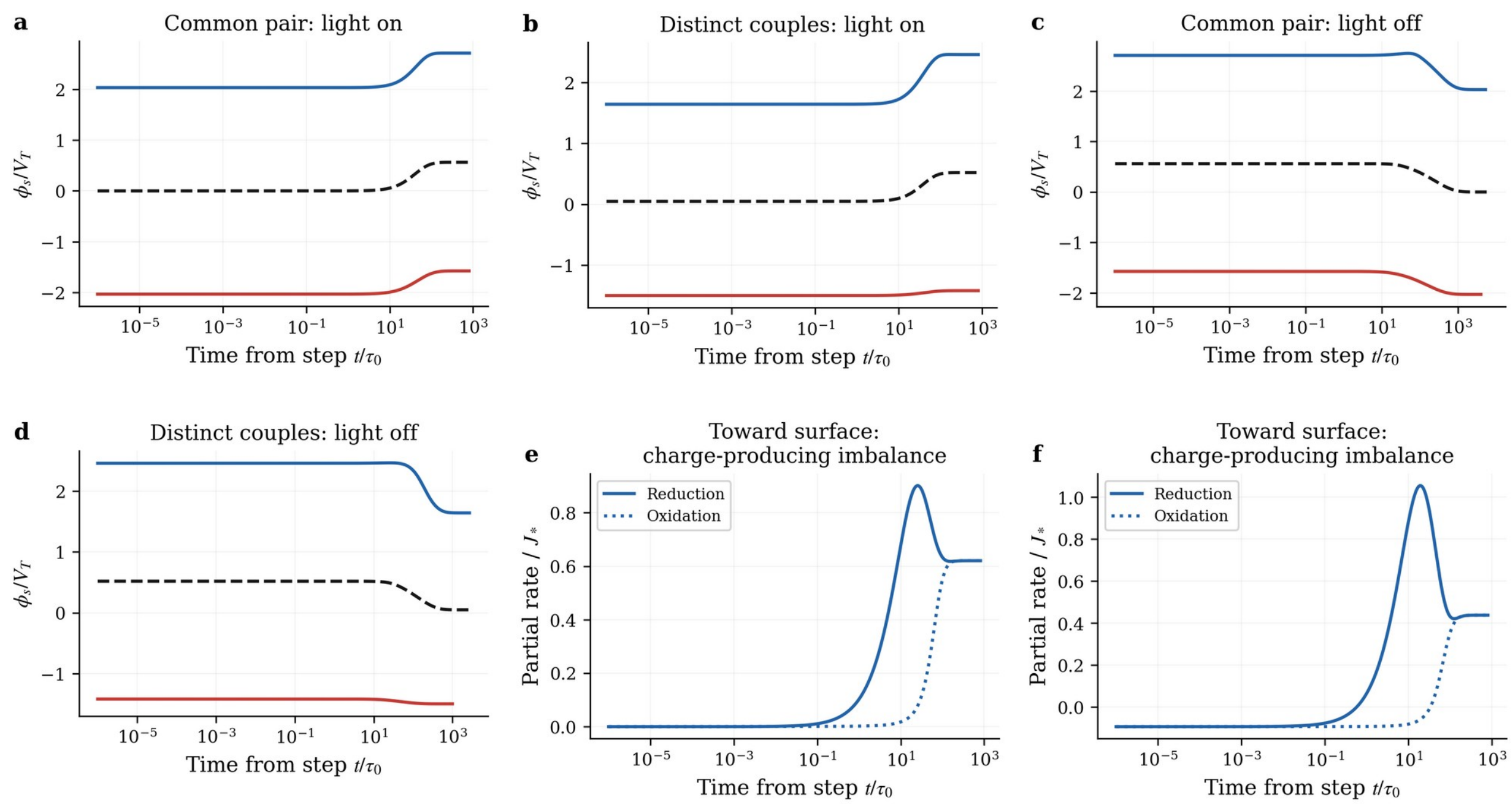

***Figure S14.*** *Light-on and light-off comparison for uniform floating-metal operation, for common-pair recycling and distinct-couple conversion. Both polarities and the passive control are retained in the voltage panels. Panels (e,f) resolve both partial rates for outward polarization after the light-on step. Their temporary imbalance changes compact interfacial charge through* $\left(c_H/\delta\right)\partial_t v_s = r_r - r_o$ *in dimensionless time; the integrated relation is verified directly. This mechanism has no finite electronic-site inventory and is distinct from the finite-site response of Case B. The rear remains a hole-equilibrating metal; it is not the inactive rear of the isolated liquid cases.*

The distinct-couple voltage spectra use the chemical gap stated in their caption. They must not be combined with common-pair transient curves as though they described one chemical system.

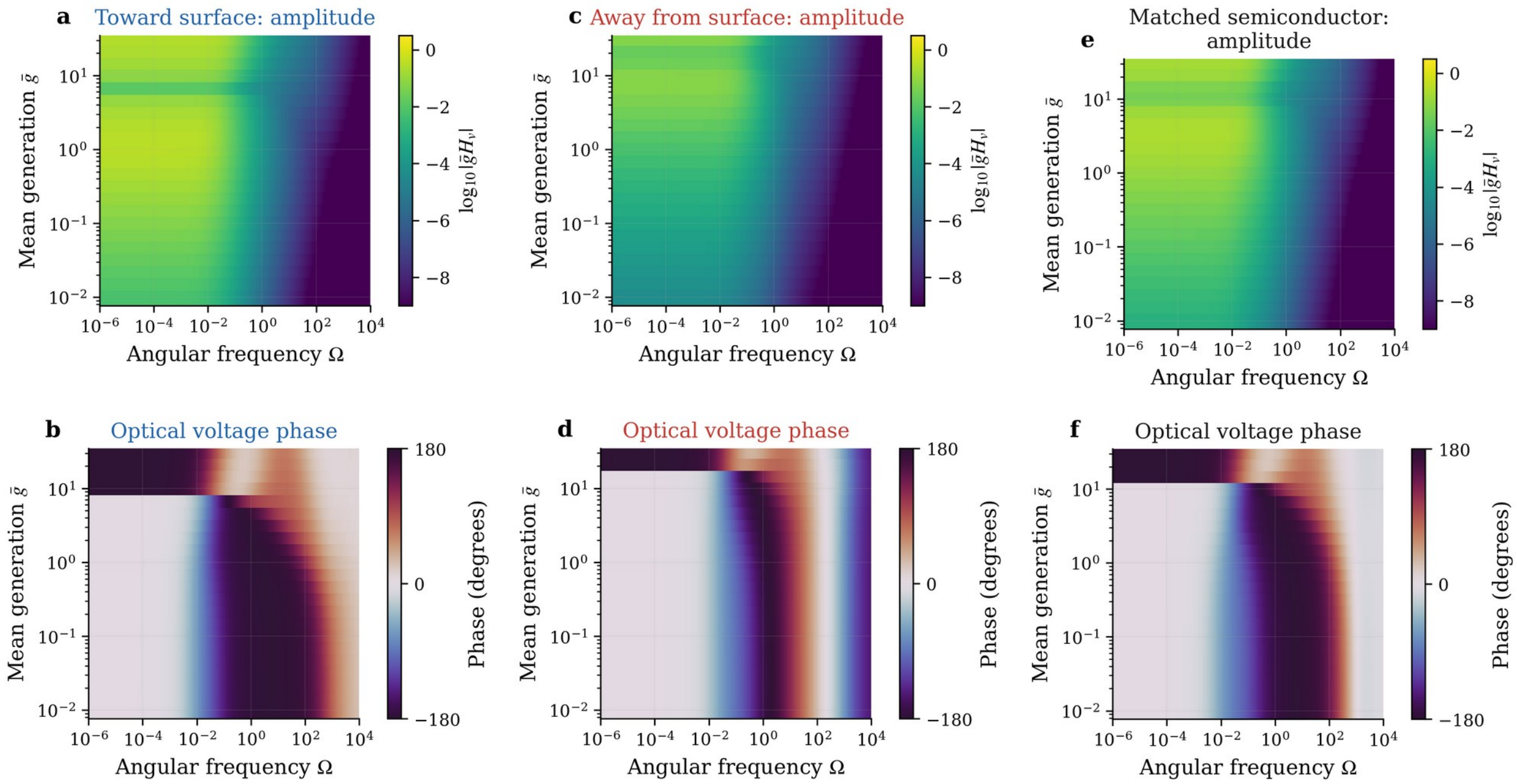

***Figure S15.*** *Evaluated optical voltage spectra for two front channels and a floating rear metal, distinct couples with gap* $4k_BT$*. Each column uses the indicated prepared polarization or nonpolar semiconductor control; amplitude is shown above its corresponding phase. The source perturbation is linearized about its own stationary illuminated state. Amplitude is* $|\bar{g}H_v|$ *and phase is relative to a positive cosine source. The 161 frequencies and 26 means are evaluated from the full finite-depth Jacobian; the maps display the positive means. White phase cells indicate* $|H_v|<10^{-13}$*, for which no phase is displayed. No local voltage is substituted for the carrier affinity. The rear metal remains hole-permeable while its total external current is zero. Equal stationary partial rates describe signed conversion between the two distinct maintained couples. Dimensionless angular frequency is* $\Omega=\omega_{\mathrm{phys}}\tau_0$*.*

Interfacial speed and carrier mobility are independent controls on collection. When both complementary transfer channels approach the available-pair ceiling, the polarity contrast in the rate can decrease even while the polarization states and potentials remain different.

Figure S16 compares the effects of interfacial transfer speed and carrier mobility on collection under fixed-bias and paired-reaction conditions.

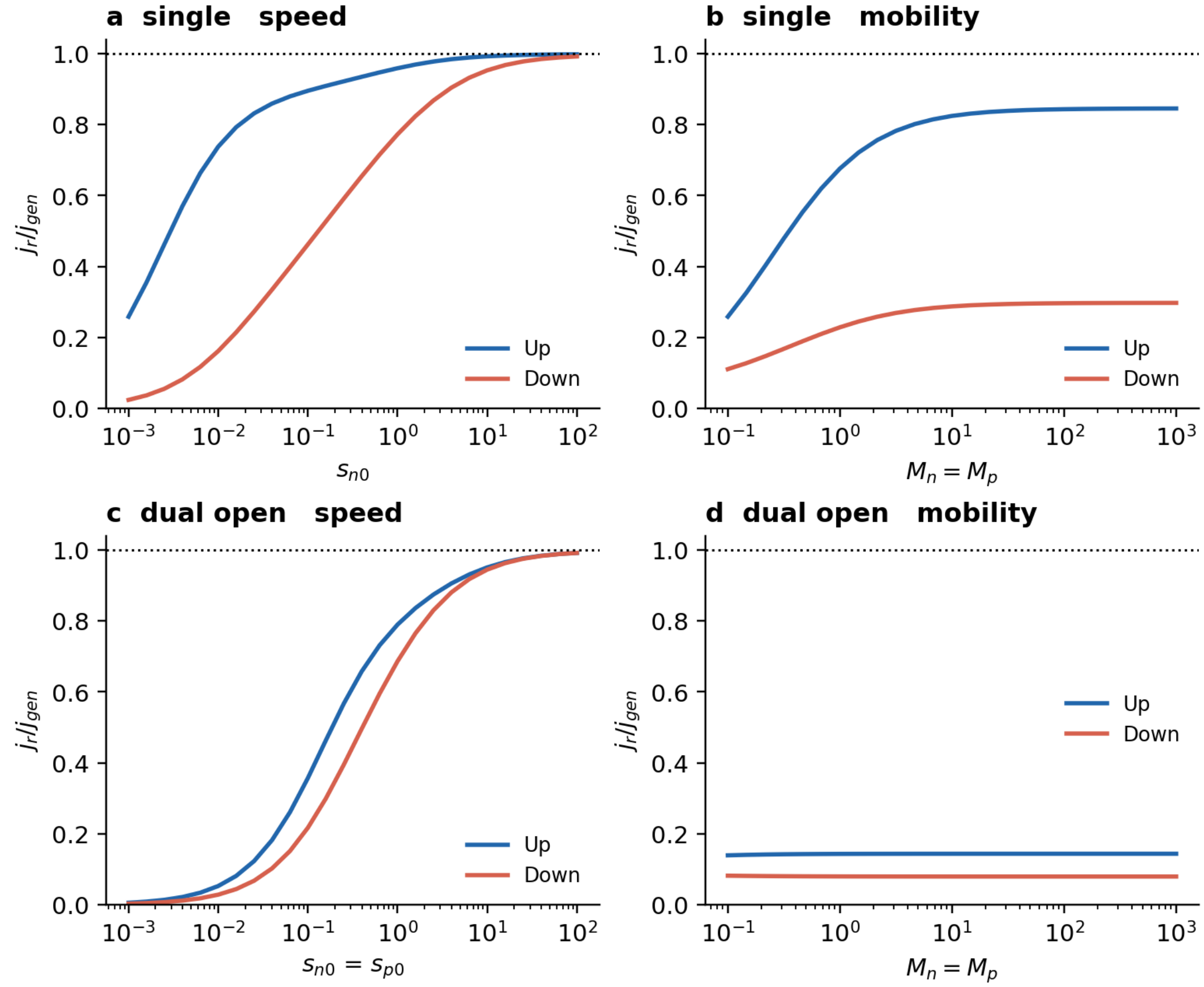

***Figure S16.*** *Competition between interfacial transfer, carrier transport, and recombination. At* $g_0 = 10$*, the upper row gives single-channel reduction at zero applied bias and the lower row gives common-pair turnover at open circuit. The left column varies interfacial transfer; both electron and hole velocities vary together in the paired case. The right column varies carrier mobility with the other inputs fixed. The dotted line is the absorbed mobile-pair supply. Increasing both paired transfer prefactors from 0.03 to 30 raises collection toward approximately 97.94% for Up and 97.85% for Down, so polarity contrast becomes small when nearly every generated pair is consumed at the surface. This ceiling applies to the stated operating conditions and current-source balance; applied-bias thermal generation or rear electron injection can supply additional current in other contact problems.*

## S5.5 Recombination timescale and static doping scope

The reactive calculation uses the instantaneous net pair-loss function $R = (np-1)/(n+p+2)$, with carrier concentrations normalized by $n_i$ and time by $\tau_0$. This is the stationary symmetric Shockley–Read form [S7]. Its incorporation into a time-dependent carrier system is a constitutive reduction and does not by itself add the dynamics or electrical charge of a bulk trap population. Charge conservation of the implemented pair-loss equations is exact because the same loss appears in both carrier equations. The physical validity of eliminating trap occupation is a separate question.

For a single electron-occupied trap fraction $f$, capture coefficients $c_n, c_p$, trap concentration $N_t$, and thermal densities $n_1, p_1$, detailed balance gives $e_n = c_n n_1$ and $e_p = c_p p_1$. The resolved occupation obeys

$$\dot{f} = c_n n(1-f) - c_n n_1 f - c_p p f + c_p p_1 (1-f), \tag{S186}$$

$$\tau_{\mathrm{occ}}^{-1} = c_n (n + n_1) + c_p (p + p_1). \tag{S187}$$

Here densities and time are dimensional. The separate electron and hole removal rates are $U_n = N_t [c_n n(1-f) - c_n n_1 f]$ and $U_p = N_t [c_p p f - c_p p_1 (1-f)]$, so $N_t \dot{f} = U_n - U_p$. The associated variable trap charge is $-qN_t f$ plus any fixed reference charge. Its derivative cancels the unequal free-carrier charging caused by $U_n$ and $U_p$. Imposing the same instantaneous scalar removal rate for both carriers therefore corresponds to eliminating this charge inventory, rather than representing it implicitly as a dynamical variable.

Setting $\dot{f}=0$ gives $f_\infty=(c_n n+c_p p_1)/[c_n(n+n_1)+c_p(p+p_1)]$ and

$$U_n=U_p=R_{\mathrm{SRH}}=\frac{np-n_i^2}{\tau_p(n+n_1)+\tau_n(p+p_1)}, \tag{S188}$$

$$\tau_n=(N_t c_n)^{-1}, \tag{S189}$$

$$\tau_p=(N_t c_p)^{-1}. \tag{S190}$$

The reduction used in the figures follows for $n_1=p_1=n_i$ and $\tau_n=\tau_p=\tau_0$. A frequency-domain interpretation then requires $\omega\tau_{\mathrm{occ}}\ll 1$ for the illuminated states contributing to the response and a controlled omission of the trap-charge contribution to electrostatics. With equal capture coefficients, the relation is $\tau_{\mathrm{occ}}/\tau_0=(N_t/n_i)/(\hat{n}+\hat{p}+2)$. Thus specifying the recombination lifetime does not independently fix the occupation time: the ratio $N_t/n_i$ is also needed. No trap concentration or dimensional frequency calibration is assigned to the illustrative reactive parameter set. The plotted full-band spectra are accordingly the spectra of the stated memoryless pair-loss model. Fast edges in the corresponding ideal source steps share the same constitutive restriction.

For the reduced doped static comparison, D adds the specified compact-layer term to the charge integral of S3.4:

$$P_{res}+Q_s(u)-C_H V_T u=0. \tag{S191}$$

This expresses how accumulation/depletion asymmetry and compact screening alter the stationary potential. It is a reduced electrostatic comparison, not a full doped continuation of the selective-contact SRH current or its optical spectrum. The contact and redox references remain essential to such a continuation. The existing intrinsic reactive results above retain their declared rate law and electrical control.

## S6. Case E: opposite domains and lateral charge return

### S6.1 Uniform floating state and stopping charge

The mean state of the analytical domain model belongs to a carrier-blocking body, not to the floating metal in Case D. Its exposed capacitance is the sole mean electrical coupling to the solution; rear compensation cancels the mean rear bound charge. The uniform mean, nonzero lateral modes, and nonlinear thick-crystal benchmark are kept separate below.

For an electrically isolated thin film with a single compact interface, the spatially uniform component obeys

$$\dot{n}=G-k_b(np-n_i^2)-r_r/L, \tag{S192}$$

$$\dot{p}=G-k_b(np-n_i^2)-r_o/L, \tag{S193}$$

$$C_0 U=\sigma_0+qL(p-n), \tag{S194}$$

$$C_0\dot{U}=q(r_r-r_o). \tag{S195}$$

When both reaction coefficients vanish, $d=p-n$ is fixed by the initial charge and $U$ is constant. With $N=(n+p)/2$, illumination changes the pair population through $\dot{N}=G-k_b[N^2-n_i^2-d^2/4]$. Its stationary value is $[n_i^2+d^2/4+G/k_b]^{1/2}$; the full Riccati transient is given in Equation (S196). The quasi-Fermi splitting can therefore increase while the uniform exposed electrostatic potential remains fixed. This conclusion depends on the stated single-exposed-capacitance geometry. Internal voltage changes and a second exterior capacitive coupling are distinct cases, treated in the common model.

With fixed $d=p-n$, set $K_d^2=n_i^2+d^2/4+G/k_b$. The blocking pair population obeys $\dot{N}=k_b\left(K_d^2-N^2\right)$, with

$$\frac{N(t)-K_d}{N(t)+K_d}=\frac{N(0)-K_d}{N(0)+K_d}e^{-2k_bK_dt}. \tag{S196}$$

The mean exposed potential is fixed by the initial charge. Neither $G$ nor the quasi-Fermi splitting is a net charging current.

For one reduction channel, with no changing site inventory, the net transferred electron-equivalent charge is exactly the change in compact charge. A stationary stopping state sets the reduction affinity to zero while generation balances recombination:

$$q\int_0^T r_r(t)\ dt=C_0\left[U(T)-U(0)\right]. \tag{S197}$$

With fixed bulk activities, write $n_r(U)=n_r^0e^{U/V_T}$ and $p_o(U)=p_o^0e^{-U/V_T}$. At a stationary stopping state $r_r=0$ and $np=K^2=n_i^2+G/k_b$. Its potential is the unique root

$$C_0U_\infty-\sigma_0=qL\left[\frac{K^2}{n_r^0e^{U_\infty/V_T}}-n_r^0e^{U_\infty/V_T}\right]. \tag{S198}$$

The left side increases with $U_\infty$ and the right side decreases. Forward reduction initially raises the body potential. Positivity of capacitance alone would not impose a finite budget on an ideal unbounded current source; the stopping condition here follows from reversible transfer and carrier balance.

Opening the complementary oxidation channel permits sustained equal partial rates. The rates and floating potential must be solved together; prescribing the potential independently would remove the electrical closure.

Opening $S_p>0$ permits a stationary paired rate $r_r=r_o=r$. For any trial $U$,

$$n=n_r(U)+r/S_n, \tag{S199}$$

$$p=p_o(U)+r/S_p, \tag{S200}$$

$$LG=r+L\,k_b\left(np-n_i^2\right). \tag{S201}$$

Define

$$A_2=\frac{L\,k_b}{S_nS_p}, \tag{S202}$$

$$B_1(U)=1+L\,k_b\left(\frac{n_r(U)}{S_p}+\frac{p_o(U)}{S_n}\right), \tag{S203}$$

$$G=LG-L\,k_b\left(n_r^0p_o^0-n_i^2\right), \tag{S204}$$

$$r(U)=\frac{2G}{B_1(U)+\sqrt{B_1(U)^2+4A_2G}}. \tag{S205}$$

This rate expression is conditional on $U$. The missing electrostatic equation is

$$H(U) \equiv C_0 U - \sigma_0 - qL[p(U) - n(U)] = 0. \tag{S206}$$

Together, Equations (S202)–(S205) and (S206) are a closed implicit analytical solution. On the physical branch, $n, p > 0$. Differentiating the carrier and reaction balances at fixed generation gives $n'(U) > 0$ and $p'(U) < 0$, hence $H'(U) > C_0 > 0$. Its limits have opposite signs, establishing a unique floating stationary state for constant positive transfer coefficients. Equilibrated electron sites add $q N_s n/(n+n_t)$ to $H$, where $n_t = k_e/k_c$, and preserve monotonicity. The admissibility argument and the population derivatives in Equations (S208)–(S209) establish these properties explicitly. Mathematical existence does not extend the thin-film or constant-capacitance assumptions beyond their range.

For fixed $U$, let $f(r) = r + L k_b (np - n_i^2) - LG$, with $n = n_r + r/S_n$ and $p = p_o + r/S_p$. On the physical interval,

$$D_r = \partial_r f = 1 + L k_b (p/S_n + n/S_p) > 0. \tag{S207}$$

At its lower endpoint at least one carrier vanishes and $f < 0$; at large positive $r$, $f \to \infty$. Thus the physical quadratic branch exists and is unique, including reverse operation. The reference product $n_r p_o$ determines the sign threshold, but not the floating voltage.

For constant transfer coefficients, put $n_U = n_r/V_T > 0$ and $p_U = -p_o/V_T < 0$, the derivatives at fixed $r$. Along the physical rate root,

$$n'(U) = \frac{n_U (1 + L k_b n/S_p) - (L k_b n/S_n) p_U}{D_r} > 0, \tag{S208}$$

$$p'(U) = \frac{p_U (1 + L k_b p/S_n) - (L k_b p/S_p) n_U}{D_r} < 0. \tag{S209}$$

Consequently $H' = C_0 + qL(n' - p') > C_0$. For equilibrated sites it acquires the additional positive term $q N_s n_t n'/(n+n_t)^2$. The limits of $H$ have opposite signs, proving uniqueness of the complete uniform state, rather than merely of a rate at a prescribed potential.

The rate changes sign at

$$G_{\text{th}} = k_b n_i^2 \left[ e^{\Delta\mu/(k_B T)} - 1 \right]. \tag{S210}$$

Below this threshold, the chemostatted chemical cycle can run backward. It is a consequence of the specified bimolecular law and homogeneous thin-film reduction. Although the electrostatic factors cancel in $n_r p_o$, they remain in the two individual carrier densities and the floating-potential equation.

Figure S17 contrasts the stopping charge for an isolated single-channel reaction with the sustained turnover enabled by opening the complementary reaction.

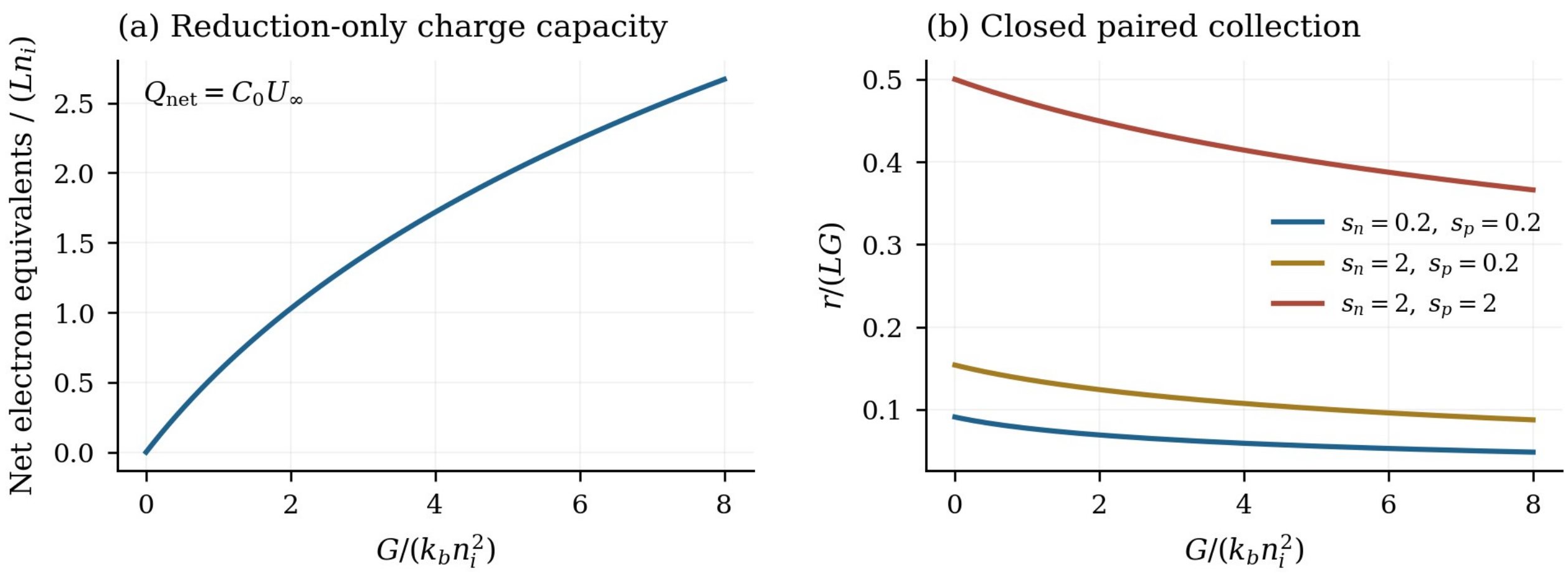

***Figure S17.*** *Uniform stopping charge and sustained paired turnover with electrical closure. Panel (a) gives the net electron equivalents transferred before an isolated reduction-only interface reaches its stopping state, with* $Q_{\mathrm{net}}=C_0 U_\infty$ *and initially zero mean potential. The rear blocks both carriers, the mean rear field is zero, and the exposed capacitance is the sole electrical coupling to the solution. The stationary condition sets the reduction affinity to zero while generation balances bimolecular recombination. Panel (b) opens both carrier-transfer channels and solves the paired rate together with the floating electrostatic condition. The reference populations are* $n_r^0=p_o^0=n_i$, $C_0 V_T/(qL n_i)=4$, *and the legend defines* $s_n=S_n/(L k_b n_i)$ *and* $s_p=S_p/(L k_b n_i)$. *The ordinate is the common partial rate divided by generated pairs per area. This establishes sustained charge-balanced turnover; a common redox pair gives recycling, while distinct product-forming channels require their specified species balances.*

The stopping-charge integral is an inventory result. The slow-charging transient additionally assumes pair relaxation is fast; it is not needed to establish the endpoint. This auxiliary uniform comparison supports the mean-state construction and is not substituted for the contacted Case D device.

### S6.2 Geometry, electrostatics, and regime of the thin-film reduction

Case E replaces the uniform exposed face by adjacent domains with polarization directed outward, $+P$, and inward, $-P$. Their residual surface charges have opposite signs after fixed chemical compensation. The domain boundary is vertical and electrically neutral: a lateral change of $P_z$ does not itself produce a bulk charge $-\nabla \cdot P$. Surface fringing fields nevertheless drive lateral carrier redistribution. We first follow the blocking photovoltage and then open both chemical channels, retaining the same film and passive dielectric backing. Neither face exchanges carriers with a rear electrode. The backing contributes an electrostatic response rather than an electronic reservoir; neglecting it requires its response to be small compared with the surface and film contributions.

The rear face is impermeable to carriers and ions and has a fixed compensating sheet that cancels its bound polarization charge. It rests on a passive dielectric of permittivity $\epsilon_{\mathrm{back}}$, with no electrode. For this analytical limit the exposed residual source $\sigma(x)$ includes the top bound polarization charge and specified fixed compensation. It excludes the mobile carriers and the exposed double layer. Fixed dopant charge can be included in its depth-integrated value.

The solution contains supporting electrolyte and two buffered redox couples. The double layer relaxes rapidly, its diffuse capacitance is large compared with the compact capacitance, and $k\lambda_D^l \ll 1$. Its differential capacitance per area is $C_0$, while $v(x)$ denotes the potential just outside it relative to the remote solution. Chemical activities there are held fixed in the leading calculation. The remote liquid has no net electrical current; nonzero lateral current modes decay into the liquid. Finite species delivery is examined separately.

The reference screening length is $\lambda_{\mathrm{sc},z}=\left[\epsilon_z V_T/\left(q\left(n_0+p_0\right)\right)\right]^{1/2}$, where $n_0, p_0$ are the uniform operating-state densities and $\epsilon_z$ is the stable normal differential permittivity. The film is thin enough that $kL\sqrt{\epsilon_x/\epsilon_z}\ll 1$, $L/\lambda_{\mathrm{sc},z}\ll 1$, and its through-thickness potential variation is small relative to $V_T$. Weak vertical extraction gradients additionally require $S_i L/D_{i,z}\ll 1$, or the corresponding load-weighted condition. We consider times longer than vertical diffusion relaxation. The depth-averaged densities then approximate the surface densities while small vertical gradients supply transfer. Normal and lateral diffusivities may differ. For nonzero wave number,

$$C_g(k)\phi_{s,k}-C_0 v_k=\sigma_k+qL\left(p_k-n_k\right), \tag{S211}$$

$$C_g(k)=C_0+\epsilon_{\mathrm{back}}|k|+\epsilon_x L k^2. \tag{S212}$$

The backing term follows from its decaying Laplace solution. Neglecting it is an electrostatic limit, not a consequence of electronic insulation. In the uniform mode it vanishes; with no mean rear field, $C_0 U=\sigma_0+qL(p-n)$, where $U=\phi_{s,0}-v_0$ and $v_0=0$ fixes the reference. The dielectric coefficient $\epsilon_x$ is the stable differential response of the frozen branch. Carrier screening is supplied by $p_k-n_k$, so it is not inserted again into $C_g$.

For a nonzero mode, the backing potential decays as $\phi_k e^{|k|(z+L)}$ away from the film. Its normal field at the rear gives $D_{f,z}(-L)=P_z-\epsilon_{\mathrm{back}}|k|\phi_k$ after accounting for the fixed rear compensation. Through-thickness integration of solid Gauss law gives

$$-\epsilon_x L\partial_x^2\phi_s+D_{f,z}(0)-D_{f,z}(-L)=qL\left(p-n+N_f\right). \tag{S213}$$

Here $N_f=N_D^+-N_A^-$ is the fixed net donor density, whose integrated charge is included in $\sigma(x)$. The exposed compact charge is $C_0(\phi_s-v)$. Applying the top displacement jump gives Equations (S211)–(S212). This derivation retains the passive backing response without supplying a carrier or Faradaic path. It also shows why the mobile-carrier capacitance belongs to the carrier solution and is absent from $C_g$.

The small parameters are $kL\sqrt{\epsilon_x/\epsilon_z}$, $L/\lambda_{\mathrm{sc},z}$, $\epsilon_{\mathrm{back}}|k|L/\epsilon_z$, and the vertical collection parameter $S_i L/D_{i,z}$. For a finite load, the sufficient last condition may be weakened to $S_n L\left|n-n_r\right|/\left(D_{n,z}n\right)\ll 1$ and $S_p L\left|p-p_o\right|/\left(D_{p,z}p\right)\ll 1$. The solution is evaluated after vertical diffusion relaxation. Lateral diffusivity can vary independently in an anisotropic film. Taking an isotropic diffusivity to zero at fixed thickness would leave this vertical collection limit; the physically blocked lateral-path limit keeps normal supply fast. Weak lateral perturbation additionally requires $\left|\phi_k-v_k\right|/V_T\ll 1$, $|\delta n|/n_0\ll 1$, and $|\delta p|/p_0\ll 1$. The electrochemical expansion must include the voltage derivative of $S_i$ if it is appreciable.

The compact-dominated limit is $C_D\gg C_0$, $k\lambda_D^1\ll 1$, with diffuse-layer relaxation fast compared with the plotted times. The liquid outside that layer is locally electroneutral and Ohmic. For a deep liquid with a mode $\phi_1(x,z)=\phi_{1,k}\cos(kx)e^{-|k|z}$,

$$j_{1,k}=\kappa_1|k|\phi_{1,k}=\phi_{1,k}/R_1(k), \tag{S214}$$

$$R_1(k)=1/\left(\kappa_1|k|\right). \tag{S215}$$

This is a modal area-specific resistance, with units $\Omega\ \mathrm{m}^2$. A mean through-current is absent; opposite modes close laterally in the liquid. A finite liquid depth or a different far boundary would change the kernel and must be specified rather than absorbed into an unexplained scalar resistance.

The two distinct redox couples have equal thermoneutral electron-energy references in the example, giving $n_*=p_*=n_i$. Their outer activities are buffered. With constant transfer velocity $S$, the reversible partial rates are

$$r_r=S\left[n_s-n_i e^{(\phi_s-\phi_1)/V_T}\right], \tag{S216}$$

$$r_o=S\left[p_s-n_i e^{-(\phi_s-\phi_1)/V_T}\right]. \tag{S217}$$

These are electron-equivalent particle fluxes in $\mathrm{m}^{-2}\mathrm{s}^{-1}$. Constant $S$ is a kinetic assumption; potential-dependent activation barriers and finite reactant supply require the additional closures in the static surface-kinetic analysis. A local Ohmic current does not by itself supply reactant molecules.

At first order, $n=N+a_k\cos kx$, $p=N-a_k\cos kx$. The conventional Faradaic current into the liquid is $j_{F,k}=q(r_{o,k}-r_{r,k})=-2qr_k$, where $r_k=S\left[a_k-n_i(\phi_{s,k}-\phi_{1,k})/V_T\right]$. Let $q_{H,k}=C_0(\phi_{s,k}-\phi_{1,k})$ denote the solid-side compact charge. The total liquid boundary current is

$$j_{1,k}=j_{F,k}+\dot{q}_{H,k}. \tag{S218}$$

Every nonzero mode carries a laterally closed liquid current. During a transient, the liquid current includes compact-layer charging as well as Faradaic current. The zero mode is determined independently by the body charge and paired reaction balance; it is not evaluated by assigning an infinite spreading resistance to a uniform through-current.

The finite-resistance Case E calculation eliminates the diffuse-layer dynamics while retaining the compact charge and the nonzero-mode current in the outer liquid. This elimination introduces a physical time condition in addition to the geometric conditions for the thin-film reduction. The calculation is a specified asymptotic model; assigning a finite liquid resistance does not by itself establish that its full frequency or step-response range is experimentally accessible. The relation between resistance and the omitted time can be checked without assigning a complete material parameter set.

For a symmetric dilute electrolyte with a common ionic diffusivity $D_i$, the bulk conductivity and charge-relaxation time satisfy $\kappa_1=\epsilon_1 D_i/\lambda_D^2$ and $\tau_M=\epsilon_1/\kappa_1=\lambda_D^2/D_i$. These are the bulk charge-relaxation quantities; the macroscopic interfacial charging time also contains geometric and compact-layer factors [S18]. With the deep-liquid modal resistance and the normalization used in Case E,

$$\rho_1=\frac{C_*}{\kappa_1 k_1 t_0}, \tag{S219}$$

$$\frac{\tau_M}{t_0}=\rho_1\frac{\epsilon_1 k_1}{C_*}=\rho_1\frac{\epsilon_1}{\epsilon_x}\frac{\epsilon_x}{C_* \mathrm{l}_0}(k_1\mathrm{l}_0)=0.6321364339\ \rho_1\frac{\epsilon_1}{\epsilon_x}. \tag{S220}$$

The final equality follows from $\epsilon_x L/(C_*\mathrm{l}_0^2)=0.008$, $L/\mathrm{l}_0=0.002$, and $k_1\mathrm{l}_0=0.1580341085$. I uses the lateral dielectric constant already present in the electrostatic kernel and does not require assigning a distinct normal dielectric constant to the liquid-time relation. The same material assignment must also satisfy

$$\frac{C_D}{C_0}=\frac{2(\epsilon_1/\epsilon_x)}{\lambda_D/\mathrm{l}_0}, \tag{S221}$$

$$\frac{D_i t_0}{\mathrm{l}_0^2}=\frac{(\lambda_D/\mathrm{l}_0)^2}{0.6321364339\,\rho_1(\epsilon_1/\epsilon_x)}, k\lambda_D\ll 1, C_D\gg C_0. \tag{S222}$$

Thus the Debye length, ionic diffusivity, dielectric ratio, and resistance cannot all be selected independently after the fixed normalized geometry has been chosen. Compact-dominated charging and appreciable Ohmic resistance are compatible in a suitable limit. Their compatibility does not guarantee that diffuse-layer relaxation is fast on every plotted time scale.

For the illustrative dielectric ratio $\epsilon_l/\epsilon_x=0.8$, the assigned parameters give $\tau_M/t_0=2.52855$ at $\rho_1=5$ and 252.855 at $\rho_1=500$. The diagnostic $\omega\tau_M\leq 0.1$ restricts the corresponding normalized frequencies to about 0.0395 and 0.000395. A late observation time does not independently validate the effect of an omitted early ionic history on integrated product. The finite-liquid curves retained here are solutions of the fast-diffuse model, not full ionic transients outside that regime.

The complete spectra extend to $\Omega=10^3$. To satisfy the same frequency diagnostic over that entire range would require $\epsilon_1/\epsilon_x\leq 3.16\times 10^{-5}$ at $\rho_1=5$, or $3.16\times 10^{-7}$ at $\rho_1=500$. Those ratios have not been assigned to a material. For the most resistive blocking case, the formal on/off extrema occur at approximately $1.991\,t_0$ and $3.547\,t_0$, respectively, before the charge-relaxation time of the illustrative liquid. The plotted extrema therefore characterize the reduced charge-allocation equations. They do not resolve an electrolyte transient at that dielectric ratio. The stationary solutions remain the long-time solutions of the declared electrostatic and kinetic boundary problem, independently of this temporal restriction.

Retaining finite diffuse-layer dynamics requires the charge mode of the linearized ionic transport problem, together with the electrostatic Laplace mode. For a half-space and lateral wave number $k$, the charge mode has inverse length $p=\left[k^2+\lambda_D^{-2}+i\omega/D_i\right]^{1/2}$, in addition to the harmonic mode with inverse length $|k|$. The boundary admittance then follows from the ion-specific flux conditions, compact-layer relation, and reaction-plane electrochemical potentials. Replacing $\kappa_1$ by $\kappa_1+i\omega\epsilon_1$ in the outer Ohmic kernel includes displacement in that outer medium but does not alone determine the diffuse ionic charge or the reactive boundary state. Such a finite-ionic-time solution is not used for the temporal figures reported here. The regime markings retain the complete limiting-model solutions while making that scope explicit.

### S6.3 A single normalized stationary and dynamic formulation

Use a length unit $l_0$, $t_0=\left(k_b n_i\right)^{-1}$, $V_T=k_B T/q$, and $C_*=2qL\,n_i/V_T$. The computational coordinate is $x/l_0$, so reported wave numbers are $k\,l_0$. Define

$$\theta=t/t_0, \tag{S223}$$

$$g=G/\left(k_b n_i^2\right), \tag{S224}$$

$$m=N/n_i, \tag{S225}$$

$$u=\phi_s/V_T, \tag{S226}$$

$$v=\phi_1/V_T, \tag{S227}$$

$$\zeta=u-v, \tag{S228}$$

$$b=a/n_i, \tag{S229}$$

$$c_0=C_0/C_*, \tag{S230}$$

$$c_g=C_g/C_*, \tag{S231}$$

$$c_f=c_g-c_0, \tag{S232}$$

$$\delta_k=\sigma_k/\left(C_* V_T\right), \tag{S233}$$

$$d_k=D_x k^2 t_0, \tag{S234}$$

$$s=S\,t_0/L, \tag{S235}$$

$$\rho_1 = R_1(k_1) C_* / t_0, \tag{S236}$$

$$h_k = t_0 / [R_1(k) C_*]. \tag{S237}$$

The symbols $c_f$ and $c_k = b_k - \zeta_k$ denote different quantities: $c_f$ is a geometric capacitance ratio, whereas $c_k$ is the dimensionless reaction-contrast amplitude. The code stores them as cf and c, respectively.

The displayed fundamental is rounded; the calculation files retain its full machine precision. The positive domain is centered at $x=0$, $-w/2<x<w/2$, and the negative domain at $x=w$; $w=\pi/k_1$. The frozen smooth source is

$$\delta(x) = \sum_{j\,\mathrm{odd}} \delta_j \cos(k_j x), \tag{S238}$$

$$\delta_j = \frac{4\delta_d}{\pi j} (-1)^{(j-1)/2} \exp\left[-(k_j a_w)^2/2\right], \tag{S239}$$

$$k_j = j k_1. \tag{S240}$$

For the deep liquid, $h_j = j/\rho_1$. Thus the conductivity, rather than each harmonic's resistance, is held fixed. The wall smoothing keeps the retained high-$k$ components compatible with a thin-sheet approximation and avoids treating an ideal discontinuity as a resolved physical wall.

The conditions for the reduction include $kL\sqrt{\epsilon_x/\epsilon_z} \ll 1$, $L/\lambda_{\mathrm{sc},z} \ll 1$, $SL/D_z \ll 1$, $\omega L^2/D_z \ll 1$, and $\omega \tau_D^1 \ll 1$. Assuming the reference $\epsilon_z = \epsilon_x$, the maximum retained $kL$ is 0.0635. At the largest mean intensity, $L/\lambda_{\mathrm{sc},z}$ is about 0.071. The largest plotted $\Omega = 10^3$ gives $\omega L^2/D_z = 0.00667$. The source amplitude remains a perturbative parameter: every first-order potential or allocation result is linear in $\delta_d$, and every reported mean-yield correction is quadratic. The actual liquid Debye time is not assigned a material value; the plotted band requires the stated fast-Debye inequality.

The fixed geometry and all transport coefficients are listed in S8. The two transfer choices are $s=0$ for blocking and $s=0.2$ for paired reactions. Both reactions are available on both domains; their spatial preference follows from the weak residual-charge pattern rather than a chemical mask.

At zeroth order $n=p=N$. Pair generation and recombination affect the neutral population; they cancel from the charge-difference equation. Dividing the depth-integrated mean carrier equation by $n_i/t_0$ gives

$$m' = g - (m^2 - 1) - s(m-1). \tag{S241}$$

Its stable illuminated root and its small-signal decay are

$$\overline{m} = 1 + \frac{2\overline{g}}{2+s+\sqrt{(2+s)^2 + 4\overline{g}}}, \tag{S242}$$

$$\gamma = 2\overline{m} + s = \sqrt{(2+s)^2 + 4\overline{g}}, \tag{S243}$$

$$H_m(z) = 1/(z+\gamma). \tag{S244}$$

For a finite step $g_i \to g_f$, let $m_i = \overline{m}(g_i)$,

$$\Gamma = \sqrt{(2+s)^2 + 4 g_f}, \tag{S245}$$

$$m_\pm = (-s \pm \Gamma)/2, \tag{S246}$$

$$\beta = (m_i - m_+)/(m_i - m_-). \tag{S247}$$

Then

$$m(\theta)=\frac{m_+-\beta m_- e^{-\Gamma\theta}}{1-\beta e^{-\Gamma\theta}}, \tag{S248}$$

$$I_m(\theta)=\int_0^\theta m(\tau)\,d\tau=m_+\theta+\log\frac{1-\beta e^{-\Gamma\theta}}{1-\beta}. \tag{S249}$$

These relations apply to light-on, light-off and nonzero-mean steps without replacing the changing carrier state by its endpoint or temporal average. When $\beta=0$, the initial state already equals the final steady state. At $g_i=0$, the leading initial population is $m_i=1$, but the patterned equilibrium has a nonzero second-order population correction, derived below.

Using Gauss law and expanding the two partial rates to first order gives

$$b_k=\delta_k-c_g u_k+c_0 v_k, \tag{S250}$$

$$\zeta_k=u_k-v_k, \tag{S251}$$

$$c_k=b_k-\zeta_k. \tag{S252}$$

The electron particle flux is $D_x\left[(N/V_T)\partial_x\phi_s-\partial_x a\right]$; the hole perturbation is its negative in the symmetric charge mode. Their difference equation and the liquid closure become

$$b_k'=d_k(m u_k-b_k)-s c_k, \tag{S253}$$

$$h_k v_k=-s c_k+c_0\zeta_k'. \tag{S254}$$

The physical mobile film charge is $-2qLn_i b_k$. In normalized form the independent balance checks are

$$-b_k'+d_k(m u_k-b_k)-s c_k=0\,(\text{film}), \tag{S255}$$

$$h_k v_k+s c_k-c_0\zeta_k'=0\,(\text{liquid}), \tag{S256}$$

$$c_f u_k'+d_k(m u_k-b_k)+h_k v_k=0\,(\text{combined}). \tag{S257}$$

The combined relation follows from the first two and the derivative of Gauss law. The instantaneous Gauss inventory may also be written $\delta_k=b_k+c_0\zeta_k+c_f u_k$. In the screening-allocation panels, $b$ is the positive screening amplitude; its physical mobile-charge sign is negative.

The $k=0$ mode is handled separately. It is not assigned an infinite-resistance version of a nonzero mode. Symmetry and the mean electrical gauge impose zero mean electrostatic potential and zero net through-current. The two globally integrated partial reactions are equal at all times in this symmetric problem, while their nonzero local difference supports a closed lateral current loop. Equal mean partial rates do not set the electron and hole quasi-Fermi levels equal.

For finite $0<\rho_1<\infty$, suppress the mode subscript and write $x=(u,v)^T$. Equations (S250)–(S254) give

$$Mx'=B-K(m)x, \tag{S258}$$

$$M=\begin{pmatrix} c_g & -c_0 \\ -c_0 & c_0 \end{pmatrix}, \tag{S259}$$

$$B=\delta\begin{pmatrix} d+s \\ -s \end{pmatrix}, \tag{S260}$$

$$K=\begin{pmatrix} (d+s)c_g+dm+s & -\left[(d+s)c_0+s\right] \\ -s(c_g+1) & h+s(c_0+1) \end{pmatrix}. \tag{S261}$$

The mass matrix contains only the retained film/backing and compact charge storage. Its determinant is $c_0 c_f$; a zero-backing/zero-thickness limit must therefore be reduced as a differential–algebraic problem instead of inverted as two independent stored voltages.

There is no independent imposed voltage source: $B$ follows from the frozen residual charge and the same carrier/chemical equations. The matrix is nonsingular for the retained passive backing/film response because $\det M = c_0 c_f > 0$. Let $A = (d+s)c_g + dm + s$. Direct expansion gives

$$\det K = hA + sd\left[c_f + m(c_0+1)\right] > 0, \tag{S262}$$

$$\operatorname{tr}\left(M^{-1}K\right) = d\left(1+\frac{m}{c_f}\right) + s\left(1+\frac{1}{c_0}\right) + \frac{h c_g}{c_0 c_f} > 0. \tag{S263}$$

The electrostatic eigenvalues therefore have positive real parts. The neutral mean population also decays with $\gamma > 0$. These results exclude an unstable linear charge mode under the stated positive coefficients. They do not require that every measured variable have a monotone step response.

For finite $h$, define

$$f = \frac{d}{d+s}, \tag{S264}$$

$$\chi = \frac{sf}{h}, \tag{S265}$$

$$\vartheta = \frac{\chi(m-1)}{1+\chi}. \tag{S266}$$

The steady solution of (S261) is

$$\bar{u} = \frac{\delta}{c_g + fm + (1-f) + \left[c_0 + (1-f)\right]\vartheta}, \tag{S267}$$

$$\bar{v} = -\vartheta \bar{u}, \tag{S268}$$

$$\bar{c} = \frac{f(m-1)}{1+\chi}\bar{u}. \tag{S269}$$

At $h \to \infty$, $\vartheta \to 0$, recovering an ideal local liquid. At $s=0$, $\bar{u} = \delta/(c_g+m)$, $\bar{v}=0$ for every finite $h$. At thermoneutral dark equilibrium $m=1$, $\bar{b} = \bar{\zeta} = \bar{u} = \delta/(c_g+1)$, $\bar{v}=0$ and every first-order reaction mode vanishes. At $\delta=0$, all spatial modes vanish at every intensity, leaving uniform paired turnover. At $d \to 0$ with $s>0$, $\bar{c} \to 0$; the local uniform paired reaction can remain nonzero because both couples are available everywhere. Thus a vanishing lateral current does not prohibit uniform local conversion of two distinct reservoirs.

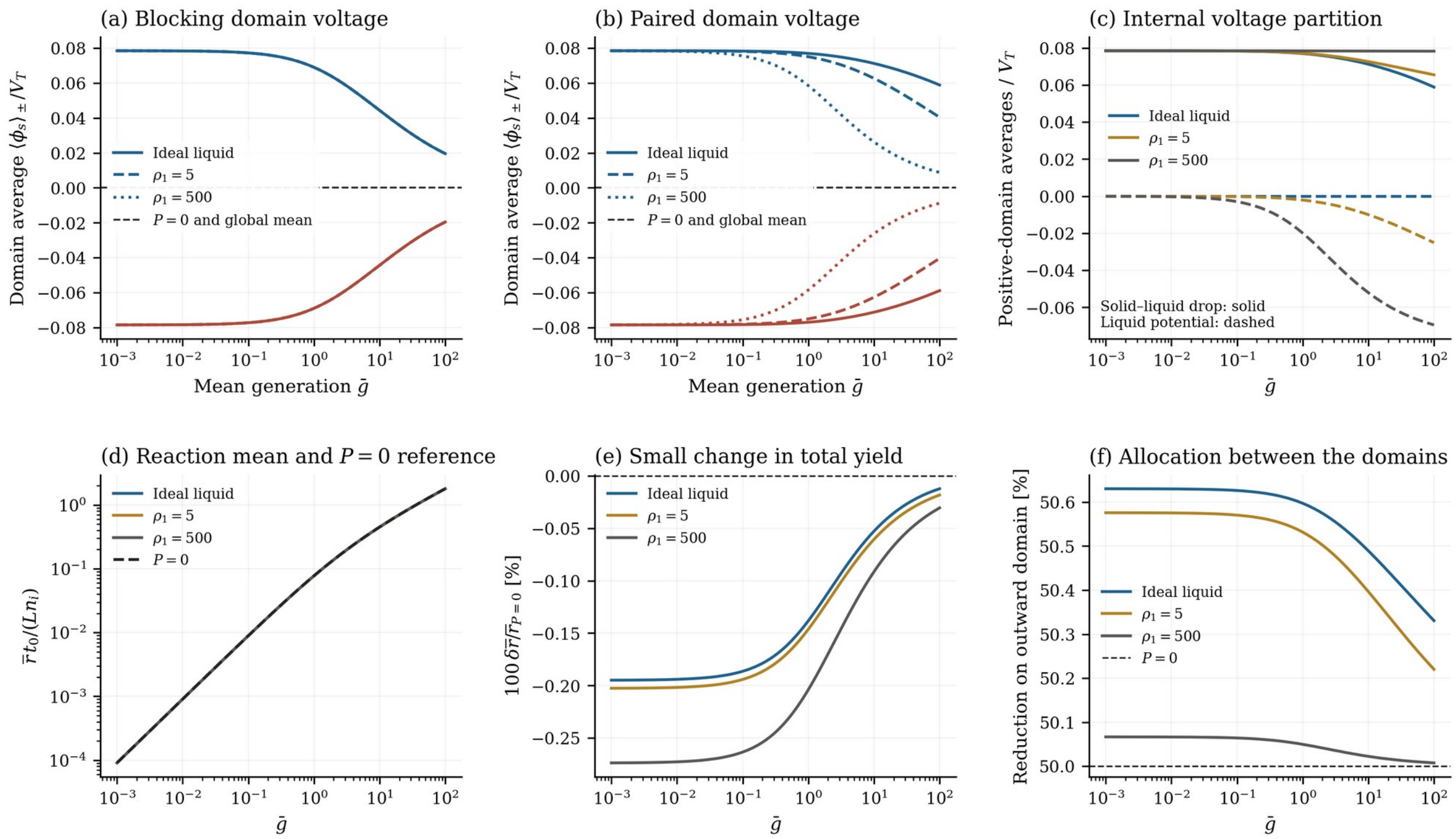

*Figure S18. Steady illumination changes domain voltage, voltage partition and reaction allocation. (a) Blocking domain-average solid potentials, with all finite liquid conductivities giving the same stationary curves. (b) Paired-reaction domain potentials for the three loads. In (a,b), blue is the outward positive-residual-charge domain, red the inward domain, and black dashed is both the matched unpatterned potential and the global mean. (c) Positive-domain compact drop, solid curves, and liquid potential, dashed curves, for the reactive case. (d) Common mean reaction rate, including its quadratic correction, against the unpatterned reference; the curves nearly coincide. (e) The small relative mean-yield correction resolved separately. (f) First-order fraction of total reduction assigned to the outward domain, with 50% for the unpatterned control. Both distinct redox couples remain available everywhere. Mean-yield corrections are second order; the domain allocation is first order.*

These stationary results separate domain voltage, compact-layer driving force, reaction allocation, and mean conversion. In the stated example the fraction of reduction assigned to the outward domain changes only from about 50.56% toward 50% as liquid resistance grows. The weak preference is not a prediction of exclusive domain selectivity.

### S6.4 Periodic illumination and finite steps

Let $g=\bar{g}+\mathrm{Re}\left(\hat{g}\,e^{z\theta}\right)$, $z=i\Omega$. The source charge, geometry and liquid conductivity are held fixed. Only $K$'s $(1,1)$ entry depends on $m$, with $\partial_m K_{11}=d$. Linearization around $\bar{x}$ therefore gives

$$\left(K+zM\right)\hat{x}=-d\,\bar{u}\;\hat{m}\begin{pmatrix}1\\0\end{pmatrix}, \tag{S270}$$

$$H_x(z)=\frac{-d\,\bar{u}}{z+\gamma}\left(K+zM\right)^{-1}\begin{pmatrix}1\\0\end{pmatrix}. \tag{S271}$$

This explicitly derives the optical forcing. Supplying an independent sinusoidal surface-voltage source would solve a different experiment. Likewise, identifying the fixed spatial harmonic $k$ with the temporal frequency $\Omega$ would not constitute a light-modulation calculation.

For explicit evaluation, the following coefficients are all taken at the same stationary mean population. No independently imposed optical voltage source is introduced.

$$A=\left(d+s\right)c_g+dm+s, \tag{S272}$$

$$B_1 = -[(d+s)c_0 + s], \tag{S273}$$

$$C_1 = -s(c_g+1), \tag{S274}$$

$$E_1 = h + s(c_0+1), \tag{S275}$$

$$\Delta(z) = (A + z c_g)(E_1 + z c_0) - (B_1 - z c_0)(C_1 - z c_0). \tag{S276}$$

$$H_u = \frac{-d\bar{u}(E_1 + z c_0)}{(z+\gamma)\Delta(z)}, \tag{S277}$$

$$H_v = \frac{d\bar{u}(C_1 - z c_0)}{(z+\gamma)\Delta(z)}, \tag{S278}$$

$$H_\zeta = H_u - H_v = \frac{-d\bar{u}(h - s c_f)}{(z+\gamma)\Delta(z)}, \tag{S279}$$

$$H_b = -c_g H_u + c_0 H_v, \tag{S280}$$

$$H_c = H_b - H_\zeta. \tag{S281}$$

The numerator of $H_\zeta$ has no $z$ term because the rapidly driven solid and liquid potentials can move together. For a single mode, $h = s c_f$ gives zero compact-drop optical response at every frequency even though the solid and liquid potentials respond. In the present reference the fundamental condition is $\rho_1 = 1487.8825$. Different stripe harmonics have different $c_f$ and $h$; they do not share one stripe-wide cancellation point.

At any finite $h$, $H_u = O(\Omega^{-2})$ and $H_\zeta = O(\Omega^{-3})$ as $\Omega \to \infty$, within the frequency range where the reduction remains valid. At ideal liquid $v = 0$,

$$\omega_q = d(1 + m/c_g) + s(1 + 1/c_g), \tag{S282}$$

$$H_u = \frac{-(d/c_g)\bar{u}}{(z+\gamma)(z+\omega_q)}. \tag{S283}$$

The ideal-liquid high-frequency limit and the finite-liquid high-frequency limit need not commute for the compact drop. This is a boundary-condition limit, not a resolved Debye process.

For a cosine stripe, the domain-average weights are $W_j = 2\sin(j\pi/2)/(j\pi)$. Hence

$$H_{\phi,+} = \sum_j W_j H_{u,j}, \tag{S284}$$

$$H_{\phi,-} = -H_{\phi,+}, \tag{S285}$$

$$H_{\Delta\phi} = 2H_{\phi,+}, \tag{S286}$$

$$H_{\text{global}} = 0. \tag{S287}$$

The gain displayed in the figures is $|H_{\phi,+}|$. The phase is unwrapped on the branch tending to $180^\circ$ at zero frequency because the static derivative is negative. A branch crossing is not silently clipped: the resistive blocking example reaches $196.45^\circ$. A true intensity sinusoid requires $|\hat{g}| \ll \bar{g}$ with $\bar{g} > 0$; the formal dark derivative is a one-sided illumination susceptibility rather than a physically admissible sinusoid of negative generation.

For ideal liquid, the nonzero-mode equation is scalar:

$$u' + \left[d + s(1 + 1/c_g) + \frac{d}{c_g} m(\theta)\right] u = \frac{d+s}{c_g}\delta. \tag{S288}$$

Define

$$\Lambda_k(\theta)=\left[d+s\left(1+1/c_g\right)\right]\theta+\frac{d}{c_g}I_m(\theta). \tag{S289}$$

The solution is

$$u_k(\theta)=e^{-\Lambda_k(\theta)}u_k(0)+\frac{(d+s)\delta_k}{c_g}\int_0^\theta e^{\Lambda_k(\tau)-\Lambda_k(\theta)}d\tau. \tag{S290}$$

The numerical quadrature uses the exponent difference directly to avoid overflow. For finite liquid, let $A_k(\theta)=M_k^{-1}K_k[m(\theta)]$, $f_k=M_k^{-1}B_k$, and let $\Phi_k(\theta,\tau)$ satisfy $\partial_\theta\Phi_k=-A_k(\theta)\Phi_k$, $\Phi_k(\tau,\tau)=I$. Then

$$x_k(\theta)=\Phi_k(\theta,0)x_k(0)+\int_0^\theta \Phi_k(\theta,\tau)f_k\,d\tau. \tag{S291}$$

This representation is exact for the reduced model. The matrices at different $m$ generally do not commute, so replacing the propagator by an exponential of an averaged coefficient is not justified. The supplied code evaluates (S291) by BDF integration with the analytic block-sparse Jacobian and the exact mean solution. This is numerical evaluation of linear modal ordinary differential equations, not time integration of the full two-dimensional drift–diffusion or ionic partial differential equations.

There is no instantaneous jump of the carrier densities or either capacitively stored potential. Since the initial charge state is stationary and $m$ is continuous, $x'(0^+)=0$, while $m'(0^+)=g_f-g_i$. Differentiating (S261) once gives, for finite liquid resistance,

$$u''(0^+)=v''(0^+)=\frac{-d\,u(0)}{c_f}(g_f-g_i), \tag{S292}$$

$$\zeta''(0^+)=0. \tag{S293}$$

For ideal liquid, $v=0$ and $u''(0^+)=-d\,u(0)(g_f-g_i)/c_g$. This short-time distinction explains the leading common motion of solid and liquid potentials in the resistive blocking example and the additional high-frequency suppression of the compact drop. It holds after the fast normal and diffuse-layer degrees of freedom omitted by the reduction have equilibrated.

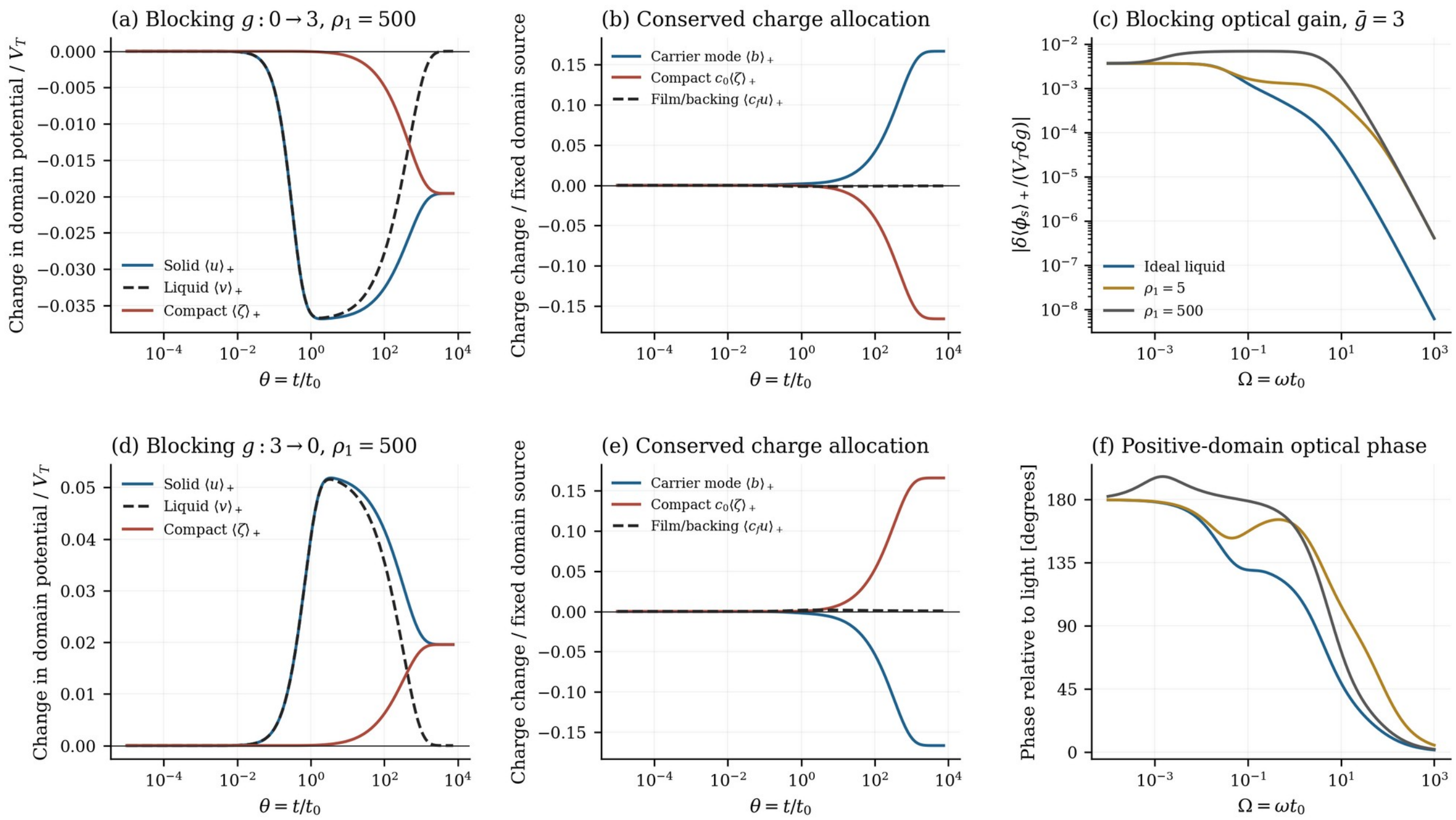

***Figure S19.*** *Blocking overshoot and optical phase arise from coupled charge inventories. (a,d) Changes of positive-domain solid potential, liquid potential and compact drop after light-on and light-off for* $\rho_1 = 500$. *(b,e) Changes of the three terms in the screening ledger* $\delta = b + c_0\zeta + c_f u$, *normalized by the fixed domain-averaged source. Their sum remains zero. Positive* $b$ *denotes the magnitude of negative mobile screening charge, so it is an algebraic screening allocation rather than the physical sign of the mobile charge. (c,f) Blocking optical gain and phase at* $\bar{g} = 3$ *for all three loads. The high-resistance phase reaches about 196.45° on the branch connected to 180° at zero frequency. Initially the solid and liquid potentials move together; later compact charging restores the common blocking endpoint. All linear decay modes are stable. The domain pattern and polarization are frozen throughout, and no trap relaxation is included.*

The state-resolved blocking comparison identifies the mechanism: the solid and liquid potentials initially move together, followed by redistribution between mobile-carrier screening and compact charge. The result uses frozen polarization and no trapping variable. Its formal early extrema at the largest liquid load must be read with the liquid-time restriction in S6.2.

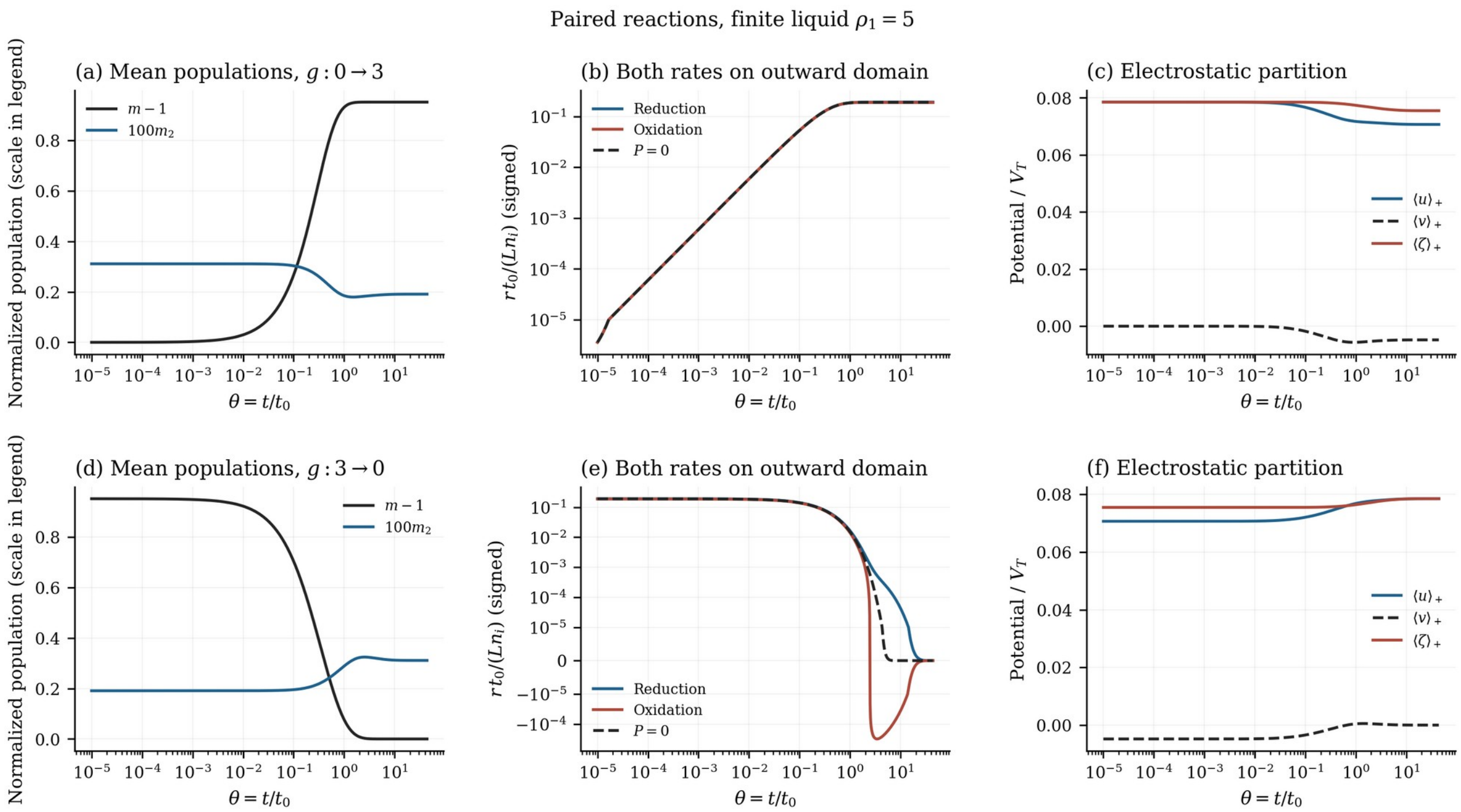

***Figure S20.*** *State-resolved paired light-on and light-off dynamics at* $\rho_1 = 5$. *The top row is* $0 \to 3$; *the bottom row is* $3 \to 0$. *(a,d) Leading mean population above equilibrium and the quadratic mean correction, with its factor of 100 displayed in the legend. (b,e) Both signed partial rates averaged over the outward domain, together with the uniform* $P=0$ *rate. The inward-domain reduction equals the outward-domain oxidation by symmetry. (c,f) Separate solid potential, liquid potential and compact drop. The small reverse partial-rate tail after light-off results from the different population and spatial-charge relaxation rates in the reversible model. It does not represent a resolved morphological dissolution process. Local contrasts are first order with the common quadratic mean included. The local contrasts in this retained comparison contain first-order harmonics and the common quadratic mean. They are not the fully completed local quadratic profiles defined in S6.5.*

## S6.5 Local quadratic chemistry, mean yield, and accumulated amount

The first-order charge solution is sufficient for voltage spectra, but a local chemical footprint requires an order-consistent neutral population. We therefore present the completed local quadratic field before defining accumulated amount. The retained state-resolved comparison above is explicitly labeled as the earlier mean-only quadratic reconstruction.

The local chemical rate in the weak-pattern model can be completed through quadratic order without changing its first-order charge equations. This completion is useful because a small omitted rate term can affect the location of a threshold near a nearly uniform product background. Let $X = x/l_0$, $D = D_x t_0 / l_0^2$, and let the residual-charge amplitude be the small parameter $\delta_d$. The symmetric carrier expansion is $n/n_i = m + b + N_2 + O(\delta_d^3)$ and $p/n_i = m - b + N_2 + O(\delta_d^3)$. Here $b$, $u$, and $v$ are first order, while $N_2(X, \theta)$ is the complete second-order neutral population, including its spatial mean. The charge potential has no even term through this order under the carrier-exchange and half-period symmetry used in the analytical reference.

The normalized local partial rates are

$$R_r = s\left[m - 1 + (b - \zeta) + N_2 - \frac{\zeta^2}{2}\right] + O(\delta_d^3), \tag{S294}$$

$$R_o = s\left[m - 1 - (b - \zeta) + N_2 - \frac{\zeta^2}{2}\right] + O(\delta_d^3), \tag{S295}$$

$$R_{r,o}=\frac{r_{r,o}t_0}{L n_i}. \tag{S296}$$

Adding the two carrier equations gives the neutral equation. The mean drift flux contains $b\ \partial_X u$, and the recombination law contains $2mN_2-b^2$. Subtracting the zeroth-order mean equation therefore gives

$$\partial_\theta N_2=D\ \partial_X^2 N_2-(2m+s)N_2-D\ \partial_X(b\ \partial_X u)+b^2+\frac{s}{2}\zeta^2. \tag{S297}$$

The drift term must be retained even though its period average is zero. It redistributes the local neutral population and therefore changes the common local contribution to reduction and oxidation. Omitting every nonzero spatial component of $N_2$, while retaining only $m_2=\langle N_2\rangle$, gives the previous local first-order reconstruction with a corrected mean. The mean equation itself remains

$$m_2{}'+(2m+s)m_2=\langle b^2\rangle+\frac{s}{2}\langle\zeta^2\rangle, \tag{S298}$$

$$\overline{R}=s\left[m-1+m_2-\frac{1}{2}\langle\zeta^2\rangle\right]. \tag{S299}$$

In darkness, $m=1$, $b=\zeta=u$, and $N_2=\zeta^2/2$. Both local partial rates then vanish. This equilibrium supplies the initial neutral field for a light-on calculation. For a step between nonzero intensities, the initial field is the full stationary solution at the initial generation. Added product amount is initialized to zero at the specified switch; the initial carrier and compact-layer inventories retain their stationary values.

Because the first-order pattern contains only odd harmonics of $k_1$, the quadratic forcing contains only even harmonics, including zero. Write $N_2=\sum_{l\geq 0}N_{2,2l}\cos(2lk_1x)$, and denote the corresponding forcing coefficient in the last three terms of the local neutral equation by $F_{2l}(\theta)$. The equations for these coefficients are independent scalar equations forced by the already determined first-order states:

$$N_{2,j}{}'+\left[2m(\theta)+s+D_x(jk_1)^2t_0\right]N_{2,j}=F_j(\theta), \tag{S300}$$

$$j=0,2,4,\dots, \tag{S301}$$

For stationary illumination, $N_{2,j}=F_j/\left[2\overline{m}+s+D_x(jk_1)^2t_0\right]$. For a step, let $I_m(\theta)=\int_0^\theta m(\tau)\,d\tau$ and $\Lambda_j(\theta)=2I_m(\theta)+\left[s+D_x(jk_1)^2t_0\right]\theta$. Then

$$N_{2,j}(\theta)=e^{-\Lambda_j(\theta)}N_{2,j}(0)+\int_0^\theta e^{\Lambda_j(\tau)-\Lambda_j(\theta)}F_j(\tau)\ d\tau. \tag{S302}$$

This representation retains the exact time-dependent mean population; replacing it by either endpoint would change the transient. The first-order charge propagator supplies $b$, $u$, and $\zeta$, after which the local quadratic population requires only a forced linear solve. The correction consequently remains an analytical modal reduction with numerical evaluation of its integrals. It is not a nonlinear time-dependent two-dimensional semiconductor–electrolyte simulation.

The complete accumulated reduction amount through this order is

$$\frac{M_r(x,T)}{Ln_i}=s\int_0^\theta[m(\tau)-1]d\tau+\sum_{j\,\mathrm{odd}}A_j^{(1)}(\theta)\cos(jk_1x)+\sum_{j\,\mathrm{even}}A_j^{(2)}(\theta)\cos(jk_1x), \tag{S303}$$

$$A_j^{(1)} = s\int_0^{\theta} c_j(\tau)\,d\tau\,, \tag{S304}$$

$$A_j^{(2)} = s\int_0^{\theta}\left[N_{2,j}(\tau) - \frac{1}{2}\left(\zeta^2\right)_j(\tau)\right]d\tau\,, \tag{S305}$$

$$\theta = T/t_0. \tag{S306}$$

The even sum includes $j=0$. This mean term is counted once; it is not added again after inserting the corrected spatial mean. Reversing the sign of the odd amount gives the oxidation distribution. Every nonzero even term integrates to zero over a full period and also over either equal-width domain because its period divides $w$. The local correction therefore preserves the previously verified mean amount and domain-integrated allocation through quadratic order. The exact ratio of domain allocation receives higher-order changes from normalization by the corrected mean, beyond the order reported here.

The zero even mode is the mean correction and is counted once. Nonzero even modes integrate to zero over either equal-width domain, so the local completion preserves the previously calculated global mean and domain allocation through this order. It can still alter a threshold crossing of a nearly uniform amount.

Let $m_2 = O\left(\delta_d^2\right)$ be the mean carrier correction. For the real cosine amplitudes,

$$B_2 = \langle b^2 \rangle = 1/2\sum_j b_j^2\,, \tag{S307}$$

$$Z_2 = \langle \zeta^2 \rangle = 1/2\sum_j \zeta_j^2. \tag{S308}$$

Expanding the spatially averaged bimolecular loss and the exponentials in (S216)–(S217) gives

$$m_2' + (2m+s)\,m_2 = B_2 + \frac{s}{2}Z_2\,, \tag{S309}$$

$$\overline{R} = s\left[m - 1 + m_2 - Z_2/2\right]. \tag{S310}$$

The correct dark initial state is $m_2(0) = Z_2(0)/2$, because $b=\zeta$ at equilibrium. Starting $m_2$ at zero would create a spurious dark chemical transient.

For steady illumination,

$$\overline{m}_2 = \frac{B_2 + s\,Z_2/2}{2m+s}\,, \tag{S311}$$

$$\delta\,\overline{R} = \frac{s}{2m+s}\left(B_2 - m\,Z_2\right). \tag{S312}$$

For one harmonic, $\overline{b}/\overline{\zeta} = 1 + f_{\text{eff}}(m-1)$, where $f_{\text{eff}} = f/(1+\chi m)$. When $m>1$, the yield correction changes sign at

$$f_{\text{eff}} = \frac{1}{1+\sqrt{m}}. \tag{S313}$$

This threshold is evaluated in the single-mode scan; it is not applied indiscriminately to a stripe sum with different transfer factors.

The complex mean-rate optical response follows by linearizing (S309)–(S310):

$$H_{m_2} = \frac{-2\,\overline{m}_2\,H_m + 2\langle \overline{b}\,H_b \rangle + s\langle \overline{\zeta}\,H_\zeta \rangle}{z+\gamma}\,, \tag{S314}$$

$$H_{\overline{R}} = s\left[H_m + H_{m_2} - \langle \overline{\zeta} H_\zeta \rangle\right]. \tag{S315}$$

The mean pair ledger through this order is

$$\left(m + m_2\right)' = g - \left[m^2 - 1 + 2\,m\,m_2 - B_2\right] - \overline{R}. \tag{S316}$$

Figure S21 shows how electronic and ionic transport change the mean-yield correction and the mean chemical optical response.

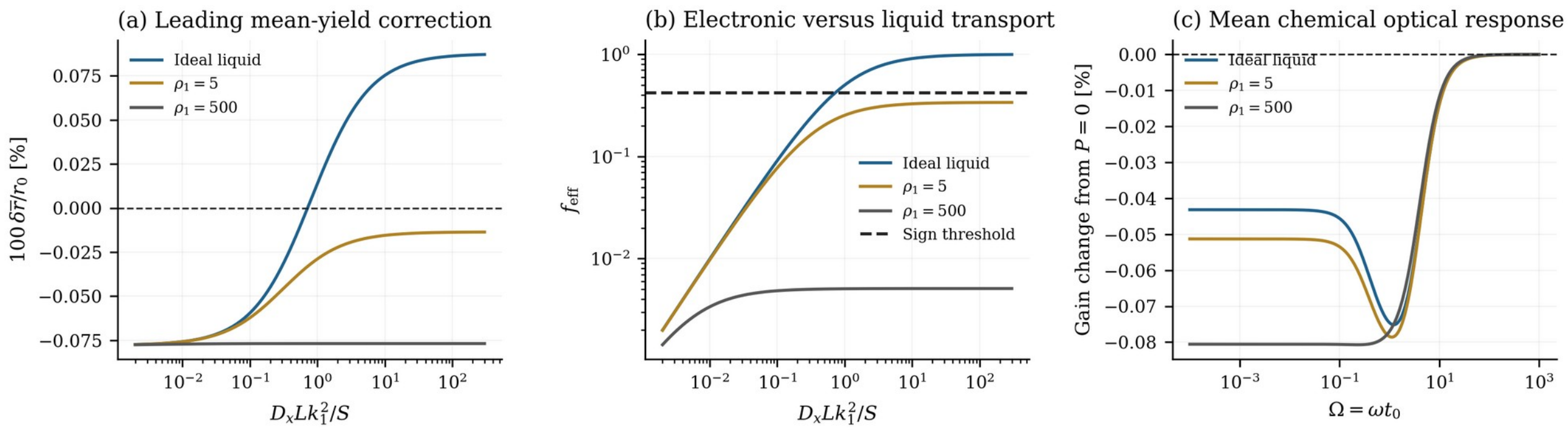

***Figure S21.*** *Electronic and ionic transport can change the sign of the first mean-yield correction. (a) Stationary quadratic mean-rate correction in a single-harmonic test as* $D_x L k_1^2/S$ *varies, at fixed* $g = 3$ *and for three liquid conductivities. (b) Corresponding effective allocation factor* $f_{\mathrm{eff}}$*, compared with the analytical sign threshold* $1/\left(1+\sqrt{\overline{m}}\right)$*. The single-mode scan holds source amplitude and wave number fixed while changing electronic diffusivity. (c) Change of the mean chemical optical gain from the corresponding* $P = 0$ *value for the full fixed stripe. The mean correction includes its own transient equation; it is not imposed quasistatically during modulation or a finite step. Panels (a,b) and (c) use the same constitutive laws but different stated spatial source shapes.*

The mean-yield correction can change sign because carrier separation and modified reverse transfer compete. This is a controlled second-order statement for the specified weak pattern, constant transfer velocities, and bimolecular recombination. It is not a continuation to near-binary nonlinear selectivity.

The amount on the opposite domain is obtained by the exact half-period symmetry, not a new fit. A $\nu$-electron product has molar amount $M_r/\left(\nu N_A\right)$. A detected footprint at threshold $M_{\mathrm{det}}$ is

$$D\left(T\right) = \left\{x : M_r\left(x, T\right) \geq M_{\mathrm{det}}\right\}, \tag{S317}$$

$$f_{\mathrm{det}} = \frac{\left|D \cap \text{one period}\right|}{2w}. \tag{S318}$$

For the illustrated light-on threshold, $M_{\mathrm{det}} = 3\,L\,n_i$. Because thresholding is discontinuous, the figure evaluates a dense grid of mean amounts by shape-preserving interpolation of the saved modal amount coefficients, then reconstructs the profile and thresholds it on 8192 spatial samples. It does not interpolate the detected fraction itself. The matched unpatterned reference has a uniform amount and therefore a threshold step when its mean reaches the threshold.

The evaluated completed local correction is only 0.150% of the uniform leading rate at its largest liquid resistance, but about 29% of the maximum first-order contrast. Relative to retaining only the quadratic mean, the detected fraction changes most near a threshold crossing; these changes do not represent equal changes in total conversion. Main-paper local rate and amount plots use the completed field. Verification and the explicit retained order are given in S7.

### S6.6 Unequal-carrier extension and spatial extent

For an unequal-carrier uniform state, first solve the uniform reaction balance and floating closure for $n_0, p_0, U_0, r_0$. The response below is linearized about that maintained state with potential-independent transfer velocities $S_n, S_p$, corresponding to $\alpha_n = \alpha_p = 0$ in Equations (S23)–(S24) and $U = \phi_s - v$. Potential-dependent transfer velocities add derivative terms, including $S_n^{'}(U_0)(n_0 - n_{r0})\delta U$ to the electron rate, and require the corresponding changes to the matrix. A finite uniform light step requires evolving mean coefficients, as in the coupled population and mode equations above. Write $n_{r0} = n_r(U_0)$, $p_{o0} = p_o(U_0)$, $a_i = D_i k^2$, $s_i = S_i / L$, $R_n = \partial R / \partial n$, and $R_p = \partial R / \partial p$ at that state. Define

$$A_n = a_n + s_n + R_n, \tag{S319}$$

$$A_p = a_p + s_p + R_p, \tag{S320}$$

$$B_n = (a_n n_0 + s_n n_{r0}) / V_T, \tag{S321}$$

$$B_p = (a_p p_0 + s_p p_{o0}) / V_T, \tag{S322}$$

$$E_n = s_n n_{r0} / V_T, \tag{S323}$$

$$E_p = s_p p_{o0} / V_T, \tag{S324}$$

$$g_* = \frac{q(S_n n_{r0} + S_p p_{o0})}{V_T}. \tag{S325}$$

For Laplace variable $s$, the explicit coefficient matrix for $y = (\delta n_k, \delta p_k, \phi_k, v_k)^{\mathrm{T}}$ is

$$A(s) = \begin{pmatrix} s + A_n & R_p & -B_n & E_n \\ R_n & s + A_p & B_p & -E_p \\ qL & -qL & C_g & -C_0 \\ q S_n & -q S_p & -(g_* + s C_0) & R_l^{-1} + g_* + s C_0 \end{pmatrix}. \tag{S326}$$

Let $\delta G_k(s)$ denote the Fourier/Laplace component of a small generation perturbation. It is zero for nonzero $k$ under spatially uniform illumination. The analytical response is

$$y(s) = A(s)^{-1} \begin{pmatrix} \delta G_k(s) + \delta n_k(0) \\ \delta G_k(s) + \delta p_k(0) \\ \sigma_k(s) \\ -C_0[\phi_k(0) - v_k(0)] \end{pmatrix}. \tag{S327}$$

The steady algebraic problem is

$$A(0)\, y_{\mathrm{st}} = (\delta G_{k,\mathrm{st}}, \delta G_{k,\mathrm{st}}, \sigma_{k,\mathrm{st}}, 0)^{\mathrm{T}}. \tag{S328}$$

This uses the steady source amplitudes, rather than substituting $s = 0$ into their Laplace transforms. For time response, the initial data must satisfy Gauss law. The fourth row includes double-layer charging and is not an independently fitted circuit element. The partial rates are

$$\delta r_r = S_n[\delta n_k - n_{r0}(\phi_k - v_k)/V_T], \tag{S329}$$

$$\delta r_o = S_p[\delta p_k + p_{o0}(\phi_k - v_k)/V_T]. \tag{S330}$$

The two electronic voltage perturbations remain separate: $\delta U_n = \phi_k - V_T \delta n_k / n_0$ and $\delta U_p = \phi_k + V_T \delta p_k / p_0$. Their conventional sheet currents are $K_i = -\Sigma_i \partial_x U_i$, with $\Sigma_n = q D_n n_0 L / V_T$ and $\Sigma_p = q D_p p_0 L / V_T$. Subtraction of the first two rows cancels generation and recombination and reproduces the charge balance. The symmetric specialization reduces to the closed coefficients in the associated model discussion. A scalar current law with a shared quasi-Fermi perturbation would require an additional compatibility condition; it is not imposed here.

This matrix is the unequal-carrier linear extension with separately closed uniform state and explicit liquid charging. It is not an additional fully nonlinear patterned or doped frequency sweep. Potential-dependent transfer coefficients require their derivatives and corresponding changes to the matrix, as stated in the preceding boundary law.

The retained ideal-liquid sharp-stripe outer approximation has

$$l_c^2 = \frac{D_x L (C_0 + C_q)}{S (C_0 + C_*)}, \tag{S331}$$

$$C_q = 2 qLN / V_T, \tag{S332}$$

$$\delta r(x) = \frac{r_0 \sigma_d}{V_T (C_0 + C_q)} \frac{\cosh[(x - w/2)/l_c]}{\cosh[w/(2 l_c)]}, 0 < x < w, \tag{S333}$$

$$r_0 = S(N - n_i). \tag{S334}$$

Its distance $d$ measured inward from each boundary contains 90% of the positive-domain excess when

$$d_{90}^{\text{excess}} = \frac{w}{2} - l_c \operatorname{asinh}\left[0.1 \sinh\left(\frac{w}{2 l_c}\right)\right]. \tag{S335}$$

The physical wall is smooth in the Fourier calculation. Equation (S333) is only the local-capacitance outer limit away from that wall and excludes the total uniform background. It cannot be called the 90% width of total reaction or the detected footprint. The finite-liquid Fourier calculation uses its actual harmonic kernels instead of imposing this hyperbolic profile.

The hyperbolic profile describes the polarization-dependent excess, not the entire reaction. The uniform background remains present. Total-rate percentile width, excess-rate width, and a thresholded amount footprint therefore measure different quantities. Optical attenuation is a depth scale, and the uniformly illuminated pattern has no lateral beam edge.

Electrical conduction and reactant supply are separate. For a dilute redox species whose migration is negligible in supporting salt, a consumption mode with a fixed-concentration reservoir at $H$ has

$$\delta c_{s,k} = - r_k K(k), \tag{S336}$$

$$K(k) = \frac{\tanh(|k| H)}{D_l |k|}, \tag{S337}$$

$$K(0) = H / D_l. \tag{S338}$$

Constant surface activities require $|r_k| K(k) \ll c_b$ and $r_0 H / D_l \ll c_b$ for sustained uniform consumption. In a closed bath the uniform inventory evolves instead. Near equilibrium, an independent one-electron redox pair has concentration resistance $R_{\text{diff}} = (V_T / q)[K_O / c_O^b + K_R / c_R^b]$. Distinct couples require their own stoichiometric terms. Away from that small-signal limit, both activity-induced reference-density shifts and transfer-coefficient shifts enter the boundary matrix. The main figures assume rapid chemical supply and do not claim a mass-transport-limited deposition rate.

### S6.7 Nonlinear finite-depth benchmark

The nonlinear stationary benchmark resolves normal carrier gradients and prescribed complementary reaction regions. Its masks are retained in the nonpolar control. It therefore tests how polarization modifies carrier collection within assigned chemical regions; it does not independently predict that only one half-reaction is available on each polarization.

The crystal occupies $z<0$ and the period is $2w$. The equations are

$$-\epsilon_f \nabla^2 \phi = q(p-n), \tag{S339}$$

$$\Gamma_n = \mu_n n \nabla \phi - D_n \nabla n, \tag{S340}$$

$$\Gamma_p = -\mu_p p \nabla \phi - D_p \nabla p, \tag{S341}$$

$$\nabla \cdot \Gamma_n = \nabla \cdot \Gamma_p = G(z) - R(n,p), \tag{S342}$$

$$G(z) = G_0 e^{\alpha z}, \tag{S343}$$

$$G_0 = \frac{\eta_g \alpha I_0}{E_\gamma}, \tag{S344}$$

$$R = \frac{2np}{\tau(n+p)}, \tag{S345}$$

$$D_i = \mu_i V_T. \tag{S346}$$

Recombination is set to zero continuously at $n=p=0$. This is a zero-background, high-injection approximation with no explicit equilibrium carriers, dopants, or traps. The shared pair loss appears identically in both continuity equations. At the exposed face,

$$\epsilon_f \partial_z \phi + C_{\text{eff}} \phi_s = \sigma_{\text{res}}(x), \tag{S347}$$

$$\Gamma_{n,z} = \chi_r S_n n_s, \tag{S348}$$

$$\Gamma_{p,z} = \chi_o S_p p_s, \tag{S349}$$

$$\chi_r = 1_{0<x<w}, \tag{S350}$$

$$\chi_o = 1 - \chi_r. \tag{S351}$$

The baseline transfer velocities are potential independent. Optional barrier laws are $S_n = S_{n0} e^{-\gamma_n \phi_s / V_T}$ and $S_p = S_{p0} e^{\gamma_p \phi_s / V_T}$; the evaluated cases use the baseline $\gamma_n = \gamma_p = 0$. Both rates are forward, high-driving approximations. Their polarity assignment is imposed through chemical availability. The calculation tests how polarization and carrier transport modify rates under those masks. The passive control sets the residual source to zero, retaining the masks and every other parameter.

The deep boundary is $z=-H$ with $\partial_z \phi = \Gamma_{n,z} = \Gamma_{p,z} = 0$, where

$$H = \max\left(3w, \frac{8}{\alpha}, 8 \max_i \sqrt{D_i \tau}\right) \tag{S352}$$

before applying the doubled-depth factor. There is no bottom carrier reservoir or electrode. This distant truncation approximates surface behavior of a thick specimen. Majority-carrier tails in the zero-background model make deep stored charge more delicate than the converged surface rates. The doubled-depth check is therefore interpreted as a test of surface observables.

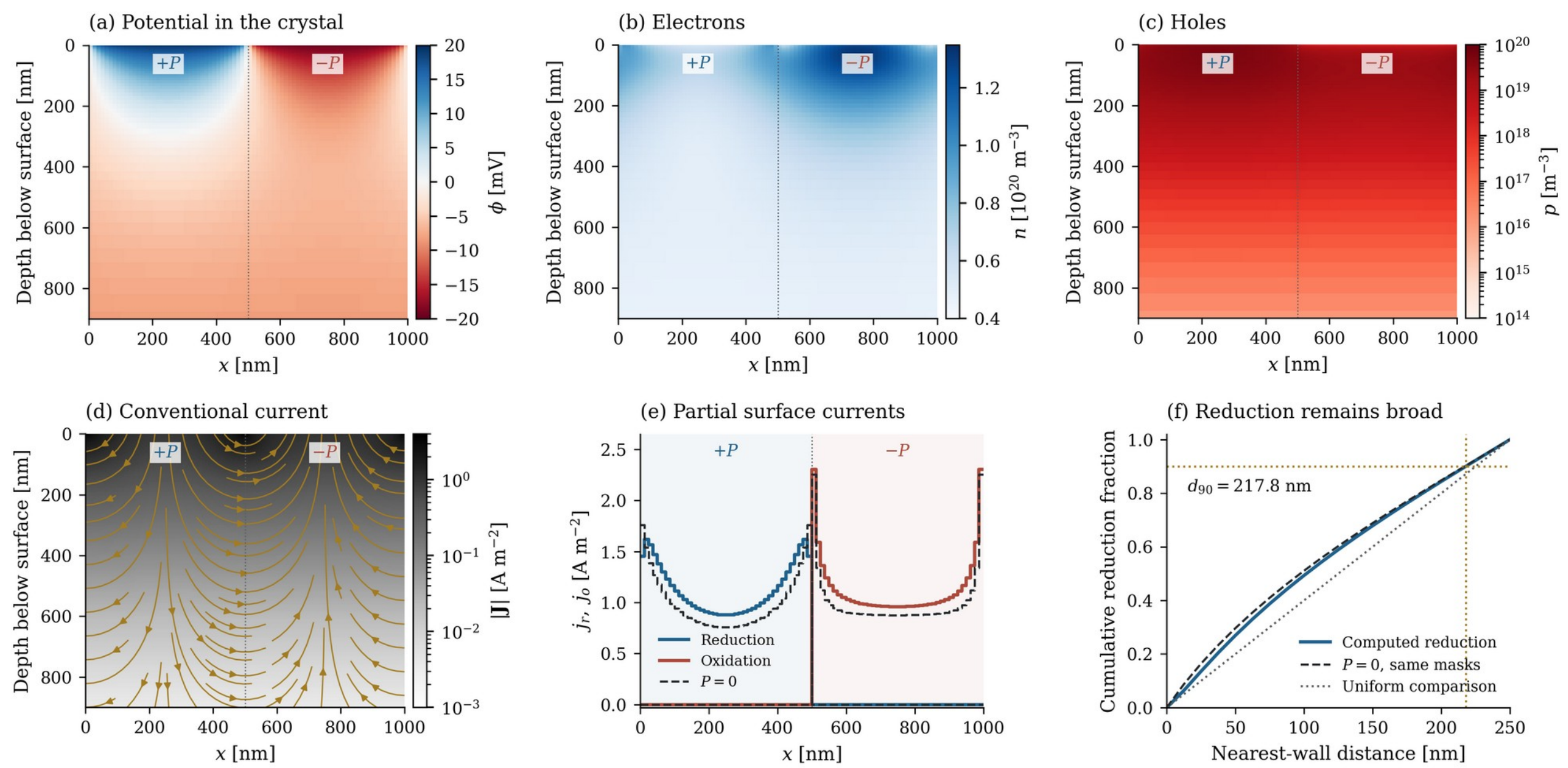

***Figure S22.*** *Nonlinear two-domain benchmark and matched passive control. (a) Electrostatic potential, (b) electron density, and (c) hole density in the crystal. The hole scale is logarithmic and the electron scale is linear. (d) Conventional-current magnitude and streamlines show the solid portions of the current paths; the ionic continuation is in the ideally conducting liquid. Maps display the first 900 nm of a 1500 nm computational depth. (e) Positive partial current densities for reduction on the outward* $+P$ *stripe and oxidation on the inward* $-P$ *stripe. Black dashed curves remove the polarization charge while preserving the same chemical masks and every transport parameter. (f) Cumulative reduction with nearest-wall distance for the polarized and passive cases, alongside a uniform profile. The polarized calculation gives* $d_{90}=217.8$ *nm, 24.80% pair collection, and a mean reduction-domain current of 1.134 A* $\mathrm{m}^{-2}$*. These* $80\times70$ *finite-volume fields are the baseline case of Table S3. The complementary reaction masks specify chemical availability and are not predictions of domain selectivity.*

The baseline has 24.80% pair collection, full-period mean reduction current 0.5668 A $\mathrm{m}^{-2}$, and a nearest-wall distance of 217.8 nm containing 90% of the reduction on the assigned stripe. Removing the residual polarization but retaining the masks gives 22.00% collection and 216.9 nm. The imposed chemical pattern remains selective in that control; the difference is the additional polarization-dependent transport effect.

For the reduction stripe, define the nearest-wall distance $d(x)=\min(x,w-x)$ and the cumulative fraction of its total reduction,

$$F(d)=\frac{\int_0^w r_r(x)\,1_{\{d(x)\le d\}}\,dx}{\int_0^w r_r(x)\,dx}, \tag{S353}$$

$$F(d_{90})=0.9. \tag{S354}$$

This definition includes the entire reduction rate within the preassigned active stripe. It differs from $d_{90}^{\mathrm{ex}}$ in the analytical two-domain derivation, which contains 90% of the polarization-dependent excess above the nonzero uniform background of the unmasked model. Uniform activity gives $d_{90}=0.45\,w$. A high immediate edge rate does not by itself make this integrated width small.

The zero-residual-source control is especially informative. It still has a reduction stripe because the masks remain imposed, and it gives 22.00% collection and $d_{90}=216.9$ nm. Thus the edge-enhanced profile in this model cannot be attributed exclusively to polarization charge or an intrinsically active domain wall. The spatial change in carrier-selective boundary conditions is itself sufficient to generate lateral collection. Residual polarization then modifies that transport and the resulting profile. In contrast, setting the pattern source to zero in the unmasked symmetric model removes its opposite rate harmonics.

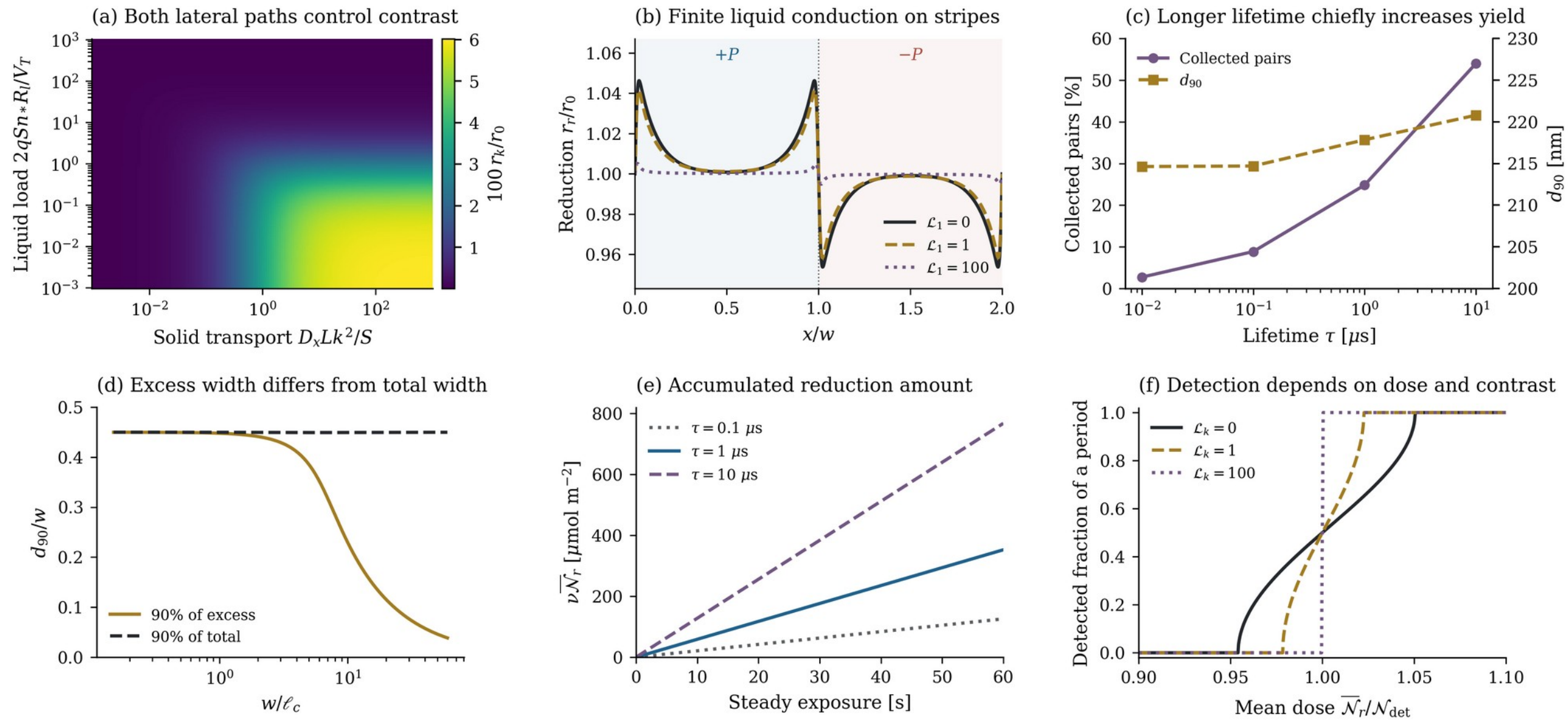

***Figure S23.*** *Transport, reaction extent, and accumulated chemical amount. (a) Harmonic partial-rate contrast versus independent solid-transport and liquid-load parameters in the analytical film. (b) Stationary reduction across the stripe pair for three liquid conductivities, expressed by the fundamental load* $L_1$*. Each harmonic uses the same conductivity, with* $R_l(k)=1/(\kappa_l|k|)$*; oxidation has the complementary first-order profile. (c) Nonlinear lifetime dependence of pair collection and the width containing 90% of reduction. Lines connect the four evaluated lifetimes. (d) Analytical widths containing 90% of the polarization-dependent excess and of the total rate; the total includes its uniform background. (e) Full-area mean reduction amount obtained by integrating the stationary nonlinear partial current. The plotted* $\nu\overline{N}_r$ *is electron-equivalent molar amount, with unit Faradaic selectivity and unchanged geometry. (f) Detectable fraction for a harmonic stationary rate as its mean accumulated dose crosses a specified threshold;* $L_k$ *is the load of that harmonic. Panels (a,b,d,f) are controlled thin-film predictions; (c,e) use the thick-crystal masked benchmark. The curves describe their stated geometries and interfacial conditions rather than a common material fit.*

Figure S23 combines two explicitly different reductions: its analytical panels concern the unmasked thin film, whereas lifetime-dependent collection and amount use the masked thick-crystal model. Increasing the lifetime from 10 ns to 10 μs raises collection from 2.71% to 53.99%, while the integrated width changes from about 214.6 to 220.8 nm. A diffusion length therefore cannot be identified directly with this reaction footprint.

*Table S3. Stationary nonlinear cases. Values are rounded for presentation; machine-readable files retain full precision.*

| **Case** | **Grid** | **Collected pairs [%]** | $d_{90}$ **[nm]** | **Mean** $j_r$ **on reduction domain [A m$^{-2}$]** |
|---|---|---|---|---|
| Coarse baseline | $40 \times 50$ | 24.6977 | 217.797 | 1.129039 |
| Baseline | $80 \times 70$ | 24.79597 | 217.83898 | 1.133530 |
| Refined baseline | $120 \times 110$ | 24.83695 | 217.83161 | 1.135403 |

| Case | Grid | Collected pairs [%] | $d_{90}$ [nm] | Mean $j_r$ on reduction domain [A $m^{-2}$] |
|---|---|---|---|---|
| Doubled depth | $80\times110$ | 24.70953 | 217.70069 | 1.129578 |
| $\tau=10$ ns | $80\times70$ | 2.70625 | 214.63088 | 0.123714 |
| $\tau=100$ ns | $80\times70$ | 8.84266 | 214.69136 | 0.404236 |
| $\tau=10$ $\mu$s | $80\times70$ | 53.98629 | 220.79109 | 2.467945 |
| Passive control, $\sigma_d=0$ | $80\times70$ | 21.99741 | 216.94099 | 1.005595 |

### S6.8 Static donor and acceptor shift

For the wide-domain paired-reaction reduction, equal electron- and hole-consuming transfer coefficients impose $n_b e^m = p_b e^{-m}$, where $m$ is the mean solid potential relative to the solution in units of $V_T$. Thus $m=1/2\ln(p_b/n_b)$. The polarization-dependent contrast is controlled by the sum of interfacial and semiconductor differential screening. Doping therefore shifts the mean potential as well as changing the contrast.

This wide-domain closure isolates a mean electrochemical shift caused by unequal populations. It neither replaces the nonzero-mode charge-return calculation nor supplies a nonlinear unmasked strong-selectivity solution. The same reduced comparison gives both polarization orientations at a specified doping; no doped frequency or finite-step family is reported for E. Its static potentials are included in S8.

## S7. Numerical methods, verification, and scope of the checks

The numerical checks distinguish algebraic consistency, numerical accuracy, and physical applicability. Conservation identities test implementation; mesh and tolerance comparisons estimate numerical error for the sampled states; asymptotic conditions delimit physical validity. These are separate tests and should not be combined into a universal accuracy claim.

### S7.1 Stationary and time-dependent solvers

A uses a scalar Poisson–Boltzmann quadrature and an independently assembled carrier–Poisson boundary-value problem. Its nonlinear source parameter and weak-potential temporal family are separately specified. C evaluates a bounded-exponential boundary solution and a conservative charge-sector finite-volume evolution. The independent nonlinear-species A/C comparison uses the same electrostatic map, so it tests the carrier/ion reduction rather than independently validating that map.

Stationary solutions use adaptive collocation for $(u, u', v, d, \ln n, \ln p, j_n, j_p)$, initialized with 151 points and permitting 20000. The normalized tolerance is $2\times10^{-7}$. Continuation in generation, bias, and chemical gap follows the selected branch. Floating operation adds $V$ and the zero rear conduction-current condition. Logarithmic densities preserve positivity. The distinct-couple crossing is refined with a scalar bracketed root and a $2\times10^{-8}$ collocation tolerance. The four principal polar redox currents reproduce their rounded record values within $3.1\times10^{-7}$.

The time-dependent implementation is independent of the stationary collocation. It uses cell-centered finite volumes, logarithmic carrier densities, exact cell-averaged optical absorption, algebraic electrostatics from integrated Gauss law, and backward differentiation with relative/absolute tolerances $2\times10^{-7}/2\times10^{-9}$. Scharfetter–Gummel interior fluxes use $B(\xi)=\xi/(e^{\xi}-1)$ and $B(0)=1$:

$$j_{n,i+1/2}=\frac{M_n}{\Delta x}\left[B(\Delta v)n_{i+1}-B(-\Delta v)n_i\right], \tag{S355}$$

$$j_{p,i+1/2}=\frac{M_p}{\Delta x}\left[B(\Delta v)p_i-B(-\Delta v)p_{i+1}\right]. \tag{S356}$$

Each surface half-cell is coupled algebraically to its capture/release or reaction law, so transport and transfer resistances enter the same flux. The Landau gradient term uses natural face conditions. Dark states are relaxed on the actual finite-volume mesh before illumination, avoiding an initial transient from interpolation of a collocation state. The principal contacted-film trajectories use 81 cells. The dry charging and reactive families retain their own pulse amplitudes and time units as specified in their case sections.

For the nonlinear domain benchmark, positive carrier populations are represented by logarithmic variables. Periodic lateral face fluxes enter neighboring volumes with opposite signs; the bottom is carrier-blocking. The surface half-cell is coupled to the reaction plane rather than identifying the top-cell density with the surface density. A sparse Newton method with damping and continuation solves the stationary system.

Let $B(\delta)=\delta/(e^{\delta}-1)$ and let a face point from cell $i$ to cell $j$, with $\Delta u=u_j-u_i$ and center separation $d_{ij}$. The dimensional face fluxes are

$$\Gamma_{n,ij}=\frac{D_n}{d_{ij}}\left[B(-\Delta u)n_i-B(\Delta u)n_j\right], \tag{S357}$$

$$\Gamma_{p,ij}=\frac{D_p}{d_{ij}}\left[B(\Delta u)p_i-B(-\Delta u)p_j\right]. \tag{S358}$$

Each integrated internal face flux enters neighboring control volumes with opposite signs. Series expansions near zero and asymptotic formulas at large $|\Delta u|$ avoid cancellation or overflow in $B$ and its derivative. Lateral faces are periodic. No carrier flux is allowed at the bottom.

The surface potential is determined at the reaction plane, rather than replaced by the top-cell value. If the top center lies a physical distance $h$ below the plane, its Robin boundary gives

$$\phi_s=\frac{\epsilon_f\phi_i+h\sigma_{\mathrm{res}}}{\epsilon_f+hC_{\mathrm{eff}}}. \tag{S359}$$

Let $\Delta u=(\phi_s-\phi_i)/V_T$ and $A_i=D_i/h$. Eliminating the surface density between the half-cell Scharfetter–Gummel flux and the reaction boundary gives

$$n_s=\frac{A_nB(-\Delta u)}{S_n+A_nB(\Delta u)}n_i, \tag{S360}$$

$$r_r=S_nn_s, \tag{S361}$$

$$p_s=\frac{A_pB(\Delta u)}{S_p+A_pB(-\Delta u)}p_i, \tag{S362}$$

$$r_o=S_pp_s. \tag{S363}$$

Here the local velocity is zero on a blocked region. The corresponding surface density approaches its Boltzmann extrapolation and the flux is zero. These half-cell relations are especially important for the fast hole reaction, whose transfer length is smaller than the first full cell.

Grid, depth, and independent half-space checks are provided for the nonlinear domain calculation. Surface observables are more robust than deep stored majority-carrier charge in its zero-background, high-injection model.

### S7.2 Optical linearization and independent temporal checks

The independently implemented finite-volume model eliminates electrostatics algebraically, retains every active carrier/polarization/site state, and uses logarithmic carrier densities. Let $\dot{y}=F(y,g)$ and let an observable be $O(y)$. At a stationary state, $J=F_y$, $b=F_g$, and $c_O=O_y$. Complex-step differentiation of the analytic flux implementation gives these derivatives without differencing a very small optical signal. The calculated response is

$$H_O(\Omega;\bar{g})=c_O(i\Omega I-J)^{-1}b, \tag{S364}$$

$$\tilde{O}=H_O\tilde{g}. \tag{S365}$$

A complex Schur decomposition evaluates the full frequency family. Negative real parts of the state eigenvalues are checked at every sampled operating point. The zero-frequency result satisfies

$$H_O(0;\bar{g})=\frac{d\bar{O}}{d\bar{g}}=-c_O J^{-1}b, \tag{S366}$$

and is checked against independently converged stationary states above and below the mean. For a sufficiently small source step $\Delta g$, the linear observable increment is

$$\Delta O(\hat{t})=c_O J^{-1}\left(e^{J\hat{t}}-I\right)b\ \Delta g. \tag{S367}$$

Both this expression and full nonlinear small-step trajectories are evaluated. Full nonlinear sinusoidal runs also test selected amplitude and phase values. A formal operator alone is therefore not treated as a completed optical spectrum.

Seven selected nonlinear sinusoidal integrations use source amplitude 0.1% of the mean. These include both B models for the toward preparation, all three primary D preparations, and the toward preparation for each floating chemistry. They test amplitude and phase independently of the stationary resolvent. The periodic source is imposed at fixed bias for primary D and under the actual floating-current closure for the floating comparisons.

The primary finite-volume grid has 81 cells. Stationary states are independently compared with adaptive collocation solutions. Spatial comparisons use 41, 81, and 161 cells at selected means and five angular frequencies, including zero. Charge-flow quadratures are integrated as auxiliary ODE states, so their accumulated currents can be compared with changing carrier/site inventory and, for floating operation, compact charge. These quadratures do not feed back into the physical dynamics.

In Case E, verification integrates carrier difference and compact charge directly rather than reusing the two-potential transfer matrix. The finite-step calculation retains the exact evolving mean population. The local quadratic field uses 101 odd charge modes and 202 even neutral modes, including zero; a 1024-point periodic transform evaluates their retained products without aliasing. This is verification of a modal reduction, not a full nonlinear ionic time-dependent calculation.

### S7.3 Consolidated verification record

Representative conservation, convergence, and independent-solution checks are collected in Table S4.

*Table S4. Representative numerical verification. Entries are specific to the stated calculations and do not establish material calibration.*

| **Calculation** | **Independent comparison or balance** | **Reported result / limitation** |
|---|---|---|
| A, nonlinear stationary | Quadrature versus full carrier–Poisson BVP | Maximum potential discrepancy $4.22 \times 10^{-11}$ in ϕ/V_T; carrier-density relative error $4.22 \times 10^{-11}$. |
| A/C, weak-charge reduction | Full electron/hole and ionic species on the same electrostatic discretization | Tests the constitutive reduction, not independence of the electrostatic map; the reported ionic overshoot persists. |
| B/D, stationary and periodic | Collocation, 41/81/161-cell refinement, stationary derivative, and direct sinusoidal integration | Seven nonlinear modulation checks; maximum complex voltage-transfer discrepancy about $4 \times 10^{-5}$ relative. See the calculation-specific scope below. |
| B/D, contact energy accounting | Sixteen stationary states; both carrier-contact laws and floating chemistries | Energy-identity discrepancy below $7.7 \times 10^{-12}$ in normalized units. |
| E, linear modal response | 90 independent stationary carrier/Gauss/liquid solves | Maximum relative coefficient discrepancy $2.87 \times 10^{-15}$. |
| E, local quadratic completion | Independent fourth-order finite-difference residual | $3.13 \times 10^{-6}$, $1.97 \times 10^{-7}$, and $1.24 \times 10^{-8}$ at 2048, 4096, and 8192 spatial points. |
| E, nonlinear surface observables | Refined mesh and doubled depth | Collection changes 0.1653% and 0.3486%; $d_{90}$ changes 0.003381% and 0.06348%, respectively. |
| E, nonlinear solver limit | Independent 1D half-space solution | 0.56909777 versus 0.56912754 A $m^{-2}$; relative difference 0.00523%. |
| Doping A/B | Zero-frequency gain versus stationary derivative | Separate reported checks are retained in the A/B doping subsections; no C–E dynamic extension is inferred. |

A separate nonlinear species comparison tests the charge-sector reduction at the small residual source used for the temporal A/C examples. Both electron and hole populations, and both ionic species in C, are evolved without linearizing their drift terms or pair recombination. The comparison uses the same discrete electrostatic charge-to-potential map as the weak-potential calculation, so it tests the constitutive reduction rather than providing an independent verification of that map. At 32 solid cells and, for C, 48 liquid cells, all species remain positive and stationary charge and Poisson residuals remain small. The maximum selected relative stationary voltage difference is $0.00203\%$ in A and $0.000446\%$ in C. The largest light-step differences are $1.7524 \times 10^{-6} V_T$ and $2.2719 \times 10^{-7} V_T$, respectively, corresponding to $0.004904\%$ and $0.003562\%$ of the full voltage changes. The total solid charge remains within $4.22 \times 10^{-15}$ in A and $5.33 \times 10^{-15}$ in C in the normalized ledger. The ionic overshoot persists. Spatial discretization error and the error from weak-potential linearization are thus assessed separately; these values should not be combined into a claim of exact agreement with a finite-potential continuum solution.

For B/D, the finite-volume spectrum study samples 390 operating points across five boundary-law configurations and three preparations. Negative eigenvalue real parts establish stability only in the retained one-dimensional state space. The separate unequal-mobility fold is not part of that all-stable spectrum sweep. The 41-to-161-cell tests and direct modulation checks concern selected operating points, not a global error bound.

Independent stationary diagnostics used tighter collocation tolerance and sixteen selected dark and illuminated states, including both rear carrier laws and common-pair and distinct-couple floating operation. The maximum absolute energy-identity discrepancy was below $7.7\times10^{-12}$ in normalized units, each integrated carrier-balance discrepancy below $3.0\times10^{-12}$, the electrostatic voltage-moment discrepancy below $8.5\times10^{-13}$, and the spatial variation of total stationary conduction current below $3.2\times10^{-15}$. These checks validate the stated equations and numerical solutions; they do not establish a material calibration or validate omitted trap and liquid dynamics.

The local quadratic correction was also checked against the previously calculated mean population and integrated amount. Its nonzero even modes leave both domain-integrated allocations unchanged through the retained order. A high-order spatial-residual convergence test supports the forcing and drift sign; it does not validate the fast-diffuse approximation outside its time range.

### S7.4 Physical validity and reproducibility

The model has four independent restrictions: nondegenerate carrier statistics; the assigned polarization/dielectric constitutive law; geometric thin-film and weak-pattern orderings; and elimination of trap or diffuse-layer dynamics. The timescale restrictions are stated with D and E, before those temporal results are interpreted. A contact, fixed electrical reference, or buffered chemical reservoir must be carried through the corresponding conservation equations.

The results are predictions of the stated illustrative constitutive sets. They have not been calibrated to one oxide by independently relabeling intensity or time. A dimensional comparison requires a jointly consistent choice of polarization scale, background permittivity, thickness, intrinsic density, transport coefficients, recombination parameters, surface capacities, and transfer velocities. The calculation contains photogeneration, ordinary drift and diffusion, polarization dynamics, and the declared boundary reactions; it contains no additional bulk photovoltaic shift-current term.

Calculation scripts, saved arrays, figure assets, and machine-readable verification ledgers are identified in the accompanying analysis archive. The parameter definitions and boundary conditions above specify the calculation represented by each figure. Reproduction must preserve the preparation, contact law, electrochemical reference, and normalization of that calculation.

## S8. Parameters, comparative doping results, and model context

### S8.1 Parameters of the retained examples

Parameter sets are retained separately because the cases do not describe one identically parameterized specimen. The isolated nonlinear benchmark, A/C weak-potential dynamics, B/D finite-depth models, E thin-film modes, and E nonlinear crystal have different controlled approximations. The table labels below refer to those families rather than redefining the physical cases. The isolated-slab and contacted-film parameter sets are given in Tables S5 and S6. The temporal-domain, additional analytical-domain, and nonlinear-domain sets are given in Tables S7, S8, and S9, respectively.

*Table S5. Parameters of the dry isolated-slab benchmark.*

| Quantity | Definition or role | Value |
|---|---|---|
| $a_0$ | $L/(2\lambda_0)$, where $\lambda_0^2=\epsilon V_T/(2qn_i)$ | 0.35 |
| $\lvert s_P\rvert$ | $\lvert P_{\mathrm{res}}\rvert L/(2\epsilon V_T)$ for each polar state | 3 |

| Quantity | Definition or role | Value |
|---|---|---|
| $P_{\text{res}}$ in passive control | Residual source removed, other parameters retained | 0 |
| $g$ in intensity curve | $G/\left(k_b n_i^2\right)$ | 0 to $10^8-1$ |
| Profile illumination | $g$ for illuminated depth profiles | $10^4$ |
| $P_{\text{ref}}$ | Nonzero residual magnitude common to the two polar films; used for all charge-profile normalization | $2\epsilon V_T(3)/L$ |
| $d_n, d_p$ in independent BVP checks | $D_n/D_{\text{ref}}, D_p/D_{\text{ref}}$ | $0.7, 2.3$ |
| $\kappa$ in independent BVP checks | $k_b N_G(L/2)^2/D_{\text{ref}}$ | 0.4 |

The mapping to dimensional inputs is joint: $n_i = 2\epsilon V_T a_0^2/\left(q L^2\right)$, $\left|P_{\text{res}}\right| = 2\epsilon V_T \left|s_P\right|/L$, and $G = g k_b n_i^2$. Absolute $\epsilon$ has units F m$^{-1}$, $n_i, N_G$ have units m$^{-3}$, $P_{\text{res}}$ has units C m$^{-2}$, and $k_b$ has units m$^3$ s$^{-1}$. Specifying only an optical intensity or a time conversion is insufficient to assign these dimensionless curves to a material. Nondegenerate carrier statistics require the local densities $N_G e^{|v_s|}$ to remain small relative to the appropriate band density of states. The constant dielectric response, fixed residual compensation, continuum length scales, and uniform generation must also remain appropriate. No material-specific calibration is claimed for this benchmark.

The weak-potential A/C dynamic reference uses $L=1$, $\epsilon_s=5$, $P=0.4$, $D_s=0.5$, and $a_s^0=20$. These values give a maximum midpoint-referenced potential of about $0.0305\,V_T$ in A. The C-specific compact, liquid, and rear-capacitance values are retained in its figure captions and governing boundary conditions.

*Table S6. Parameter sets for cases B and D.*

| Parameter | B: dry two-carrier rear contact | D: electrochemical front and selective rear |
|---|---|---|
| $\epsilon^* = \epsilon_b V_T/\left(P_* L\right)$ | 0.15 | 0.12 |
| $\delta = q n_i L/P_*$ | 0.08 | 0.005 |
| $M_n = D_n \tau_*/L^2$ | 10 | 100 |
| $M_p = D_p \tau_*/L^2$ | 10 | 100 |
| $A$ | $-5$ | $-30$ |
| $B$ | 5 | 30 |
| $C$ | 0 | 0 |
| $\kappa$ | 0.03 | 0.04 |
| $\gamma$ | 1 | 0.2 |
| $\beta = \alpha_{\text{abs}} L$ | 2 | 2 |
| Recombination $R$ | $np-1$ | $(np-1)/(n+p+2)$ |

| **Parameter** | **B: dry two-carrier rear contact** | **D: electrochemical front and selective rear** |
|---|---|---|
| Site capacities $s_+ = s_-$ | $1.5$ with sites; states absent without sites | No site states |
| Site capture ratios $K_+ = K_-$ | $0.03$ when sites are active | Inactive |
| Site emission times $T_+ = T_-$ | $100$ when sites are active | Inactive |
| Compact capacitance $c_H$ | Not used | $0.35$ |
| Reduction prefactor $s_{n0}$ | Not used | $0.03$ |
| Oxidation prefactor $s_{p0}$ | Not used | $0$ in primary D; $0.03$ in two-channel comparisons |
| Activation fractions $\alpha_n, \alpha_p$ | Not used | $1/2, 1/2$ |
| Maintained redox energies $(\rho_r, \rho_o)$ | Not used | $(0, 0)$ for common pair; $(2, -2)$ for distinct couples |

The first-order Landau comparisons use $(-2, -6, 8)$ for B and $(-12, -24, 36)$ for D. Both the first- and second-order choices have zero-field minima at $u = \pm 1$. Additional charging sweeps use equal mobilities 1, 10, and 100; asymmetric mobilities $(30, 3)$; or site activities $(0.06, 0.015)$. These changes are specified in the corresponding figure captions. A dimensional mapping must supply mutually consistent $L, n_i, P_*, \tau_*, \epsilon_b$, and absorption parameters. In particular,

$$I_{\text{inc}} = \frac{g_0 n_i h\nu}{\tau_* \eta_{\text{pair}} (1 - R_{\text{opt}}) \alpha_{\text{abs}}}, \tag{S368}$$

$$j_{\text{gen}} = J_* g_0 \frac{1 - e^{-\beta}}{\beta}. \tag{S369}$$

*Table S7. Fixed geometry and constitutive inputs for the temporal domain calculations.*

| **Fixed input** | **Value** | **Meaning** |
|---|---|---|
| $c_0$ | 2 | Compact capacitance divided by $C_*$ |
| $\epsilon_{\text{back}} / (C_* \mathrm{l}_0)$ | 0.02 | Passive backing response |
| $\epsilon_x L / (C_* \mathrm{l}_0^2)$ | 0.008 | Lateral film electrostatics |
| $D_x t_0 / \mathrm{l}_0^2$ | 0.6 | Lateral carrier diffusion |
| $s$ | 0 or 0.2 | Blocking or paired front transfer |
| $\delta_d$ | 0.24 | Residual stripe source amplitude |
| $k_1 \mathrm{l}_0$ | 0.1580341085 | Fundamental, unchanged with illumination |
| $a_w / w$ | 0.01 | Gaussian wall-smoothing width |
| $L / \mathrm{l}_0$ | 0.002 | Film thickness |

| Fixed input | Value | Meaning |
|---|---|---|
| $D_z t_0 / l_0^2$ | 0.6 | Normal-equilibration validity check |
| Odd harmonics | $1, 3, \dots, 201$ | 101 retained cosine modes |
| $\rho_1$ | 0, 5, 500 | Ideal, intermediate and resistive liquid |

The residual-charge pattern, wall smoothing, and mode-dependent liquid loads are defined in S6.3. The table's parameters remain fixed during the indicated illumination and liquid-resistance comparisons.

*Table S8. Parameters of the additional analytical domain examples in Figure S23(a,b,d,f). These dimensionless examples are distinct from the nonlinear thick-crystal baseline.*

| Quantity or check | Value |
|---|---|
| $G/(k_b n_i^2)$ | 3 |
| $S/(L k_b n_i)$ | 0.2 |
| $D_x k_0^2/(k_b n_i)$ | 0.6 |
| $C_0/C_*$ | 2 |
| $\sigma_d/(C_0 V_T)$ | 0.12 |
| $\epsilon_{\text{back}} k_0/C_0$ | 0.01 |
| $\epsilon_x L k_0^2/C_0$ | 0.004 |
| $k_0 L$ | 0.002 |
| $C_0 L/\epsilon_z$ | 0.001 |
| $w/l_c$ for reconstructed stripes | 10 |
| $a_w/w$ | 0.01 |
| Reactive $N_\infty/n_i$ | 1.95182845 |
| Blocking $N_\infty/n_i$ | 2 |

The reference code units are $q=V_T=n_i=n_*=k_b=1$, $L=0.002$, $k_0=1$, $C_0=0.008$, $S=0.0004$, $D_x=0.6$, $G=3$, $\sigma_d=0.00096$, $\epsilon_x=\epsilon_z=0.016$, and $\epsilon_{\text{back}}=0.00008$. They give $L/\lambda_{\text{sc},z}=0.03124$ at the reactive mean and $SL/D_z=1.33\times10^{-6}$ for $D_z=0.6$. The stripe fundamental is $k_1=0.1580341$ in those units. Retained high harmonics are suppressed by the Gaussian wall factor; the thin-film and small-amplitude assumptions remain satisfied over the displayed profiles. Figure S23(d) uses the separate local-capacitance outer limit at the same excess amplitude $\varepsilon_c$, whereas the stripe plots retain all three contributions to $C_g(k)$.

*Table S9. Parameters of the nonlinear baseline.*

| Quantity | Value | Role |
|---|---|---|
| One-domain width $w$ | 500 nm | Period $2w=1$ $\mu$m |
| Temperature | 300 K | $V_T=25.852$ mV |

| Quantity | Value | Role |
|---|---|---|
| Electron and hole mobilities | $3\times10^{-7}$ m$^2$ V$^{-1}$ s$^{-1}$ | Equal illustrative transport inputs |
| Lifetime $\tau$ | 1 $\mu$s | Pair lifetime when $n=p$ |
| Relative permittivity | 100 | Fixed-domain effective value |
| Residual-charge amplitude | 0.002 C m$^{-2}$ | Compensated source, not spontaneous polarization |
| Interface capacitance | 0.1 F m$^{-2}$ | Fast ionic response |
| Absorption coefficient | $10^7$ m$^{-1}$ | Optical depth 100 nm |
| Photon energy | 3.5 eV | Above-gap benchmark |
| Incident intensity | 10 W m$^{-2}$ | Uniform laterally |
| Mobile-pair optical yield | 0.8 | Multiplies absorbed photon rate |
| $S_n, S_p$ | 0.1, 10 m s$^{-1}$ | Complementary masks |
| Wall smoothing length $a$ | 10 nm | Tanh length near a wall |
| Baseline mesh | $80\times70$ | Lateral by normal cells |
| First normal cell | 1 nm | Geometric stretching toward the bottom |
| Baseline computational depth | 1500 nm | 3000 nm in the depth check |

The mobility gives $D=7.756\times10^{-9}$ m$^2$ s$^{-1}$ and $\sqrt{D\tau}=88.07$ nm. The transfer lengths are $D/S_n=77.56$ nm and $D/S_p=0.7756$ nm, requiring a resolved reaction-plane half cell. $w^2/D=32.24$ $\mu$s is a lateral diffusion time, not an automatically applicable lifetime or reaction width.

### S8.2 Cross-case stationary doping and the A/B dynamic extension

Figure S24 and Figure 15a of the main text evaluate these relations for r_i=10⁻⁶, a normalized polarization source σ/Q_0=2 in the contacted-space-charge cases, equal normalized front and rear passive capacitances in C, and unit compact-layer capacitance in D and E. Signed dopant density is swept continuously from p-type to n-type. The values are dimensionless mechanism tests. A material calibration must specify the density scale, dielectric response, band density of states, dopant ionization, defect compensation, thickness, and the contact and redox references together.

***Table S10.*** *Static potentials in the stationary doping extension at g=1 for (N_D−N_A)/N_ref=−10, 0, and +10. Values are in V_T and correspond to the reduced case-specific boundaries described above.*

| Case | p-type +P | p-type −P | intrinsic +P | intrinsic −P | n-type +P | n-type −P |
|---|---|---|---|---|---|---|
| A | 0.5769 | -0.5769 | 0.8611 | -0.8611 | 0.5769 | -0.5769 |
| B | 0.6968 | -0.5680 | 1.3170 | -1.3170 | 0.5680 | -0.6968 |
| C | 0.2885 | -0.2885 | 0.4305 | -0.4305 | 0.2885 | -0.2885 |
| D | 0.5062 | -0.4497 | 0.8151 | -0.8151 | 0.4497 | -0.5062 |
| E | 2.7894 | 1.8355 | 0.8284 | -0.8284 | -1.8355 | -2.7894 |

The dark populations are evaluated in cancellation-free form. For $d \geq 0$, $n_0 = \frac{1}{2}[d + \sqrt{d^2 + 4 r_i^2}]$ and $p_0 = r_i^2 / n_0$; for $d < 0$, $p_0 = \frac{1}{2}[-d + \sqrt{d^2 + 4 r_i^2}]$ and $n_0 = r_i^2 / p_0$. Evaluating $n_0$ first for strongly *p*-type material loses most of its significant digits at $r_i = 10^{-6}$ and overestimates $p_0$ by about 0.5% at $d = -10$. The values in Table S10 use the stable form, so the *p*-type and *n*-type entries are exact mirror images under exchange of the two polarization orientations.

***Table S11.*** *Static optical sensitivity* $|H_V(0)|$ *and corner frequency* $\Omega_c$ *(in units of* $k_b N_{ref}$*) at* $g = 1$ *for* |d|=10. *For p-type material the two orientations of Case B are interchanged.*

| State | $\lvert H_V(0) \rvert$ | $\Omega_c$ |
|---|---|---|
| **A, intrinsic** | $5.79 \times 10^{-2}$ | 1.51 |
| **A, *n*- or *p*-type** | $4.09 \times 10^{-3}$ | 4.26 |
| **B, intrinsic, ±P** | 0.289 | 1.43 |
| **B, *n*-type, +P (accumulation)** | $4.18 \times 10^{-3}$ | 7.82 |
| **B, *n*-type, −P (depletion)** | $9.59 \times 10^{-3}$ | 6.14 |

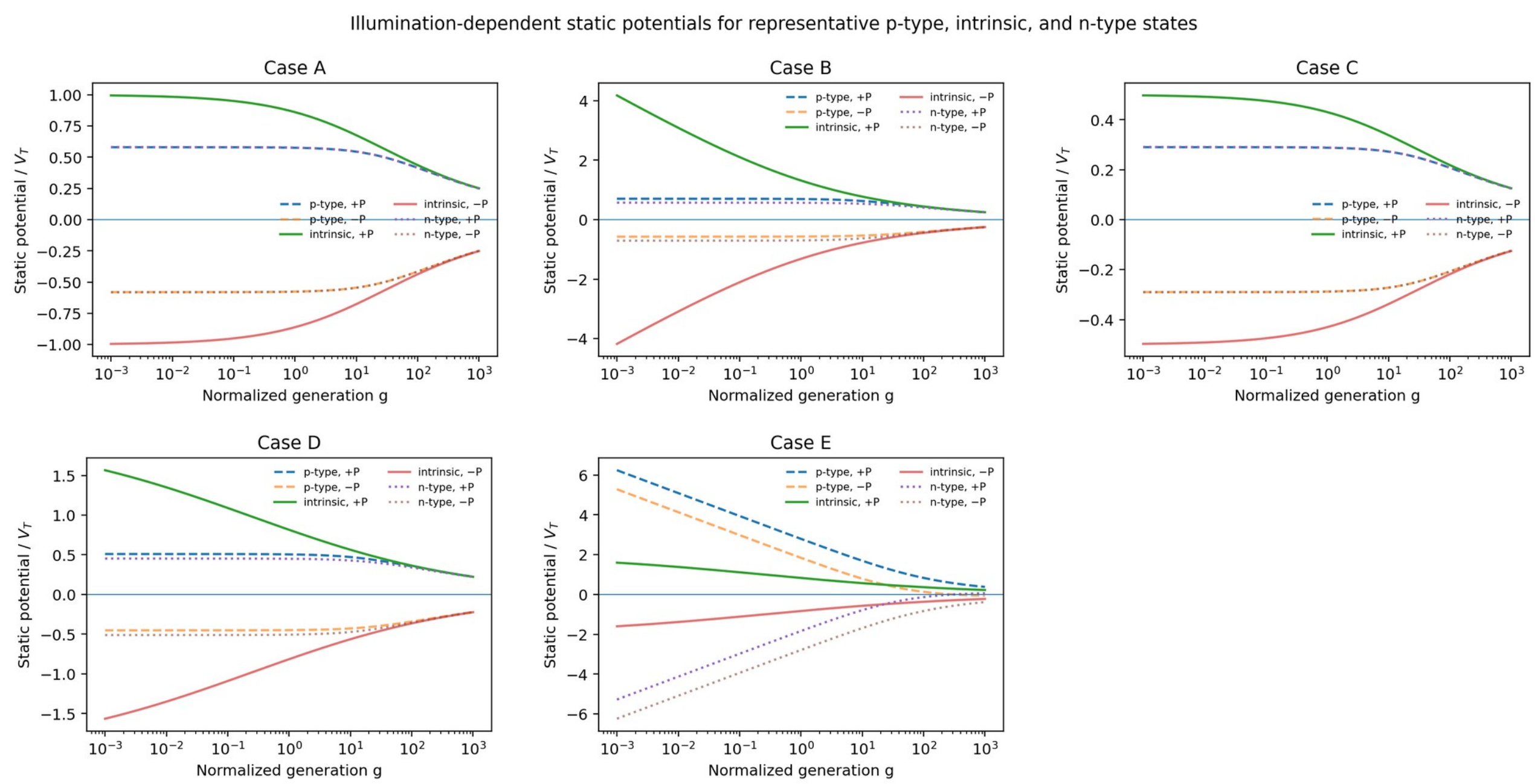

***Figure S24.*** *Static potential versus normalized illumination for representative p-type, intrinsic, and n-type states in the five boundary cases. Dynamic doping responses for A and the bare fixed-polarization B variant are specified in S2.4 and S3.4 and compared in Figure 15 of the main text.*

The numerical values above are mechanism tests at the specified normalized inputs. A/C inherit donor–acceptor symmetry of their linear blocking reduction. B/D show accumulation/depletion asymmetry in their space-charge closures. E adds a mean-potential shift to the domain contrast. Only A and the explicitly fixed-polarization B variant have an additional dynamic doping calculation; its gains do not replace the intrinsic Landau/site responses elsewhere in the study.

### S8.3 Literature context relevant to the model choices

Classical semiconductor photoelectrode theory separates carrier generation, transport, loss, and collection. The depletion and collection limits of Gärtner, Butler, Reichman, and the nanowire treatment of Foley et al. supply useful boundary-specific benchmarks [S25–S28]. Interfacial electron-transfer work motivates rate laws containing both semiconductor carrier populations and chemical activities [S12–S14]. Intensity-modulated photocurrent theory explicitly combines transfer, recombination, and charge storage; it does not identify every measured time constant with one elementary reaction [S22–S24].

The closest coupled continuum precedents solve semiconductor carrier transport and electrolyte reaction/transport together under specified solid contacts and liquid reservoirs [S15, S16]. Ferroelectric surface-compensation models couple polarization to an electrical or chemical environment [S10, S11]. Coated-ferroelectric calculations already treat one-dimensional and multidomain reaction currents with declared electrical boundaries [S29, S30]. These sources motivate a comparison of operating conditions rather than a claim that quantitative ferroelectric reaction theory was previously absent.

Oxides illustrate why spatially selective photochemistry is not unique to ferroelectrics. Facets and terminations of $TiO_2$, $SrTiO_3$, and $BiVO_4$ can differ in charging, adsorption, and reaction activity. Surface protonation changes the electrostatic boundary, and defects affect carrier localization, mobility, and loss. [S31–S34] The coexistence of SrO and $TiO_2$ terminations has been shown to be important for water splitting on $SrTiO_3$, with distinct contributions to oxidation chemistry and band alignment. Spatial separation of hydrogen- and oxygen-evolution sites is also observed on the stepped surfaces of La-doped $NaTaO_3$. [S35, S36] Selective cocatalyst placement in $SrTiO_3$:Al has enabled high-efficiency overall water splitting, but its interpretation involves both carrier delivery and utilization at the interface. A long-lived optical population can be a reactive population or a stored population; an efficient extraction pathway can also shorten a luminescence lifetime. [S37–S41]

The material law and the measured observable must therefore be specified separately. Transport studies on $BiVO_4$ identify low carrier mobility and small-polaron conduction, while time-resolved microscopy distinguishes the spatial evolution of electrons and holes. Operando measurements in liquid additionally reveal facet- and defect-dependent potential responses that vary with electrolyte pH. [S42–S45] The chemical detection volume can extend beyond the solid interface: in $TiO_2$ suspensions, mobile hydroxyl radicals generated through an oxygen-reduction pathway can react with species in solution. [S46] These results motivate retaining separate descriptions of the solid carriers, the interfacial reaction, and transport of chemical species. The same distinctions remain relevant when polarization supplies an additional, electrically reversible source of surface asymmetry.

Domain-selective Ag/Pb marker reactions and electrically written patterns establish programmable local reactivity [S47–S49]. Other observations identify wall-adjacent patterns and dependence on screening, wavelength, concentration, and illumination [S50–S52]. Single-domain and contacted photocatalytic studies further connect polarization to charge collection and chemical output, while retaining different complementary pathways [S53–S58]. These are distinct experimental endpoints and should be compared to the appropriate potential, current, or amount observable.

Surface chemistry can modify activity independently of simple carrier separation. Selective mediator adsorption and interfacial defect control in ferroelectric photocatalysis illustrate why charge availability, transfer, and productive utilization are separate [S57–S61]. Local electrochemistry similarly demonstrates that the compensating path can extend beyond the locally excited region [S20, S21, S62]. The five-case formulation keeps these paths explicit; a microscopic adsorption mechanism, exceptional wall conductivity, or an intrinsic optical current requires its own constitutive terms.

The Supplementary References are numbered independently of the main article.

**Supplementary References**